\documentclass[12pt,a4paper]{book}

\usepackage{a4wide}
\usepackage{amssymb}
\usepackage{amsmath}
\usepackage{graphicx}
\usepackage{xcolor}
\usepackage[titletoc]{appendix}
\usepackage[nouppercase]{frontespizio}

\newcommand{\beq}{\begin{equation}}
\newcommand{\eeq}{\end{equation}}

\newcommand{\be}{\begin{equation}}
\newcommand{\ee}{\end{equation}}
\newcommand{\ba}{\begin{eqnarray}}
\newcommand{\ea}{\end{eqnarray}}
\newcommand{\nn}{\nonumber}
\newcommand{\q}{\theta}
\newcommand{\m}{\mathcal}
\newcommand{\we}{\varepsilon}

\def\[{\left[}
\def\]{\right]}
\def\({\left(}
\def\){\right)}
\def\[{\left[}

\begin{document}



\begin{frontespizio}
\Istituzione {Alma Mater Studiorum - Universit\'a di Bologna}
\Divisione {\vspace{0.6cm} DOTTORATO DI RICERCA IN}
\Scuola {FISICA \vspace{0.4cm} \\  \noindent   Ciclo XXIX}
\Titolo {Supersymmetric $4d$ gauge theories and Integrability}
\Sottotitolo {Settore Concorsuale di Afferenza: 02/A2 \vspace{0.2cm} \\  \noindent Settore Scientifico Disciplinare: FIS/02 \vspace{0.4cm} \\  \noindent Coordinatore Dottorato: Prof. Silvia Arcelli }
\NCandidato {Presentata da}
\Candidato {Alfredo Bonini}
\NRelatore {Supervisore}{}
\Relatore {Prof. Francesco Ravanini}
\NCorrelatore {Correlatore}{}
\Correlatore {Dott. Davide Fioravanti}
\Piede {Esame Finale anno 2018}
\end{frontespizio}

\tableofcontents

\newpage

\section*{Abstract}
\addcontentsline{toc}{chapter}{Abstract}

This thesis is devoted to some particular aspects of integrability in $4d$ SUSY gauge theories. Taking advantage of the integrable structures emergent in the theory, non-local observables such as null polygonal Wilson loops are studied in $4d$ planar $\mathcal{N}=4$ Super Yang-Mills. Their duality with the $4d$ gluon scattering amplitudes makes the analysis even more interesting. The so-called \emph{Pentagon approach}, an application of the Operator Product Expansion (OPE) method to the null polygonal Wilsol loops, makes possible a non-perturbative evaluation of these objects. They are recast as an OPE series over the $2d$ GKP flux-tube excitations, a description reminescent of the QCD flux-tube stretching between quarks. The integrability of the flux-tube allows us to evaluate the series, in principle, for any value of the coupling constant. From this analysis, several results have been obtained. In the strong coupling regime we reproduced the TBA-like equations expected from the minimal area problem in string theory, finding agreement with the $AdS/CFT$ prediction. In this respect, of fundamental importance is the emergence of effective bound states between elementary fermionic excitations. Along the way, we encountered some intriguing analogies between these null polygonal Wilson loops and the Nekrasov instanton partition function $\mathcal{Z}$ for $\mathcal{N}=2$ theories. Furthermore, a new non-perturbative enhancement of the classical string argument has been confirmed, stemming from the dynamics of the string on the five sphere $S^5$ and described by the non-linear $\sigma$-model $O(6)$. Some properties of a fundamental building block of the OPE series, the $SU(4)$ structure of the form factors for a specific twist operator $\hat{P}$, have been analysed. This $SU(4)$ matrix part is given a representation in terms of rational functions, organized in a Young diagrams pattern.

\newpage

\section*{Preface}
\addcontentsline{toc}{chapter}{Preface}

The work presented in this thesis mainly consists of a collection of the results obtained during the PhD at the University of Bologna, with my scientific advisor Dr. Davide Fioravanti and collaborators Dr. Simone Piscaglia and Dr. Marco Rossi. The research resulted in a series of papers \cite{BFPR,BFPR2,BFPR3,BFPR4,BFPR5}, whose content will be elaborated in the second part of the text. In the first part we give a pedagogical survey of some tools of integrability and how they can be applied to $4d$ supersymmetric gauge theories, helping the reader to contextualize the rest of the work.

\chapter*{Introduction and Overview}
\addcontentsline{toc}{chapter}{Introduction and Overview}
\label{Ch1}

In this brief introduction we describe the motivations behind the theoretical work displayed in the text. Furthermore, we highlight the results and give a plan of the thesis.

\paragraph{Motivations}

Our universe is currently described by the Standard Model, which is a quantum field theory and where all the interactions but gravity are described by the gauge principle, with the group $SU(3)_c \otimes SU(2)_I \otimes U(1)_Y$. Combined with General Relativity, the relativistic theory of gravity, the Standard Model successfully predicts all the experiments we can do on Earth. However, there are still some observations, expecially at the cosmological level, to account for and an extension of the contemporary models is necessary. On the other hand, on a purely theoretical ground, the road to unification is yet to be concluded and merging gravity with the other interactions, in a fully consistent quantum description, is probably the major theoretical issue of modern physics. Therefore, there are both experimental and theoretical reasons to search for completions and/or extensions of the current theories. The main candidate for the unification is String Theory, which has the nice feature of including gravity in a consistent framework and implies the presence of, in some extent, supersymmetry in Nature. Putting aside gravity and remaining in the realm of quantum field theories, a less ambitious road involves the supersymmetric extension of the Standard Model. These models are the simplest candidates to partially solve some of the observational and theoretical puzzles above and are currently tested in the accelerators.

Even within the Standard Model, there are some phenomena that call for a better description. These belong to the class of the non-perturbative effects, relevant when the coupling constant is not small. The paradigmatic case is the low-intermediate energy QCD, where the coupling is of order $\alpha_s \sim 1$ and the usual perturbative expansion breaks down. The only tool available so far is the lattice numerical approach, which yielded some positive results but is limited by the computational power. Therefore, analytic tools are very welcome and would enlarge our knowledge of the Standard Model and of Quantum Field Theory in general. An exact evaluation of the non-perturbative effects is possible in some special models of QFT, mainly in two spacetime dimensions. This is due to the presence of a special feature called \emph{integrability}. 

Integrability itself has a long history, starting from the Liouville definiton for the classical systems and culminating with the two dimensional ($1+1$ Minkowski) integrable QFTs. In a nutshell, it appears when the conserved charges equal the degrees of freedom in number: the dynamics is constrained and the model is solvable. The application of integrability techniques allowed to obtain exact results for many observables, even at the non-perturbative level. Until recently, integrability thought to be confined in the realm of two dimensional systems and thus to be irrelevant for $4d$ gauge theories. The seminal work \cite{MinZar} proved otherwise, showing the emergence of some integrable structures in the extended supersymmetric theory $\mathcal{N}=4$ Super Yang-Mills when computing the anomalous dimension of the gauge invariant operators. This fact opened a whole new field of research.
Later on, more integrability based methods have been applied to compute other observables in $\mathcal{N}=4$ SYM. Furthermore, some integrable structures appeared in other theories as well, like $\mathcal{N}=2$ SYM and $\mathcal{N}=6$ $3d$ Chern-Simons.

Another leivmotiv of modern theoretical physics is duality. The discovery that string theory and gauge theory might describe, in some peculiar cases, the same physical system has been probably the most important theoretical development of the last two decades \cite{Mal,Wit,GKP1}. To be specific, the so-called $AdS/CFT$ correspondence states that a string theory living in a space $AdS_{d+1}\times X$, containing a $d+1$-dimensional $AdS$ part, is dual to a quantum field theory with conformal invariance ($CFT$ side) which dwells on the boundary of $AdS_{d+1}$, namely the $d$-dimensional Minkowski spacetime. The gauge  theory living on the bounday usually contains more than one supercharges $\mathcal{N}>1$. This is a remarkable statement because it relates two very different frameworks like String Theory and Quantum Field Theory, living in two \emph{different} spacetime dimensions. In fact, it is a practical realisation of the holographic principle \cite{Hol} proposed long ago, which states that the informations on the gravitational degrees of freedom in the bulk are encoded in the boundary. From the practical point of view, this connection may be used to gain more insight on both sides of the correspondence. Being a weak/strong duality, it connects the weak coupling regime of one side to the dual strongly coupled dynamics. This feature makes its verification difficult, as we cannot compare the perturbative computations of both sides. However, if we trust the correspondence, it may give us precious non-perturbative informations from relatively simple perturbative methods. The most famous and well-studied case is the duality between $\mathcal{N}=4$ SYM and IIB String Theory on $AdS_5\times S^5$. There is a considerable interplay between integrability and $AdS/CFT$, as the emergent integrability in $\mathcal{N}=4$ played a major role to test the duality. In fact, also the string side $AdS_5\times S^5$ shown some integrable features. Therefore, integrability in $AdS/CFT$ is an independent area of research by itself. 

Although these methods involve only gauge theories with extended SUSY, which are not directly relevant to describe our universe, they are worth of consideration as they give us additional insight on these mathematical models and might help us to understand better some non-perturbative features of QFTs in full generality. In addition, even though they are different from the Standard Model, in some regimes they coincide or show many similarities with the QCD. For instance, the deep inelastic scattering in QCD is well-described by its supersymmetric cousin $\mathcal{N}=4$ Super Yang-Mills. The asymptotic freedom present in QCD shows up in many supersymmetric models with $\mathcal{N}=2$ as well. Summarizing, this purely theoretical work is very important, as it could give us new tools and insight to understand some phenomena taking place in the real world.

\paragraph{Results and contents}

The results obtained throughout my PhD, which culminated in a series of papers \cite{BFPR,BFPR2,BFPR3,BFPR4,BFPR5}, are here briefly highlighted. They will be described, in great details, in the Chapters \ref{ChMat},\ref{ChCla},\ref{ChSca}. Also, a plan of the thesis is outlined.

The central topic of the PhD has been the application of integrability techniques in $4d$ SUSY gauge theories. In particular, we focused on the maximally supersymmetric $\mathcal{N}=4$ gauge theory, with gauge group $SU(N)$, in the planar limit $N\to\infty$. Integrability is emergent  at the level of the gauge invariant operators, when the spectrum of the anomalous dimension is considered. The Operator Product Expansion tool, usually applied to products of local operators in a CFT, works for the null polygonal Wilson loops as well \cite{Anope} and can be employed through the so-called \emph{Pentagon approach} \cite{BSV1}. They are the simplest among the non-local operators and, remarkably, they are dual to the $4d$ gluons scattering amplitudes \cite{AM-amp,DKS,BHT}. This is a unique feature of the special theory under investigation. Therefore, the OPE method provides tools for a non-pertubative evaluation of the $4d$ amplitudes.

In this setup, we computed the Wilson loop in the strong coupling limit, where two different regimes are present. On the one side, the prediction from (classical) string theory is reproduced considering only the fermionic and gluonic sectors of the gauge theory. The typical Thermodynamic Bethe Ansatz (TBA)-like equations \cite{TBuA,YSA} for the scattering amplitudes are obtained within the OPE framework. The fermions coalesce, in the strong coupling limit, into effective bound states \cite{FPR2,BSV3}, playing a major role for the emergence of the TBA description. In this respect, some intriguing analogies with the Nekrasov function in $\mathcal{N}=2$ theories \cite{Nekrasov} are discussed. On the other hand, a previously proposed \cite{BSV4} non-perturbative contribution is shown to appear from the scalar sector which is described by the $O(6)$ non-linear $\sigma$-model \cite{AM}. This correction is, surprisingly, of the same order as the classical one.
In addition, we investigated the $2d$ integrable system underlying the $4d$ theory, in particular the pentagon field (twist) operator appearing in the OPE series and its form factors. A coupling independent part, called the matrix part, characterizes the structure under the $SU(4)$ $R$-symmetry and it is given by a multiple integral over auxiliary rapidites of the $SU(4)$ spin chain \cite{BSV4,BSVagosto,BFPR}. These integrals have been computed and given a representation in terms of Young diagrams.

The content of the thesis is organized as follows. The first chapter is meant as a general introduction to integrability and $AdS/CFT$ duality. Its purpose is to give an overview of the main concepts and tools, introducing the reader to the rest of the work. We discuss integrability, both by itself and applied to the $4d$ SUSY theories and $AdS/CFT$. Moreover, we give a brief introduction of string theory and SUSY gauge theories. 
In Chapter \ref{ChOPE} we focus on the Wilson loops in $\mathcal{N}=4$ Super Yang-Mills, discussing their duality with the scattering amplitudes. A glimpse of the strong coupling regime and the results known in literature is given. The most important part is the OPE method, which will be the main tool from which the results have been obtained. Chapter \ref{ChMat} is devoted to a particular feature of the $2d$ integrable system underlying $\mathcal{N}=4$ SYM: the matrix part of the form factors of the twist operators $\hat{P}$. The classical string result is discussed in Chapter \ref{ChCla}, where we analysed the fermions and gluons contributions. The formation of bound states between the former and the subsequent resummation is thoroughly analysed. In the fifth chapter, the scalars are studied and their quantum correction discussed in details, both for the hexagon and the polygons with $n>6$. To conclude, many technical details are organized in several appendices.

\chapter{Integrability and $4d$ gauge theories}
\label{Ch2}

In this chapter we introduce some basic concepts and tools that will help to understand and contextualize the work described in the thesis. We mainly focus on integrability, which is the main responsible for the results obtained in the thesis. We briefly discuss supersymmetric gauge theories and string theory, showing how the are deeply intertwined through the $AdS/CFT$ correspondence. The last part focus on one of the most important discovery in theoretical physics of the new century: the emergence of integrable structures in some $4d$ supersymmetric models. This fact has also been of paramount important to give strong evidence of the validity of the duality above. Although, we must mention that the integrability techniques leading to most results of the thesis are more complicated than those described in this chapter. However, an introduction to the basis of quantum integrability is useful nonetheless, as it helps to grasp the physical picture behind the various methods and computations in the maix text.

\section{Quantum Integrability}
\label{Integrability}

Integrabiliby itself is an old concept, for which many different definitions and formulations exist. However, the physics behind integrable systems is more or less the same. The main feature is the presence of as many conserved charges as there are degrees of freedom: this fact constaints the dynamics so much that it becomes exactly solvable. First developed in the realm of classical systems by Liouville, it has been discovered in the quantum systems by Bethe and later developed as a coherent and complete set of tools. The same physical picture works also for continuum models, both classical and quantum. This fact led to the theory of $2d$ integrable quantum field theories, characterized by an infinite set of local conserved charges. In this section we focus on the quantum theory of integrability, reviewing the Bethe Ansatz technique and discussing the integrable features in $2d$ quantum field theory. Some important extensions and applications are also discussed, among them the celebrated Thermodynamic Bethe Ansatz.

\subsection{The Bethe Ansatz}

In this paragraph we briefly describe the technique going under the name of Bethe Ansatz \cite{Bethe}. We follow history and discuss the system to which the Bethe Ansatz was applied for the first time, the Heisenberg spin chain. 

\subsubsection{Heisenberg spin chain}

Let us consider a $SU(2)$ spin chain of lenght $L$ with periodic boundary condition

\be\label{HSpin}
H=-J\sum_{i=1}^L \vec{S}_i\cdot\vec{S}_{i+1}, \quad  \vec{S}_{L+1}=\vec{S}_1
\ee

where the interaction acts between nearest neighbours. The Hilbert space on which the system lives is the tensor product

\be
\mathcal{H}=\underbrace{\mathbb{C}^2\otimes\cdots\otimes \mathbb{C}^2}_{L}
\ee

If $J>0$, called the ferromagnetic case, the spins tend to align and the ground state is doubly degenerate with all spins pointing in the same direction. This is called ferromagnetic vacuum and we choose

\be 
|0\rangle = |\uparrow\cdots\uparrow\rangle \equiv |\uparrow\rangle\otimes\cdots\otimes |\uparrow\rangle
\ee

As the Hamiltonian commutes with $S_z$, we can find a set of common eigenvectors for the two operators $H,S_z$. The vacuum is the highest eigenstates for the latter

\be 
S_z|0\rangle = \frac{L}{2}|0\rangle, \quad H'|0\rangle = 0
\ee

where we shifted the Hamiltonian (\ref{HSpin}) such that the vacuum energy is set to zero.

In $1931$, Bethe pushed forward a proposal for the general solution, translating the eigenvalue problem into a set of relatively simple algebraic equations. To begin with, let us start with the states with only one spin reversed. We define

\be\label{nstate} 
|n\rangle = |\uparrow\cdots\uparrow\downarrow_n\uparrow\cdots\uparrow\rangle
\ee

where the spin down is located in the $n$-th spot. It is an eigenvector of the total spin 

\be 
S_z|n\rangle = \left(\frac{L}{2}-1\right)|n\rangle 
\ee

but not of the Hamiltonian (\ref{HSpin}). There are $L$ different states of the type (\ref{nstate}), labelled by the integer $n=1,...,L$, thus we expect to find $L$ suitable linear combinations which diagonalize $H$

\be 
|\psi\rangle =\sum_{n=1}^L \psi(n)|n\rangle
\ee

We think of a spin down as a (pseudo)particle propagating along the chain, which gives us the guess for the wave function $\psi(n)=e^{ipn}$

\be 
|p\rangle = \sum_{n=1}^{L}e^{ipn}|n\rangle
\ee

where $p$ represents the momenta of the particle moving around the chain, called \emph{magnon}. The periodic boundary conditions (PBC) impose the quantization of the momenta

\be 
p=\frac{2\pi k}{L}, \quad k=0,\cdots , L-1
\ee

from which we recover the expected number $L$ of eigenvalues with one spin down.

The energy of the single magnon depends on the momenta  

\be 
\epsilon(p)=2J\sin\frac{p^2}{2}
\ee  

so that the eigenvalue problem for one spin down is solved as

\be 
H'|p\rangle = 2J\sin\frac{p^2}{2} |p\rangle, \quad S_z|p\rangle = \left(\frac{L}{2}-1\right)|p\rangle 
\ee

The non trivial physics starts when we consider two spins down, where our state is labelled by two integer numbers $n_1$, $n_2$: there are $\frac{L(L-1)}{2}$ different states, thanks to the symmetry $|n_1,n_2\rangle = |n_2,n_1\rangle$. This time, the eigenstates are

\be 
|\psi\rangle = \sum_{n_1<n_2}^{L}\psi(n_1,n_2)|n_1,n_2\rangle
\ee
 
with the educated guess for the wave function

\be\label{BetheWave2} 
\psi(n_1,n_2) = e^{ip_1 n_1}e^{ip_2 n_2}  + S(p_1,p_2)e^{ip_1 n_2}e^{ip_2 n_1}
\ee

where the interpretation of the unknown function $S(p_1,p_2)$ is clear: it represents the scattering phase between two magnons, as it appears once we exchange the momenta $p_1$ and $p_2$. 

Now, it is helpful to introduce the Bethe rapidity parameter $u$, related to the momentum through

\be 
e^{ip}=\frac{u+i/2}{u-i/2}, \quad u=\frac{1}{2}\cot\frac{p}{2}
\ee

from which the energy of a single magnon is given by

\be\label{e(u)} 
\epsilon(u)=\frac{J}{u^2+1/4}
\ee

After some algebraic calculations, using (\ref{HSpin}), the S-matrix in terms of the rapidities reads

\be\label{Smagnons} 
S(u_1,u_2)=\frac{u_1-u_2-i}{u_1-u_2+i}=S(u_1-u_2)
\ee

which depends only on the difference of the rapidities. 

The allowed momenta are determined by the PBC

\be 
e^{ip_1L}S(p_1,p_2)=e^{ip_2L}S(p_2,p_1)=1
\ee

which, expressed in terms of the rapidities, assume the simple algebraic form

\be 
\left(\frac{u_i+i/2}{u_i-i/2}\right)^L = \frac{u_i-u_j+i}{u_i-u_j-i}, \quad j\neq i=1,2
\ee

A remarkable fact is that, even though the magnons scatter non trivially, the total energy is still given by the sum of the individual energies

\be 
H'|\psi(p_1,p_2)\rangle = E(p_1,p_2)|\psi(p_1,p_2)\rangle  \quad E(p_1,p_2) = \epsilon(p_1) + \epsilon(p_2)
\ee

which means that the interaction acting between magnons only affects the allowed momenta whereas the total energy is still that of free systems, \emph{i.e.} additive. 

Importantly, this picture does not change when we move to the general case, with $m>2$ excitations: it turns out that the multiparticle scattering phase is factorised in terms of the two-particle one $S(p_1,p_2)$. As we will see in details later, this is one of the main features of an integrable system. The general eigenstate

\be\label{BetheWave} 
|\psi(p_1,...,p_m)\rangle=\sum_{n_1<....<n_m}^L\psi(n_1,...,n_m)|n_1,.....,n_m\rangle
\ee

is given by the wave function, generalization of (\ref{BetheWave2})

\be\label{BetheWave1} 
\psi(n_1,...,n_m)=\sum_{P}e^{i\sum_{i=1}^m p_{P(i)}n_i+i/2\sum_{i<j}\delta(p_{P(i)},p_{P(j)})}
\ee

where the scattering matrix has been parametrized as $S(p_i,p_j)=e^{i\delta(p_i,p_j)}$.

Again, by imposing the PBC we obtain the celebrated set of \emph{Bethe equations}

\be 
e^{ip_iL}\displaystyle\prod_{j\neq i}^M S(p_i,p_j)=1
\ee

which, in terms of the rapidity parameter, assumes the inspiring form

\be\label{Bethe} 
\left(\frac{u_i+i/2}{u_i-i/2}\right)^L = \displaystyle\prod_{j\neq i}^M\frac{u_i-u_j+i}{u_i-u_j-i}
\ee

The eigenvalues are simply the sum of the individual energies

\be\label{Energy}
E(\lbrace p_i\rbrace) = \sum_{i=1}^M \epsilon(p_i) 
\ee

This description in terms of particles whose scattering is factorizable is at the core of quantum integrability and it holds for integrable $2d$ QFTs as well.

\paragraph{Inhomogenuities, $SU(2)$ symmetry and the Nested Bethe Ansatz} The $SU(2)$ spin chain description can be emergent in physical systems endowed with $SU(2)$ symmetry\footnote{For instance, an integrable $2d$ QFT with particle transforming as the fundamental of $SU(2)$. }. 

To begin with, we define the Bethe equations adding, at each site, the inhomogenuities $u_i$

\be\label{BetheIn} 
\displaystyle\prod_{l}^L\frac{w_i-u_l+i/2}{w_i-u_l-i/2} = \displaystyle\prod_{j\neq i}^M \frac{w_i-w_j+i}{w_i-w_j-i}
\ee

where, for future convenience, we named the Bethe rapidities of the magnons $w_i$.

In a QFT model with $SU(2)$ symmetry, particles come in two different flavours. They are labelled by the relativistic rapidties $\theta_i\equiv \frac{\pi u_i}{2}$, where energy and momentum are

\be 
E(u)=m\cosh \frac{\pi u}{2}, \quad p(u)=m\sinh \frac{\pi u}{2}
\ee

The scattering matrix $S(u_i-u_j)$ acts between the physical particles is and produce the Bethe equations for the rapidities $u_i$

\be\label{SU2}
e^{ip_jL}\displaystyle\prod_{m\neq j}^N S(u_j-u_m)\displaystyle\prod_{k=1}^M\frac{u_j-w_k+i/2}{u_j-w_k-i/2}=1
\ee

Now, some explanations are due. Here, $N$ is the total number of particles and we see the physical scattering matrix acting between them. $M$ is the number of particles, out of $N$, with $SU(2)$ degree of freedom reversed: we see that there is a fictitious rapidity $w_i$ associate to them. Equations (\ref{SU2}) must be supplemented with that of these rapidities (\ref{BetheIn}), also called isotopic or auxiliary roots. Therefore, we see that a $SU(2)$ spin chain emerge for the system, \emph{i.e.} there are $M$ magnons propagating on a $SU(2)$ spin chain of lenght $N$ . These magnons, from the point of view of the quantum field theory under investigation, do not carry energy and momentum but play an indirect role, by determining the allowed physical rapidities $u_i$.

This argument generalizes to any symmetry group and it goes under the name of \emph{Nested Bethe Ansatz} \cite{Nested}. The name means that, depending on the symmetry group, there are several levels of spin chains whose inhomogenuities are the magnons of another chain or the physical particles of a QFT model. For our purpose, in the text we will encounter the $SU(4)$ spin chain emergent from a physical system enjoying that symmetry, namely the GKP vacuum. Sticking to the special unitary group, a physical system requiring the nesting procedure could be either a spin chain with $SU(N>2)$ symmetry or a QFT with particles transforming under $SU(N\geq 2)$. 

\paragraph{Bound states}

The Bethe equations (\ref{Bethe}) may also describe bound states between the particles. This happens when we consider complex solutions of the Bethe equations. The picture simplifies in the large size $L$ limit, where the complex rapidities arrange themselves in strings in the complex plane with a real center \cite{Tak}, which can be thought as the rapidity of the composite particle. This goes under the name of \emph{string hypothesis}.

To give a skecth of how it works, we stick to the $SU(2)$ spin chain and consider a set of rapidities solution of the Bethe equations (\ref{Bethe}), allowing one of them to be complex $u_1$ with positive imaginary part\footnote{With a negative imaginary part, the argument is specular and still applies.}. Considering the Bethe equation for $i=1$

\be 
\left(\frac{u_1+i/2}{u_1-i/2}\right)^L = \displaystyle\prod_{j> 1}^M\frac{u_1-u_j+i}{u_1-u_j-i}
\ee

from which we see that the LHS, as the modulus of $\frac{u_1+i/2}{u_1-i/2}$ is bigger than one, goes to infinity for $L\to\infty$. Accordingly, an infinite must occur on the RHS: this is possible only if another complex rapidity $u_2$ exists such that $u_2=u_1-i$. In the simplest case, $u_2$ has a negative imaginary part and the requirement of total real momenta tells us the roots are complex conjugate. Therefore they form a 2-string in the complex plane with real $\tilde{u}$ center

\be 
u_1=\tilde{u} + i/2, \quad u_2=\tilde{u} - i/2
\ee 

If the imaginary part of $u_2$ is still positive\footnote{If it lies on the real axis, an additional rapidity in the lower half plane is necessary.}, the argument can be iterated to obtain the general configuration. In the end, the bound state is given by the $l$-string 

\be 
u_a=\tilde{u}+\frac{i}{2}(l+1-2a), \quad a=1,...,l
\ee

centered around $\tilde{u}$ and composed by $l$ rapidities spaced by $i$ in the complex plane. 

The interpretation as a bound states is confirmed by the analysis of the wave function and the energy. The former is shown to decrease exponentially as the separation between the magnons grows. For the energy, let us consider a $2$-string with real center $\tilde{u}$ whose energy is given

\be 
E_2(\tilde{u})=\epsilon(\tilde{u}+i/2) + \epsilon(\tilde{u}-i/2)
\ee  

with (\ref{e(u)}). The total energy reads

\be 
E_2(\tilde{u})=\frac{2J}{\tilde{u}^2+1}
\ee

which is less than any two particle state with real momenta. Generalizing, a $Q$-string has energy

\be  
E_Q(\tilde{u})=\frac{JQ}{\tilde{u}^2+\frac{Q^2}{4}}
\ee

It is possible to write down the Bethe equations directly involving the strings, by defining the S-matrix between these composite particles. This procedure is called \emph{fusion}: the scattering matrix between a magnon and a $Q$-string is given by

\be 
S^{1Q}(u-w_{Q})=\displaystyle\prod_{w_j\in \lbrace w_Q\rbrace} S^{11}(u-w_j)
\ee

where $S^{11}$ is the original scattering phase between single magnons (\ref{Smagnons}). In the same way the scattering between two composite objects is defined. 

\paragraph{Algebraic Bethe Ansatz}

It is worth to mention that a more formal and general approach has been developed, in the second half of the 20th century. It applies to the models with a finite number of degrees of freedom and goes under the name of \emph{Algebraic Bethe Ansatz}, or quantum inverse scattering method\footnote{For a beautiful review, see \cite{Fad1}.}. This method allows to find the commuting conserved charges and construct the eigenstates in a straightforward way. The diagonalisation of the Hamiltonian is obtained in an algebraic fashion, employing tools such as the Lax operator and the transfer matrix, from which the Bethe equations result as a consistency condition. This approach emerged as an extension to the quantum world of some techinques belonging to classical integrability.

\subsection{Integrability in $2d$ Quantum Field Theory}
\label{Int2QFT}

The Bethe Ansatz technique exposed above works also in the realm of $2d$ QFTs, where integrability emerge when the theory is endowed with an infinite number of local conserved charges. In a general quantum field theory, the effect of those charges depends drastically on the number of dimensions where the model dwells. For $d>2$, an infinite number, it implies a trivial scattering, \emph{i.e.} a free theory. This is a consequence of the Coleman-Mandula theorem \cite{ColMan}. On the other hand, for two dimensional systems the consequences are much more intriguing: an integrable $2d$ QFT is characterized by the following important features \cite{Parke}

\begin{description}
  \item[$\bullet$] The number of particles is conserved during the scattering: production and annihilation processes are forbidden.
  \item[$\bullet$] Strict conservation of momentum: the set of momenta does not change during the process, the scattering is therefore \emph{elastic}.
  \item[$\bullet$] Factorisation of the scattering: multiparticle processes are described in terms of the two-particle S-matrix $S(p_i,p_j)$ only.
\end{description}

A non-trivial consistency condition for the last statement is the famous \emph{Yang-Baxter equation}, which assumes the sketchy form $S_{12}S_{13}S_{23}= S_{23}S_{13}S_{12}$. It follows from the fact that the scattering can be achieved in many different sequences of two-body processes. It is obviously satisfied when only one kind of particle is present in the spectrum, namely the S-matrix is just a function of the momenta, while it is severely constraining the matrix structure of $S_{a,b}^{c,d}$ if we have internal degrees of freedom.

Now we are going to convince the reader that the features above hold, using an argument expesed in \cite{DoreyS}. The proof makes use of the action of the conserved charges $Q_s$, where $s$ is the Lorentz spin, on the eigenstates of momentum 

\be 
Q_s|p\rangle_a = q^{(a)}_s(p)|p\rangle_a
\ee

where $a$ represents the other internal quantum numbers. The locality of the charges implies the additivity when we have a multiparticle state

\be 
Q_s|p_1,...,p_n\rangle_{a_1,...,a_n} = (q^{(a_1)}_s(p_1)+ ... + q^{(a_n)}_s(p_n))|p_1,...,p_n\rangle_{a_1,...,a_n}
\ee

Let us consider a scattering of $n$ particles into a final state consisting of $m$ particles, with the set of initial and final momenta $\lbrace p_i\rbrace_n$, $\lbrace p'_j\rbrace_m$. The conservation of the charge $Q_s$ requires the equality

\be\label{conscharge} 
\sum_{i=1}^n q^{(a_i)}_s(p_i) = \sum_{j=1}^m q^{(a_j)}_s(p'_j)
\ee

which is valid for any of the infinite values of $s$, the spin index labelling the charges. The set of constraints (\ref{conscharge}) is so strong that it fixes $m=n$ and the set of momenta $p_i=p'_j$. It follows that the particles cannot be produced or annihilated and they can only exchange their momenta.

The factorisability can be shown with the action of the charges on the wave function. Interestingly, the argument also show the triviality of the theory for $d>2$. Suppose we have a gaussian wave packet

\be\label{wavapacket}
\psi(x) = \int dp e^{-a(p-p_0)^2}e^{ip(x-x_0)}
\ee 

representing a particle in $x_0$ with momentum distribution centered around $p_0$. The effect on the wavefunction of $e^{icQ_s}$ is a momentum dependent shift, the new center of the wave packet being $x_1=x_0 - scp_0^{s-1}$. 

Consider a scattering process with three particles, all described by a wave packet like (\ref{wavapacket}). The initial state is given by $p_1>p_2>p_3$, therefore the particle 1 is the leftmost, 2 is in the middle and 3 is the rightmost. The scattering may happen in several ways, all of them are obtained by a suitable application of the operator $e^{icQ_s}$, as it shifts differently the different wave packets. Since $[Q_s,S]=0$, the related amplitudes are equivalent and the factorisation follows. On the other hand, if we have $d>2$, the application of the charge $Q_s$ may shift the particles in such a way that they never interact, \emph{i.e.} the amplitudes is vanishing implying a free theory. 

To summarize, the main quantity to be determined in a $2d$ integrable QFT is the two-particle S-matrix $S(p_i,p_j)$. From the knowledge of the S-matrix, it is possible to find the spectrum of the theory by employing the Bethe ansatz method depicted in the previous subsection. This works when the theory is confined in a large volume $L$, which allows us to impose the usual periodic bounday conditions. This volume has to be much larger than the inverse mass scale of the theory\footnote{If there is more than one mass, we need to consider the smallest one.}, namely $L\gg m^{-1}$.  The two-particle S-matrix will be the topic of the next part, devoted to the so-called \emph{S-matrix theory}.

\subsubsection{The S-matrix theory}

Here we analyse the two-particle S-matrix $S(p_1,p_2)$ for a relativistic $2d$ model. We will see that many physical contraints \cite{ZamS1,ZamS2} can be imposed, limiting the possible form of the two-particle S-matrix, a fact that sometimes lead to its complete determination. 

The two-particle process we want to describe is

\be 
p_1+p_2\to p_3+p_4
\ee

and involves four $2d$ momenta which satisfy the obvious conservation law $p_1+p_2\to p_3+p_4$. In general the particles may be labelled by additional indices, representing some internal symmetry. The strict conservation of momenta tells us that either $p_3=p_2$, $p_4=p_1$ or $p_3=p_1$, $p_4=p_2$. In both cases, the S-matrix depends on the incoming momenta only and, once we add the possible internal symmetry, it reads $S_{a,b}^{c,d}(p_1,p_2)$ and describes the process

\be 
p^{(a)}_1+p^{(b)}_2\to p^{(c)}_1+p^{(d)}_2
\ee

In a relativistic theory the S-matrix depends only on the Lorentz invariant combinations of $p_i$, namely the Mandelstam variables $s,t,u$. In two dimensions only one of them is independent, we choose $s$ which represent the center of mass energy and reads

\be\label{stheta}  
s=(p_1+p_2)^2=m_1^2+m_2^2+2m_1m_2\cosh\theta_{12}, \quad \theta_{12}\equiv \theta_1-\theta_2
\ee

where we introduced the relativistic rapidity\footnote{Note that a Lorentz transformation is just a shift in the rapidity space $\theta\to \theta + a$.} according to

\be 
p^{(0)}=m\cosh\theta , \quad p^{(1)}=m\sinh\theta
\ee

This relativistic invariance implies that the S-matrix depends only of the differences of the rapidities, \emph{i.e.} $S_{a,b}^{c,d}(\theta_{12})$.

The S-matrix theory consists in a series of constraints to the analytic form of $S(\theta)$, coming from various physical arguments. Here we breifly list and discuss those properties. 
 
We ask that our function $S(\theta)$ satisfies the \emph{Hermitean analiticity condition}

\be 
S_{a,b}^{c,d}(\theta)=[S_{a,b}^{c,d}(-\theta^*)]^*
\ee

which implies that it is an analytic function in the region $0<\textit{Im}\theta<\pi$, called the \emph{physical strip}. 

An important physical contraint comes from the conservation of probability, also called unitary in quantum mechanics, which requires our function to satisfy

\be 
S_{a,b}^{c,d}(\theta)S_{c,d}^{e,f}(-\theta)=\delta_a^e\delta_b^f
\ee

Moreover, the relativistic invariance implies the crossing symmetry, which relates the $s$-channel to the $t$-channel: we can interchange an incoming particle with an outgoing one and obtain

\be 
S_{a,b}^{c,d}(\theta)=S_{a\bar{d}}^{c\bar{b}}(i\pi-\theta)
\ee

where the bar operation means that we are considering the antiparticle. 

Let us not forget the additional condition from the Yang-Baxter equation, which, once considering all the indices, reads

\be 
S_{a_1,a_2}^{c_1,c_2}(\theta_{12})S_{c_1,a_3}^{b_1,c_3}(\theta_{13})S_{c_2,c_3}^{b_2,b_3}(\theta_{23})=S_{a_2,a_3}^{c_2,c_3}(\theta_{23})S_{a_1,c_3}^{c_1,b_3}(\theta_{13})S_{c_1,c_2}^{b_1,b_2}(\theta_{12})
\ee

\paragraph{Bound States}

As learned from basic quantum mechanics, poles in the S matrix may correspond to bound states of the theory. Here the case is no different, and our S matrix could be endowed with poles in the complex plane, symptom of a possible bound state. To be precise, let us stick with the case of diagonal scattering $S_{a,b}$ and consider a simple pole on the imaginary axis $\theta=iu_{ab}^c$ for the S-matrix $S_{ab}(\theta)$, where the quantity $u_{ab}^c$ is called \emph{fusing angle}. The bound state emerges from the process $a+b\to c$ and formula (\ref{stheta}) gives us the mass of the composite object

\be 
m_c^2=m_a^2+m_b^2+2m_am_b\cos u_{ab}^c
\ee

Crossing symmetry tells us that we can think of any particle $a,b,c$ as a bound states of the remaining two: the relation between the fusing angles is quite constraining

\be 
u_{ab}^c+u_{bc}^a+u_{ac}^b=2\pi
\ee

The fusion procedure gives us the S-matrix involving a bound state, as depicted also for the Bethe ansatz equations previously: the scattering between the excitation $c$, composed by $a,b$, and a particle $d$ is given by the product

\be 
S_{cd}(\theta)=S_{ad}(\theta+i\bar{u}_{a\bar{c}}^b)S_{bd}(\theta-i\bar{u}_{b\bar{c}}^a)
\ee

which is called the bootstrap equation and is at the core of the so-called \emph{bootstrap approach} \cite{ZamS1}.

\subsection{Form factors}

In this paragraph we discuss the form factors, which are key quantities to compute in an integrable QFT. As an important application, they can be used to evaluate the correlation functions and in particular their UV limit. The form factors theory closely follows that of the S matrix, as many physical requirements hold here as well. We define the form factor of a specific operator $\mathcal{O}$ as the matrix element between the $n$ particle state with definite rapidities and the vacuum as

\be\label{FFO} 
F^{\mathcal{O}}_{\vec{a}}(\theta_1,\cdots ,\theta_n)\equiv \langle 0|\mathcal{O}(0)|\theta_1,\cdots,\theta_n \rangle^{\textit{in}}_{\vec{a}}
\ee

where $\vec{a}=(a_1,...,a_n)$ is the set of internal indices labelling the particles. The rapidities are ordered decreasingly $\theta_1>....>\theta_n$: the general form factor can be obtained through a process of analitical continuation.

The form factors $F_n$ are required to satisfy certain properties, which leads to their partial, or even complete in some peculiar cases, determination. The first is the so-called \emph{Watson's equation}: it simply states that if we exchange two particles nearby, the S matrix between them appears

\be\label{Watson} 
F_n(\theta_1,....,\theta_n) = S(\theta_i-\theta_{i+1})F_n(\theta_1,.,\theta_{i+1},\theta_{i},...,\theta_n)
\ee

where, for simplicity's sake, we omitted the obvious dependence on the internal quantum numbers. 
Another constraint concerns the behaviour in the complex plane: the physical rapidities are real but we can consider $F_n$ as an analytic function of all its variables. Shifting by $2\pi i$ a rapidity does not change the physical value of the momentum but has the effect of transporting the particle on the other side of the list: this is called \emph{monodromy} and reads

\be\label{Monodromy} 
F_n(\theta_1+2\pi i,....,\theta_n)=F_n(\theta_2,....,\theta_n,\theta_1)
\ee
So far, our relations involved form factors with the same number of particles. An interesting formula relates $F_n$ to the form factors with two particles less. General considerations lead us to find that there are poles in the complex plane whenever $\theta_{ij}=i\pi$ for some $i,j$: that means that two particles have the opposite momentum and build a single particle configuration. These singularities do not depend on the specific model but are quite general, they are called \emph{kinematic poles}. The residue of this single pole is related to the form factor $F_{n-2}$ through

\be\label{Kinematic} 
\textit{Res}_{\theta_{12}=i\pi}F_n(\theta_1,....,\theta_n)=2iF_{n-2}(\theta_3,....,\theta_n)\left[1-\displaystyle\prod_{i=3}^n S(\theta_2-\theta_i)\right]
\ee

Until now, we did not mention the theory whose form factor we were computing: in fact, the properties (\ref{Watson}, \ref{Monodromy}, \ref{Kinematic}) are general and independent of the specific model under investigation. A model dependent physical constraint involves the possible bound states present in the theory, emergent whenever the S-matrix is endowed with single poles on the imaginary axis in the physical strip. The residue of the form factors for these poles yield

\be\label{BoundS} 
\textit{Res}_{\theta_{12}=iu_{12}^{(12)}}F_{\vec{a}}(\theta_1,...\theta_n) = \sqrt{2iR_{(12)}}\Gamma_{12}^{(12)}F_{a_{(12)},a_3,..,a_n}(\theta_{(12)},\theta_3,...,\theta_n)
\ee
where $\theta_{(12)}\equiv \frac{\theta_1+\theta_2}{2}$ and $R_{(12)}$ is the residue of the S-matrix at $iu_{12}^{(12)}$.


The general solution of the properties above can be written as

\be 
F^{\mathcal{O}}_n(\theta_1,....,\theta_n) = K^{\mathcal{O}}_n(\theta_1,....,\theta_n)\displaystyle\prod_{i<j}F_{min}(\theta_{ij})
\ee
where $F_{min}$ satisfies the monodromy and the Watson's for $n=2$ and is chosen to be free of poles and zeroes in the physical sheet. The involved structure is carried by the function $K^{\mathcal{O}}_n$, which contains also the informations on the operator $\mathcal{O}$. It is symmetric and periodic with $2\pi i$.

\subsubsection{Correlation functions and conformal limit}

An important application of the form factors consists in the study of the correlation functions for the operator $\mathcal{O}$. Let us consider the two-point function $\langle 0|\mathcal{O}(x)\mathcal{O}(0)|0\rangle$ and insert the identity as a sum over states

\be 
\mathbf{1} = \sum_{n=0}^{\infty}\frac{1}{n!}\int\displaystyle\prod_{i=1}^n \frac{d\theta_i}{2\pi}\sum_{\vec{a}}|\theta_1,\cdots,\theta_n \rangle_{\vec{a}} {}_{\vec{a}}\langle \theta_1,\cdots,\theta_n|
\ee

We shift the operator

\be 
\mathcal{O}(x) = e^{i\hat{p}x}\mathcal{O}(0)e^{-i\hat{p}x}
\ee

and apply the momentum $\hat{p}=(p^{(0)},p^{(1)})$ to the eigenstates, so that we obtain the form factor series for the two-point function

\be\label{CorrFF}
\langle 0|\mathcal{O}(x)\mathcal{O}(0)|0\rangle = \sum_{n=0}^{\infty}\frac{1}{n!}\int\displaystyle\prod_{i=1}^n\left( \frac{d\theta_i}{2\pi} e^{-mr\cosh\theta_i}\right)G_n(\theta_1,...,\theta_n)
\ee 

where the $G_n$ represents the squared form factor, summed over the internal symmetry indices

\be 
G_n(\theta_1,...,\theta_n)=\sum_{\vec{a}}|\langle 0|\mathcal{O}|\theta_1,...\theta_n\rangle|^2
\ee

and the Lorentz invariant distance $r=\sqrt{x_{\mu}x^{\mu}}$ appears.

The series (\ref{CorrFF}) is well-suited to study the short-distance behaviour $r\to 0$ where the theory becomes conformal. In fact, the asymptotic behaviour must be dictated by the scaling dimension of the operator $\Delta_{\mathcal{O}}$

\be 
\langle 0|\mathcal{O}(x)\mathcal{O}(0)|0\rangle \sim \frac{1}{r^{2\Delta_{\mathcal{O}}}}
\ee 

There is an interesting formula of $\Delta_{\mathcal{O}}$, coming from the expansion (\ref{CorrFF}). We follow the procedure developed in \cite{Smirnov}, for a review see \cite{BabKar2}. To begin with, we consider the logarithm of the two-point function

\be 
F^{\mathcal{O}}(x)\equiv \log \langle 0|\mathcal{O}(x)\mathcal{O}(0)|0\rangle
\ee

for which it is a well-known fact that its expansion contains the connected version $g_n$ of the functions $G_n$, \emph{i.e.} we have

\be 
F^{\mathcal{O}}(x)=\sum_{n=1}^{\infty}\frac{1}{n!}\int\displaystyle\prod_{i=1}^n\left( \frac{d\theta_i}{2\pi} e^{-mr\cosh\theta_i}\right)g_n(\theta_1,...,\theta_n)
\ee

The important point is the asymptotic factorisation of the G functions when $k$ rapidities are sent far away from the others

\be 
G_n\to G_{n-k}G_k
\ee

which entails that the $g_n$ go to zero whenever one or more rapidities is well-separated from the others.
As the function $g_n$ depends only on the differences $\alpha_i\equiv \theta_{i+1}-\theta_1$, we can integrate over $\theta_1$ to get

\be\label{IntK0}
F^{\mathcal{O}}(x)=\sum_{n=1}^{\infty}\frac{1}{n!}\int\displaystyle\prod_{i=1}^{n-1}\frac{d\alpha_i}{2\pi}g_n(\alpha_1,....,\alpha_{n-1})K_0(mr\xi)
\ee

where we have introduced the modified Bessel function of the third kind 

\be 
K_0(z)\equiv \frac{1}{2}\int d\theta e^{-z\cosh\theta}
\ee

and the function of the remaining rapidities

\be 
\xi^2=\left(\sum_{i=2}^{n}\cosh\theta_i +1\right)^2-\left(\sum_{i=2}^{n}\sinh\theta_i\right)^2
\ee

At this point, we can expand the Bessel function

\be 
K_0(z)=-\log z + \gamma_{E} -\log 2 + O(z^2\log z)
\ee
  
inside the integral to get the expression for the conformal weight, which is simply the series

\be\label{DeltaFF} 
\Delta_{\mathcal{O}}=\frac{1}{2}\sum_{n=1}^{\infty}\frac{1}{n!}\int\displaystyle\prod_{i=1}^{n-1}\frac{d\alpha_i}{2\pi}g_n(\alpha_1,....,\alpha_{n-1})
\ee

We see that the necessary condition for this method to work is that the integral of the connected functions over their $n-1$ rapidities is finite. The same method would not have been possible for the correlator, as the integral of the functions $G_n$ is not well-defined. The expansion (\ref{DeltaFF}) is usually fastly convergent and constitutes an useful method to compute the scaling dimension in an integrable QFT. It is noteworthy the fact that it is a UV-IR connection, as the form factors are defined only in the infinite volume limit whereas the short-distance regime is described by the high energy dynamics. 

This technique will be fundamental in  Chapter \ref{ScalarHex}, where we analyse the scalar contribution to the hexagonal Wilson loop which is a correlator in the $O(6)$ sigma-model.

\subsection{Thermodynamic Bethe Ansatz}

An important application of the Bethe ansatz technique concerns the physics in the thermodynamic limit for systems at finite temperature. We will see that the free energy density can be computed in terms of a non-linear integral equation, called Thermodynamic Bethe Ansatz, referred as TBA, equation. In addition, with a double Wick rotation, it applies also to the ground state energy of finite size systems. The latter version, developed for the first time by Zamolodchikov \cite{Zam}, found important applications in the field of $AdS/CFT$.  

\subsubsection{Yang-Yang TBA}

In this paragraph we consider an integrable system, for instance a spin chain or a $1+1$ dimensional QFT in a large volume, at finite temperature. The thermodynamic limit is taken as we send to infinity the lenght of the chain together with the number of excitations. Therefore, the density (occupation number) of states becomes the main physical quantities we want to study.  The techniques described here were first developed by Yang and Yang \cite{YangYang}.

Consider a set of Bethe equations for $n$ excitations, in an integrable system defined on a circle of size $L$

\be\label{Bethe}
e^{ip_iL}\displaystyle\prod_{j\neq i}^n S(p_i,p_j)=1, \quad i=1,...,n
\ee

which, we recall, are nothing but the periodic boundary condition once we take into account the scattering between the particles. Taking the logarithm we get

\be 
ip_iL + \sum_{j\neq i}\log S(p_i,p_j) = 2\pi i I_i
\ee

The integer numbers $I_i$ identify the particular solution to the Bethe equations. To convince the reader of this fact, for a free theory $S=1$ and 

\be 
I_i = \frac{p_i L}{2\pi}
\ee

For an interacting theory the meaning is the same but the relation betweem the set of $I_i$ and the momenta is no longer an easy one. Let us define the counting function $c(p)$ through

\be 
Lc(p) = \frac{Lp}{2\pi} + \frac{1}{2\pi i}\sum_j \log S(p-p_j)
\ee

depending on the set of quantum numbers $I_j$. Notice that $Lc(p_i)=I_i$, \emph{i.e.} it assumes integer values when the momentum coincides with a Bethe root, \emph{i.e.} a solution of the Bethe equations. The other values of momenta for which $Lc(p^h_i)=J_i$ is an integer are called holes.

Defining the two densities of states $\rho^r(p)$, $\rho^h(p)$, respectively for roots and holes, we find

\be 
\rho^r(p) + \rho^h(p) = \frac{dc(p)}{dp}\equiv \rho(p) 
\ee

For a free system we would have simply a constant density of states $\rho(p)=1/2\pi$. The difference here is that the density of states depends on the actual excitations, therefore the spectrum is self-determined by the presence of the interaction. 

In the thermodynamic limit the number of excitations goes to infinity and we can replace sum over Bethe roots with an integral $\sum_j f(p_j) \to \int dp \rho^r(p)f(p)$, thus the continuum version of the Bethe equations become (we take the derivative with respect to $p$)

\be\label{Bethecont}
L\rho(p) = \frac{L}{2\pi} + \frac{1}{2\pi i}\int dp' \rho^r(p')K(p,p')
\ee  

where the kernel is defined as

\be 
K(p,p')\equiv \frac{d}{dp}\log S(p,p')
\ee

The integral equation (\ref{Bethecont}) is the continuum version of the logarithm of the Bethe equations (\ref{Bethe}). This is the starting point for the thermodynamic considerations. The energy density reads

\be 
e[\rho^r] = \int dp \rho^r(p)E(p)
\ee

where $E(p)$ is the single particle energy. The equilibrium is determined by the minimum of the free energy density

\be\label{Free} 
f[\rho^r,\rho] = e[\rho^r] - Ts[\rho^r,\rho]
\ee

where $T$ is the temperature of the system.

The free energy is a functional of two densities, among which the constraint (\ref{Bethecont}) is present\footnote{In principle we could use (\ref{Bethecont}) and express everything in terms of $\rho(p)$ only.}. The entropy density, for a fermionic statistic, reads

\be 
s = \int dp\left[\rho(p)\log\rho(p)-\rho^r(p)\log\rho^r(p)-\rho^h(p)\log\rho^h(p)\right]
\ee

Extending the case of free systems, it is natural to define the pseudoenergy as

\be 
\frac{\rho^h(p)}{\rho^r(p)} \equiv e^{\frac{\epsilon(p)}{T}}
\ee

from which the minimum condition $\delta f=0$ becomes the famous TBA equation

\be\label{TBA}
\epsilon (p) = E(p) -T\int dp' K(p,p')\log\left(1+ e^{-\epsilon(p')/T}\right)
\ee

The critical value of the functional (\ref{Free}), \emph{i.e.} the actual free energy, reads

\be 
f=-T\int\frac{dp}{2\pi}\log\left(1+e^{\epsilon(p)/T}\right)
\ee

It is clear from (\ref{TBA}) that for a free particle, as $K=0$, the pseudoenergy reduces to the single particle energy $E(p)$. 

We must mention that if different types of excitations are present, for instance labelled by some internal indices or even bound states, the TBA becomes a set of (coupled) non-linear integral equations of the type (\ref{TBA}), with a matrix kernel $K_{a,b}(p,p')$.

\subsubsection{Finite size energy and the central charge}

Having studied the thermodynamic in the large volume limit, which culminated in a set of non-linear integral equations of the type (\ref{TBA}), it is time to employ the trick, due to Zamolodchikov, and apply the results to the finite (\emph{i.e.} not large) size systems.

The realm of relativistic theories is the natural framework to use the method, but it can be extended also to any integrable system with some modifications. Let us consider an euclidean $2d$ QFT, defined on a torus with dimensions $L$ and $R$, which we consider both finite for now. 
Thanks to the relativistic nature of the theory, we can consider two different quantization schemes, depending on which variable we take as time (imaginary, as we are in the euclidean version), then we send $L$ to infinity. 

The first scheme, with $L$ as time and $R$ as the size of the space (with PBC), yields for the partition function in the large $L$ limit

\be 
Z(L,R)\simeq e^{-L E_0(R)}
\ee

which is dominated by the ground state energy $E_0(R)$ of the theory for finite size $R$.

In the other scheme, $L$ plays the role of the size and $R$ of the inverse temperature of our theory. The partition function is

\be 
Z(L,R)\simeq e^{-RLf(1/R)}
\ee

where $f(1/R)$ is the density of free energy at temperature $T=1/R$. 

Thus, as the infinite volume $L$ drops out in the equivalence, we get the remarkable relation

\be\label{Ef} 
E_0(R)=R f(1/R)
\ee

which relates the finite size $R$ ground state energy to the free energy density, at infinite volume, for temperature $1/R$.

Thus we can apply the TBA above, adapted for a relativistic system, where

\be 
E(p)=m\cosh\theta, \quad p=m\sinh\theta
\ee

The non-linear integral equation reads

\be\label{TBAr} 
\epsilon (\theta) = mR\cosh\theta - \int\frac{d\theta'}{2\pi}\varphi(\theta-\theta')\log\left(1-e^{-\epsilon(\theta')}\right)
\ee

where the kernel is

\be 
\varphi(\theta)\equiv -i\frac{d\log S(\theta)}{d\theta}
\ee

Equation (\ref{TBAr}) allows us to find the dispersion relation for a finite temperature and the (critical) free energy is

\be\label{freenergy}
f(R) = m\int\frac{d\theta}{2\pi}\cosh\theta\log\left(1+e^{-\epsilon(\theta)}\right)
\ee

which, by means of (\ref{Ef}), gives the finite size energy $E_0(R)$. It can be parametrised as

\be 
E_0(R)=-\frac{\pi c(r)}{6R}, \quad r=mR
\ee

where $c(r)$, called \emph{scaling function}, is the finite size generalisation of the central charge governing the Casimir energy in the conformal limit

\be 
E_c(R)=-\frac{\pi c}{6R}
\ee

Summarizing, the scaling function reads

\be 
c(r)=-\frac{6R}{\pi}r\int\frac{d\theta}{2\pi}\cosh\theta\log\left(1+e^{-\epsilon(\theta)}\right)
\ee

whose $r$ limit can be addressed to obtain the central charge of the associated CFT, namely $c(0) = c$, see \cite{Zam}.

\paragraph{The Y-system}

It is worth to mention that a set of coupled Thermodynamic Bethe Ansatz equations like (\ref{TBA}), can be recast in an inspiring set of functional equations, the so-called Y-system \cite{ZamY}. Defining the Y functions as

\be 
Y_a(\theta)\equiv e^{\epsilon(\theta)}
\ee

it is possible to show that, upon analytical continuation, they satisfy a set of functional equations, schematically of the form

\be\label{Ysystem} 
Y_a(\theta+i\pi/h)Y_a(\theta-i\pi/h)=\displaystyle\prod_{b}[1+Y_b(\theta)]^{l_{ab}}
\ee

Some comments are due. The incidence matrix $l_{ab}$ tells us how the different nodes of the TBA equations, once recast in the so-called \emph{universal form}, are connected \cite{Dynkin}. The equations (\ref{Ysystem}) reflect the structure of Dynkin diagram associated to the algerba of the theory. Along the process we lose some informations: for instance, the driving term, \emph{i.e.} the dispersion relation, does not appear anymore in the equations. This means that the Y-system has a higher degree of universality that the TBA equations. On the other hand, a solution of the Y-system do not necessarily solve the TBA equations: in order to assure that, some additional constraints such as the asymptotic properties are needed.

What is important is that (\ref{Ysystem}), despite being obtained from the TBA describing the ground state energy, can be applied to \emph{any} excited state of the theory. This is possible thanks to the analytic continuation of (\ref{TBA}) in the complex plane \cite{DoreyTateo}. Therefore, (\ref{Ysystem}) is the starting point for the analysis of the finite size spectrum of an integrable model. This has important application in the context of $AdS/CFT$ as well.

\section{Supersymmetry and AdS/CFT duality}
\label{SUSYADSCFT}

In this section we discuss supersymmetric gauge theories and their duality with string theory, through the $AdS/CFT$ correspondence. In particular we focus on the case of interest for this work, the duality between the maximally supersymmetric $\mathcal{N}=4$ SYM and IIB string theory living on  $AdS_5\times S^5$ spacetime.  We will review the main features of those models, giving the reader the minimum background necessary to undestand the framework of the work.

\subsection{Supersymmetric gauge theories}

Supersymmetry is a spacetime symmetry which extends the Poincar\'e group. In addition to the momentum generator $P_{\mu}$ and the angular (Lorentz) momentum $J_{\mu\nu}$, there are a number $\mathcal{N}$ of Weyl (left and right) spinor charges $Q^{A}_{\alpha},\bar{Q}^{A}_{\dot{\alpha}}$, where $A=1,...,\mathcal{N}$. Given its spinorial nature, it has the effect of turning a boson into a fermion and viceversa, schematically

\be 
Q|B\rangle = |F\rangle , \quad  Q|F\rangle = |B\rangle
\ee

thus changing the spin of the particle. That is why it is a spacetime symmetry and not just an internal one. The main consequence is that a supersymmetric theory contains the same number of bosonic and fermionic degrees of freedom: therefore, each particle has its own superpartner whose spin differs by $1/2$. 

Supersymmetry has been introducted, in the 70's, for several reasons. From a purely theoretical point of view, it represents the minimal extension of the standard QFTs bases on Poincar\'e symmetry: allowing the charges to be fermionic, the maximal symmetry group becomes, from the $\textit{Poincar\'e}\times\textit{Internal}$ allowed by the Coleman-Mandula theorem, to the 

\be 
\textit{SuperPoincar\'e}\times\textit{Internal}
\ee

that is, the more general symmetry group is the Poincar\'e group with supersymmetry, called\emph{ SuperPoincar\'e}, times a possible internal symmetry which does not mix with spacetime. As some charges are fermionic, the algebra of the generators contains also some anticommutation relations. Furthermore, as $\lbrace Q,\bar{Q}\rbrace \sim P$, supersymmetry is also related to general relativity: if we promote supersymmetry to be local, we find a model invariant under general coordinate transformations, obtaining the theory called Supergravity. Supersymmetry, relating fermions with bosons, provides a further unifications involving matter and radiation, usually considered and described in a different way. In addition, it is also one of the main prediction of string theory. Another reason to hope for supersymmetry is that helps the three different gauge couplings of the Standard Model to unify at a certain energy scale $M_{gut}$.

On the experimental point of view, there are some reasons to believe that supersymmetry might be a property of Nature. There are several puzzles that are not explained by the Standard Model of particle physics. The biggest one is the clash between the zero-point energy predicted by the current models and the cosmological constant. One of the advantage of supersymmetry is that it addresses the problem of the zero-point energy of the vacuum, as fermions and bosons contribute with the opposite sign to it. However, to be relevant for the real world, supersymmetry has to be broken somewhere along the way down to our energy scale. This is due to the fact that it is not manifest in the spectrum of particles we observe. This fact gives a non-zero value for the zero-point energy, still too large but is much lower than the previous one, so the issue is reduced by many orders of magnitude. Another unsolved problem is the presence of dark matter, which could be explained within supersymmetry by a particular weakly interacting superpartner of the known particles, for instance the neutralino.  In addition, the naturalness and the hierarchy problem are partially addressed by the supersymmetric models.

The simplest relevant model for the phenomenology contains one supercharge $\mathcal{N}=1$ and it is called Minimal Supersymmetric Standard Model (MSSM), which is currently under investigation at the LHC. Other phenomenologically relevant models, always with $\mathcal{N}=1$, are proposed. 
The number of supercharges can be increased, up to $\mathcal{N}=4$ if we do not consider supergravity. The extended SUSY theories $\mathcal{N}>1$ are not directly relevant for the phenomenology, as they do not possess chiral matter. However, in some regimes they behave in the same way as the relevant models, for instance theories with $\mathcal{N}=2,4$ may be thought as cousins of the QCD. Interestingly, the large amount of symmetries gives them some very interesting features: in particular, the non-perturbative regime can be dealt with sometimes exactly and non trivial results are obtained. This is partially due to an interplay with integrabily, as we will see in more details in the rest of the text. They could be useful toy models to develop new tools and gain additional insight in the non-perturbative structure of QFTs. Hopefully, this might help to study the non-perturbative regimes in the real theories, like for instance confinement in QCD.

\paragraph{Supersymmetric algebra} Here we breifly review the SuperPoincar\'e algebra, showing how the various multiplest appear. As mentioned earlier, a supersymmetric field theory is obtained from an ordinary Quantum Field Theory by enhancing the symmetry group, in particular adding some spinor supercharges $Q,\bar{Q}$ to the generators of the Poincar\'e group. This enlarge the symmetry to the so-called \emph{Super-Poincar\'e} group. The simplest case with only one supercharge, divided in left and right Weyl spinors, extend the Poincar\'e group by adding, to the usual ones involving the generators $P_{\mu},J_{\mu\nu}$, the following (anti)commutation relations

\be 
[Q_{\alpha},P_{\mu}]=0, \quad [Q_{\alpha},J_{\mu\nu}]=(\sigma_{\mu\nu})^{\beta}_{\alpha} Q_{\beta}, \quad \lbrace Q_{\alpha}, \bar{Q}_{\dot{\beta}}\rbrace = 2 \sigma ^{\mu}_{\alpha\dot{\beta}} P_{\mu}
\ee

where the relations for the barred spinor are analogous.

Enhancing the number of spinor charges $Q_{\alpha}^A,\bar{Q}_{\dot{\alpha}}^A$, with $A=1,..,\mathcal{N}$, the commutation relations are modified to

\ba 
&&[Q_{\alpha}^A,P_{\mu}]=0, \quad [Q_{\alpha}^A,J_{\mu\nu}]=(\sigma_{\mu\nu})^{\beta}_{\alpha} Q_{\beta}^A,  \nn \\
&&\lbrace Q_{\alpha}^A, \bar{Q}_{\dot{\beta}}^B\rbrace = 2 \sigma ^{\mu}_{\alpha\dot{\beta}} P_{\mu}\delta^{AB}, \quad \lbrace Q_{\alpha}^A, Q_{\beta}^B\rbrace = \epsilon_{\alpha\beta}Z^{AB}
\ea 

where $Z^{AB}$ is called the central charge of the $SU(\mathcal{N})$ algebra rotating the supercharges. 
This last subgroup is particularly important, expecially in $\mathcal{N}=4$ SYM, as the particles of the theory transform as representations of $SU(\mathcal{N})$\footnote{We stress that it is not a local symmetry, but only a global one.}.

In the same way as the Poincar\'e algebra determines the possible multiplets of the theory, the extended algebra here considered is amenable for the same treatment. The main difference is that, within the same multiplet, particles with different spin are present. For instance, the \emph{chiral multiplet} contains a complex scalar and its associated superpartner, a Weyl spinor. On the other hand, the \emph{vector supermultiplet} describes a gauge field whose superpartner is a Weyl fermion. Only for the extended algebras, the so-called hypermultiplet represents the matter field, particularly important in $\mathcal{N}=2$. These multiplets are separated when we have only one supercharge but merge together when more charges are added to the algebra. In the special case $\mathcal{N}=4$ all the particles belong to the same multiplet. The appropriated formalism to describe those group of particles makes use of a generalized version of the quantum field, the superfields involving Grassmann variables. 

We conclude with an important remark, valid in four dimensions. It is possible to show that, for $\mathcal{N}$ supercharges, the particle with the highest spin has $s\geq \frac{\mathcal{N}}{4}$. Therefore, if we do not want gravity to be involved\footnote{In that case, the highest number of supercharge would be $\mathcal{N}=8$.}, the largest number of allowed supercharges is $\mathcal{N}=4$, which is the theory we will mainly focus on in the following. 

\subsubsection{$\mathcal{N}=4$ Super Yang-Mills}

As we have seen above, $\mathcal{N}=4$ is the maximum number of allowed supercharges in a $4d$ theory without gravity. The associated gauge theory, with gauge group $SU(N)$, plays a rather special role in the realm of SUSY theories. The large number of symmetries imposes many constraints on the Lagrangian, which drastically reduce the number of free parameters. The same does not happen for theories with lowest number of supercharges. For instance, all the particles belong to the same multiplet: that implies, as the spectrum contains the gauge massless bosons, that all the particles have mass zero and transform under the adjoint representation of $SU(N)$. The Lagrangian reads

\be\label{N=4}
\mathcal{L}= \textit{Tr}\left[-\frac{1}{2g_{YM}^2}F_{\mu\nu}F^{\mu\nu} - D_{\mu}\Phi^i D^{\mu}\Phi^i - i \bar{\psi}_{A}\sigma^{\mu}D_{\mu}\psi^A + \text{interactions} \right]
\ee

Some comments on the formula above are due. The trace is taken over the gauge group $SU(N)$ indices. The only free parameter, beside the number of colors $N$, is the dimensionless coupling constant $g_{YM}$: it appears also in the definition of the field strenght $F_{\mu\nu}$ and the covariant derivative $D_{\mu}$.
The field content in (\ref{N=4}) consists of the vector supermultiplet $(A_{\mu},\psi_{\alpha}^A,\bar{\psi}_{\dot{\alpha}}^{\bar{A}},\Phi^i)$, where $A,\bar{A}=1,...,4$ and $i=1,....,6$. 

A quick glance to (\ref{N=4}) tells us that the Lagrangian is scale invariant. This fact, however, does not imply that the theory mantains the symmetry at the quantum level. There are many examples of conformally invariant Lagrangians, whose symmetry is spoiled by renormalisation effects. For instance, pure QCD (no quarks) is conformal invariant at the classical level but the process of renormalisation introduce a scale $\Lambda_{QCD}$. Therefore, the theory is not fixed by the coupling constant, which is running, but by a particular energy scale.

In $\mathcal{N}=4$ SYM, remarkably, the $\beta$-function which describes the renormalisation of the coupling constant is vanishing at any loop order, namely

\be 
\beta(\mu)=\mu\frac{\partial g_{YM}}{\partial\mu}=0
\ee

where $\mu$ is the energy scale. The consequences are striking: the coupling constant does not run and it is the true parameter describing the theory. In addition, as there is not an emergent energy scale in the theory, the conformal symmetry is preserved at the quantum level and we are dealing with a $4d$ CFT.

Joining together conformal symmetry, the $\mathcal{N}=4$ supercharges and the Poincar\'e group, the full symmetry group of the theory becomes $PSU(2,2|4)$, which contains two bosonic subgroups. The first one represent the conformal $4d$ symmetry, isomorphic to $SU(2,2)\sim SO(2,4)$. The other is $SU(4)\sim SO(6)$, representing the $R$-symmetry which rotates the four supercharges into each other. The fields of the theory transform under the adjoint of $SU(N)$, whose indices have been omitted. All the particles belong to a specific representation of the $SU(4)$ $R$-symmetry. The gauge boson is the singlet, as the are no $SU(4)$ indices. The left and right Weyl fermions are respectively the fundamental $\bold{4}$ and the antifundamental $\bar{\bold{4}}$, while the real scalars transform as the antisymmetric of $SU(4)$ or the fundamental vector of $SO(6)$. Adding the $32$ odd supercharges to the set of the bosonic generators, we build the whole group $PSU(2,2|4)$.

The bosonic subgroup has rank six, therefore we have the set of the six charges $(\Delta,S_1,S_2,J_1,J_2,J_3)$ labelling the states/operators of the theory.  The first three are associated to the spacetime symmetry, the conformal group: $\Delta$ represents the scaling dimension, eigenvalue of the dilatation operator $\mathcal{D}$, whereas $S_1,S_2$ defines the representation of the Lorentz group. The three charges $J_i$ are a sort of $SU(4)$ angular momentum in the space of the supercharges.

\subsection{String theory and the AdS/CFT duality}

In this part we briefly introduce string theory and discuss the important duality with some SUSY gauge theories, the $AdS/CFT$ correspondence. We will focus on the special case $AdS_5/CFT_4$, where the maximal supersymmetric $\mathcal{N}=4$ is involved.

\subsubsection{String Theory}

String theory is the main candidate for a full quantum theory of all the known interactions, gravity included. Originally, it was born to describe some properties of the strong force, when QCD was still under construction. In particular, the relation between mass and spin for some group of hadrons is well explained imagining a quark and an antiquark connected by a relativistic rotating string. However, after the predictive success of QCD, this project was abandoned. However, the discovery of the graviton in its spectrum and the absence of internal inconsistencies gave string theory a new light in the 80's. 

The basic objects are tiny strings, whose different vibration modes give rise to particles. In this description, different particles come from the same foundamental object. According to the two revolutions of the last century, the strings are relativistic and quantized, in the language of first quantization. 

The are two appealing reasons to think that string theory could actually describe our universe. First, gravity is included in a full consistent quantum framework, a feature that is still missing in QFT. In addition, string theory is unique, as there are no free adimensional parameters, so everything we see should be, in principle, fixed by the dynamics of the model. The only dimensional parameter is the string scale (or lenght), which is thought to be the order of the inverse Plank mass $\sim M_{p}^{-1}\sim l_p$. To compare, in the Standard Model there are around 30 free parameters whose values are determined by experiments. In addition, a consistent model of strings includes naturally the concept of supersymmetry. 

Looking at the downsides, a significant complication is that the theory is, upon quantisation, consistently defined only in $9+1$ spacetime dimensions. To make up for the six missing dimensions, we need to bend the extra ones into a small compact dimensions. Unfortunately, there are about $10^{500}$ different solutions and the way we bend them affects drastically the low-energy properties of our world. This is one of the main obstacle faced by the researcher who try to get some predictions out of the theory.  

On the experimental level, being the string scale very large compared with the energies currently tested in our accelerators, finding a phenomenological prediction is a very challenging task. However, there are some proposals, expecially at the cosmological level, to test some low-energy effects of the theory. This is the field called String Phenomenology.

\paragraph{The string action} 

Here we briefly introduce the starting object from which string theory is developed, namely the string action. We consider only the bosonic subsector to avoid unnecessary technical complications: we stress, however, that also fermionic degrees of freedom have to be included. The theory is bases on the language of first quantisation, therefore the string action is the generalisation to a one-dimensional object of the worldline formalism for a relativistic particle. In the same way as a particle is described by a worldline, a string sweeps a $2d$ spacetime surface called \emph{worldsheet}, described by the functions $X^{\mu}(\tau,\sigma)$. The action is the area of the worldsheet and is given by the Nambu-Goto action

\be\label{NambuGoto} 
S=-T\int d^2\sigma \sqrt{-\dot{X}^2 X'^2 + (\dot{X}\cdot X')^2}, \quad \dot{X}^{\mu}\equiv \frac{\partial X^{\mu}}{\partial\tau}, \quad X'^{\mu}\equiv \frac{\partial X^{\mu}}{\partial\sigma}
\ee

where $T\equiv \frac{1}{2\pi\alpha'}\equiv \frac{1}{2\pi l_s^2}$ is the string tension and the integral is over the two worldsheet coordinates $\tau,\sigma$. 

The equation of motion from (\ref{NambuGoto}) are very involved, making the quantization difficult. Fortunately, there is an alternative action available, which is equivalent to the Nambu-Goto one. It introduces an additional metric on the worldsheet $g^{\mu\nu}$ and reads

\be\label{Polyakov} 
S_p=-\frac{T}{2}\int d^2\sigma \sqrt{-g}g^{\alpha\beta}\partial_{\alpha}X^{\mu}\partial_{\beta}X^{\nu}\eta_{\mu\nu}
\ee

where $\eta_{\mu\nu}$ is the metric of the spacetime in which the string dwells. 

The Polyakov action (\ref{Polyakov}) is the starting point for the quantisation, which can be achieved in several alternative methods. From the analysis of the symmetries and the requirement that they are preserved at the quantum level\footnote{In particular, the Weyl symmetry.}, it follows that the only consistent number of spacetime dimensions is $D=26$ for the bosonic string and $D=10$ for the full superstring theory involving fermions as well.

For our purpose, we point out that the action (\ref{Polyakov}) describes a $2d$ quantum field theory living on the worldsheet span by the coordinates $(\tau,\sigma)$. The fields of this QFT $X^{\mu}$ are nothing but scalar fields whose internal indices are the Lorentz indices $\mu=1,..,D$ of the original string theory. Therefore, for a non-trivial fixed background geometry, the worldsheet theory becomes a $2d$ non-linear $\sigma$-model.  

The interaction between strings is still described by the action (\ref{Polyakov}), but considering worldsheets with different topologies. The interaction series is organized as a topological expansion, weighted by the string coupling $g_s$, of the path integral of the Polyakov action (\ref{Polyakov}). The tree-level amplitudes are described by the surfaces with zero handles, equivalent to a cilinder. Strictly speaking, the $\sigma$-model description works only when the tree-level contributions are considerer, \emph{i.e.} for free strings. 

We mention that, in principle, both the coupling between strings and the background spacetime are determined by the dynamics of the strings. However, it is meaningful to study approximations with fixed background and coupling, as we will see in the case of the $AdS/CFT$.

\subsubsection{The AdS/CFT duality}

The idea of a connection between strings and gauge theory was first proposed by 't Hooft, who studied the large $N$ expansion \cite{largeN} in a $SU(N)$ gauge theory. The Feynman diagrams are classified in a topological fashion: the are weighted by a factor $N^{2-2g}$, where $g$ is the genus of the associated two-dimensional surface\footnote{The surface on which the diagram can be drawn without any self-intersection of lines.}. The genus corresponds, roughly speaking, to the number of handles of the surface. For instance, a sphere and a torus have respectively $g=0$ and $g=1$. The expansion recall that of string theory, where the parameter is the string coupling $g_s$ and the different worldsheets are classified by the same topological invariant. The identification between the two expansion parameters is $g_s \sim 1/N$. In the large $N$ limit of the gauge theory, the only relevant diagrams are the planar ones, which can be drawn on a plane without self-intersections. For this reason, the large $N$ limit is also referred to as the planar limit. In that regime, in order to have a finite theory, wa also send to zero the Yang-Mills coupling: the only parameter becomes the 't Hooft coupling $\lambda\equiv g_{YM}^2 N$. Another hint which suggested the duality was the holographic principle \cite{Hol}. The area law for the entropy of a black hole $S \sim A$ suggests that, in a gravitational theory, the information stored in a given volume $V$ can be described by some degrees of freedom living on the surface $\partial V$ enclosing the volume.

The $AdS/CFT$ correspondence \cite{Mal,Wit,GKP1} realizes practically the ideas described above. It states a duality between a string theory, describing gravity, living in a spacetime containing an $AdS_{d+1}$ subspace, and a conformal gauge theory located at the boundary of $AdS_{d+1}$, which is the $d$-dimensional Minkowski spacetime $M^{d}$. The holography comes from the fact that the quantum field theory on the boundary describes the gravitational dynamics in the bulk. It is thus quite remarkable and very different from many other dualities, as it relates two theories with different degrees of freedom (only one side has gravity) and different spacetime dimensions. 

In addition, another and probably the most important feature, expecially from the computational point of view, is the weak/strong nature of the correspondence. It means that the strong coupling on one side is mapped to the weakly coupled dual theory. This is, at the same time, a good and a bad thing. On the one hand, a proof of the duality is very hard, as we cannot compare perturbative calculations on both sides and non-perturbative tools, for instance integrability, are needed. On the other hand, however, it could give us hint about the nature of the strongly coupled dynamics by simply looking at the dual theory in the weak coupling regime.  

Several examples of $AdS/CFT$ duality are known nowadays. The most famous and best known case is the $AdS_5/CFT_4$, which is actually the one relevant for this thesis and the only one that will be referred to in the following.

\subsubsection{$\mathcal{N}=4$ SYM and IIB $AdS_5\times S^5$ string theory}

Here we discuss the basic features of the $AdS/CFT$ duality we are interested in: it concerns the $4d$ conformal SUSY gauge theory $\mathcal{N}=4$ Super Yang-Mills and the type $IIB$ string theory living on the spacetime $AdS_5\times S^5$. 

The gauge theory is defined by two dimensionless parameters, the Yang-Mills coupling $g_{YM}$ and the number of colors $N$. On the other side, strings are coupled by $g_s$ and have lenght $l = \sqrt{\alpha}$; in addition, the sphere $S^5$ and the $AdS_5$ part share the common radius $R$. Actually, the last two only appear in the dimensionless combination $(R/l)$.

The are different versions of the duality, depending on whether some particular limits are taken or not.
The stronger version related the two theories for any value of the parameters, once identified as

\be\label{AdSCFT} 
\lambda = g_{YM}^2N=\left(\frac{R}{l}\right)^4, \quad  g_{YM}^2=4\pi g_s
\ee 

A slightly weaker statement, which is the one we are actually interested in and in which most results have been obtained, regards the large $N$ limit. Importantly, the string coupling goes to zero: we are thus dealing with \emph{free strings}, namely with worldsheets with the simples topology. In this case, the string theory is described by the $1+1$ QFT on the worldsheet. The correspondence states that the two theories, in the planar limit $N\to\infty$, $g_s\to 0$, are equivalent, \emph{i.e.} they describe the same physics from a very different perspective.

From the first relation in (\ref{AdSCFT}), we can see the strong/weak nature of the duality. The strong coupling side of the gauge theory corresponds to $R\gg l$, \emph{i.e.} a weakly curved spacetime form the point of view of the string, whose worldsheet QFT $\sigma$-model is thus weakly coupled. On the other hand, for small $\lambda$ the string feels the whole spacetime metric and the strongly interacting $\sigma$-model takes over.

Having discussed the symmetries of $\mathcal{N}=4$ SYM previously, now we analyse the string theory and check that the symmetries of the two sides match. The string describes a $1+1$ QFT on the worldsheet, in particular a non-linear $PSU(2,2|4)$ $\sigma$-model, which corresponds to the coset $\frac{PSU(2,2|4)}{SO(1,4)\times SO(5)}$. The bosonic part of the action gives, respectively, the non-linear $\sigma$-models  $SO(2,4)$ and $SO(6)$. The first corresponds to the dynamics on $AdS_5$, whereas the latter describes the string on the five-sphere $S^5$. The coupling the two subsectors is carried by the fermions.

The duality also should give us a map between the observables, of which the main example is the following. On the gauge side we have the gauge invariant (single trace only, in the planar limit) operators, labelled by the charges $(\Delta,S_1,S_2,J_1,J_2,J_3)$. They are mapped to the string states: more precisely, $\Delta$ is related to the energy of the string, while the spins $S_1,S_2$ represent the dynamics of the string moving in $AdS_5$ and the triplet of $J_i$ is related to the string angular momentum on the sphere $S^5$.

\section{Integrable structures in $4d$ SUSY field theories}
\label{IntSUSY}

The standard integrability features highlighted for quantum field theories in Section \ref{Integrability} are confined to two dimensional models, as a consequence of the Coleman-Mandula theorem \cite{ColMan}. However, in the last two decades, the study of extended $4d$ SUSY models unveiled some integrable features as well. In this section we discuss this surprising emergence of integrability in the theories $\mathcal{N}=2,4$, with particular focus on $\mathcal{N}=4$ SYM which will be the main topic of the thesis. For the latter, the authors \cite{MinZar} found an application of the Bethe Ansatz in the planar limit, giving rise to a new field of research. At the same time, integrability found applications on the string dual model as well \cite{BPR}.

\subsection{Integrability in planar $\mathcal{N}=4$ SYM}

Despite the lack of an infinite set of local commuting charges, which is the necessary condition for standard integrability in a $2d$ theory, $\mathcal{N}=4$ SYM still enjoys some integrability features. This happens when the planar limit $N\to\infty$ is taken. The motivations behind this emergent property are not completely understood yet, although it might be related to the so-called Yangian symmetry, an infinite set of \emph{non-local} commuting charges characterizing the theory. Specifically, an integrable description emerged for the first time when dealing with the spectral problem of the anomalous dimensions, which we are going to depict in details below.

\subsubsection{The spectral problem} 

In a conformal field theory, thanks to the radial quantisation and the state/operator map, we can diagonalise the dilatation operator $\mathcal{D}$ in place of the Hamiltonian to solve the spectral problem of the theory, which is defined as follows. The two-point function of a single trace\footnote{In the planar limit, they are the only relevant ones.} gauge invariant operator $\mathcal{O}$ behaves schematically as

\be\label{2pDelta}
\langle\mathcal{O}(x)\mathcal{O}(0)\rangle\sim\frac{1}{x^{2\Delta(g)}}, \quad \Delta(g)^{\mathcal{O}}=\Delta_0^{\mathcal{O}} + \gamma^{\mathcal{O}}(g)
\ee

where $\Delta^{\mathcal{O}}(g)$ is the scaling dimension, eigenvalue of the dilatation operator $\mathcal{D}$. The scaling dimension plays the role of the energy, while the gauge invariant operators can be thought of as the (eigen)states of the theory.

The scaling dimension consists of two parts. The bare one, which is integer of half-integer and it is nothing but the engineering dimension of the operator $\mathcal{O}$, it is the tree level contribution to the correlation function. The quantum corrections are switched on with the coupling and are contained in the so-called anomalous dimension $\gamma^{\mathcal{O}}(g)$. 

The operator $\mathcal{O}$ can be represented as a string of fields

\be\label{OTr} 
\mathcal{O}\sim \textit{Tr}\left[A_1A_2......A_n\right]
\ee

where $A_i$ is a generic gauge covariant field of the theory and the trace is taken over the $SU(N)$ indices of the gauge group. The gauge covariant fields we can insert to compose the operator $\mathcal{O}$ are: the field streght $F_{\mu\nu}$, the (anti)fermion $\psi_{\alpha}^A,\bar{\psi}_{\dot{\alpha}}^{\bar{A}}$, the scalars $\Phi^i$ and the covariant derivative $D_{\mu}$. The quantum corrections $\gamma^{\mathcal{O}}(g)$ are due to the renormalisation effects, necessary to keep the loop contributions finite. The main contribution from \cite{MinZar} was to find a spin chain description of the problem. 

We stress that an operator like (\ref{OTr}) is labelled by the six charges $(\Delta,S_1,S_2,J_1,J_2,J_3)$ of which only the first is corrected at the quantum level. Now an important remark about (\ref{2pDelta}) is due: the formula is valid, strictly speaking, only for some linear combinations of single trace operators. More precisely, the renormalisation process introduces mixing between operator with the same charges $(J_1,J_2,S_1,S_2,S_3)$, so that the anomalous part is actually a matrix $\Gamma$ acting in the space of operators, with eigenvalues $\gamma$.

\paragraph{The $SU(2)$ sector} 

Although the authors studied (at one-loop) the $SO(6)$ sector, made up by the six real scalars $\Phi^i$, to give a sketch of the method we analyse the simplest $SU(2)$ sector, with two complex scalars built out of $\Phi^i$ according to

\be 
Z=\Phi_1+i\Phi_2, \quad X=\Phi_3+i\Phi_4
\ee

In this case, a gauge invariant operator has the sketchy form

\be 
\mathcal{O} \sim \textit{Tr}\left[ZXZZX.....XZZXXZ\right]
\ee

which already shows an interesting analogy with the spin chain, once we identify $(Z,X)$ with the two spin states $(\uparrow,\downarrow)$. 

The anomalous dimension matrix can be expanded in loop contributions $\Gamma=g^2\Gamma^{(2)} + O(g^3)$, where we parametrize the 't Hooft coupling as $\lambda = 16\pi^2 g^2$. The main achievement was to show that, at one-loop level, the anomalous dimension matrix of the $SU(2)$ sector is given by the Hamiltonian of the Heisenberg spin chain, namely

\be 
\Gamma^{(2)}=g^2\mathcal{H}_{SU(2)}, \quad (X,Z)\Leftrightarrow (\downarrow,\uparrow)
\ee

where the spin states are identified with the two scalars, as depicted before. 

Therefore, the vacuum state is represented by the operator $\textit{Tr}[Z^L]$ and the excitations corresponds to the insertion of a field $X$. The operator with definite one-loop anomalous dimension $g^2\gamma^{(2)}$ are linear combination of terms schematically of the type $\textit{Tr}[Z^{L-M}X^M]$, where $M$ is the number of ''magnons'' propagating along the chain/operator. An additional constraint comes from the ciclicity of the trace: we need to consider only those states which respect translational invariance. For instance, with one magnon the only state allowed has momentum $p=0$ whereas in general the total momentum must vanish.

\paragraph{The full $PSU(2,2|4)$ sector: the \emph{Asymptotic} Bethe Ansatz equations}

The procedure outlined above can be pushed forward, by enlarging the sector and considering more than one loop. A complete non-perturbative evaluation is still missing so far. However, a set of equations, describing a long-range spin chain, has been proposed  \cite{BSABA,BES12} to account for any loop order for the full $\mathcal{N}=4$ SYM theory, \emph{i.e.} a $PSU(2,2|4)$ chain. It is described by the Beisert-Staudacher, or \emph{Asymptotic} Bethe Ansatz (ABA), equations

\ba\label{ABA}
1&=&\displaystyle\prod_{j\neq k}^{K_2}\frac{u_{1,k}-u_{2,j}-i/2}{u_{1,k}-u_{2,j}+i/2}\displaystyle\prod_{j=1}^{s}\frac{1-\frac{g^2}{2x_{1,k}x^-_{4,j}}}{1-\frac{g^2}{2x_{1,k}x^+_{4,j}}}\nonumber \\ 
1&=&\displaystyle\prod_{j\neq k}^{K_2}\frac{u_{2,k}-u_{2,j}+i}{u_{2,k}-u_{2,j}-i}\displaystyle\prod_{j=1}^{K_1}\frac{u_{2,k}-u_{1,j}-i/2}{u_{2,k}-u_{1,j}+i/2}\displaystyle\prod_{j=1}^{K_3}\frac{u_{2,k}-u_{3,j}-i/2}{u_{2,k}-u_{3,j}+i/2}\nonumber \\
1&=&\displaystyle\prod_{j=1}^{K_2}\frac{u_{3,k}-u_{2,j}-i/2}{u_{3,k}-u_{2,j}+i/2}\displaystyle\prod_{j=1}^{s}\frac{x_{3,k}-x^-_{4,j}}{x_{3,k}-x^+_{4,j}}\nonumber \\ 
1&=& \left(\frac{x^-_{4,k}}{x^+_{4,k}}\right)^L\displaystyle\prod_{j\neq k}^s \frac{x^-_{4,k}-x^+_{4,j}}{x^+_{4,k}-x^-_{4,j}}\frac{1-\frac{g^2}{2x^+_{4,k}x^-_{4,j}}}{1-\frac{g^2}{2x^-_{4,k}x^+_{4,j}}}\sigma^2(u_{4,k},u_{4,j})\times  \nonumber \\
&\times &\displaystyle\prod_{j=1}^{K_3}\frac{x_{4,k}^+-x_{3,j}}{x_{4,k}^--x_{3,j}}\displaystyle\prod_{j=1}^{K_5}\frac{x_{4,k}^+-x_{5,j}}{x_{4,k}^--x_{5,j}}\displaystyle\prod_{j=1}^{K_1}\frac{1-\frac{g^2}{2x_{1,j}x^+_{4,k}}}{1-\frac{g^2}{2x_{1,j}x^-_{4,k}}}\displaystyle\prod_{j=1}^{K_7}\frac{1-\frac{g^2}{2x_{7,j}x^+_{4,k}}}{1-\frac{g^2}{2x_{7,j}x^-_{4,k}}} \nonumber \\
1&=&\displaystyle\prod_{j=1}^{K_6}\frac{u_{5,k}-u_{6,j}-i/2}{u_{5,k}-u_{6,j}+i/2}\displaystyle\prod_{j=1}^{s}\frac{x_{5,k}-x^-_{4,j}}{x_{5,k}-x^+_{4,j}}\nonumber \\ 
1&=&\displaystyle\prod_{j\neq k}^{K_6}\frac{u_{6,k}-u_{6,j}+i}{u_{6,k}-u_{6,j}-i}\displaystyle\prod_{j=1}^{K_7}\frac{u_{6,k}-u_{7,j}-i/2}{u_{6,k}-u_{7,j}+i/2}\displaystyle\prod_{j=1}^{K_5}\frac{u_{6,k}-u_{5,j}-i/2}{u_{6,k}-u_{5,j}+i/2}\nonumber \\
1&=&\displaystyle\prod_{j\neq k}^{K_6}\frac{u_{7,k}-u_{6,j}-i/2}{u_{7,k}-u_{6,j}+i/2}\displaystyle\prod_{j=1}^{s}\frac{1-\frac{g^2}{2x_{7,k}x^-_{4,j}}}{1-\frac{g^2}{2x_{7,k}x^+_{4,j}}}
\ea

where the function $x(u)$ is the Jukovsky map

\be 
x(u)=\frac{u}{2}\left[1+\sqrt{1-\frac{2g^2}{u^2}}\right], \quad x^{\pm}(u)\equiv x\left(x\pm\frac{i}{2}\right)
\ee

Note that in (\ref{ABA}) only the fourth node carries momentum and energy, thus the anomalous dimension is expressed in terms of the roots $u_4$ as

\be 
\gamma = ig^2\sum_{k=1}^s \left[\frac{1}{x_{4,k}^+}-\frac{1}{x_{4,k}^-}\right]
\ee

The equations (\ref{ABA}) are asymptotic in the sense that they are valid for long operators $L$ only.
The range of the interaction increases as the number of loops grows, so that the equations solve the problem only up to $\lambda^{2L}$: for higher loops, some wrapping effects appear as the interaction goes along the whole chain.

The wrapping effects are estimated to be exponential $O(e^{-L})$ and can be captured for finite $L$ by an ingegnious refinement of the Thermodynamic Bethe Ansatz procedure \cite{TBA1,TBA3,TBA2} outline before. The $AdS_5/CFT_4$ played a main role in the complete solution of the spectral problem, as the mirror rotation\footnote{The procedure that swaps space and time.} can be done on the dual $2d$ QFT theory on the worldsheet. It is worth to mention a more complete description of the spectrum, which is encapsulated in the so-called Quantum Spectral Curve \cite{QSC} formalism, which enjoys some computational advantages with respect the TBA setup. It is important to mention that these tools can be applied also to other cases of gauge-gravity duality, for instance $AdS_4/CFT_3$ \cite{BFTAdS4CavFio,GLMFFPT}.

\paragraph{The $sl(2)$ sector}

A noteworthy sector, closed under renormalization, is composed by operators of the type

\be\label{SL2} 
\textit{Tr}[D_+^S Z^L] + \cdots
\ee
where the dots indicate that the light-cone covariant derivatives $D_+^S$ can act in different positions the string of scalar fields. The operator (\ref{SL2}) has Lorentz spin $S$ and classical scaling dimension $\Delta_0=S+L$ so that

\be 
\Delta(g,L,S)=S+L+\gamma(g,L,S)
\ee
is the full scaling dimension, where $\gamma$ represents the anomalous part.

This sector has been the subject of several investigations \cite{FGR12,FGR34,FRS,BGK}, expecially in the large spin limit $S\to\infty$ whose behaviour is described, on the dual string side, by the so-called GKP string solution \cite{GKP2}. It describes a long spinning string with angular momentum $S,L$ respectively in $AdS_5$ and $S^5$. 

Interestingly, in such a limit the anomalous dimension is given, at the leading order, by the Sudakov factor

\be\label{Suda} 
\gamma(g,L,S)=f(g)\ln S + O((\ln S)^0)
\ee
where $f(g)$ is the \emph{scaling function}. 

We remark that the function $f(g)$ is more general, as it also appears to describe the cusp anomalous dimension for light-like cusped Wilson loops. The scaling function has been carefully studied in both weak \cite{BES12} and strong coupling limit \cite{Scaling}. An additional reason to investigate the $sl(2)$ sector is its applications to QCD, for instance to study the deep anelastic scattering. For our purpose, the excitations over the GKP string are of fundamental importance in the computation of the null polygonal Wilson loops and will be described in some more details in the next chapter.

\subsection{Integrability in $\mathcal{N}=2$}

Some integrable structures appear also in less supersymmetric gauge theories as well, for instance $\mathcal{N}=2$ models. These theories are more similar to the QCD than $\mathcal{N}=4$, as there are massive matter fields and they usually enjoy asymptotic freedom. Of course, they are not conformal invariant. Even though less symmetric than $\mathcal{N}=4$ SYM, exact results are still possible to obtain and some of the techniques employed to obtain them show deep and interesting connections with the language of integrable systems. 

\paragraph{Seiberg-Witten curve and Classical Integrability} The first connection appeared in the 90's, in particular with the the theory of classical integrable systems. For $\mathcal{N}=2$ theories, Seiberg and Witten \cite{SW1} managed to compute the exact low-energy effective action, encoding all the necessary informations in the so-called \emph{Seiberg-Witten curve}, which is a complex elliptic curve, \emph{i.e.} a surface in $\mathbb{C}^2$. Eventually, it yields the prepotential $\mathcal{F}_{SW}$, namely the logatirhm of the low-energy effective action which is a fuction in the moduli space, \emph{i.e.} the space of VEVs $\vec{a}$ for the scalar fields. Immediately thereafter, a connection with classical integrable systems appeared, as the SW curve turned out to be the classical spectral curve characterizing a classical integrable system \cite{SWInt}.

\subsubsection{The Nekrasov function $\mathcal{Z}$ and Quantum Integrability}

A quantum version of the correspondence above has been proposed recently \cite{NekSha}. A fundamental quantity is the so-called Nekrasov instanton partition function $\mathcal{Z}$, developed in \cite{Nekrasov} to compute the instantons effect to the partition function in $\mathcal{N}=2$ gauge theories. A deformed spacetime, called the $\Omega$-background, is necessary to regularize the sum over the instantons and it is parametrised by $\epsilon_1,\epsilon_2$. Combined with the classical and one loop contribution, the Nekrasov function yields the Seiberg-Witten prepotential through the limit procedure

\be\label{FSW} 
\mathcal{F}_{SW}(\vec{a},q)=\lim_{\epsilon_1,\epsilon_2\to 0}\epsilon_1 \epsilon_2\log\left[\mathcal{Z}_{tree}\mathcal{Z}_{loop}\mathcal{Z}(\vec{a},q,\epsilon_1,\epsilon_2)\right]
\ee

where the dependence on the instanton parameter $q$ and the VEVs is highlighted.

The quantisation of the classical integrable system associated to the SW curve has been proposed to be given by switching on one of the parameter, say $\epsilon_1$, which plays the role of the Planck constant. The limit $\epsilon_2\to 0$ is called the Nerkasov-Shatashvili (NS) limit and it provides, throught the Nekrasov function, a quantized version of the aforementioned duality. Defining the superpotential according to

\be 
\mathcal{W}(\vec{a},q,\epsilon_1)=\lim_{\epsilon_2\to 0}\epsilon_2\log\left[\mathcal{Z}_{tree}\mathcal{Z}_{loop}\mathcal{Z}(\vec{a},q,\epsilon_1,\epsilon_2)\right]
\ee

the proposal is that the equations

\be 
\exp\left(\frac{\partial\mathcal{W}(\vec{a})}{\partial a_i}\right)=1, \quad i=1,...,r
\ee

are nothing but the Bethe equations for a quantum integrable system. 

A different relation with the integrability language emerges as we focus on the instanton contribution 

\be 
\mathcal{W}_{inst}(\vec{a},q,\epsilon_1)=\lim_{\epsilon_2\to 0}\epsilon_2\log\mathcal{Z}(\vec{a},q,\epsilon_1,\epsilon_2)
\ee

which turns out ot be described in terms of a TBA-like equation. In particular, its value is given by a critical Yang-Yang functional $YY_c$, coming from a saddle point procedure whose equation of motion assumes the same form as the TBA equation (\ref{TBA}).

The Nekrasov function will be extensively studied in the Appendix \ref{NekApp}, focusing on the NS limit and deriving the TBA-like equation. Remarkably,  this limit will show a striking analogy with the strong coupling regime for the null polygonal Wilson in $\mathcal{N}=4$, for which an analogous result holds as well. Additional insight on the emergence of these integrable features in $\mathcal{N}=2$ models has been obtained in the recent papers \cite{BoFN2}.

\chapter{$\mathcal{N}=4$ null polygonal Wilson loops and the OPE series}
\label{ChOPE}

In this chapter we set the stage for the second part of the thesis, containing the results obtained during the PhD. As mentioned in the introduction, integrability appears in a subtle and peculiar way in $\mathcal{N}=4$ SYM gauge theory, first applied to solve the spectral problem and later to other observables as well. Instead of the two-point functions and their anomalous dimensions, we focused on non-local observables in $\mathcal{N}=4$ Super Yang-Mills, the null polygonal Wilson loops. This theory enjoy another remarkable duality, \emph{i.e.} that between null polygonal Wilson loops and the $4d$ gluon scattering amplitudes. For the Wilson loops, a non-perturbative approach has been proposed in \cite{BSV1}, building on an earlier idea of \cite{Anope} and ispired by the interesting spin chain description in \cite{Bel-Qua}. The idea is borrowed form the QCD and it is to describe the loop in terms of the $2d$ the flux-tube, which is integrable in $\mathcal{N}=4$ SYM. Therefore, it provides a non-perturbative tool to study the scattering amplitudes in a four dimensional interacting theory.

\section{$\mathcal{N}=4$ Wilson loops and 4d scattering amplitudes}
\label{WLAmp}

In this section we consider planar $\mathcal{N}=4$ Super Yang-Mills and deal with another important duality, involving two very different kind of observables. On the one hand, there are the null polygonal Wilson loops, which are non-local gauge invariant operators. On the other side, the $4d$ gluon scattering amplitudes are on-shell quantities. A correspondence between them was proposed \cite{AM-amp,BHTDKS} and then successfully tested both at weak and strong coupling \cite{WLAmp}. The aim for this section is to briefly describe both quantities and provide a dictionary of the duality.

\subsection{The Wilson loop}

The Wilson loops are among the simplest non-local observables in a gauge theory. For a non-SUSY theory, they are defined as the vacuum expectation value of a gauge invariant operator, describing the parallel transport of a quark along a closed path $\mathcal{C}$. The quark acquires a phase factor

\be 
\Psi(x+\mathcal{C})=\mathcal{W}(\mathcal{C})\Psi(x)
\ee

which is represented by the operator $\mathcal{W}(\mathcal{C})$. For a $SU(N)$ gauge theory, the Wilson loop operator is defined by the path ordered exponential

\be 
W(\mathcal{C})\equiv \frac{1}{N}\textit{Tr} \mathcal{W}(\mathcal{C}) = \frac{1}{N}\textit{Tr}\left(\mathcal{P} e^{i\oint_{\mathcal{C}} ds \dot{x}_{\mu}A^{\mu}}\right)
\ee

where the trace is taken over the fundamental representation of the gauge group. The symbol $\mathcal{P}$ means the path ordering operation, necessary as we are dealing with a non-abelian theory.
Historically, they were introduced to address the confinement problem in QCD. For instance, consider a rectangular path with two timelike sides $T$ and two spacelike $R$: the large $T$ limit gives the quark-antiquark potential through

\be 
V(R)=-\lim_{T\to\infty}\frac{\log\langle W(R,T)\rangle}{T}
\ee

From now on, we omit the symbol $\langle...\rangle$ and suppose the operation of expectation value to be taken, so that $W(\mathcal{C})$ is no longer an operator but a number, more precisely a functional of the path $\mathcal{C}$. 

Moving to $\mathcal{N}=4$ SYM, the definition of Wilson loop needs a refinement, as there are no massive fundamental quarks in the theory. To this purpose, the $AdS/CFT$ duality furnishes us a picture to solve the problem: we think of $\mathcal{N}=4$ SYM as the low-energy worldvolume theory of open strings ending on a stack of $N$ D3-branes.  We introduce an additional D3 brane, separated from the others in the transverse directions $x^{i+3}$, $i=1...,6$, giving an VEV to the scalars $\phi^i$. The separation is given by $x^{i+3}=Mn^i$, where $M\gg 1$ and the vector is unitary $\delta_{ij}n^i n^j=1$. 
On the gauge side, the procedure consists of breaking the $SU(N+1)$ symmetry down to $SU(N)\times U(1)$, giving birth to massive fundamental particles. That said, the simplest Wilson loops (bosonic) in $\mathcal{N}=4$ SYM reads

\be\label{BosonicWL} 
W(\mathcal{C})\equiv \langle 0|\textit{Tr}\mathcal{P} e^{i\oint_{\mathcal{C}} ds \left(\dot{x}_{\mu}A^{\mu} + |\dot{x}|\phi_i n^i\right)}|0\rangle
\ee

where, beside the usual gauge connection, the scalar fields also appear. The unit vector $n^i$ identifies a particular direction in the five-sphere $S^5$. 
The definition (\ref{BosonicWL}) can be generalised by including the fermions along the path, so that we speak of charged, under the $SU(4)$ R-symmetry, Wilson loops. In our work we stick with the simplest case where only scalar fields are involved. 

\paragraph{Null polygonal contour}

Of particular interest is the class of Wilson loops defined on a null polygonal contour. Thus, instead of a functional, we have a function of $n$ vertices $W_n(x_1,...,x_n)$. 
The presence of the cusps along the contour introduces UV divergences, so that the Wilson loop needs to be renormalised in order to get a finite quantity. The process of renormalisation breaks the conformal symmetry of the theory. This breaking is completely captured by the so-called BDS ansatz, originally proposed for the scattering amplitudes \cite{BDS}, which fixes the finite part of the loop up to $n=5$.

Adding more sides, a non-trivial correction called the remainder function appears. It is a conformal invariant quantity, function of the $4d$ cross ratios $\tau_i,\sigma_i,\phi_i$, $i=1,...,n-5$ which encodes the geometric informations on the loop as they depend on the vertices $x_i$. 

\subsection{Wilson loop/amplitudes duality}

As mentioned previously, the null polygonal Wilson loops are thought to be dual to the four dimensional scattering amplitudes between gluons (gauge particles) in $\mathcal{N}=4$ SYM. The purpose of this part is to explain a little more this statement and provide a dictionary of the duality.  

A generic amplitude depends on the initial and final states, containing momenta, helicities (which can be $\pm 1$) and colour degrees of freedom. We can strip off the last dependence by defining the so-called color-ordered amplitudes. The discrete helicity degrees of freedom identify the type of the amplitude we are dealing with. In this work we focus only on the MHV amplitude, where the name stands for Maximal Helicity Violation. Considering all the momenta incoming, it means that $n-2$ gluons have a particular helicity and $2$ the opposite: the cases where they have the same helicity or only one differ from the others are shown to be vanishing. Therefore, we consider a function only of the momenta $\mathcal{A}^{MHV}(p_1,...,p_n)$, where the conservation\footnote{All momenta are considered incoming.} implies $\sum_i p_i=0$.

Omitting the MHV for the sake of simplicity, the amplitude consists in two factors, the tree-level contribution and the loops effect

\be 
\mathcal{A}(p_1,...,p_n)=\mathcal{A}^{\textit{tree}}(p_1,...,p_n)\mathcal{A}^{L}(p_1,...,p_n)
\ee

We remark that, as the theory contains massless particles, we have IR divergences that have to be cured by a renormalisation process. On the other hand, the theory is UV-divergence free. The finite part of the amplitudes can be captured by the BDS ansatz for the loop part $\mathcal{A}^L$ and works fine up to five gluons, thus describe the amplitudes with $n=4,5$ exactly at any loop order\footnote{The three gluon amplitude is vanishing.}.

The duality reads

\be\label{WL/Amp} 
\mathcal{A}^L(p_1,...,p_n)=W(x_1,...,x_n), \quad p_i=x_{i+1}-x_i
\ee

once the sides of the loop are identifyied with the momenta of the gluons and the IR-UV divergencies are removed. The null sides are due to the masslessness of the particles and the contour is closed due to the momentum conservation $\sum_i p_i=0$.

Although this fact has been proposed by strong coupling computation through the $AdS/CFT$ duality, it received many checks at weak coupling by means of pure gauge theory techniques, \emph{i.e.} loop computations. Therefore it is now thought no longer as a conjecture, but as a well-established fact. However, a mathematical proof is still missing. We mention that the duality extends to any helicity configuration by considering the $SU(4)$ charged Wilson loops.

\subsection{Strong coupling limit}
\label{StCou}

In this part we anticipate some strong coupling features of these Wilson loops/amplitudes. There are two main contributions to that regime. Using the string picture, one comes from the classical dynamics in $AdS_5$ \cite{AM-amp} and it has been the first hint to the aforementioned duality. This contribution will be analysed below. On the other hand, the pentagonal OPE method, which will be introduced later, suggested \cite{BSV4} a correction of the same order coming from the non-perturbative dynamics of the string on the five sphere $S^5$ and described by a non-linear $\sigma$-model $O(6)$ \cite{AM}, where the Wilson loop becomes a correlator of a specific twist operator. The particles of this QFT are the two-dimensional version of the scalars of the gauge theory. In particular, as the mass of the scalars goes to zero exponentially with the coupling \cite{AM,BKBFO6,FGR12}

\be 
m\sim e^{-\sqrt{\lambda}/4}
\ee

we find ourselves in the short-distance regime and the typical $\sqrt{\lambda}$ behaviour follows. This contribution is discussed in Chapter \ref{ChSca}.

The full strong coupling behaviour, depicted here for the hexagon for simplicity, reads

\be\label{Wfactorised}
W_6 \simeq W_{AdS_5}W_{S^5}\left[1+O\left(\frac{1}{\sqrt{\lambda}}\right)\right] \, ,
\ee

where the two terms take the form

\be\label{strongW}
W_{AdS_5}\simeq C_{AdS_5}(\tau,\sigma,\phi)e^{-\frac{A_6(\tau,\sigma,\phi)}{2\pi}\sqrt{\lambda}}, \quad  \, W_{S^5} \simeq C_{S^5}(\tau,\sigma)\lambda^B e^{\sqrt{\lambda}A} \, ,
\ee

The coefficients for the second contribution will be computed in Chapter \ref{ChSca}, whereas the first term is described below and in more details in the Appendix \ref{TBApp}.

\paragraph{Minimal area and TBA}

Here we analyse the first contribution to (\ref{strongW}), by means of the $AdS/CFT$ duality which allows us to compute the scattering/amplitudes Wilson loops \cite{AM-amp}. In this picture, the gluons are open strings ending on D-branes and their amplitude is computed by the minimal area of the string whose worldsheet ends on the contour with lightlike segments. The same string yields, in the strong coupling limit, the expectation value of the null polygonal Wilson loop. This was the first hint of the aforementioned duality. Thus, the amplitude/Wl is given by the classical string minimal area in the $AdS_5$ space

\be 
W \sim e^{-S_{\textit{min}}}
\ee

where the symbol $\sim$ means the equality holds at strong coupling up to a prefactor.

The quantity $S_{\textit{min}}$ is the minimum value of the worldsheet action (saddle point) for a string moving in $AdS_5$ attached to a null polygon at the boundary. The polygon is given by the momenta of the gluons involved in the scattering process. 

Remarkably, as first shown in \cite{TBuA} for $n=6$ and later extended in \cite{YSA}, the minimal area problem reduces to a set of non-linear integral equations, whose form recalls very much that of Thermodynamic Bethe Ansatz. This gives $S_{min}$ in terms 

\be\label{SYYc}
S_{min}=\frac{\sqrt{\lambda}}{2\pi}YY_c=\frac{\sqrt{\lambda}}{2\pi}A
\ee

where $YY_c$ is the critical value of the Yang-Yang functional of the TBA setup and coincide with the regularised area of the worldsheet $A$. Some details, which will be compared to our computations on the gauge side, are listed in the appendix \ref{TBApp}. We mention that the prefactor $C_{AdS_5}(\tau,\sigma,\phi)$ should follow from the one-loop correction to the classical approximation, which has not been computed yet on the string side. This could constitute an interesting project to pursue in the future. 

An interesting remark on the two contributions in (\ref{strongW}) concerns the collinear limit $\tau\to\infty$: in this regime, the classical contribution goes to zero, as the area is exponentially suppressed $A\sim e^{-\sqrt{2}\tau}$. Therefore, the only non-vanishing effect is given by the scalars, as long as we remain in the short-distance regime $z\simeq m\tau \ll 1$.

\section{The OPE for Wls and the Pentagon approach}
\label{SecOPE}

In this section we introduce a non-perturbative method to study the null polygonal Wilson loops. It will be the main framework from which the results of the thesis are obtained, in Chapters \ref{ChMat},\ref{ChCla},\ref{ChSca}. It is based on the concept of Operator Product Expansion, usually employed in a conformal field theory to the product of local operators, expanding it in series of powers of the distance. Building on earlier ideas \cite{AM}, the authors \cite{Anope} showed that an analogue technique works for the null polygonal Wilson loops as well in a conformal field theory as $\mathcal{N}=4$ SYM. To be more precise, the method compute the finite conformal invariant ratio $W_n$, once all the divergencies are removed. This quantity $W_n$ depends on $3(n-5)$ conformal invariant cross ratios $\tau_i,\sigma_i,\phi_i$ encoding the geometry of the loop\footnote{To make a paraller with the usual formulation of the OPE, they play the role of the distance.} and starts to be non-trivial for $n>5$\footnote{In fact, the method described below yields $W_4=W_5=1$ by definition.}.  In the work \cite{BSV1}, the authors pushed forward a computational method of this OPE series, called \emph{Pentagon approach} for reasons that wil be clear later. The method has been explained in details in \cite{BSV2} and successfully tested against both weak \cite{BSV3,BSV5,DPHPap} and strong coupling computations \cite{BSV3,Bel1509,FPR2}.

Following \cite{BSV1,BSV2}, the physical picture behind the method involves the two-dimensional flux-tube dynamics. The Wilson loop is seen as a series of free evolutions and transitions of this flux-tube. We decompose the $n$-polygon in consecutive pentagons, whose overlap are middle squares. The latter corresponds to the free evolution of the flux-tube, described by a phase which couples the cross ratios to the charges (energy, momentum and angular momentum) of the flux-tube. The transition is represented by a pentagon and is due to the cusp the flux-tube encounters during its propagation.

Mathematically, the picture described above emerges as follows: we think of the Wilson loop as a correlation function

\be\label{WLCorr} 
\langle 0|\hat{P} e^{-\tau_{n-5}\hat{H}+i\sigma_{n-5}\hat{p}+i\phi_{n-5}\hat{J}}\hat{P}.......\hat{P}e^{-\tau_1\hat{H}+i\sigma_1\hat{p}+i\phi_1\hat{J}}\hat{P}|0\rangle
\ee
where the operator $\hat{P}$ acts in the two dimensional space where the flux-tube lives, as well as its charges $\hat{H},\hat{p},\hat{J}$. It sort of represents the effect of a cusp during the propagation. The series can be obtained by inserting $n-5$ identities inside the correlator (\ref{WLCorr}): we consider the sum over all the flux-tube states $\psi$

\be 
\bold{1}=\sum_{\psi}|\psi\rangle\langle \psi|
\ee
to obtain the \emph{OPE series} for the null polygonal Wilson loop

\be\label{OPEsum} 
W_n=\sum_{\left\{\psi\right\}}\displaystyle\prod_{i=1}^{n-5}e^{-E_{\psi_i}\tau_i+ip_{\psi_i}\sigma_i+im_{\psi_i}\phi_i}P(0|\psi_1)P(\psi_1|\psi_2)\cdots P(\psi_{n-5}|0)
\ee
where we used the short-hand notation

\be 
P(\psi_1|\psi_2)\equiv \langle\psi_2|\hat{P}|\psi_1\rangle
\ee
to indicate the matrix element of the operator $\hat{P}$. In the original work \cite{BSV1}, they have been referred to as the pentagon transitions. In the paraller with the more common OPE series, these quantities are the analogue of the structure constants and therefore carry the main informations on the dynamics of the theory.

The series (\ref{OPEseries}) is an expansion around the collinear limit $\tau_i\to\infty$, where two consecutive sides become collinear. The effect of creating a new side, namely a cusp, can be mimicked by the insertion of the operator $\hat{P}$. In the collinear limit, only the first terms of the series contribute as they become more and more suppressed by the exponential $e^{-\tau E}$ in the propagation phase.

So far, we have not employed the integrability whatsoever, as the expansion (\ref{OPEsum}) is valid for any four dimensional CFT. For this series to be of any use we need to know the states on which sum over.  Here is where integrability pops out, as the flux-tube turns out to be integrable. Therefore, this method is non-perturbative in $g$: all the building blocks of the series can be, in principle, studied at any coupling. We mention that the series (\ref{OPEsum}), upon little modifications, is valid also for the $SU(4)$ charged Wilson loops, for which the operator $\hat{P}$ acquires a non-trivial fermionic component.

\subsection{The GKP flux-tube}
\label{SecGKP}

In order to study the OPE series (\ref{OPEsum}), we need to analyse the flux-tube and its excitations $\psi$. There are two main description of the flux-tube, depending on the side of the $AdS/CFT$ we are considering. On the string side, it is represented by the GKP string \cite{GKP2}, a long spinning string in $AdS^5$ with large spin $S\to\infty$. Its gauge dual is a special kind of single trace operator, composed by a large number $S$ of (light-cone) covariant derivative $D_+$. The GKP flux-tube vacuum is thus represented by the operator

\be\label{GKP} 
\textit{Tr}[ZD_+......D_+Z], \quad S\to\infty
\ee
belonging to the so-called $sl(2)$ sector of $\mathcal{N}=4$ SYM, already introduced in Section \ref{IntSUSY}. 

The GKP vacuum preserves the $SU(4)$ R-symmetry: this means that its excitations are representation of this group. In the gauge theory picture, an excitation corresponds to an insertion of a field in the sea of covariant derivatives (\ref{GKP}). They can be gluons\footnote{With two different helicity states.}, fermions, antifermions and scalars, which are respectively $\bold{1},\bold{4},\bold{\bar{4}},\bold{6}$ of $SU(4)$. 

The analysis of the excitation energies above the vacuum (\ref{GKP}) is still based on the Asymptotic Bethe Ansatz equations (\ref{ABA}), although some important modifications are required. Being asymptotic, they are valid only up to the order $\lambda^{2L}$, where $L$ is the number of field, excluding covariant derivatives, present in the operator. In principle, then, wrapping effects should be present when considering a small number of insertions in (\ref{GKP}). Fortunately, the large spin limit brings some simplifications as the wrapping effects have been shown to be suppressed for $S\to\infty$.

Equations (\ref{ABA}) directly apply to the excitations over the so-called string BMN vacuum \cite{BMN}, on the gauge side composed by a series of (complex) scalars $\textit{Tr}[ZZ...ZZ]$, while on the other hand we need to study how our excitations behave in the sea of covariant derivatives in (\ref{GKP}).
The procedure to pass from the (\ref{ABA}) to the Bethe equations for the flux-tube particles follows the technique of the non-linear integral equation \cite{NLIE}\footnote{For a review, see for instance \cite{FR1081}.} and it was fruitfully employed in \cite{Basso,FPR1}\footnote{On the string side, the worldsheet S matricex has been evaluated in \cite{BB12}.}. In the limit $S\to\infty$, what emerges is a set of Bethe equations for a spin chain with lenght $2\ln S$, which allows the study of the finite corrections to the anomalous dimension

\be 
\gamma - f(g)\ln S = \sum_{j}E(u_j)
\ee
as the excitation energies $E(u_j)$ of this long chain or, according to the string description, of the GKP string.

To sum up, the flux-tube states $\psi$ are single and multi-particle Bethe states where the particles are gluons and bound states thereof, fermions, antifermions and scalars. The excitations satisfy a set of Bethe equations, see for instance \cite{FPR2} for a detailed discussion. The physical quantities characterising these states are $2d$ the scattering matrices $S_{a,b}(u_a,u_b)$ and the dispersion relations $E_a(u),p_a(u)$. The remaining quantities to determine are the pentagon transitions 

\be\label{Pentagon} 
P_{a_1,..,a_n|b_1,..,b_m}(u_1,...,u_n|v_1,...,v_m)
\ee
where the indices $a_i,b_j$ label the particles including possible internal indices. 

For the quantities (\ref{Pentagon}), a set of axioms \cite{BSV1,BSV5}, largely inspired by the constraints following the form factor interpretation, has been proposed. The solution of those axioms furnish the pentagon transitions, in principle for any value of the coupling. 

Therefore, the underlying integrability in $\mathcal{N}=4$, appearing in the computation of the anomalous dimensions, can be transported to study the flux-tube and allows a non-perturbartive determination of the dynamical quantities in the OPE series (\ref{OPEsum}). The sum over states actually involves a sum over the particles along with integrals over the momenta of the excitations. For the hexagon, it schematically reads

\be\label{OPEseries} 
W=\sum_{n}\frac{1}{n!}\sum_{a}S(a)\int\displaystyle\prod_{i=1}^n \frac{du_i}{2\pi}\mu(u_i)e^{-E_i(u_i)\tau + ip_i(u_i)\sigma + im_i\phi}P(0|u_1,....,u_n)P(-u_n,...-u_1|0)
\ee

where the sum $\sum_a S(a)$ is a short-hand for the sum over the several kind of particles with its associated symmetry factor $S(a)$. The quantity $\mu_i(u_i)$ is the integration measure, appearing due to the parametrisation of the momenta $p(u)$ through the rapidity $u$.

\subsection{The operator $\hat{P}$ and the $SU(4)$ matrix part} 
\label{SecMat}

In this part we analyse the operator $\hat{P}$ appearing in the OPE series (\ref{OPEsum}) through its form factors. The pentagon transitions $P(\psi_1|\psi_2)$, being form factors, have to satisfy the set of constraints exposed in the first chapter. However, there is a significant difference with respect the usual case, as here we are dealing with a twist operator \cite{BSV1,Twist}. In particular, its pentagonal nature affects the monodromy property, which differs from (\ref{Monodromy}): for the operator $\hat{P}$, we need five mirror transformation to get the original particle, instead of the customary four. Therefore (\ref{Monodromy}) is modified into\footnote{We neglected the internal indices for simplicity.}

\be\label{PMonodromy} 
P(u_1^{5\gamma},..,u_n)=P(u_2,....,u_n,u_1)
\ee

where the symbol $\theta^{\gamma}$ is used to indicate the operation of mirror rotation, for instance in relativistic theories $\theta^{\gamma} = \theta + i/2$. This property can be pictorially understood as follows: the original excitation lives on the bottom side of the pentagon and the effect of a mirror rotation is to move the particle around the pentagon. Therefore, we need five rotations to move the particle along the whole pentagon. A secondary difference concerns the recursive relation (\ref{Kinematic}) from the kinematic poles, for which the S-matrix part on the RHS does not appear when the twist operator is considered. 

A very interesting simplication occurs when considering the $SU(4)$ structure of this special operator: this is the topic of the following part and of the third chapter of the thesis.

\subsubsection{The $SU(4)$ matrix part}

The most general form factor/pentagon transition is

\be\label{PentFF}
\langle 0|\hat{P}|A^{(1)}_{a_1}(\theta_1).....A^{(n)}_{a_n}(\theta_n)\rangle = P_{A^{(1)}_{a_1},...,A^{(n)}_{a_n}}(\theta_1,...\theta_n)
\ee

where $A^{(i)}_{a_i}$ labels a particle among gluons, fermions and scalars and $a_i$ refers to its potential R-symmetry index, present only for scalars and fermions. 

In the OPE series, what appears is a product of several pentagon transitions like (\ref{PentFF}) and possibly a sum over the internal indices. In the following we focus on the simplest case, the hexagonal Wilson loop, for which in the OPE series (\ref{OPEseries}) we have the factor

\be 
\sum_{\vec{a}}|\langle 0|\hat{P}|A^{(1)}_{a_1}(\theta_1).....A^{(n)}_{a_n}(\theta_n)\rangle|^2
\ee

In general, such a quantity does not admit any simplification. However, for the operator $\hat{P}$, the authors \cite{BSV1} proposed the following formula

\be\label{FFsquare}
\sum_{\vec{a}}|\langle 0|\hat{P}|A^{(1)}_{a_1}(\theta_1).....A^{(n)}_{a_n}(\theta_n)\rangle|^2 = \Pi^{dyn}_{A^{(1)}_{a_1},...,A^{(n)}_{a_n}}\Pi^{mat}_{A^{(1)}_{a_1},...,A^{(n)}_{a_n}}
\ee

where the upperscript \textit{dyn} and \textit{mat} stand for dynamical and matrix. The former encodes the complete dependence on the coupling constant $\lambda$ and it is two-body factorizable

\be 
\Pi^{dyn}_{A^{(1)}_{a_1},...,A^{(n)}_{a_n}} = \frac{1}{\displaystyle\prod_{i<j}P_{A^{(i)},A^{(j)}}(\theta_i|\theta_j)P_{A^{(j)},A^{(i)}}(\theta_j|\theta_i)}
\ee

where the functions appear at the denominator thanks to a special property \cite{BSV1,BSV5} satisfied by the pentagon transitions. Note that it does not contain the internal indices, whose effect is completely inside the matrix part which does not depend on $\lambda$, but only on differences of rapidities $\theta_i-\theta$. We stress that this is a remarkable simplification, as the knowledge of the two particles pentagon transitions and the matrix part would allow to find the general form factor squared. We can say that the dynamics, containing the coupling, and the $SU(4)$ group structure do not couple to each other.

\paragraph{Integral representation}

The matrix part has been given an integral formula \cite{BSVagosto}, where the auxiliary rapidities of the  $SU(4)$ spin chain are involved. In fact, the equations determining the flux-tube excitations must be supplemented by those for the auxiliary $SU(4)$ degrees of freedom. This is the extension to the $SU(4)$ symmetry of what we have seen for the $SU(2)$ QFT in the first chapter, where the spin chain with inhomogenuities emerges. The equations for the $SU(4)$ spin chain are

\ba\label{SU4roots}
1&=&\prod _{j=1}^{N_f}\frac{u_{a,k}-u_{f,j}-i/2}{u_{a,k}-u_{f,j}+i/2} \prod _{j\not=k}^{K_a}\frac{u_{a,k}-u_{a,j}+i}{u_{a,k}-u_{a,j}-i} \prod _{j=1}^{K_b}\frac{u_{a,k}-u_{b,j}-i/2}{u_{a,k}-u_{b,j}+i/2} \nn \\
1&=&\prod _{h=1}^H \frac{u_{b,k}-u_h-i/2}{u_{b,k}-u_h+i/2} \prod _{j=1}^{K_a}\frac{u_{b,k}-u_{a,j}-i/2}{u_{b,k}-u_{a,j}+i/2} \prod _{j=1}^{K_c}\frac{u_{b,k}-u_{c,j}-i/2}{u_{b,k}-u_{c,j}+i/2} \prod _{j\not=k}^{K_b}\frac{u_{b,k}-u_{b,j}+i}{u_{b,k}-u_{b,j}-i}  \nn \\
1&=&\prod _{j=1}^{N_{\bar f}}\frac{u_{c,k}-u_{\bar f,j}-i/2}{u_{c,k}-u_{\bar f,j}+i/2} \prod _{j\not=k}^{K_c}\frac{u_{c,k}-u_{c,j}+i}{u_{c,k}-u_{c,j}-i} \prod _{j=1}^{K_b}\frac{u_{c,k}-u_{b,j}-i/2}{u_{c,k}-u_{b,j}+i/2} \nn
\ea

where the auxiliary roots are of three different types $u_{a,k}, u_{b,k}, u_{c,k}$ and correspond to the nodes of the associated $SU(4)$ Dynkin diagram. The physical rapidities $u_f,u_{\bar{f}},u_h$, respectively for fermions, antifermions and scalars, play the role of the inhomogenuities for this chain. 
Inspired by the Bethe equations (\ref{SU4roots}), the general formula for the matrix part has been proposed
\small 
\be\label{GenMat} 
\Pi_{mat}(\bold{u},\bold{v},\bold{s})=\frac{1}{K_a!K_b!K_c!}\int\displaystyle\prod_{i=1}^{K_a}\frac{da_i}{2\pi}\displaystyle\prod_{j=1}^{K_b}\frac{db_j}{2\pi}\displaystyle\prod_{l=1}^{K_c}\frac{dc_l}{2\pi}\frac{g(\bold{a})g(\bold{b})g(\bold{c})}{f(\bold{u}-\bold{a})f(\bold{v}-\bold{c})f(\bold{s}-\bold{b})f(\bold{a}-\bold{b})f(\bold{b}-\bold{c})}
\ee
\normalsize

where we used the vector notation for the rapidities and the functions

\ba  
f(\bold{a}-\bold{b})&=&\displaystyle\prod_{i,j}f(a_i-b_j), \quad f(x)=x^2+1/4 \nn \\
g(\bold{a})&=& \displaystyle\prod_{i<j}g(a_i-a_j), \quad g(x)=x^2(x^2+1)
\ea

The number of auxiliary rapidities is determined by the equations

\ba\label{AuxRoots}  
N_f-2K_a+K_b&=&0 \nn \\
N_{\bar{f}}-2K_c+K_b&=&0 \nn \\
H+K_a+K_c-2K_b&=&0
\ea

and the requirement that the solutions $K_i$ are integer yields the constraint

\be\label{SingletPimat}
N_f+2H+3N_{\bar{f}} = 4n, \quad n\in \mathbb{Z}
\ee

which is the singlet condition for the intermediate multiparticle state $\psi$. 
We remark that the formula has been proposed to be valid also for charged Wilson loops, \emph{i.e.} non-singlet states, where the charges are encoded in the RHS of (\ref{AuxRoots}) which can be either $0$ or $1$. 
The matrix factor defined in (\ref{GenMat}) it is a rational function of the differences $u_{ij}$ and it gets complicated as the number of particles grows.. However, as we are going to show in the next chapter, thanks to the integral representation, it is possible to employ a computational procedure based on sums over Young diagrams. We will do it for two particular cases, $2n$ scalars and $n$ couples $f\bar{f}$. 

It is worth to mention that a formula of the type (\ref{GenMat}) does not come unexpected, as an integral representation over the auxiliary variables of the form factors is known through the so-called off-shell Bethe Ansatz, see for instance \cite{BabKar1,BabKar2} for the applications to the $SU(4)$ and $O(6)$ symmetries. However, the case here is different as the integrals in (\ref{GenMat}) compute part of the square in (\ref{FFsquare}) where we got rid of the internal indices. A concluding remark concerns the generalization of formula (\ref{FFsquare}) to the case with more edges $n>6$. An integral representation along the line of (\ref{GenMat}) is still unknown and it would be interesting to find one. On the other hand, the form factors of the operator $\hat{P}$ (\ref{Pentagon}), thus including their internal indices, have been computed in \cite{Bel1607} direclty by solving the axioms.

\chapter{Form factors: the $SU(4)$ matrix part}
\label{ChMat}

This part of the work concerns the form factors of the twist operator $\hat{P}$, also called pentagon transitions. They are the main building blocks of the OPE series (\ref{OPEseries}) introduced in  the previous chapter. In the expansion over the flux-tube excitations, the product of such form factors appears along with a sum over the $SU(4)$ indices when the fermions and scalars are involved. 

The prototype for the application of the OPE is the hexagonal Wilson loop, for which the form factor part reads

\be\label{FFsum} 
\sum_{\vec{a}}|\langle 0|\hat{P}|\Phi_{a_1}(\theta_1).....\Phi_{a_{2n}}(\theta_{2n})\rangle|^2 \equiv \Pi_{mat}^{(2n)}(\theta_1,....,\theta_{2n})\Pi^{(2n)}_{dyn}(\theta_1,....,\theta_{2n})
\ee

\emph{i.e.} the square form factor, after the sum over the internal indices $\vec{a}$, factorizes \cite{BSV1} in two parts, the dynamical and matrix factors, as discussed in Section \ref{SecMat}. The latter is present only when $SU(4)$ indices appear and is coupling independent. 
The matrix factor enjoys the multiple integral representation introduced in Section \ref{SecMat}, where the auxiliary rapidities of the $SU(4)$ spin chain are considered and integrated over. 

In this chapter, taking advantage of some analogies with the $\mathcal{N}=2$ instanton partition function, we manage to solve systematically the integrals and recast the matrix part as a sum over rational functions. An interesting classification in terms of Young diagrams is given. The methods portrayed here are published in \cite{BFPR3,BFPR5}. In the Appendix \ref{Pol} some useful properties of several polynomial functions appearing during the procedure are highlighted.

\section{The scalars}
\label{ScaMat}

The six scalars $\Phi^i$ of $\mathcal{N}=4$ SYM transform as the antisymmetric representation of $SU(4)$, namely the fundamental of $SO(6)$, so that we will refer to their $SO(6)$ matrix structure, which we call $\Pi_{mat}^{(2n)}$ for brevity. As we are interested in the MHV amplitudes, the singlet constraint (\ref{SingletPimat}) specialized to scalars give $H=2n$, \emph{i.e.} the number of excitations is even.

The general formula (\ref{GenMat}) applied to the scalar case \cite{BSV4} reads
\ba\label{ScaPiMat}
\Pi_{mat}^{(2n)}(u_1,\ldots,u_{2n}) &=& \frac{1}{(2n)!(n!)^2}\int_{-\infty}^{+\infty}
\prod_{k=1}^{n}\frac{da_k}{2\pi}
\prod_{k=1}^{2n}\frac{db_k}{2\pi}\prod_{k=1}^{n}\frac{dc_k}{2\pi} \cdot \\
&\cdot& \frac{\displaystyle\prod_{i<j}^{n} g(a_i-a_j) \prod_{i<j}^{2n} g(b_i-b_j) \prod_{i<j}^{n} g(c_i-c_j)}
{\displaystyle \prod_{j=1}^{2n} \left(\prod_{i=1}^{n} f(a_i-b_j) \prod_{k=1}^{n} f(c_k-b_j)
\prod_{l=1}^{2n} f\left (u_l-b_j \right)\right)} \ , \nn
\ea
where we recall the functions $f(x)=x^2+\frac{1}{4}$ and $g(x)=x^2(x^2+1)$.

In this section we compute explicitly the multiple integrals by residues, employing the symmetries by a method based on Young diagrams. 

The variables $a,c$ in (\ref{ScaPiMat}) do not couple to each other and we can recast into
\ba\label{Pin_mat}
\Pi_{mat}^{(2n)}(u_1,\ldots,u_{2n}) =
\frac{1}{(2n)!(n!)^2}\int \prod_{k=1}^{2n}\frac{db_k}{2\pi}\,\left[\mathcal{D}_{2n}(b_1,\ldots,b_{2n})\right]^2\,
\frac{\displaystyle \prod_{i<j}^{2n} g(b_i-b_j) }
{\displaystyle \prod_{k=1}^{2n} \prod_{l=1}^{2n} f(u_l-b_k)} \, ,
\ea
where the symmetric function $\mathcal{D}_{2n}$ encodes the integrals on $a,c$
\be\label{int_a}
\mathcal{D}_{2n}(b_1,\ldots,b_{2n})\equiv
\int_{-\infty}^{\infty}
\prod_{k=1}^{n}\frac{da_k}{2\pi}
\frac{\displaystyle\prod_{i<j}^{n} g(a_i-a_j)}
{\displaystyle \prod_{j=1}^{2n} \prod_{i=1}^{n} f(a_i-b_j)} \ .
\ee
The function (\ref{int_a}) can be evaluated by multiple residues and given the expression
\be\label{int_a2}
\mathcal{D}_{2n}(b_1,\ldots,b_{2n})=
\sum_{\alpha_1=1}^{2n}\ldots\sum_{\alpha_n=1}^{2n}
\frac{\displaystyle\prod_{i<j}^n g(b_{\alpha_i}-b_{\alpha_j})}
{\displaystyle\prod_{k=1}^n \prod^{2n}_{\gamma_k=1 \, ,\,\gamma_k\neq\alpha_k} f(b_{\alpha_k}-b_{\gamma_k}+\frac{i}{2})} \, .
\ee
where $S_{\vec{\alpha}}=\{\alpha_1,\ldots,\alpha_n\}$  indicates a partition of labels $\alpha_k$ (with $\alpha_k\in \{1,\ldots,2n\}$) and we also introduced the complementary set
$\bar S_{\vec{\alpha}}=\{1,\ldots,2n\} - \{\alpha_1,\ldots,\alpha_n\}$. Equipped with these notations, we can write (\ref{int_a2}) as
\be
\mathcal{D}_{2n}(b_1,\ldots,b_{2n})=2n \frac{\delta _{2n}(b_1,\ldots,b_{2n})}{\prod \limits_{\stackrel {i,j=1}{i<j}} ^{2n} [ (b_i-b_j)^2+1]} \, ,
\ee
where we introduced the symmetric function
\be
\delta _{2n}(b_1,\ldots,b_{2n}) \equiv \frac{n!}{2n} \sum _{\alpha _1<\alpha _2< \dots  < \alpha _{n}=1}^{2n} \left (  \prod _{\stackrel {i\in S_{\vec{\alpha}},
j\in S_{\vec{\alpha}}, i<j} {i\in \bar S_{\vec{\alpha}},
j\in \bar S_{\vec{\alpha}}, i<j}} [ (b_i-b_j)^2+1] \right ) \prod _{k=1}^n \prod _{\beta \in  \bar S_{\vec{\alpha}}} \frac{b_{\alpha _k}-b_{\beta}-i}{b_{\alpha _k}-b_{\beta}} \, .
\label{def-delta}
\ee
The symmetry under $b_i\leftrightarrow b_j$ tells us that the function defined above is a polynomial, since single poles for $b_i=b_j$ are forbidden and double poles do not appear. Some useful properties of the polynomials $\delta_{2n}$ are listed in the Appendix \ref{Pol}. What is relevant to us is that the matrix factor assumes the inspiring form
\be\label{Nekr-scal}
\Pi_{mat}^{(2n)}(u_1,\ldots,u_{2n})=\frac{4n^2}{(2n)!(n!)^2}\int \displaystyle\prod_{i=1}^{2n}\frac{db_i}{2\pi}\frac{[\delta_{2n}(b_1,\ldots,b_{2n})]^2}{\displaystyle\prod_{i,j}^{2n}f(u_i-b_j)}\displaystyle\prod_{i<j}\frac{b_{ij}^2}{(b_{ij}^2+1)}
\, , \quad b_{ij}\equiv b_i -b_j \, ,
\ee
which shows striking similarities with the Nekrasov instanton partition function in $\mathcal{N}=2$ theories. In details, (\ref{Nekr-scal}) is compared to $\mathcal{Z}_{U(2n)}^{(2n)}$, the $2n$-instanton contribution to the partition function of a $U(2n)$ theory, in which the rapidities $u_i$ play the role of the VEVs $a_i$ of the scalar fields and the instanton coordinates $\phi_i$ are represented by the auxiliary rapidities $b_i$. For $\mathcal{Z}$, an evaluation by residues, which results in a sum over Young tableaux configurations \cite{Nekrasov2}, is well-known. This connection allows us to push forward a computation of $\Pi_{mat}^{(2n)}$ by the same procedure, classifying the contributions in Young tableaux. 

\subsection{Young diagrams method}

As said, inspired by the analogy with the Nekrasov function, we are going to evaluate (\ref{Nekr-scal}) by residues. We must remark, though, some differences with respect to the Nekrasov partition function. In our case the polar part is simpler, as we do not have the two deformation parameters $\epsilon_1,\epsilon_2$ which are present in $\mathcal{Z}_{U(2n)}^{(2n)}$,, but only one\footnote{In $\mathcal{Z}$ each VEV $a_i$ has its associated Young tableaux, while here we have a column associated to any $u_i$ and the Young tableaux description appears once we symmetrize $u_i\leftrightarrow u_j$.} fixed to $\pm i$. The polynomial $\delta_{2n}$, on the other hand, is absent in the Nekrasov function and is source of some important effects. However, the key features of the multiple integrals are shared by $\Pi_{mat}^{(2n)}$, $\mathcal{Z}_{U(2n)}^{(2n)}$ and are basically as follows:
\begin{itemize}
\item The poles $b_i=u_j+\frac{i}{2}$, relating the residues positions to the physical rapidities;
\item The double zeroes $b_{ij}^2$, which have the effect of canceling the contributions when two or more residues evaluated at the same point. For instance, if we take the first residue in $b_1=u_k+\frac{i}{2}$, the poles in $b_{j\neq 1}=u_k+\frac{i}{2}$ are no longer present when we integrate over $b_j$;
\item The polar part $\frac{1}{b_{ij}^2+1}$, whose role is to arrange the residues in strings, displaced by $+i$, in the complex plane. Considering the example before, the first residue in $b_1=u_k+\frac{i}{2}$ generates poles of the type $b_{j\neq 1}=u_{k}+\frac{3i}{2}$.
\end{itemize}
As a consequence, a particular configuration is represented by the $2n$ coordinates of the residues, all different and arranged in strings in the complex plane starting from $u_i+\frac{i}{2}$ and displaced by $+i$. The procedure will be clarified later with a detailed analysis of the cases $n=1,2$.
For $2n$ scalars, the procedure culminates in the formula
\be\label{PiMat-Young}
\Pi_{mat}^{(2n)}(u_1,\ldots,u_{2n})=\sum_{l_1+\ldots +l_{2n}=2n, l_i<3,l_{i+1}\leq l_i}(l_1,\ldots,l_{2n})_s=\sum_{|Y|=2n,l_i<3}(Y)_s \, .
\ee
which is expressed as a sum over Young tableaux. Some explanations about (\ref{PiMat-Young}) are due. The symbol $(l_1,\ldots,l_{2n})$ represents the contribution of a particular residue pattern: there are $l_i$ residues with real rapidities $u_i$ arranged in a string in the upper half plane. The constraint $\sum_{i=1}^{2n}l_i=2n$ follows from the fact that we have $2n$ integrations. On the other hand, the constraint $l_i<3$ comes from the property

\be\label{zerodelta} 
\delta_{2n}(u_1,u_1+i,u_1+2i,u_4,.....,u_n)=0
\ee 
which cancels the configurations with strings of three or more rapidities in the complex plane. 
The index $s$ in $(l_1,\ldots,l_{2n})_s$ stems for the sum over permutation of inequivalent rapidities, namely
\be
(l_1,\ldots,l_{2n})_s\equiv (l_1,\ldots,l_{2n}) + \textrm{permutations of $l_1,\ldots,l_{2n}$} \, .
\ee
In the following we will sometimes use the symbol $Y$ (with $|Y|=\sum _i l_i$) as a shorthand for $(l_1,\ldots ,l_{2n})$.

The fundamental building block of (\ref {PiMat-Young}) is the diagram $(l_1,\ldots,l_{2n})$, which represents the contribution when $l_i$ residues arrange around the real rapidiy $u_i$. Its value, once taking into account an additional $(2n)!$ factor from the permutation of the rapidities, reads

\footnotesize
\ba\label{diagr}
&&(l_1,\ldots,l_{2n})= \frac{4}{[(n-1)!]^2}\frac{1}{\displaystyle\prod_i^{2n}(l_i!)^2}\frac{1}{\displaystyle\prod_{k=1}^{2n}\prod_{j\neq k}\prod_{m=1}^{l_k}(u_k-u_j+(m-1)i)(u_k-u_j+mi)}\cdot\nn\\
&\cdot &\displaystyle\prod_{i<j}^{2n}\prod_{m=1}^{l_i}\prod_{k=1}^{l_j}\frac{(u_i-u_j+(m-k)i)^2}{(u_i-u_j+(m-k)i)^2+1}\delta ^2_{2n}(Y)\equiv \frac{4}{[(n-1)!]^2}\frac{1}{\displaystyle\prod_i^{2n}(l_i!)^2}\delta ^2_{2n}(Y)[l_1,\ldots ,l_{2n}] \, .
\ea
\normalsize
which, for convenience, has been divided in two parts, one due to $\delta_{2n}(Y)$\footnote{This short-hand notation stands for $\delta _{2n}$ computed on the residues pattern $Y=(l_1,\ldots ,l_{2n})$.} and the rest, indicated as $[l_1,\ldots ,l_{2n}]$.

Actually, the formula (\ref{diagr}) can be specialized for $l_i\leq 2$, as they are the only non-vanishing diagrams. The most generic contribution contains $k$ columns with  $l_i=2$, $2(n-k)$ with $l_i=1$ and $k$ with $l_i=0$, giving a total number of $n+1$ different Young tableaux.\\
To begin with, we start with the simplest case
\be
(1,1,\ldots,1,1)_{2n}=\frac{4}{[(n-1)!]^2}\delta_{2n}^2(u_1,\cdots ,u_{2n})[1,1,\ldots,1,1]_{2n} \, ,
\ee
which, thanks to
\be
[1,1,\ldots ,1,1]_{2n}=\displaystyle\prod_{i<j}^{2n}\frac{1}{(u_{ij}^2+1)^2} \, ,
\ee
and the Pfaffian formula\footnote{I am very grateful to Ivan Kostov and Didina Serban for pointing out this interesting formula.} of $\delta_{2n}$
\be
\delta_{2n}(b_1,\ldots ,b_{2n})=
\frac{n!}{2n}2^n\displaystyle\prod_{i<j}\frac{b_{ij}^2+1}{b_{ij}}\textrm{Pf} \, D \, , \quad D_{ij}=\left(\frac{b_{ij}}{b_{ij}^2+1}\right) \, ,
\ee
elaborated in the Appendix \ref{Pol}, turns into
\be\label{1111}
(1,1, \ldots ,1,1)_{2n}=2^{2n}\textrm{Det}\left(\frac{u_{ij}}{u_{ij}^2+1}\right)\displaystyle\prod_{i<j}^{2n}\frac{1}{u_{ij}^2} =
\frac{4n^2}{(n!)^2} \prod _{i<j}^{2n} \frac{1}{(1+u_{ij}^2)^2} \, \delta^2_{2n}(u_1,\ldots ,u_{2n})
\, .
\ee
where the subscript $2n$ in $(1,1,\ldots,1,1)_{2n}$ highlights the fact that there are $2n$ variables, which will be a necessary distinction in the following. The polynomial computed in a configuration of the type $(2,\ldots ,2,0,\ldots ,0)$ combined with the fact
\be
[2,\ldots ,2,0,\ldots ,0]_{2n}=\displaystyle\prod_{i<j}^{n}\frac{1}{(u_{ij}^2+1)^2(u_{ij}^2+4)^2}\frac{1}{\displaystyle\prod_{i=1}^{n}\displaystyle\prod_{j=n+1}^{2n}u_{ij}
(u_{ij}+i)^2(u_{ij}+2i)} \, ,
\ee
gives the residues contribution for the other extremal case
\be\label{2200}
(2, \ldots ,2,0, \ldots , 0)_{2n}=\frac{1}{\displaystyle\prod_{i=1}^{n}\displaystyle\prod_{j=n+1}^{2n}u_{ij}(u_{ij}+i)^2(u_{ij}+2i)} \, .
\ee
We have now everything to write a general formula: a particular configuration, up to a permutation of the rapidities, contains $k$ columns of height two, $k$ of height  zero and $2n-2k$ with height one.

The relation (\ref{delta-fact}) in Appendix \ref {Pol} gives the polynomial $\delta _{2n}$ computed in the configuration $(2, \ldots ,2,0, \ldots , 0_{2k},1, \ldots ,1)_{2n}$ in terms of the two special cases described before. The intermediate subscript $2k$ means that the first $2k$ columns are of the type $(2,\ldots, 2,0,\ldots ,0)_{2k}$, while the remaining $2n-2k$ contains $1$.

Using this formula for the polynomials $\delta_{2n}$, combined with (\ref{diagr}) applied to $[l_1,\ldots ,l_{2n}]$
\ba
&&[2, \ldots ,2,0, \ldots , 0_{2k},1, \ldots ,1]_{2n}=[2, \ldots ,2,0, \ldots , 0]_{2k}\cdot [1_{2k+1},1, \ldots ,1,1]_{2n} \cdot \nn\\
&& \cdot \displaystyle\prod_{j=2k+1}^{2n}\displaystyle\prod_{i=1}^{k}
\frac{1}{u_{ij}(u_{ij}-i)(u_{ij}^2+1)(u_{ij}+2i)^2}\displaystyle\prod_{l=k+1}^{2k}\frac{1}{u_{lj}(u_{lj}-i)} \, .
\ea

we can write the general contribution as follows
\ba\label{general}
&&(2, \ldots ,2,0, \ldots , 0_{2k},1, \ldots ,1)_{2n}=(2, \ldots ,2,0, \ldots , 0)_{2k}\cdot (1_{2k+1},1, \ldots ,1,1)_{2n} \cdot \nn\\
&& \cdot \displaystyle\prod_{j=2k+1}^{2n}\displaystyle\prod_{i=1}^{k}\frac{1}{u_{ij}(u_{ij}+i)}\displaystyle\prod_{l=k+1}^{2k}\frac{1}{u_{lj}(u_{lj}-i)}= \frac{1}{\displaystyle\prod_{i=1}^{k}\displaystyle\prod_{j=k+1}^{2k}u_{ij}(u_{ij}+i)^2(u_{ij}+2i)}\cdot \nn\\
&& \cdot 2^{2n-2k}\textrm{Det}_{(i,j)=2k+1}^{2n}\left(\frac{u_{ij}}{u_{ij}^2+1}\right)\displaystyle\prod_{i<j=2k+1}^{2n}\frac{1}{u_{ij}^2} \displaystyle\prod_{j=2k+1}^{2n}\displaystyle\prod_{i=1}^{k}\frac{1}{u_{ij}(u_{ij}+i)}\displaystyle\prod_{l=k+1}^{2k}\frac{1}{u_{lj}(u_{lj}-i)} \, ,
\ea
where $(1_{2k+1},1,\ldots,1,1)_{2n}$ is the contribution of the type (\ref{1111}) involving the variables $u_{2k+1},\ldots,u_{2n}$ only. The same thing for the determinant in (\ref {general}), whose matrix elements are $u_{ij}/(u_{ij}^2+1)$, with $2k+1\leq i,j\leq 2n$.

Recalling (\ref{PiMat-Young}), we must sum over the Young configurations, which are in turn given by permuting the (inequivalent) rapidities in (\ref{general}). We symmetrize the Young diagram by summing over the $(2n)!$ permutations $P$, thus divide by the overcounting factor $(k!)^2(2n-2k)!$, leading to
the final expression for the matrix part
\ba
&&\Pi_{mat}^{(2n)}(u_1, \ldots , u_{2n})=\sum_{k=0}^{n}\frac{2^{2(n-k)}}{(2n-2k)!(k!)^2}\sum_{P}\frac{1}{\displaystyle\prod_{i=1}^{k}\displaystyle\prod_{j=k+1}^{2k}u_{P_i P_j}(u_{P_i P_j}+i)^2(u_{P_i P_j}+2i)}\cdot \nn\\
&& \label{MatYoung} \\
&& \cdot \textrm{Det}_{(i,j)=2k+1}^{2n}\left(\frac{u_{P_i P_j}}{u_{P_i P_j}^2+1}\right)\displaystyle\prod_{i<j=2k+1}^{2n}\frac{1}{u_{P_i P_j}^2}\displaystyle\prod_{j=2k+1}^{2n}\displaystyle\prod_{i=1}^{k}\frac{1}{u_{P_i P_j}(u_{P_i P_j}+i)}\displaystyle\prod_{l=k+1}^{2k}\frac{1}{u_{P_l P_j}(u_{P_l P_j}-i)} \nn \, .
\ea

Formula (\ref{MatYoung}), which is the main achievement of this section, it is a more explicit version of (\ref{PiMat-Young}) and it represents the matrix factor as a finite sum of rational functions. The drawback, however, is that the polar structure of $\Pi_{mat}^{(2n)}$ is somehow hidden in that expression, as there are many fictitious poles that cancel once we sum over all the configurations. The polar structure of $\Pi_{mat}^{(2n)}$ will be analysed in the following, taking advantage of another feature, the asymptotic factorisation. This will be extensively analysed in Section \ref{ScalarHex}.

\medskip
In order to elucidate the Young tableaux method, we outline the computations for the simplest cases, \emph{i.e.} two and four scalars.

\medskip

\noindent\textbf{$\bullet$\ Two scalars ($n=1$):}

\medskip

For a couple of scalars, the integral formula (\ref{Pin_mat}) reads
\ba
\Pi_{mat}^{(2)}(u_1,u_{2})=\frac{1}{2}\int\frac{da\,dc}{(2\pi)^2}\,\frac{db_1\,db_2}{(2\pi)^2}\,
\frac{g(b_1-b_2)}{f(u_1-b_1)f(u_1-b_2)f(u_2-b_1)f(u_2-b_2)}\cdot \nn\\
\cdot \frac{1}{f(a-b_1)f(a-b_2)f(c-b_1)f(c-b_2)} \ ,
\ea
which, after an evaluation by residues of the integrals over $a$ and $c$, turns to
\be\label{b1b2int}
\Pi_{mat}^{(2)}(u_1,u_{2})=2\int\frac{db_1\,db_2}{(2\pi)^2}\,
\frac{1}{f(u_1-b_1)f(u_1-b_2)f(u_2-b_1)f(u_2-b_2)}\,\frac{(b_1-b_2)^2}{(b_1-b_2)^2+1} \ .
\ee
We see that our polynomial is trivial for two particles, {\it i.e.} $\delta_2=1$. The contour integrals over $b_1$, $b_2$ can be easily performed without any Young tableaux technique (we have just $3\times 2=6$ residues to evaluate) and we obtain
\be\label{2scalars}
\Pi_{mat}^{(2)}(u_1,u_{2})=
\frac{6}{[(u_1-u_2)^2+1][(u_1-u_2)^2+4]} \ .
\ee
However, it is useful to solve the $n=1$ case within the Young tableaux framework, in order to give a simple sketch of the procedure. Afterwards, we will deal with $n=2$, the first non trivial case.\\
We start from the double integral (\ref{b1b2int}), closing the contour in the upper half plane. Therefore the integral over $b_1$ gets contributions from the poles in $b_2+i$, $u_1+i/2$ and $u_2+i/2$, leading to 
\be
\Pi_{mat}^{(2)}(u_1,u_{2})=\int\frac{db_2}{2\pi}\frac{A+B+C}{(b_2-u_1-i/2)(b_2-u_1+i/2)(b_2-u_2-i/2)(b_2-u_2+i/2)}  \, ,
\ee
where we defined the three different contributions
\ba
A&&=\frac{-1}{(b_2-u_1+3i/2)(b_2-u_1+i/2)(b_2-u_2+3i/2)(b_2-u_2+i/2)} \nn\\
B&&=\frac{2}{(u_1-u_2)(u_1-u_2+i)}\frac{(b_1-u_1-i/2)^2}{(b_1-u_1-3i/2)(b_1-u_1+i/2)} \nn\\
C&&=\frac{2}{(u_2-u_1)(u_2-u_1+i)}\frac{(b_1-u_2-i/2)^2}{(b_1-u_2-3i/2)(b_1-u_2+i/2)}
\ea

These come from, respectively, the poles $b_1=b_2+i$, $u_1+i/2$ and $u_2+i/2$. In the integral over $b_2$, each term contains two poles, thus in total we have $3\times 2=6$ residues. The various contributions can be classified by the position of the poles of the auxiliary roots $(b_1,b_2)$: they are $(u_1+i/2,u_1+3i/2)$, $(u_1+3i/2,u_1+i/2)$, $(u_1+i/2,u_2+i/2)$, $(u_2+i/2,u_1+i/2)$, $(u_2+i/2,u_2+3i/2)$ and $(u_2+3i/2,u_2+i/2)$. The key feature is that the residues are invariant under the exchange $b_1\leftrightarrow b_2$ and only three terms are truly different: we represent them by an array of two numbers $(l_1,l_2)$ with $l_1+l_2=1$, where $l_i$ labels the number of roots in the string with real position $u_i$. We define them as follows:
\ba
(2,0)&&\equiv(u_1+i/2,u_1+3i/2) + (u_1+3i/2,u_1+i/2)= 2\times (u_1+i/2,u_1+3i/2) \nn\\
(0,2)&&\equiv(u_2+i/2,u_2+3i/2) + (u_2+3i/2,u_2+i/2)= 2\times (u_2+i/2,u_2+3i/2)\nn\\
(1,1)&&\equiv(u_1+i/2,u_2+i/2) + (u_2+i/2,u_1+i/2) = 2\times (u_1+i/2,u_2+i/2)\, ,
\ea
which are nothing but the $n=1$ version of $(l_1,\ldots, l_{2n})$. In the end, the total matrix part amounts to
\be
\Pi_{mat}^{(2)}(u_1,u_2)=(1,1)+(2,0)+(0,2)
\ee
with
\ba
(1,1)&&=\frac{4}{[(u_1-u_2)^2+1]^2}\nn\\
(2,0)&&=\frac{1}{(u_1-u_2)(u_1-u_2+i)^2(u_1-u_2+2i)}\nn\\
(0,2)&&=\frac{1}{(u_2-u_1)(u_2-u_1+i)^2(u_2-u_1+2i)} \, ,
\ea
in agreement with (\ref{2scalars}).
The last step is the symmetrisation: we note that $(2,0)$ and $(0,2)$ are related by the exchange $u_1\leftrightarrow u_2$ and thus we define the symmetric function $(2,0)_s=(2,0)+(0,2)$, which we call Young tableaux, and write the final form
\be
\Pi_{mat}^{(2)}(u_1,u_2)=(1,1)_s+(2,0)_s \, ,
\ee
Note that we do not need to symmetrize the other contribution, as $(1,1)_s\equiv (1,1)$ already.

\medskip

\noindent\textbf{$\bullet$\ Four scalars ($n=2$):}

\medskip

For $n=2$, formula (\ref{Pin_mat}) becomes:
\be
\Pi_{mat}^{(4)}(u_1,\ldots,u_{4})= \frac{1}{6}\int
\frac{db_1 db_2 db_3 db_4}{(2\pi)^4}\,
\frac{[\delta_4(b_1,\ldots,b_4)]^2}
{\displaystyle \prod_{i,j=1}^{4}f(u_l-b_j)}
\,\prod_{i<j}\frac{(b_i-b_j)^2}{(b_i-b_j)^2+1} \, .
\ee

The total number of residues to take into account is $7\times 6\times 5\times 4=820$, as each integration lowers the number of residues by one. Therefore, a brute force approach would be very inefficient.
The Young tableaux expansion helps us, employing two symmetries: the permutations of isotopic rapidities $b_i$ (bringing a factor $4!=24$) and that of $u_i$ (which symmetrizes the residue contributions), which gives us only 5 different Young tableaux: $(1,1,1,1)_s$, $(2,1,1,0)_s$, $(2,2,0,0)_s$, $(3,1,0,0)_s$ and $(4,0,0,0)_s$. Each of them is a sum over the permutations in $(l_1,\ldots,l_4)$, which are respectively $1,12,6,12,4$. As a combinatorial check, $(1+12+6+12+4)\times 24=840$, which is the total number of residues stated before. The method employs the fact that many of them are either equal (by exchanging the $b_i$) or related by permutations of $u_i$. From (\ref{zerodelta}) follow

\be 
\delta_4(u_1,u_1+i,u_1+2i,u_1+3i)=\delta_4(u_1,u_1+i,u_1+2i,u_2)=0
\ee

which means that the latter two diagrams actually vanish. Therefore, the $n=2$ matrix part is given by
\be
\Pi_{mat}^{(4)}(u_1,u_2,u_3,u_4)=(1,1,1,1)_s + (2,1,1,0)_s + (2,2,0,0)_s \, ,
\ee
where, according to (\ref{general}), we have
\ba
(1,1,1,1)_s&=&(1,1,1,1)=16 \, \textrm{Det}\left(\frac{u_{ij}}{u_{ij}^2+1}\right)\displaystyle\prod_{i<j}^{4}\frac{1}{u_{ij}^2} \, ,  \\
(2,2,0,0)&=&\frac{1}{\displaystyle\prod_{i=1}^2\prod_{j=3}^4 u_{ij}(u_{ij}+i)^2(u_{ij}+2i)} \, , \\
(2,0,1,1)&=&\frac{1}{u_{12}(u_{12}+i)^2(u_{12}+2i)}
\frac{4}{(u_{34}^2+1)^2}\frac{1}{\displaystyle\prod_{j=3}^{4}u_{1j}(u_{1j}+i)u_{2j}(u_{2j}-i)} \,
\ea
and the symmetrisation is obtained as follows
\small
\ba
(2,2,0,0)_s &=& (2,2,0,0) + (2,0,2,0) + (2,0,0,2) + (0,2,2,0) + (0,2,0,2) + (0,0,2,2)\nn\\
(2,1,1,0)_s&=& (2,1,1,0) + (2,1,0,1)+ (2,0,1,1)+(1,2,1,0)+(1,1,2,0)+(1,0,1,2) + \nn\\
&+& (1,2,0,1) + (1,0,2,1) + (1,1,0,2) + (0,2,1,1) + (0,1,1,2) + (0,1,2,1) \, .
\ea
\normalsize

\subsubsection{Recursion relation}

An interesting application of the method outlined above is a sort of recursion relation for the matrix factor, which relates the residue of $\Pi_{mat}^{(2n)}$ in $u_i=u_j+2i$ to the matrix part with $2$ scalars less 
\be\label{ResPimatSca}
-2i \textit{Res}_{u_2=u_1+2i} \Pi_{mat}^{(2n)}(u_1,\cdots ,u_{2n})= \frac{\Pi_{mat}^{(2n-2)}(u_3,\cdots ,u_{2n})}{\displaystyle\prod_{j=3}^{2n}u_{1j}(u_{1j}+i)^2 (u_{1j}+2i)} \, .
\ee

We remark that (\ref{ResPimatSca}) has a clear physical origin, as $\Pi_{mat}^{(2n)}$ is part of the squared form factor of $\hat{P}$ and, as a consequence, must satisfy certain axioms. One of them concerns the kinematic poles, in particular their residues, relating them to the form factor with two particles less. They are those in $u_i=u_j+2i$, therefore (\ref{ResPimatSca}) is just consequence of the form factor interpretation of the pentagonal transitions.

We can prove (\ref{ResPimatSca}) by means of the sum over Young diagrams (\ref{PiMat-Young}): we note that the pole in $u_2=u_1+2i$ is present only in the terms of the type $(2,0,l_3,\cdots ,l_{2n})$, where $\sum_{i=3}^{2n}l_i=2n-2$. The sum on the RHS is that of $\Pi_{mat}^{(2n-2)}(u_3,\cdots ,u_{2n})$. To go further we work out the expression of $(2,0,l_3,\cdots ,l_{2n})$, splitting it in three different contributions
\be\label{split}
(2,0,l_3,\cdots ,l_{2n})=(2,0)\cdot (l_3,\cdots ,l_{2n})M_{\lbrace l_i \rbrace}
\ee
where $M_{\lbrace l_i \rbrace}$ is the mixed term and depends on the specific configuration ${\lbrace l_i \rbrace}$ and all the rapidities $u_i$. 

The pole for $u_2=u_1 +2i$, with residue $i/2$, is contained in $(2,0)$ only and the quantity $M_{\lbrace l_i \rbrace}$, when evaluated in $u_2 = u_1 + 2i$ (which we call $M^*$), does no longer depends on ${\lbrace l_i \rbrace}$. As a result, it becomes a prefactor multiplying the sum and the matrix part for fewer scalars is recovered
\be
\textit{Res}_{u_2=u_1+2i}\Pi_{mat}^{(2n)}(u_1,\cdots ,u_{2n}) = -\frac{1}{2i} M^*(u_1,u_3,\cdots ,u_{2n}) \Pi_{mat}^{(2n-2)}(u_3,\cdots ,u_{2n})
\ee
As a final step, we use (\ref{2200}) and (\ref{general}) and write the mixed contribution for the configuration $(2_3,\cdots ,2_{k+1},0,\cdots ,0_{2k},1,\cdots ,1)$ as
\be
\frac{1}{\displaystyle\prod_{j=3}^{k+1}u_{1j}(u_{1j}+i)^2(u_{1j}+2i)\prod_{j=k+2}^{2k}u_{2j}(u_{2j}-i)^2(u_{2j}-2i)\prod_{j=2k+1}^{2n}u_{1j}(u_{1j}+i)u_{2j}(u_{2j}-i)}
\ee
where other are obtained by a suitable permutation. Identifying $u_2=u_1 +2i$ we get
\be
M^*(u_1,u_3,\cdots ,u_{2n})=\frac{1}{\displaystyle\prod_{j=3}^{2n}u_{1j}(u_{1j}+i)^2(u_{1j}+2i)}
\ee
which finally proves the claim.

\subsection{Polar structure and polynomials}

Here we discuss the polar structure of the matrix factor. For this purpose, we use the asymptotic factorisation discussed in Section \ref{ScalarHex}. 

Starting from the known case $n=1$, formula (\ref{2scalars}), where the poles are explicit, we will prove that the matrix part can be written as follows
\be\label{P2n}
\Pi_{mat}^{(2n)}(u_1,\cdots , u_{2n})=\frac{P_{2n}(u_1,\ldots,u_{2n})}{\displaystyle\prod_{i<j}^{2n}(u_{ij}^2+1)(u_{ij}^2+4)} \, ,
\ee
where $P_{2n}$ is a symmetric polynomial depending on the differences $u_{ij}$.

The argument goes as follows: when two arbitrary rapidities $u_p$, $u_q$ get large, the results of Section \ref{ScalarHex} state that
\be\label{fact}
\Pi_{mat}^{(2n)}(u_1, \ldots, u_{p}+\Lambda,\ldots ,u_{q}+\Lambda, \ldots, u_{2n})\simeq \Lambda^{-8(n-1)}\Pi_{mat}^{(2)}(u_p,u_q)\Pi_{mat}^{(2n-2)}(u_1, \ldots, \underline{u_p},\ldots ,\underline{u_q}, \ldots, u_{2n}) \, ,
\ee
where the notation $\underline{u_k}$ means the omission of the rapidity $u_k$.

We remember that $\Pi_{mat}^{(2n)}(u_1,\ldots, u_{2n})$ depends only on the differences and may show singularities when $u_{ij}$ pick particular values. Of course, a singular value of $\Pi_{mat}^{(2n)}$ for the particular difference $u_{pq}$ is unchanged by the shifts in the LHS of (\ref {fact}). The RHS tells us where the singularities occur: $u_{pq}=\pm i, \pm 2i$. As the argument does not depend on the particular couple of rapidities we choose, the structure (\ref {P2n}) follows.

In conclusion, the only unknown in the matrix part (\ref {P2n}) are the polynomials $P_{2n}$. The simplest cases ($n=1,2$) are reported in Appendix \ref{Pol}: for $n\geq 3$ expressions for $P_{2n}$ get rapidly involved and a simple formula is not known. However, the residue formula (\ref{ResPimat}) gives us some contraints on $P_{2n}$: when evaluated in a specific configuration, it is related to a smaller polynomial, see Appendix \ref{Pol}. The degree of the polynomial $P_{2n}(u_1,\ldots, u_{2n})$ may be found here by comparing (\ref{P2n}) to (\ref{Nekr-scal}). The degree of $\Pi_{mat}^{(2n)}(u_1,\ldots, u_{2n})$ is found to be equal to $-4n^2$ by using integral representation (\ref{Nekr-scal}) and the fact that the degree of $\delta _{2n}(u_1,\ldots, u_{2n})$ is $2n(n-1)$. It then follows that the degree of $P_{2n}(u_1,\ldots, u_{2n})$ is $-4n^2+4\frac{2n(2n-1)}{2}=4n(n-1)$. Other properties of these polynomials, for instance the general form of their highest degree, are discussed in Appendix \ref{Pol}.

\section{The fermions}
\label{FerMat}

The method of the previous section can be adapted to study the fermion-antifermion matrix part. The results discussed here are subject of the paper \cite{BFPR5}. For the purpose of our work, we are interested in the case with $n$ couples $f\bar{f}$: the general formula (\ref{GenMat}) becomes \cite{BSV4,BFPR}

\ba\label{Pi_mat^ff}
\Pi_{mat}^{(n)}(\{u_i\},\{v_j\})=\frac{1}{(n!)^3}\int
\prod_{k=1}^n\left(\frac{da_k db_k dc_k}{(2\pi)^3}\right)
\,\frac{\displaystyle\prod_{i<j}^n g(a_i-a_j) g(b_i-b_j) g(c_i-c_j)}
{\displaystyle\prod_{i,j}^n f(a_i-b_j) f(c_i-b_j) \prod_{i,j}^n f(u_i-a_j) f(v_i-c_j)} \nn \, , \\
\,
\ea
where, as usual, the integrations are on the whole real axis.

The variables $a$ and $c$ appear symmetrically in (\ref{Pi_mat^ff}) and do not couple to each other, allowing us to simplify the formula, obtaining

\ba
\Pi_{mat}^{(n)}(\{u\},\{v\}) = \frac{1}{(n!)^3}\int\prod_{k=1}^n\frac{db_k}{2\pi}\prod_{i<j}^n g(b_{ij})
\mathcal{D}_{2n}(b_1,\dots,b_n,u_1,\dots,u_n)\mathcal{D}_{2n}(b_1,\dots,b_n,v_1,\dots,v_n)  \,.\nn
\ea

where the function $\mathcal{D}$ has been defined in the previous section for the scalars. Passing to the function $\delta_{2n}$, we get the inspiring form
\ba\label{ff-int}
\Pi_{mat}^{(n)}(\{u_i\},\{v_j\}) &=&
\frac{4n^2}{(n!)^3}\frac{1}{\displaystyle\prod_{i<j}^n(u_{ij}^2+1)(v_{ij}^2+1)}
\int\prod_{k=1}^n\frac{db_k}{2\pi} \prod_{i<j}^n\left(\frac{b_{ij}^2}{b_{ij}^2+1}\right)\cdot \nn\\
& \cdot &
\frac{\delta_{2n}(b_1,\dots,b_n,u_1,\dots,u_n)\delta_{2n}(b_1,\dots,b_n,v_1,\dots,v_n)}{\displaystyle\prod_{i,j=1}^n[(b_i-u_j)^2+1][(b_i-v_j)^2+1]}
\ea

which sets the stage for a systematic evaluation by residues. Indeed, following the strategy already carried out for scalars, these configurations can be classified in diagrams: even though they are not exactly Young diagrams, with a little abuse of notations we use the same word for them.

\subsection{Young diagrams}

Along the same line for the scalars, our method relies on the following features of (\ref{ff-int}):
\begin{enumerate}
	\item the double zeroes for coinciding variables prevent singularities for coinciding $b_i=b_j$;
	
	\item poles due to the factors $\frac{1}{b_{ij}^2+1}$ play no role, due to the properties of $\delta_{2n}$: this is different form the scalars case;
	\item poles in $b_i=u_k(v_k)+i$, relating the residues to the physical rapidities.
\end{enumerate}

Therefore one must evaluate the residues for poles such as $b_k-u_j=i$ or $b_k-v_j=i$ only and at most once for a given physical rapidity: for instance, if we compute a residue for $b_1=u_j+i$, poles at $b_{k\neq 1}=u_j+i$ or $b_{k\neq 1}=u_j+2i$ do not occur.

The remarks $(2)$ is manifest in formula (\ref{ff-int}): half the entries of the $\delta_{2n}$ polynomials correspond to the $n$ integration variables $b_j$, while the remaining $n$ are fermionic rapidities, $i.e.\ u_k$ or $v_k$, and in addition to that, the property $\delta_{2n}(u_1,u_1+i,u_1+2i,u_2,u_3,\dots)=0$ proves our claim.

Recalling the diagrammatic language of the previous section, it means that one needs to consider diagrams with $n$ boxes, each one corresponding to the contribution of a single pole, which are arranged into an array with $2n$ entries, \emph{i.e.} related to the $2n$ rapidities $u_i,v_i$. This represents a fundamental difference with respect to the scalar case, as the number of rapidities is twice the number of integrations. Under these prescriptions, the matrix factor (\ref{ff-int}) reads
\be\label{ff-Y}
\Pi_{mat}^{(n)}(\{u_i\},\{v_j\})=\frac{4n^2}{(n!)^2 2^n}\frac{1}{\displaystyle\prod_{i<j}(u_{ij}^2+1)(v_{ij}^2+1)}Y_n(\{u_i\},\{v_j\})
\ee
where we factorised a constant (independent of the diagram) function and $Y_n$ denotes the sum
\be\label{sumY}
Y_n(\{u_i\},\{v_j\})=\sum_{l_1+\cdots +l_{2n}=n, l_i=0,1} (l_1,\cdots,l_{2n}) \,;
\ee
where, in each diagram $(l_1,\cdots,l_{2n})$, the first half entries corresponds to the fermion rapidities $u_i$ and the remaining $n$ to antifermions $v_i$. 
The total number of diagrams appearing in (\ref{sumY}) amounts to $(2n)!/(n!)^2$, which are the non-equivalent permutations of the $l_k$'s entries in $(l_1,\cdots,l_{2n})$.

For fermions, all the diagrams can be obtained from a fundamental single one, we choose
\small
\ba\label{Perm}
(1,\cdots ,1,0,\cdots ,0)= \frac{\delta_{2n}(u_1+i,\cdots ,u_n+i,u_1,\cdots ,u_n)\delta_{2n}(u_1+i,\cdots ,u_n+i,v_1,\cdots ,v_n)}{\displaystyle\prod_{i<j}^n(u_{ij}^2+1)(u_{ij}^2+4)\displaystyle\prod_{i,j=1}^n (u_i-v_j)(u_i-v_j+2i)} \ .
\ea
\normalsize

The other can be obtained by considering a suitable permutation of the $2n$ variables $u_i$ and $v_i$. To be specific, we need to change the positions of some $1$-entries in the array, permuting the rapidities accordingly. Whenever a $1$ is moved from the position $i\leq n$ to $n+j$ ($1\leq j\leq n$), we swap the rapidities $u_i$ and $v_j$ in (\ref{Perm}), with the \textit{caveat} that the second half of arguments in the $\delta_{2n}$-polynomials are held fixed.

We move some $1$-entries to the antifermionic positions to get the diagram
\be\label{diag_gen}
(\boldsymbol{1}_k,\boldsymbol{0};\boldsymbol{0},\boldsymbol{1}_{n-k})\equiv (1,\cdots , 1_k,0,\cdots ,0_n,0,\cdots ,0_k,1,\cdots ,1)
\ee
which, thanks to the recursion formula for $\delta_{2n}$ shown in the Appendix \ref{Pol}, assumes the explicit form
\ba\label{Genk}
(\boldsymbol{1}_k,\boldsymbol{0};\boldsymbol{0},\boldsymbol{1}_{n-k}) &=& \frac{\delta_{2k}(u_{1}+i,\cdots ,u_{k}+i,v_1,\cdots ,v_k)\delta_{2n-2k}(u_{k+1},\cdots ,u_n,v_{k+1}+i,\cdots ,v_n+i)}{\displaystyle\prod_{i,j=1}^k(u_i-v_j)(u_i-v_j+2i)\displaystyle\prod_{i,j=k+1}^n(u_i-v_j)(u_i-v_j-2i)}\cdot \nn\\
&\cdot & \displaystyle\prod_{i=1}^k\displaystyle\prod_{j=k+1}^n\frac{(u_{ij}-i)(v_{ij}+i)}{u_{ij}v_{ij}} \cdot 2^n\frac{[(n-1)!]^2}{(k-1)!(n-k-1)!} \ ,
\ea
which is valid for any $k=1,\cdots,n-1$.

Finally, the matrix factor (\ref{ff-Y}) comes out as a sum of rational functions: for fixed $k$, we need to be add together
$\binom{n}{k}^2$ contributions: they follow from applying to (\ref{Genk}) a permutation $P$ of the variables $u_i$ and a permutation $Q$ to the antifermionic ones $v_i$, up to a normalisation factor to avoid over-counting.
The matrix factor $\Pi^{(n)}_{mat}$ eventually assumes the appealing form
\ba\label{Pimatfinal}
&&\Pi^{(n)}_{mat}(u_1, \cdots ,u_n,v_1,\cdots ,v_n) = \frac{4}{\displaystyle\prod_{i<j}^n(u_{ij}^2+1)(v_{ij}^2+1)}\sum_{k=0}^n\frac{1}{[(n-k)!(k)!]^2(k-1)!(n-k-1)!}\cdot \nn\\
&\cdot &\sum_P \sum_{Q}\frac{\delta_{2k}(u_{P_1}+i,\cdots ,u_{P_k}+i,v_{Q_1},\cdots ,v_{Q_k})\delta_{2n-2k}(u_{P_{k+1}},\cdots ,u_{P_n},v_{Q_{k+1}}+i,\cdots ,v_{Q_n}+i)}{\displaystyle\prod_{i,j=1}^k(u_{P_i}-v_{Q_j})(u_{P_i}-v_{Q_j}+2i)\displaystyle\prod_{i,j=k+1}^n(u_{P_i}-v_{Q_j})(u_{P_i}-v_{Q_j}-2i)} \cdot \nn\\
&\cdot & \displaystyle\prod_{i=1}^k\displaystyle\prod_{j=k+1}^n\frac{(u_{P_i}- u_{P_j}-i)(v_{Q_i}-v_{Q_j}+i)}{(u_{P_i}- u_{P_j})(v_{Q_i}-v_{Q_j})} \ .
\ea

where, in order for the $k=0$ and $k=n$ terms to make sense, one must substitute the factorial $(-1)!$ with $2$.

As a simple application, the cases $i.e.\ n=1$ and $n=2$, are portrayed below.\\
\medskip\\
\textbf{$\bullet $ One couple $f\bar{f}$ ($n=1$):}\\
For $n=1$ we need to consider only two diagrams, $(1,0)$ and $(0,1)$:  the former takes into account the residue for the pole $b-u_1=i$,
$$
(1,0)=\frac{1}{(u-v)(u-v+2i)} \,,
$$
the latter for $b-v_1=i$,
$$
(0,1)=\frac{1}{(v-u)(v-u+2i)} \,.
$$
The expression (\ref{ff-Y}) simply reads
$
\Pi_{mat}^{(1)}(u,v)=2Y_1(u,v)=2\left[(1,0)+(0,1)\right]
$,
which returns the already known two-particle matrix factor 
\be
\Pi_{mat}^{(1)}(u,v)=\frac{4}{(u-v)^2+4} \,.
\ee
\medskip\\
\textbf{$\bullet $ Two couples $f\bar{f}$ ($n=2$):}\\
The $n=2$ case may be more clarifying, being less trivial. In
\be\label{Pi-Y}
\Pi_{mat}^{(2)}(u_1,u_2,v_1,v_2)=\frac{1}{(u_{12}^2+1)(v_{12}^2+1)}Y_2(u_1,u_2,v_1,v_2)
\ee
we have six distinct contributions, namely
\ba
Y_2(u_1,u_2,v_1,v_2)=(1,1,0,0)+(1,0,1,0)+(1,0,0,1)+(0,1,1,0)+(0,1,0,1)+(0,0,1,1) \ .\nn\\
\ea
The first one, from (\ref{Perm}) and upon using (\ref{delta20}), results
\ba\label{1100}
(1,1,0,0) &=&
\frac{\delta_{4}(u_1+i,u_2+i,u_1,u_2)\delta_{4}(u_1+i,u_2+i,v_1,v_2)}{\displaystyle(u_{12}^2+1)(u_{12}^2+4)\displaystyle\prod_{i,j}^2 (u_i-v_j)(u_i-v_j+2i)}= \\
&=& \frac{2\delta_4(u_1+i,u_2+i,v_1,v_2)}{\displaystyle\prod_{i,j=1}^2(u_i-v_j)(u_i-v_j+2i)} \nn
\ea

whereas $(0,0,1,1)$ is obtained from (\ref{1100}) through the substitution $(u_1,u_2)\leftrightarrow (v_1,v_2)$.
The diagram $(1,0,1,0)$ can be retrieved, instead, by exchanging $u_1\leftrightarrow v_1$ in the first line of (\ref{1100}) (keep in mind the \textit{caveat} about the variables of $\delta_4$):
\ba
(1,0,1,0) &=&
\frac{\delta_{4}(u_1+i,v_1+i,u_1,u_2)\delta_{4}(u_1+i,v_1+i,v_1,v_2)}{\displaystyle[(u_{1}-v_1)^2+1][(u_{1}-v_1)^2+4]}\cdot \\
&\cdot &\frac{1}{u_{12}(u_{12}+2i)v_{12}(v_{12}+2i)(u_1-v_2)(u_1-v_2+2i)(v_1-u_2)(v_1-u_2+2i)}= \nn\\
&=&\frac{4(u_1-u_2-i)(v_1-v_2-i)}{(u_1-v_2)(v_1-u_2)(u_1-u_2)(v_1-v_2)(u_1-v_2+2i)(v_1-u_2+2i)} \,;
\ea

Now, the remaining ones straightforwardly follow after a suitable permutation: $(1,0,0,1)$ is obtained from $v_1 \leftrightarrow v_2$, $(0,1,1,0)$ from $u_1\leftrightarrow u_2$ and $(0,1,0,1)$ after the exchange of both $(u_1,v_1)\leftrightarrow (u_2,v_2)$.
Summing up the six pieces, the matrix factor (\ref{Pi-Y}) amounts to
\be
\Pi_{mat}^{(2)}(u_1,u_2,v_1,v_2)=\frac{1}{((u_{1}-u_2)^2+1)((v_{1}-v_2)^2+1)}\frac{P^{(2f+2\bar f)}(u_1,u_2,v_1,v_2)}{\displaystyle\prod_{i,j=1}^2((u_i-v_j)^2+4)} \ ,
\ee

where the polynomial $P^{(2f+2\bar f)}$ is listed in the Appendix \ref{Pol}. This result agrees with the previous finding by \cite{BFPR}. Hence, we illustrated how the Young tableaux approach provides an efficient way to the compute the fermion matrix factors, in the same way as it does for the scalars.

\subsubsection{Recursion formula}

An analogue recursion relation is satisfied by the matrix factor for the fermions. It relates the residues to a matrix factor with lower number of particles and it was also proposed in \cite{Bel1509} for the case with $n_f=n_{\bar{f}}+1$.

The residue of the matrix factor $\Pi_{mat}^{(n)}$ in the kinematic pole $v_j=u_i+2i$, is given by
\be\label{ResPimat}
i \textit{Res}_{v_1=u_1+2i} \Pi_{mat}^{(n)}(u_1,\cdots ,u_n,v_1,\cdots ,v_n)= \frac{\Pi_{mat}^{(n-1)}(u_2,\cdots ,u_n,v_2,\cdots ,v_n)}{\displaystyle\prod_{j=2}^n(u_{1j}+i)u_{1j}(u_1-v_j+2i)(u_1-v_j+i)} \ .
\ee

The proof follows from the sum over Young diagram (\ref{ff-Y}), by considering the single diagram contribution (\ref{Genk}). The pole $v_1=u_1+2i$ appears only in diagrams of the type $(1,\lbrace l_{i=2}^n \rbrace , 0, \lbrace l_{n+1}^{2n} \rbrace)$, which we split into
\be
(1,\lbrace l_{i=2}^n \rbrace , 0, \lbrace l_{n+1}^{2n} \rbrace)=(1,0)\cdot (\lbrace l_{i=2}^n \rbrace , \lbrace l_{n+1}^{2n} \rbrace)\cdot M_{\lbrace l_i \rbrace} 
\ee

Only the two-particle diagram $(1,0)$ contains the pole and $(\lbrace l_{i=2}^n \rbrace , \lbrace l_{n+1}^{2n} \rbrace)$ represents a diagram in the sum defining $\Pi_{mat}^{(n-1)}$. The mixing term $M_{\lbrace l_i \rbrace}$, whose expression follows from (\ref{Genk}) after some algebraic manipulations, depends on both sets of variables and, in general, it changes according to the specific diagram ${\lbrace l_i \rbrace}$. However, when evaluating the residue around $v_1=u_1+2i$, this dependence on the diagram drops out and, after summing over the set ${\lbrace l_i \rbrace}$, the matrix factor with a decreased number of particles $\Pi_{mat}^{(n-1)}$ is reproduced, multiplied by the correct factor to give (\ref{ResPimat}).\\

As said for the scalar case, this kind of relation does not come unexpected from a physical ground, as it clearly alludes to the axiom relating the form factor\footnote{Even though, strictly speaking, we are dealing with some square modulus of a form factor.} of an operator to itself, once two particles are 'annihilated' via the evaluation of the residues for kinematic poles, $v_i=u_j+2i$. This recursion formula will find the main application in the next chapter, with very important consequences.

\subsection{Fermion polynomials}

As in the case involving scalars, it is possible to explicit the polar structure which, for a system of $n$ couples $f\bar{f}$, reads

\be\label{Mat-fer}
\Pi_{mat}^{(n)}(\{u_i\},\{v_j\})=\frac{P^{(n)}(u_1,\dots,u_n,v_1,\dots,v_n)}
{\displaystyle\prod_{i<j}^n[(u_i-u_j)^2+1]\prod_{i<j}^n[(v_i-v_j)^2+1]\prod_{i,j=1}^n[(u_i-v_j)^2+4]} \,,
\ee
where $P^{(n)}(u_1,\dots,u_n,v_1,\dots,v_n)$ is a $2n(n-1)$ degree polynomial: the simplest cases are reported in the dedicated appendix.

The proof for the general polar structure (\ref{Mat-fer}) relies on the asymptotic factorisation of $\Pi_{mat}^{(n)}$. In fact, when $k$ rapidities $u_i$ and $v_i$ get shifted by a large quantity $\Lambda$, the matrix part factorises into the product matrix factors involving two disjoint proper subgroups of particles (up to a power of the shift):
\be\label{FactPi}
\Pi_{mat}^{(n)}(\{u_{i=1}^{k} +\Lambda , u_{i=k+1}^n \},\{v_{i=1}^{k} +\Lambda , v_{i=k+1}^n  \}) \simeq \Lambda^{-4k(n-k)}\Pi_{mat}^{(k)}(\{u_{i=1}^k\},\{v_{i=1}^k\})\Pi_{mat}^{(n-k)}(\{u_{i=k+1}^n\},\{v_{i=k+1}^n\}) \ .
\ee
Formula (\ref{FactPi}) can be proven directly from the integral representation (\ref{Pi_mat^ff}): the shift by $\Lambda $ on $k$ fermion and antifermion rapidities can be re-absorbed into $k$ integration variables $a,\,b,\,c$, then the limit $\Lambda\rightarrow\infty$ leads to (\ref{FactPi}).

The polar structure (\ref{Mat-fer}) is then a direct consequence of factorisation, a sketchy\footnote{The proof does not differ from the scalar case, for which a more detailed explanation is available in \cite{BFPR3}.} proof is provided: if one shifts by
$\Lambda\gg 1$ two fermion and two antifermion rapidities, say $u_1,\,u_2,\,v_1,\,v_2$ without loss of generality, (\ref{FactPi}) for $k=2$ becomes
\be
\Pi_{mat}^{(n)}(u_1+\Lambda ,u_2+\Lambda, \{u_{i=3}^n\},v_1+\Lambda ,v_2+\Lambda, \{v_{i=3}^n\})
\simeq \Lambda^{-8(n-2)} \,\Pi_{mat}^{(2)}(u_1,u_2,v_1,v_2)
\Pi_{mat}^{(n-2)}(\{u_{i=3}^n\},\{v_{i=3}^n\}) \ .
\ee
The two-particle factor $\Pi_{mat}^{(2)}(u_1,u_2,v_1,v_2)$ satisfies the structure (\ref{Mat-fer}) and exhibits poles for $u_1-u_2=\pm i$,
$v_1-v_2=\pm i$ and $u_i-v_j=\pm 2i$. Since $\Pi_{mat}^{(n)}$ is invariant under permutations of the $u$'s and of the $v$'s, the same reasoning must hold for any arbitrary $4$-plet $\{u_i,u_j,v_k,v_l\}$, thus structure (\ref{Mat-fer}) follows. \\

Many additional features of the polynomials, expecially those due to the recursion formula (\ref{ResPimat}), are listed in the Appendix \ref{Pol}. They will turn out to be fundamental for the computation of the fermion contribution $W_f$ to the null polygonal Wilson loop, object of the next chapter.

\chapter{Fermions and gluons: the classical contribution}
\label{ChCla}

In this chapter we analyse the OPE series, for instance (\ref{OPEseries}) in the hexagon case, focusing on the contribution from fermions, antifermions and gluons in the strong coupling regime. On a general ground, we have seen in subsection \ref{StCou} that there are two different regimes in that limit. The aim of the chapter is to reproduce the first one, given by the classical string theory in $AdS_5$. More precisely, the null polygonal Wilson loop is given by the (exponential of) minimal worldsheet area of the string, moving in $AdS_5$ and attached to the polygon on the boundary. The solution of the problem takes the form of a set of non-linear integral equations which recalls the TBA systems. The excitations responsible for that are gluons and (anti)-fermions. In \cite{FPR2}, the classical contribution for the hexagon has been reproduced by means of the following hypothesis: the contribution from fermions-antifermions can be thought of as coming from effective bound states $f\bar{f}$ between them \cite{BSV3}. On the physical ground of Bethe equations, they are not real bound states, for the do not appear in the spectrum. However, for strictly infinite coupling we can think of them as physical particles and sum, instead over fermions, on these so-called \emph{mesons}. In turn, they form effective bound states among themselves and, upon summation, reproduce the central node of the TBA system. The resummation of the OPE series is possible since these composite objects are $SU(4)$ singlets. Here we prove that this hypothesis is indeed correct, following an argument outlined in \cite{BFPR4}. The chapter is structured as follows. The first section deals with the hexagon, for which we show the emergence of the meson (and bound states thereof) directly from the OPE series for the fermions \cite{BFPR,BFPR4}, thus confirming the validity of the meson hypothesis pushed forward in \cite{FPR2}. Interestingly, along the way we found some analogies with the instanton contribution to the partition function in $\mathcal{N}=2$ theories, encoded in the Nekrasov function $\mathcal{Z}$. In particular, it resembles the series over mesons appearing in the hexagon case. This analogy will help to understand the mechanism by which mesons form bound states, leading to the emergence of the TBA-like equation. In the second part, assuming the meson hypothesis, we sum over gluons and mesons for the polygonal Wilson loop with $N>6$, finding agreement with the classical string computation and reproducing the TBA and the Y-system for scattering amplitudes \cite{Anope,YSA} directly from the OPE setup. An important remark about the relation between the TBA and the bound states is due. In the usual TBA formulation, the presence of bound states generates additional TBA nodes, \emph{i.e.} there is a node for any different bound state. In our case, instead, all the tower of bound states is necessary to reproduce, upon resummation, the TBA node. This is the main difference with respect to the usual TBA formulation of integrable systems.

\section{The hexagon: fermions}
\label{FerHex}

The hexagon, under the meson hypothesis, has been resummed in \cite{FPR2} and the TBA-like equations for the amplitudes have been reproduced. The purpose of this part is to prove the validity of the procedure, by analysing the fermion-antifermion coontribution. In the process we disregard the gluons, as they are singlet under $SU(4)$ and their inclusion is straightforward \cite{BFPR5}.

\subsection{Fermions contribution and bound states}

The fermionic part of (\ref{OPEseries}) reads

\ba\label{OPEFer}
W_f=\sum_{n=0}^{\infty} W_f^{(n)} = && \sum_{n=0}^{\infty}\frac{1}{n! n!}\int _{\m{C}}\prod_{k=1}^n \left[\frac{du_k}{2\pi}\frac{dv_k}{2\pi}
\,\mu_f(u_k)\mu_f(v_k)\,e^{-\tau E_f(u_k)+i\sigma p_f(u_k)}\cdot \right.\\
&& \left. \cdot e^{-\tau E_f(v_k)+i\sigma p_f(v_k)}\right]
\Pi_{dyn}^{(n)}(\{u_i\},\{v_j\})\,\Pi_{mat}^{(n)}(\{u_i\},\{v_j\})
\nn \ .
\ea

where we adapted the singlet condition (\ref{SingletPimat}) to the case without scalars, which becomes $N_f=N_{\bar{f}}\pm 4m$. We considered only the terms with $m=0$, \emph{i.e.} with the same number of fermions and antifermions, as it has been shown to yield the main contribution to the Wilson loop.

The integration contour $\m{C}_S$ is open and restricted to the small fermion sheet, as the large fermions are damped by an exponential factor in the strong coupling limit. More details on the contour can be found in \cite{BSV3,BFPR}. The (anti)-fermionic rapidities $\{u_k\}$, $\{v_k\}$ parametrize the energy and momentum $E_f(u)$, $p_f(u)$, which couple to the cross ratios $\tau$ and $\sigma$ in the propagation phase. The other cross ratio $\phi$ couples only to the helicity of the gluons and does not appear here. The product of pentagon transition in (\ref{OPEseries}), or form factor squared, factorizes into the product of a dynamical and a matrix part \cite{BSV4}, where the latter is coupling independent. The dynamical part is two-body factorisable

\small
\be\label{dyn}
\Pi_{dyn}^{(n)}(\{u_i\},\{v_j\}
) = \displaystyle\prod_{i<j}^n\frac{1}{P(u_i|u_j)P(u_j|u_i)}
\frac{1}{P(v_i|v_j)P(v_j|v_i)}\displaystyle\prod_{i,j=1}^n \frac{1}{\bar P(u_i|v_j)\bar P(v_j|u_i)}
\ee
\normalsize 
where $P$ stands for the transition between particles of the same type ($i.e.$ fermion-fermion or anti-fermion-anti-fermion) and $\bar{P}$ for the transition between a fermion and an anti-fermion. The function $P(u|v)$ is endowed with a single pole for coinciding rapidities $v=u$, whose residue determines the measure $\mu_f(u)$ \cite{BSV1}: $\mbox{Res}\,_{v=u}\,P(u|v)= i/\mu_f(u)$. 

On the other hand, the factor $\Pi^{(n)}_{mat}$, encoding the $SU(4)$ matrix structure, has an integral representation \cite{BSV4} in terms of the auxiliary variables $a,\,b,\,c$, introduced in the previous chapter where a Young tableaux method \cite{BFPR5} has been exposed. Eventually, it led to the polar form

\be\label{Mat-fer}
\Pi_{mat}^{(n)}(\{u_i\},\{v_j\})=\frac{P^{(n)}(u_1,\dots,u_n,v_1,\dots,v_n)}
{\displaystyle\prod_{i<j}^n[(u_i-u_j)^2+1]\prod_{i<j}^n[(v_i-v_j)^2+1]\prod_{i,j=1}^n[(u_i-v_j)^2+4]} \,.
\ee

where $P^{(n)}(u_1,\dots,u_n,v_1,\dots,v_n)$ is a degree $2n(n-1)$ polynomial in the $u_i,\,v_j$\,.

The polar structure (\ref{Mat-fer}) and the properties of the polynomials $P^{(n)}$ play a main role to the emergence of the meson. In turn, they bound up to form composite states themselves, thus reproducing the (central node of) TBA-like equations for the amplitudes. First, we see how couples $f\bar{f}$ coalesce into bound states. Later, we will address the interaction between these meson and witness the emergence of their bound states. Once againg, we stress that on the ground of the Bethe ansatz equations, the aforementioned mesons do not show up in the spectrum of particles at finite coupling, as they lie outside of the physical sheet \cite{BSV3}. On the contrary, they come into existence at $g=\infty$ and start contributing to the OPE in place of (unbounded) fermions and antifermions, whose contribution is subdominant.

To begin with, making use of (\ref{dyn},\ref{Mat-fer}), we reformulate (\ref{OPEFer}) into the appealing
\be\label{Meson}
W^{(n)}=\frac{1}{n!}\int_{C_S}\displaystyle\prod_{i=1}^n\frac{du_i}{2\pi}I_n(u_1,\cdots ,u_n)\displaystyle\prod_{i<j}^np(u_{ij}) \ ,
\ee
where, with some hindsight, we factorised the factor responsible for the bound states between mesons, which we call short-range (meson-meson) potential
\be\label{pol-pot}
p(u_{ij})\equiv \displaystyle\frac{u_{ij}^2}{u_{ij}^2+1}
\ee
and enclosing the integrals on the antifermionic rapidities $v_j$ inside the function
\small
\be\label{I_n}
I_n(u_1, \cdots , u_n)\equiv\frac{1}{n!}\int_{C}\displaystyle\prod_{i=1}^n\frac{dv_i}{2\pi} R_n(\left\lbrace u_i\right\rbrace,\left\lbrace v_i\right\rbrace )P^{(n)}(\left\lbrace u_i\right\rbrace,\left\lbrace v_i\right\rbrace)\displaystyle\prod_{i,j=1}^n h(u_i-v_j)\displaystyle\prod_{i<j}^n p(v_{ij}) \ ,
\ee
\normalsize
in which the short-range potential acting between a couple $f\bar{f}$ is present

\be\label{shortff}
h(u_i-v_j)=\frac{1}{(u_i-v_j)^2+4}
\ee

The regular part $R_n$, with no poles nor zeroes in the rapidities ($u_i,\,v_i\,$), is related to the dynamical factor (\ref{dyn}) via the definition
\be\label{R_n}
R_n(u_1,\cdots ,u_n,v_1,\cdots ,v_n)\displaystyle\prod_{i<j}^n u_{ij}^2v_{ij}^2 \equiv
\Pi_{dyn}^{(n)}(u_1,\cdots ,u_n,v_1,\cdots ,v_n)\displaystyle\prod_{i=1}^n\hat{\mu}_f(u_i)\hat{\mu}_f(v_i) \ ,
\ee
where the measure and the propagation phase are combined into
\be
\hat{\mu}_f(u)\equiv \mu_f(u)e^{-\tau E_f(u) + i\sigma p_f(u)} \ .
\ee

The emergence of mesons is an outcome of the integrals over the antifermionic variable $v_j$ in (\ref{I_n}), in particular due to the short-range potential (\ref{shortff}). As explained, the integration can be safely restricted to the small-fermion sheet section $\m{C}_S$. In turn, it is useful to split $\m{C}_S$ in two parts: a closed contour $\m{C}_{HM}$, entirely lying in the lower half plane and the segment $\mathcal{I}=[-2g,+2g]$, oppositely oriented \cite{BFPR}\footnote{We mention that the interval $\mathcal{I}$ contains unphysical values of the momentum. However, they cancel with an opposite contribution present in the closed part $\m{C}_{HM}$.}. This choice allows us to evaluate part of the integrals by residue. The choice of the contour $\m{C}_S$ does not depend on the coupling constant $g$, nevertheless in the strong coupling regime we benefit from a crucial simplification. In fact, we decompose the function $I_n$ into the sum $I_n=I_n^{closed}+I_n^r$, where $I_n^{closed}$ corresponds to the expression (\ref{I_n}) with the contour $\m{C}_S$ replaced by $\m{C}_{HM}$. The remainder $I_n^r$ turns out to be subdominant at large coupling \cite{BFPR}. This can be understood as, when $\lambda\to\infty$, the rapidities get rescaled according to $u=2g\bar{u}$ and the interval $\mathcal{I}$ becomes fixed in the $\bar{u}$: in this new variables the poles are very close to the integration contour and the leading order is given by the residues of such poles. A more detailed explanation of this effect is given in \cite{BFPR}, where the simplest cases $n=1,2$ have been thoroughly analysed.

In order to illustrate how to approach the integrals (\ref{I_n}), we analyse the simplest non-trivial case, $n=2$, so to clarify our notation before moving to the general case. We treat it in a more elegant way than \cite{BFPR}, amenable for an easy generalisation to any $n$ \cite{BFPR4}. \\

\textbf{$\bullet$ $n=2$ case}\\

We want to evaluate the integral

\footnotesize
\be\label{I_2}
I_2^{closed}(u_1,u_2)=\frac{1}{2}\int_{C_{HM}}\frac{dv_1 dv_2}{(2\pi)^2}\frac{R_2(u_1,u_2,v_1,v_2)P^{(2)}(u_1,u_2,v_1,v_2)}{\left[(u_1-v_1)^2+4\right]\left[(u_1-v_2)^2+4\right]\left[(u_2-v_1)^2+4\right]\left[(u_2-v_2)^2+4\right]}\frac{v_{12}^2}{v_{12}^2+1}
\ee
\normalsize
by computing the residues. Fortunately, after the warm up of the previous chapter, we can apply a similar method to this multiple integrals. One observes that the poles arrange themselves in strings in the complex plane, with real rapidities $u_j$: indeed, if the first pole to be considered is $v_i=u_j-2i$\,, any further residue around the same real rapidity is placed below the previous one at a distance $-i$\,, $i.e.$ the general form $v_i=u_j-(2+\kappa)i$ (with $\kappa =0,1,\dots $) is found. Sticking on (\ref{I_2}), three independent residue configurations occur, namely $(u_1-2i,u_2-2i)$, $(u_1-2i,u_1-3i)$ and $(u_2-2i,u_2-3i)$, each one with a multiplicity $2!$ owing to the symmetry of $I_2^{closed}$ under permutation of the integration variables $v_i$: these configurations are respectively denoted in the following as $(1,1)$, $(2,0)$ and $(0,2)$, for compactness. The explicit form of $P^{(2)}$ entails that $(2,0)$ and $(0,2)$ actually give no contribution, since\footnote{See the Appendix \ref{Pol} for the properties of the polynomials $P^{(n)}$.}
\be\label{nullYoung}
P^{(2)}(u_1,u_2,u_1-2i,u_1-3i)=0 \ ,
\ee
while the only configuration that matters is $(1,1)$, for which $P^{(2)}$ takes the simple form
\be\label{Young-2}
P^{(2)}(u_1,u_2,u_1-2i,u_2-2i)=16\left[(u_{12}^2+16)(u_{12}^2+1)\right] \ :
\ee
as a result,
\be
I_2^{closed}(u_1,u_2) = R_2(u_1,u_2, u_1-2i,u_2-2i) \ ,
\ee
which on a physical ground hints that fermions and antifermions (with rapidities differing by the purely imaginary quantity $2i$) form a bound state.
\ \medskip\ \\

\textbf{$\bullet$ Arbitrary $n$:}\\

\vspace{0.2cm}

As already noticed for $n=2$, also in the general case for arbitrary $n$ the residue configuration $(1,1,\cdots,1)$, $i.e.$ when only poles of the kind $v_i=u_j-2i$ are involved, turns out to be the sole to contribute to $I_n$. In fact, the strong constraint
\be
P^{(n)}(u_1,\cdots ,u_n, u_1-2i, u_1-3i , v_3, \cdots , v_n)=0 \ ,
\ee
resulting from the relation (\ref{Pfer=0}) in the appendix, rules out all the residue configurations except for $(1,1,\cdots,1)$ so that, in a quite formal fashion, one can assert that
\be\label{P(Y)=0}
P^{(n)}(Y\neq (1,1\cdots ,1,1))=0 \ .
\ee
Taking into account the proper combinatorial factor, the contribution arising from the aforementioned configuration reads
\be\label{11111}
(1,1,\cdots ,1,1)=\frac{(-1)^n}{4^n}\frac{P^{(n)}(u_1,\cdots ,u_n,u_1-2i,\cdots ,u_n-2i)R_n(u_1,\cdots ,u_n,u_1-2i,\cdots ,u_n-2i)}{\displaystyle\prod_{i<j}^n(u_{ij}^2+16)(u_{ij}^2+1)}
\ee
where $P^{(n)}$ can be given the explicit form (\ref{fundP}), allowing us to retrieve for (\ref{I_n}) an expression
\be\label{I_n-clo}
I_n^{closed}(u_1,\cdots ,u_n)=(-1)^n R_n(u_1,\cdots ,u_n, u_1-2i,\cdots ,u_n-2i)
\ee
which highlights how fermion and antifermion rapidities pair up to form complex two-string, with spacing $2i$. A comparison with (\ref{dyn}), (\ref{R_n}) suggests these two-strings to find an interpretation as bound states \cite{BSV3,BFPR}, named mesons, whose energy is the sum of the energies of the single components, as well as their momentum,
\be
E_M(u)\equiv E_f(u+i)+E_f(u-i), \quad p_M(u)\equiv p_f(u+i)+p_f(u-i) \ ,
\ee
while their pentagon transition amplitude can be recognised in the expression
\be
P^{MM}(u|v) = -(u-v)(u-v+i) P(u+i|v+i)P(u-i|v-i)|\bar{P}(u-i|v+i)\bar{P}(u-i|v+i) \ ,
\ee
although, for later purposes, it is worth to introduce the regular function $P^{MM}_{reg}$, without poles nor zeroes, related to $P^{MM}$ via\footnote{In the Appendix \ref{ApSca} these and other related formulae are discussed in more details.}
\be\label{PMMreg}
P^{MM}(u|v)
= \frac{u-v+i}{u-v}\,P^{MM}_{reg}(u|v) \ .
\ee
Accordingly, the (hatted) measure can be coherently traced in the formula
\be
\hat{\mu}_M(u) \equiv \mu_M(u)e^{-\tau E_M(u) + i\sigma p_M(u)}
= -\frac{\hat{\mu}_f(u+i)\hat{\mu}_f(u-i)}{\bar{P}(u+i|u-i)\bar{P}(u-i|u+i)} \ ,
\ee
while the measure is
\be
\mbox{Res}\,_{v=u}\,P^{MM}(u|v)=\frac{i}{\mu_M(u)} \ .
\ee

These identifications lead us to recast (\ref{I_n-clo}) into the expression
\be\label{conj}
I_n^{closed}(u_1,\cdots ,u_n) 
 = \frac{\displaystyle\prod_{i=1}^n\hat{\mu}_M(u_i-i)}{\displaystyle\prod_{i<j}^n P^{MM}_{reg}(u_i-i|u_j-i)P^{MM}_{reg}(u_j-i|u_i-i)} \ ,
\ee
that, once plugged into (\ref{Meson}), legitimises us to reformulate the fermionic contribution to the hexagon Wilson loop (\ref{OPEFer}) in the strong coupling limit in terms of these novel bound states, into the series $W=W^{(M)}+\dots$

\small
\be\label{SingMes}
W^{(M)} = \sum_{n=0}^{\infty}\frac{1}{n!}\int_{C_S}\displaystyle\prod_{i=1}^n\frac{du_i}{2\pi}\hat{\mu}_M(u_i-i)\displaystyle\prod_{i<j}^n\frac{1}{P^{MM}_{reg}(u_i-i|u_j-i)
P^{MM}_{reg}(u_j-i|u_i-i)}\displaystyle\prod_{i<j}^n p(u_{ij}) 
\ee
\normalsize
where the dots are to remind that some terms, coming from the integrations on the interval $\mathcal{I}$ and originally included in $I_n$, were discarded\footnote{In order to fully reconstruct the fermionic contribution to the Wl at finite coupling, the integrations performed along the contours $C_L$ must be added too.} while considering the leading contribution $I_n^{closed}$.

To conclude with a few comments, formula (\ref{SingMes}) means that in the large coupling regime unpaired fermions and antifermions give the way to the formation of mesons. Nevertheless, we want to stress that at finite coupling the formula (\ref {SingMes}) still makes sense: indeed one can still recognise a contribution ascribable to these effective particles, and associate them (at least formally) a pentagon amplitude and a measure.\\
In the next part, some attention shall be paid to the short-range interaction $p(u_{ij})$ inside the integrand of (\ref{SingMes}). In the strong coupling limit, it accounts for the formation of bound states between mesons, following a mechanism similar to that responsible for the coalescence of instantons into bound states \cite{MenYang} in the Nekrasov partition function. This will represent the main aim of the following subsection.

\subsection{Mesons bound states and TBA}

Now we are going to analyse in details the series $W^{(M)}$ in (\ref{SingMes}) which, we recall, is the strong coupling limit of the fermionic sector of the OPE. It is a sum over effective bound states $f\bar{f}$ which we called mesons. They are singlets under $SU(4)$, so that we do not have to deal with the involved $SU(4)$ matrix structure. We employ an elegant method \cite{BFPR4}, combining the path integral approach and the Fredholm determinant formula, from which the strong coupling limit follows straightforwardly. This will give a rigorous proof of the meson hypothesis by reproducing the result of \cite{FPR2}, which was the starting point for the resummation of the OPE series. 

An interesting analogy with the Nekrasov instanton partition function $\mathcal{Z}$ \cite{Nekrasov} emerges, which depends on two deformation parameters $\epsilon_1,\epsilon_2$ defining the $\Omega$-background. The correspondence with the meson series $W^{(M)}$ becomes noteworhty expecially when the strong coupling limit of the latter is considered. On the $\mathcal{N}=2$ side, it corresponds to the so-called Nekrasov-Shatashvili regime, where one parameter is sent to zero whereas the other is kept finite. A description in terms of TBA-like equation is present for the Nekrasov function in that limit \cite{MenYang}: with this in mind, the strong coupling limit for the meson series should follow straightforwardly from this analogy. In particular, for large coupling $\lambda\to\infty$, mesons bound together following the same pattern of instantons in $\mathcal{Z}$, giving rise to the typical dilogarithm function. The set of techniques we are going to apply here to $W^{(M)}$ works for the partition function $\mathcal{Z}$ as well. For the latter, they are described in the Appendix \ref{NekApp}, where we also give more details on the Nekrasov function and discuss its similarities with the meson series $W^{(M)}$. We mention that the emergence of bound states between mesons in (\ref{SingMes}) can be equivalently shown employing the cluster expansion, which was the way they have been originally found for the Nekrasov function \cite{MenYang}. This method is worked out in the dedicated appendix. However, the approach in this section is much more elegant and straightforward.

The procedure goes as follows: along the line of \cite{FPR2,BFPR4}, we introduce a quantum field $X(u)$ whose propagator is

\be
e^{\left\langle X(u_i)X(u_j)\right\rangle} \equiv \frac{1}{P^{MM}_{reg}(u_i-i|u_j-i)
P^{MM}_{reg}(u_j-i|u_i-i)}\equiv e^{G_M(u_i,u_j)}
\ee

so that we can write the Wilson loop as an average

\be
W^{(M)} \simeq \left\langle \sum_{n=0}^{\infty}\frac{1}{n!}\int_{C_S}\displaystyle\prod_{i=1}^n\frac{du_i}{2\pi }\hat{\mu}_M(u_i-i)e^{X(u_i)}\displaystyle\prod_{i<j}^n p(u_{ij})\right\rangle
\ee

thanks to the path integral idendity\footnote{Here we neglected the diagonal term from the Gaussian identity, due to the propagator evaluated in $u_i=u_j$, as it is subleading in the strong coupling regime.}

\be 
\displaystyle\prod_{i<j}e^{\left\langle X(u_i)X(u_j)\right\rangle}=\left\langle\displaystyle\prod_{i}e^{X(u_i)}\right\rangle 
\ee

which is an extension of the usual gaussian integrals.
The Cauchy identity allows us to recast the meson-meson potential as

\be
\displaystyle\prod_{i<j}^n p(u_{ij})=\frac{1}{i^n}\det\left(\frac{1}{u_i-u_j-i}\right)
\ee

which, through the definition of the matrix $M$

\be
M(u_i,u_j)\equiv \frac{\left[\hat{\mu}_M(u_i-i)e^{X(u_i)}\hat{\mu}_M(u_j-i)e^{X(u_j)}\right]^{1/2}}{u_i-u_j-i}
\ee

allows us to find the beautiful representation

\be\label{WFre}
W^{(M)} \simeq \left\langle\det\left(1 + M\right)\right\rangle = \left\langle \exp\left[\sum_{n=1}^{\infty}\frac{(-1)^{n+1}}{n}\textit{Tr}M^n\right]\right\rangle
\ee

where, inside the average, we have the Fredholm determinant of an integral operator defined by the kernel $M(x,y)$. The trace of $M^n$ is defined as

\be
\textit{Tr}M^n \equiv \int_{C_s}\displaystyle\prod_{i=1}^n\frac{du_i}{2\pi i}\hat{\mu}_M(u_i-i)e^{X(u_i)}\displaystyle\prod_{i=1}^n\frac{1}{u_i-u_{i+1}-i}, \quad u_{n+1}\equiv u_1
\ee

This method is extensively discussed in the Appendix \ref{NekApp}, where it is applied to the Nekrasov partition function $\mathcal{Z}$.

Now we want to perform the strong coupling limit. As stated before, it corresponds to the $\epsilon\to 0$ limit for the Nekrasov function, so that we can use the result in the Appendix \ref{NekApp}: the leading order is obtained by evaluating the residues inside the closed contour $C_{HM}$

\be
\textit{Tr}M^n \simeq \frac{(-1)^{n-1}}{n}\int_{C_s}\frac{du}{2\pi}\hat{\mu}^n_M(u-i)e^{nX(u)}\simeq \frac{(-1)^{n-1}}{n}\int_{C_s}\frac{du}{2\pi}\hat{\mu}^n_M(u)e^{nX(u)}
\ee

where the shifts have been neglected, as the rapidities $u$ get rescaled in the large $\lambda$ limit. The sum over $n$ in (\ref{WFre}) gives rise to the expected dilogarithm: in the end, the strong Wilson loop reads

\be\label{Wpath}
W^{(M)}\simeq\left\langle\exp\left[\int_{C_s}\frac{du}{2\pi} Li_2\left[\hat{\mu}_M(u) e^{X(u)}\right]\right]\right\rangle
\ee

where the hatted measure is

\be
\hat{\mu}_M(u) = \mu_M(u)e^{-\tau E_M(u) + i\sigma p_M(u)}
\ee

The strong coupling measure is given by $\mu_M(u)\simeq -1$, so that (\ref{Wpath}) turns into

\be\label{WLi2}
W^{(M)}\simeq\left\langle\exp\left[-\int_{C_s}\frac{du}{2\pi}\mu_M(u) Li_2\left[-e^{-\tau E_M(u)+i\sigma p_M(u)} e^{X(u)}\right]\right]\right\rangle
\ee

which agrees with (11.53) of \cite{FPR2}, which was obtained by means of the meson hypothesis, \emph{i.e.} starting from the OPE series containing mesons and bound states thereof

\be\label{BoundMes}
W^{(M)}_{bound}=\sum_{N=0}^{\infty}\frac{1}{N!}\sum_{a_1=1}^{\infty}\cdots\sum_{a_N=1}^{\infty}\int_{C_S}\displaystyle\prod_{i=1}^N\frac{du_i}{2\pi}\frac{\hat{\mu}_M(u_i)^{a_i}}{a_i^2}\displaystyle\prod_{i<j}^N\left[\frac{1}{P^{MM}_{reg}(u_i|u_j) P^{MM}_{reg}(u_j|u_i)}\right]^{a_i a_j}
\ee

This means that, in the strong coupling regime, we have the equivalence\footnote{This requires also the explicit expression in the large $\lambda$ limit of the pentagon transitions, displayed in Appendix \ref{ApSca}.} $W^{(M)}\simeq W^{(M)}_{bound}$. Note that the measure factor $1/a^2$, for a bound state of $a$ mesons, is fundamental to reproduce the dilogarithm function and appears, in the OPE setup, thanks to the short-range potential $p(u_{ij})$. On the other hand, it was proposed in \cite{FPR2} from the analysis of the Bethe equations. 

Equation (\ref{WLi2}) is a path integral representation of the meson contribution to the Wilson loop, valid in the strong coupling limit only. What makes the method useful is that the action is large $\sim\sqrt{\lambda}$ and the saddle point technique, depicted in details in \cite{FPR2}, is appliable. This will yield the TBA-like equation previously found for the scattering amplitudes. Including also gluons and their bound states, for which the original OPE series already looks like (\ref{BoundMes}) and the dilogarithm appears naturally, the classical string result is reproduced \cite{FPR2}. The main contribution of this section was to prove the claim that, altough not present in the spectrum, effective bound states between $f\bar{f}$ emerge, along with an infinite tower of bound states between them, at infinite coupling and their contribution reproduces the central node of the TBA-like equations. Note that these are not bound states in the usual sense in the TBA framework, as they do not constitute additional nodes in the equations\footnote{The same argument works for the bound states between gluons, which are physical bound states in the Bethe equations setup but not TBA bound states.}: their resummation, on the other hand, reproduces only one node of the TBA. The emergence of the interaction between mesons and gluons can be shown easily within the same framework  \cite{BFPR5}, by considering the OPE series with both gluons and (anti)-fermions.

\section{Extension to any polygon}
\label{ResumPol}

In this section we show how the resummation of the OPE series, accounting for fermions and gluons, extends  to bigger polygons $n>6$. In doing that, we take advantage of the meson hypothesis proven for the hexagon case $n=6$. A more complete derivation would involve considering the general gluons-fermions contribution to $W_n$. However, it is rather involved and for several physical/mathematical reasons we can say with a high degree of confidence that the meson hypothesis should be trusted for any number of edges. The main achievement consists in the agreement with the string theory calculations, namely the TBA-like equations depicted in Appendix \ref{TBApp}. The content of this section is mainly taken from the paper \cite{BFPR}.

\subsection{Mesons and gluons: OPE series resummation}

Here, by means of the meson hypothesis, we sum the OPE series for mesons and gluons at strong coupling $\lambda\to\infty$, in the case of bigger polygons $n>6$. Since we are interested in the strong coupling analysis, throughout the section we use the hyperbolic rapidities $\theta$ to parametrize the energy and momenta of gluons and mesons. They are related to the Bethe rapidities through $u_M=2g\coth\theta$, $u_g=2g\tanh\theta$, respectively for mesons and gluons. This parametrization is well-suited to compare the OPE results with the TBA system for the amplitudes.

The most general pentagonal amplitude, involving two states with respectively $N$ and $M$ excitations, is denoted as

\be
P_{\vec{A} \vec{B}}(\theta_1,\ldots, \theta_N|\theta'_1,\ldots, \theta'_M)
\ee

The notation $\vec{A}=(A_1,...,A_N)$, where $A_i=a_{\alpha _i}$, means that the $i$-th excitation is a bound state of $a_{\alpha _i}$ particles of type $\alpha _i$: $\alpha _i=1,3 $ stands for gluons with positive and negative helicity and $\alpha _i=2$ stands for mesons.

The excitations are singlets, so that we do not have a matrix part and the transition factorizes as \cite{BSV1}
\be
P_{\vec{A} \vec{B}}(\theta_1,\ldots, \theta_N|\theta'_1,\ldots, \theta'_M) = \frac{\prod \limits _{i,j}P_{A_i B_j}(\theta_i|\theta'_j)}{\prod \limits _{i>j}P_{A_i A_j}(\theta_i|\theta_j) \prod \limits _{i<j}P_{B_i B_j}(\theta'_i|\theta'_j)} \, , \label {P-singl}
\ee
where $P_{A_i B_j}(\theta_i|\theta'_j)$ are the elementary transitions involving two particles. At strong coupling we have the important property
\be
P_{A_iA_j}(\theta|\theta')=[P_{\alpha _i,\alpha _j}(\theta|\theta')]^{a_{\alpha _i} a_{\alpha _j}} \, , \label
{P-strong}
\ee
where the 'fundamental' $P_{\alpha , \beta} (\theta, \theta ')$, acting between unbound mesons and gluons, are listed in Appendix \ref {ApSca}.
In the same way, for $\lambda\to\infty$ energy and momentum of bound states are given by the sum of energy and momentum of their constituents, namely
\be
E_{A_i}(\theta )=a_{\alpha _i}E_{\alpha _i}(\theta) \, , \quad p_{A_i}(\theta )=a_{\alpha _i}p_{\alpha _i}(\theta) \, ,
\ee
where the dispersion becomes relativistic 
\be
E_1(\theta)=E_3(\theta)=\sqrt{2}\cosh \theta \, , \quad E_2(\theta)=2\cosh \theta \, ; \quad
p_1(\theta)=p_3(\theta)=\sqrt{2}\sinh \theta \, , \quad p_2(\theta)=2\sinh \theta \label {en-mom} \, .
\ee

We note that the gluons have mass $\sqrt{2}$, whereas the meson mass is $2$. This agrees with the string theory perturbative analysis. 

We first deal with the heptagon, which helps to fix all the notations, then we study the most general case.

\subsubsection{Heptagon}

The OPE series at strong coupling, employing the properties discussed before, reads
\small
\ba
{\cal W}_{7}&=&\sum _{N=0}^{\infty} \sum _{M=0}^{\infty}\frac{1}{N!}\frac{1}{M!}\sum _{\alpha_1=1}^3...\sum _{\alpha_N=1}^3 \sum _{\beta_1=1}^3...\sum _{\beta_M=1}^3
\sum _{a_{\alpha _1}}...\sum _{a_{\alpha _N}} \sum _{b_{\beta _1}}...\sum _{b_{\beta _M}} \int \prod _{i=1}^N d\hat \theta_i (\tau _1, \sigma _1, \phi _1) \cdot \nonumber \\
&\cdot & \prod _{j=1}^M d\hat \theta'_j (\tau _2, \sigma _2, \phi _2) \frac{\prod \limits _{i=1}^N \prod \limits _{j=1}^M [P _{\alpha_i \beta_j}(-\theta_i|\theta'_j)]^{a_{\alpha _i}b_{\beta _j}}}{\prod \limits _{\stackrel {i,j=1}{i\not=j}}^{N} [P_{\alpha_i \alpha_j} (\theta_i|\theta_j)]^{a_{\alpha _i}a_{\alpha _j}} \prod \limits _{\stackrel {i,j=1}{i\not=j}}^{M} [P_{\beta_i \beta_j} (\theta'_i|\theta'_j)]^{b_{\beta _i}b_{\beta _j}}} \, , \label {Whep}
\ea
\normalsize

where we used a compact notation for the integration measure, including the propagation phases

\ba
d\hat \theta_i (\tau , \sigma , \phi )&=&e^{-\tau a_{\alpha _i}E_{\alpha_i}(\theta_i)+i\sigma a_{\alpha _i} p_{\alpha_i}(\theta_i)+i a_{\alpha _i} \phi  (2-\alpha _i)}\frac{\mu _{\alpha _i}(\theta_i)}{\left (a_{\alpha _i} \right )^2}(-1)^{a_{\alpha _i}-1} \frac{d\theta_i}{2\pi} \, , \nonumber \\
d\hat \theta'_j (\tau , \sigma , \phi )&=&e^{-\tau b_{\beta _j}E_{\beta_j}(\theta'_j)+i\sigma b_{\beta _j} p_{\beta_j}(\theta'_j)+i b_{\beta _j} \phi  (2-\beta _j)}\frac{\mu _{\beta _j}(\theta'_j)}{\left (b_{\beta _j} \right )^2}(-1)^{b_{\beta _j}-1} \frac{d\theta'_j}{2\pi} \, , \label {dhat} \\
\mu _1(\theta)&=&\mu _3(\theta)= - \frac{\sqrt{\lambda}}{2\pi} \frac{2}{\cosh ^2 2\theta} \, , \quad
\mu _2(\theta)=  \frac{\sqrt{\lambda}}{2\pi} \frac{2}{\sinh ^2 2\theta} \, , \nonumber
\ea

It is worth to remark the typical $1/a^2$ behaviour of the measure factor, for both gluons and mesons: although their origin is different\footnote{The gluons and bound states are physical particles.} the OPE quantities behave in the exact same way. 

Now we employ the same path integral method of the previous section, although in a more general fashion. From the Gaussian integrals, the following identity holds
\be
\langle e^{\sum \limits _{s,\alpha} X_{\alpha, 1}^{(s)} J_{\alpha,1}^{(s)}}e^{\sum \limits _{s,\alpha} X_{\alpha,2}^{(s)} J_{\alpha,2}^{(s)}}...e^{\sum \limits _{s,\alpha} X_{\alpha,d}^{(s)} J_{\alpha,d}^{(s)}} \rangle = e^{\frac{1}{2}\sum \limits _{s,s',\alpha, \beta}\sum \limits _{i,j=1}^d J_{\alpha,i}^{(s)} G_{\alpha,\beta,i,j}^{(s,s')}J_{\beta,j}^{(s')}} \label {identity} \, ,
\ee

of which we consider the limit $d\rightarrow \infty$. The average symbol $\langle \cdot \cdot \cdot \rangle $ means functional integration with respect to the fields $X_{\alpha}^{(s)}(\theta)$ in the presence of a source term with propagator
\be
\langle X_{\alpha}^{(s)}(\theta) X_{\beta}^{(s')}(\theta ') \rangle = G_{\alpha ,\beta}^{s,s'}(\theta , \theta ') \, .
\ee

which we choose to be given by the following

\small
\be
G^{(1,1)}_{\alpha, \beta}(\theta,\theta')=G^{(2,2)}_{\alpha, \beta}(\theta,\theta')=-\ln [P_{\alpha, \beta}(\theta|\theta')P_{\beta , \alpha}(\theta'|\theta)] \, , \quad G^{(1,2)}_{\alpha , \beta}(\theta,\theta')=G^{(2,1)}_{\beta ,\alpha}(\theta',\theta)=-\ln [P_{\beta ,\alpha}(\theta'|\theta)] \label {Gchoice}
\ee
\normalsize

and the currents are
\be
J^{(1)}_{\alpha}(\theta)=\sum _{i=1}^N a_{\alpha _i} \delta _{\alpha, \alpha _i}\delta (\theta-\theta_i)   \, , \quad J^{(2)}_{\beta }(\theta')=-\sum _{j=1}^M b_{\beta _j} \delta _{\beta, \beta _j}\delta (\theta'+\theta'_j) \label {Jchoice} \, .
\ee
We remark that the $s$-th currents $J^{(s)}_{\alpha}(\theta)$ have support on the straight line $\gamma _s$ defined by the condition
\be
\textrm{Im} \theta  = \hat \varphi _{s} - \frac{\pi}{4} \frac{1+(-1)^{s}}{2} \, ,
\ee
and that the kernels $G^{(s,s')}_{\alpha, \beta}(\theta ,\theta ')$ have support on $\gamma _s \times \gamma _{s'}$.
Thanks to $P_{\alpha, \beta}(-\theta|-\theta')=P_{\beta, \alpha}(\theta'|\theta)$, we obtain
\ba
&& \exp \left [ \frac{1}{2}\sum _{s,s', \alpha, \beta}\int _{\gamma _s}d\theta \int _{\gamma _{s'}}d\theta' J^{(s)}_{\alpha}(\theta) G^{(s,s')}_{\alpha, \beta}(\theta,\theta') J^{(s')}_{\beta }(\theta') \right ]=\exp \Bigl [\frac{1}{2}\sum _{i,j=1}^N a_{\alpha _i} a_{\alpha _j} G^{(1,1)}_{\alpha _i, \alpha _j}(\theta_i,\theta_j) + \nonumber \\
 && + \frac{1}{2}\sum _{i,j=1}^M b_{\beta _i} b_{\beta _j} G^{(2,2)}_{\beta _i, \beta _j}(-\theta'_i,-\theta'_j)- \frac{1}{2}\sum _{i=1}^N \sum _{j=1}^M \Bigl (a_{\alpha _i} b_{\beta _j} G^{(1,2)}_{\alpha _i, \beta _j}(\theta_i,-\theta'_j)+a_{\alpha _i} b_{\beta _j} G^{(2,1)}_{\beta _j, \alpha _i}(-\theta'_j,\theta_i) \Bigr ) \Bigr ] = \nonumber \\
 && = \frac{\prod \limits _{i=1}^N \prod \limits _{j=1}^M [P _{\alpha_i \beta_j}(-\theta_i|\theta'_j)]^{a_{\alpha _i}b_{\beta _j}}}{\prod \limits _{\stackrel {i,j=1}{i\not=j}}^{N} [P_{\alpha_i \alpha_j} (\theta_i|\theta_j)]^{a_{\alpha _i}a_{\alpha _j}} \prod \limits _{\stackrel {i,j=1}{i\not=j}}^{M} [P_{\beta_i \beta_j} (\theta'_i|\theta'_j)]^{b_{\beta _i}b_{\beta _j}}} \, .
\ea
which, using (\ref {identity}) and (\ref {Jchoice}), returns the relation we were looking for
\ba
&& \frac{\prod \limits _{i=1}^N \prod \limits _{j=1}^M [P _{\alpha_i \beta_j}(-\theta_i|\theta'_j)]^{a_{\alpha _i}b_{\beta _j}}}{\prod \limits _{\stackrel {i,j=1}{i\not=j}}^{N} [P_{\alpha_i \alpha_j} (\theta_i|\theta_j)]^{a_{\alpha _i}a_{\alpha _j}} \prod \limits _{\stackrel {i,j=1}{i\not=j}}^{M} [P_{\beta_i \beta_j} (\theta'_i|\theta'_j)]^{b_{\beta _i}b_{\beta _j}}}=
\langle \exp \left [ \sum _{s,\alpha} \int _{\gamma _s}du X_{\alpha}^{(s)}(\theta) J_{\alpha}^{(s)}(\theta) \right ] \rangle = \nonumber \\
&&  = \langle \exp \left [ \sum _{i=1}^N a_{\alpha _i} X_{\alpha _i}^{(1)}(\theta_i)- \sum _{j=1}^M b_{\beta _j} X_{\beta _j}^{(2)}(-\theta'_j) \right ] \rangle 
\ea
Therefore, under functional integration, the integrands in (\ref {Whep}) factorizes in one-particle terms, allowing us to perform the sums over $a_{\alpha _i}$ and $b_{\beta _j}$, eventually getting the dilogarithm
\ba
{\cal W}_{7}&=&\sum _{N=0}^{\infty} \sum _{M=0}^{\infty}\frac{(-1)^{N+M}}{N!M!} \langle \int _{\gamma _1}\prod _{i=1}^N \frac{d\theta_i}{2\pi}\left [ \sum _{\alpha} \mu _{\alpha}(\theta_i) \textrm{Li}_2 \left (-e^{-\tau _1 E_{\alpha}(\theta_i) +i\sigma _1 p_{\alpha} (\theta_i) +i\phi _1(2-\alpha ) +X^{(1)}_{\alpha}(\theta_i)} \right ) \right ] \cdot \nonumber \\
&\cdot & \int _{\gamma _2}\prod _{j=1}^M \frac{d\theta'_j}{2\pi}\left [ \sum _{\beta} \mu _{\beta}(\theta'_j) \textrm{Li}_2 \left (-e^{-\tau _2 E_{\beta}(\theta'_j) +i\sigma _2 p_{\beta} (\theta'_j) +i\phi _2(2-\beta ) -X^{(2)}_{\beta}(-\theta'_j)} \right )\right ] \rangle = \\
&=& \langle \exp \left [ - \int _{\gamma _1}\frac{d\theta}{2\pi} \left [ \sum  _{\alpha} \mu _{\alpha}(\theta) \textrm{Li}_2 \left (-e^{-\tau _1 E_{\alpha}(\theta) +i\sigma _1 p_{\alpha} (\theta) +i\phi _1(2-\alpha ) +X^{(1)}_{\alpha}(\theta)} \right ) \right ] \right ] \cdot \nonumber \\
&\cdot & \exp \left [ - \int _{\gamma _2}\frac{d\theta'}{2\pi} \left [ \sum  _{\beta} \mu _{\beta}(\theta') \textrm{Li}_2 \left (-e^{-\tau _2 E_{\beta}(\theta') -i\sigma _2 p_{\beta} (\theta') +i\phi _2(2-\beta) -X^{(2)}_{\beta}(\theta')} \right ) \right ] \right ] \rangle
\ea
This can be compactly written as a partition function
\be
{\cal W}_{7}=\int  \prod  _{\alpha , \beta =1}^3 {\cal D}X^{(1)}_{\alpha } {\cal D}X^{(2)}_{\beta } e^{-S[X^{(1)},X^{(2)}]} \, ,
\label {WXX}
\ee
with the action
\ba
S[X^{(1)},X^{(2)}]&=&\frac{1}{2} \sum _{s,s',\alpha ,\beta} \int _{\gamma _{s}} d\theta \int _{\gamma _{s'}} d\theta '  X^{(s)}_{\alpha}(\theta ) T^{(s,s')}_{\alpha, \beta}(\theta , \theta ') X^{(s')}_{\beta}(\theta ') +  \\
&+& \sum _{s,\alpha} \int _{\gamma _{s}} \frac{d\theta}{2\pi} \mu _{\alpha}(\theta) \textrm{Li}_2 \left ( -e^{-\tau _s E_{\alpha}(\theta) + i(-1)^{s+1}\sigma _s p _{\alpha}(\theta) +i\phi _s (2-\alpha)+ (-1)^{s+1}X^{(s)}_{\alpha}(\theta )} \right ) \nonumber \label {SXX} 
\ea
Notice that $S$ is proportional to the large coupling $\sqrt{\lambda}$. Thus, the customary saddle point technique gives the 'equation of motion' for $X^{(s)}_{\alpha}$ as extremisation of (\ref {SXX}):
\ba
&& X^{(s)}_{\alpha}(\theta)+ \sum _{s'=1}^2 \sum _{\alpha '=1}^3 (-1)^{s'} \cdot \label {Xeq}\\
&\cdot & \int _{\gamma _{s'}}\frac{d\theta '}{2\pi} \mu _{\alpha '}(\theta ') G^{(s,s')}_{\alpha, \alpha '}(\theta , \theta ')
\ln \left [ 1+e^{-\tau _{s'} E_{\alpha '}(\theta ' ) +i(-1)^{s'-1}\sigma _{s'} p_{\alpha '}(\theta ')+i\phi _{s'}(2-\alpha ') +(-1)^{s'+1}X^{(s')}_{\alpha '}(\theta ')} \right ]=0 \nonumber \, .
\ea

which, by defining the pseudoenergies according to
\be
\varepsilon ^{(s)}_{\alpha} \left ( \theta -i\hat \varphi _s +i\frac{\pi}{4}\frac{1+(-1)^s}{2} \right ) = \tau _s E_{\alpha }(\theta ) -i (-1)^{s-1}\sigma _s p_{\alpha }(\theta ) +(-1)^s X^{(s)}_{\alpha }(\theta ) \, ,
\ee
turn into the TBA setup
\small
\ba
&&\varepsilon ^{(s)}_{\alpha } ( \theta -i\hat \varphi _s ) = \tau _s E_{\alpha} \left (\theta -i\frac{\pi}{4}\frac{1+(-1)^s}{2} \right ) -i (-1)^{s-1}\sigma _s p_{\alpha }\left (\theta - i\frac{\pi}{4}\frac{1+(-1)^s}{2} \right )  - \nonumber \\
&& - \sum _{s'=1}^{2} \sum _{\alpha '=1}^3(-1)^{s+s'}\int _{\textrm{Im} \theta '=\hat \varphi _{s'}} d\theta ' \frac{\mu _{\alpha '}\left (\theta '-i\frac{\pi}{4}\frac{1+(-1)^{s'}}{2} \right )}{2\pi} G^{(s,s')}_{\alpha , \alpha '}\left (\theta -i\frac{\pi}{4}\frac{1+(-1)^s}{2}  , \theta '-i\frac{\pi}{4}\frac{1+(-1)^{s'}}{2} \right) \cdot \nonumber \\
 &&\cdot  \ln \left ( 1+e^{-\varepsilon ^{(s')}_{\alpha '} \left ( \theta '-i\hat \varphi _{s'} \right )}e^{i\phi _{s'}(2-\alpha ')} \right ) \, .
\ea
\normalsize
Finally, the saddle point yields ${\cal W}_{7}=\exp (-S )$, with the critical action $S$ given by
\small
\ba
S&=&\frac{1}{2} \sum _{s,s'=1}^2 \sum _{\alpha, \alpha '=1}^3 \int _{\textrm{Im} \theta =\hat \varphi _s} \frac{d\theta}{2\pi}
\int _{\textrm{Im} \theta '=\hat \varphi _s'} \frac{d\theta '}{2\pi} (-1)^{s+s'}\mu _{\alpha} \left ( \theta - \frac{i\pi}{4}\frac{1+(-1)^s}{2} \right )  \mu _{\alpha '} \left ( \theta ' - \frac{i\pi}{4}\frac{1+(-1)^{s'} }{2} \right ) \cdot \nonumber \\
&\cdot & G^{(s,s')}_{\alpha , \alpha '}\left (\theta -i\frac{\pi}{4}\frac{1+(-1)^s}{2}  , \theta '-i\frac{\pi}{4}\frac{1+(-1)^{s'}}{2} \right) \cdot \nonumber \\
&\cdot & \ln \left ( 1+e^{-\varepsilon ^{(s)}_{\alpha } \left ( \theta -i\hat \varphi _{s} \right )}e^{i\phi _{s}(2-\alpha )} \right )\ln \left ( 1+e^{-\varepsilon ^{(s')}_{\alpha '} \left ( \theta '-i\hat \varphi _{s'} \right )}e^{i\phi _{s'}(2-\alpha ')} \right ) + \nonumber \\
&+& \sum _{s=1}^2 \sum _{\alpha=1}^3 \int _{\textrm{Im} \theta =\hat \varphi _s} \frac{d\theta}{2\pi} \mu _{\alpha}\left ( \theta - \frac{i\pi}{4}\frac{1+(-1)^s}{2} \right ) \textrm{Li}_2 \left ( -e^{-\varepsilon ^{(s)}_{\alpha } \left ( \theta -i\hat \varphi _{s} \right )+i\phi _s (2-\alpha) }\right ) \, . \label {YY-7}
\ea
\normalsize

\paragraph{General polygon $n>7$}

For the $n$-gon, we have to deal with the product of transitions

\footnotesize
\ba
&& P_{\vec{A}^{(1)}}(0|\theta_1^{(1)},\ldots , \theta_{N^{(1)}}^{(1)}) P_{\vec{A}^{(1)}\vec{A}^{(2)}}(-\theta_{N^{(1)}}^{(1)},\ldots, -\theta_1^{(1)}|\theta_1^{(2)},\ldots, \theta_{N^{(2)}}^{(2)}) P_{\vec{A}^{(2)}\vec{A}^{(3)}}(-\theta_{N^{(2)}}^{(2)},\ldots, -\theta_1^{(2)}|\theta_1^{(3)},\ldots, \theta_{N^{(3)}}^{(3)})\ldots  \nonumber \\
&&\ldots P_{\vec{A}^{(n-6)}\vec{A}^{(n-5)}}(-\theta_{N^{(n-6)}}^{(n-6)},\ldots, -\theta_1^{(n-6)}|\theta_1^{(n-5)},\ldots, \theta_{N^{(n-5)}}^{(n-5)}) P_{\vec{A}^{(n-5)}}(-\theta_{N^{(n-5)}}^{(n-5)},\ldots, -\theta_1^{(n-5)}|0) \, .
\ea
\normalsize

which, once applying the factorisation (\ref {P-singl}), becomes

\footnotesize
\ba
&& P_{\vec{A}^{(1)}}(0|\theta_1^{(1)},\ldots , \theta_{N^{(1)}}^{(1)}) P_{\vec{A}^{(1)}\vec{A}^{(2)}}(-\theta_{N^{(1)}}^{(1)},\ldots, -\theta_1^{(1)}|\theta_1^{(2)},\ldots, \theta_{N^{(2)}}^{(2)}) P_{\vec{A}^{(2)}\vec{A}^{(3)}}(-\theta_{N^{(2)}}^{(2)},\ldots, -\theta_1^{(2)}|\theta_1^{(3)},\ldots, \theta_{N^{(3)}}^{(3)})\ldots  \nonumber \\
&&\ldots P_{\vec{A}^{(n-6)}\vec{A}^{(n-5)}}(-\theta_{N^{(n-6)}}^{(n-6)},\ldots, -\theta_1^{(n-6)}|\theta_1^{(n-5)},\ldots, \theta_{N^{(n-5)}}^{(n-5)}) P_{\vec{A}^{(n-5)}}(-\theta_{N^{(n-5)}}^{(n-5)},\ldots, -\theta_1^{(n-5)}|0) \label {P-prop}\\
&& = \frac{\prod \limits _{s=1}^{n-6}\prod \limits _{i^{(s)}=1}^{N^{(s)}} \prod \limits _{i^{(s+1)}=1}^{N^{(s+1)}} P_{A_{i^{(s)}}^{(s)}A_{i^{(s+1)}}^{(s+1)}}(-\theta_{i^{(s)}}^{(s)}|\theta_{i^{(s+1)}}^{(s+1)})}{\prod \limits _{s=1}^{n-5} \prod \limits _{\stackrel {i^{(s)},j^{(s)}=1}{i^{(s)}\not=j^{(s)}}}^{N^{(s)}} P_{A_{i^{(s)}}^{(s)}A_{j^{(s)}}^{(s)}}(\theta_{i^{(s)}}^{(s)}|\theta_{j^{(s)}}^{(s)}) } =
\frac{\prod \limits _{s=1}^{n-6}\prod \limits _{i^{(s)}=1}^{N^{(s)}} \prod \limits _{i^{(s+1)}=1}^{N^{(s+1)}} \left [ P_{\alpha ^{(s)}_{i^{(s)}} \alpha ^{(s+1)}_{i^{(s+1)}}}(-\theta_{i^{(s)}}^{(s)}|\theta_{i^{(s+1)}}^{(s+1)})\right ]^{a^{(s)}_{\alpha _i^{(s)}}a^{(s+1)}_{\alpha _i^{(s+1)}}}}{\prod \limits _{s=1}^{n-5} \prod \limits _{i^{(s)},j^{(s)}=1}^{N^{(s)}} \left [ P_{\alpha ^{(s)}_{i^{(s)}}\alpha ^{(s)}_{j^{(s)}} }(\theta_{i^{(s)}}^{(s)}|\theta_{j^{(s)}}^{(s)})\right]^{a^{(s)}_{\alpha _i^{(s)}} a^{(s)}_{\alpha _j^{(s)}}}}
\nonumber
\ea
\normalsize

in which the strong coupling approximation (\ref {P-strong}) has been employed.

As we did in the hexagon case, in view of introducing the functional integration, we define the currents
\be
J^{(s)}_{\alpha}(\theta )=(-1)^{s+1}\sum _{i=1}^{N^{(s)}}a^{(s)}_{\alpha _i^{(s)}}\delta _{\alpha, \alpha _i^{(s)}}\delta \left (\theta+(-1)^s \theta _i^{(s)}\right ) \, , \quad s=1,...,n-5 \, .
\ee
where the $s$-th current is supported on the straight line $\gamma _s$ such that
\be
\textrm{Im} \theta  = \hat \varphi _{s} - \frac{\pi}{4} \frac{1+(-1)^{s}}{2} \, .
\ee
The propagators $G^{(s,s')}_{\alpha, \beta}(\theta ,\theta ')$, living on $\gamma _s \times \gamma _{s'}$, are defined as
\ba\label{green-kernel}
G^{(s,s)}_{\alpha, \beta}(\theta ,\theta ') &=& -\ln [P_{\alpha, \beta}(\theta |\theta ')P_{\beta ,\alpha}(\theta '|\theta )] \, , \quad s=1,...,n-5 \, , \nn\\
G^{(s,s+1)}_{\alpha, \beta}(\theta ,\theta ') &=&-\ln P _{\alpha, \beta}\left((-1)^s \theta  | (-1)^s \theta ' \right ) \, , \quad s=1,...,n-6 \, , \\
G^{(s,s-1)}_{\alpha, \beta}(\theta ,\theta ') &=& -\ln P _{\alpha, \beta}\left ((-1)^s \theta  | (-1)^s \theta ' \right ) \, , \quad s=2,...,n-5 \, , \nn
\ea
where all the other 'matrix' elements are vanishing. Then, for a polygon with $n$ edges, the expectation value

\ba
&& \exp \left [ \frac{1}{2} \sum _{s,s'=1}^{n-5} \sum _{\alpha ^{(s), \alpha ^{(s')}}=1}^3 \int _{\gamma _s}d\theta  \int _{\gamma _{s'}}d\theta ' J^{(s)}_{\alpha ^{(s)}}(\theta ) G^{(s,s')}_{\alpha ^{(s)}, \alpha ^{(s')}}(\theta ,\theta ') J^{(s')}_{\alpha ^{(s')}}(\theta ') \right ]= \nn \\
&& =\langle \exp \left [ \sum _{s=1}^{n-5}\sum _{\alpha ^{(s)}=1}^{3}\int _{\gamma _s}d\theta  J^{(s)}_{\alpha ^{(s)}}(\theta ) X ^{(s)}_{\alpha ^{(s)}} (\theta ) \right ] \rangle
\ea

equals
\be
\frac{\prod \limits _{s=1}^{n-6}\prod \limits _{i^{(s)}=1}^{N^{(s)}} \prod \limits _{i^{(s+1)}=1}^{N^{(s+1)}} \left [ P_{\alpha ^{(s)}_{i^{(s)}} \alpha ^{(s+1)}_{i^{(s+1)}}}(-\theta_{i^{(s)}}^{(s)}|\theta_{i^{(s+1)}}^{(s+1)})\right ]^{a^{(s)}_{\alpha _i^{(s)}}a^{(s+1)}_{\alpha _i^{(s+1)}}}}{\prod \limits _{s=1}^{n-5} \prod \limits _{i^{(s)},j^{(s)}=1}^{N^{(s)}} \left [ P_{\alpha ^{(s)}_{i^{(s)}}\alpha ^{(s)}_{j^{(s)}} }(\theta_{i^{(s)}}^{(s)}|\theta_{j^{(s)}}^{(s)})\right]^{a^{(s)}_{\alpha _i^{(s)}} a^{(s)}_{\alpha _j^{(s)}}}}
\nonumber \, ,
\ee
i.e. it coincides with (\ref {P-prop}).

Now, generalising the summation described before for the heptagon, we get
\be
{\cal W}_{n}= \langle \exp \left [ -\sum _{s=1}^{n-5}\sum _{\alpha ^{(s)}=1}^3 \int _{\gamma _s}\frac{d\theta}{2\pi} \mu _{\alpha ^{(s)}} (\theta)  \textrm{Li}_2 \left ( -e^{-\tau _s E(\theta) +
i\sigma _s p(\theta) +i\phi _s (2-\alpha ^{(s)})+ (-1)^{s+1}X^{(s)}_{\alpha ^{(s)}}((-1)^{s+1}\theta )} \right ) \right ] \rangle \, ,
\ee
which is a partition function
\be
{\cal W}_n=\int \prod _{s=1}^{n-5} \prod  _{\alpha ^{(s)}=1}^3 {\cal D}X^{(s)}_{\alpha ^{(s)}} e^{-S[X^{(1)}...X^{(n-5)}]} \, ,
\ee
where the action is proportional to $\sqrt{\lambda}$ and given by

\small
\ba
S[X^{(1)}...X^{(n-5)}]&=&\frac{1}{2} \sum _{s,s'=1}^{n-5} \sum _{\alpha ^{(s)}, \alpha ^{(s')}=1}^3\int _{\gamma _{s}} d\theta \int _{\gamma _{s'}} d\theta '  X^{(s)}_{\alpha ^{(s)}}(\theta ) T^{(s,s')}_{\alpha ^{(s)}, \alpha ^{(s')}}(\theta , \theta ') X^{(s')}_{\alpha ^{(s')}}(\theta ') + \label {act} \\
&+& \sum _{s=1}^{n-5} \sum _{\alpha ^{(s)}=1}^3\int _{\gamma _{s}} \frac{d\theta}{2\pi} \mu _{\alpha ^{(s)}}(\theta) \textrm{Li}_2 \left ( -e^{-\tau _s E_{\alpha ^{(s)}}(\theta) +
i(-1)^{s+1}\sigma _s p_{\alpha ^{(s)}}(\theta) +  i \phi _s (2-\alpha ^{(s)})+(-1)^{s+1}X^{(s)}_{\alpha ^{(s)}}(\theta )} \right ) \nonumber \, .
\ea
\normalsize

The saddle point equations for $X^{(s)}_{\alpha ^{(s)}}$, descending from the minimisation of the functional $S[X^{(1)}...X^{(n-5)}]$,
are
\ba
\label{eq_moto}
&& X^{(s)}_{\alpha ^{(s)}}(\theta)+ \sum _{s'=1}^{n-5} \sum _{\alpha ^{(s')}=1}^3(-1)^{s'} \int _{\gamma _{s'}}\frac{d\theta '}{2\pi} \mu _{\alpha ^{(s')}}(\theta ') G^{(s,s')}_{\alpha ^{(s)}\alpha ^{(s')}}(\theta , \theta ')
\cdot \\
&\cdot &
\ln \left [ 1+e^{-\tau _{s'} E_{\alpha ^{(s')}}(\theta ' ) +i(-1)^{s'-1}\sigma _{s'}p_{\alpha ^{(s')}}(\theta ') +i\phi _{s'}(2-\alpha ^{(s')})+(-1)^{s'-1}X^{(s')}_{\alpha ^{(s')}}(\theta ')} \right ]=0 \nonumber \, .
\ea
Again, we use the pseudoenergy
\be
\varepsilon ^{(s)}_{\alpha} \left ( \theta -i\hat \varphi _s +i\frac{\pi}{4}\frac{1+(-1)^s}{2} \right ) = \tau _s E_{\alpha }(\theta ) -i (-1)^{s-1}\sigma _s p_{\alpha }(\theta ) +(-1)^s X^{(s)}_{\alpha }(\theta ) \, .
\ee
whose equation of motion reads

\small
\ba
&&\varepsilon ^{(s)}_{\alpha ^{(s)}} \left ( \theta -i\hat \varphi _s +i\frac{\pi}{4}\frac{1+(-1)^s}{2} \right ) = \tau _s E_{\alpha ^{(s)}}(\theta ) -i (-1)^{s-1}\sigma _s p_{\alpha^{(s)} }(\theta ) -  \\
&& - \sum _{s'=1}^{n-5} \sum _{\alpha ^{(s')}=1}^3(-1)^{s+s'}\int _{\gamma _{s'}} d\theta ' \frac{\mu _{\alpha ^{(s')}}(\theta ')}{2\pi} G^{(s,s')}_{\alpha ^{(s)}, \alpha ^{(s')}}(\theta , \theta ')  \ln \left ( 1+e^{-\varepsilon ^{(s')}_{\alpha ^{(s')}} \left ( \theta '-i\hat \varphi _{s'} +i\frac{\pi}{4}\frac{1+(-1)^{s'}}{2} \right )}e^{i\phi _{s'}(2-\alpha^{(s')})}  \right ) \label {eq-m} \nonumber \, .
\ea
\normalsize

or, alternatively

\footnotesize
\ba \label{epsbsv}
&&\varepsilon ^{(s)}_{\alpha ^{(s)}} ( \theta -i\hat \varphi _s ) = \tau _s E_{\alpha^{(s)} } \left (\theta -i\frac{\pi}{4}\frac{1+(-1)^s}{2} \right ) -i (-1)^{s-1}\sigma _s p_{\alpha^{(s)} }\left (\theta - i\frac{\pi}{4}\frac{1+(-1)^s}{2} \right )  - \nonumber \\
&& - \sum _{s'=1}^{n-5} \sum _{\alpha ^{(s')}=1}^3(-1)^{s+s'}\int _{\textrm{Im} \theta '=\hat \varphi _{s'}} d\theta ' \frac{\mu _{\alpha ^{(s')}}\left (\theta '-i\frac{\pi}{4}\frac{1+(-1)^{s'}}{2} \right )}{2\pi} G^{(s,s')}_{\alpha ^{(s)}, \alpha ^{(s')}}\left (\theta -i\frac{\pi}{4}\frac{1+(-1)^s}{2}  , \theta '-i\frac{\pi}{4}\frac{1+(-1)^{s'}}{2} \right) \cdot \nonumber \\
 &&\cdot  \ln \left ( 1+e^{-\varepsilon ^{(s')}_{\alpha ^{(s')}} \left ( \theta '-i\hat \varphi _{s'} \right )}e^{i\phi _{s'}(2-\alpha^{(s')})} \right )
\ea
\normalsize

\medskip

Finally, for the Wilson loop with $n$ edges we have the obvious extension of (\ref {YY-7}) we gave in the heptagon case: ${\cal W}_{n}=\exp (-S )$, where $S$ is the critical action (\ref {act}):
\small
\ba
S&=&\frac{1}{2} \sum _{s,s'=1}^{n-5} \sum _{\alpha, \alpha '=1}^3 \int _{\textrm{Im} \theta =\hat \varphi _s} \frac{d\theta}{2\pi}
\int _{\textrm{Im} \theta '=\hat \varphi _s'} \frac{d\theta '}{2\pi} (-1)^{s+s'}\mu _{\alpha} \left ( \theta - \frac{i\pi}{4}\frac{1+(-1)^s}{2} \right )  \mu _{\alpha '} \left ( \theta ' - \frac{i\pi}{4}\frac{1+(-1)^{s'} }{2} \right ) \cdot \nonumber \\
&\cdot & G^{(s,s')}_{\alpha , \alpha '}\left (\theta -i\frac{\pi}{4}\frac{1+(-1)^s}{2}  , \theta '-i\frac{\pi}{4}\frac{1+(-1)^{s'}}{2} \right) \cdot \nonumber \\
&\cdot & \ln \left ( 1+e^{-\varepsilon ^{(s)}_{\alpha } \left ( \theta -i\hat \varphi _{s} \right )}e^{i\phi _{s}(2-\alpha )} \right )\ln \left ( 1+e^{-\varepsilon ^{(s')}_{\alpha '} \left ( \theta '-i\hat \varphi _{s'} \right )}e^{i\phi _{s'}(2-\alpha ')} \right ) + \nonumber \\
&+& \sum _{s=1}^{n-5}\sum _{\alpha=1}^3 \int _{\textrm{Im} \theta =\hat \varphi _s} \frac{d\theta}{2\pi} \mu _{\alpha}\left ( \theta - \frac{i\pi}{4}\frac{1+(-1)^s}{2} \right ) \textrm{Li}_2 \left ( -e^{-\varepsilon ^{(s)}_{\alpha } \left ( \theta -i\hat \varphi _{s} \right )+i\phi _s (2-\alpha) }\right ) \, . \label {YY-n}
\ea
\normalsize

\subsection{Comparisons}

In this part we compare the results of the re-summation with the classical string minimisation exposed in Appendix \ref{TBApp}. For what concerns crossing ratios, we identify
\be
\ln y_{2,s}=-2\tau _s \, , \quad  \ln\frac{y_{2,s}}{y_{1,s}y_{3,s}}=2\sigma _s \label {cross1} \, ,
\ee
\be
e^{i\phi _{s}}=\sqrt{\frac{y_{1,s}}{y_{3,s}}} \, , \quad s=4k+1, 4k+2 \, ;
\ e^{i\phi _{s}}=\sqrt{\frac{y_{3,s}}{y_{1,s}}} \, , \quad s=4k+3, 4k+4 \label {cross2} \, .
\ee
of which the first one brings ${\cal E}_s(\theta )$ to the form
\be
{\cal E}_s(\theta )= -2\tau _s \cosh \theta + 2i (-1)^{s-1}\sigma _s \sinh \theta \, .
\ee
The measure reads
\be
\mu _{\alpha} \left (\theta - \frac{i\pi}{4}\frac{1+(-1)^s}{2} \right )= \frac{\sqrt{\lambda}}{2\pi}\frac{2}{\sinh ^2 \left [2\theta -\frac{i\pi}{2}\frac{1+(-1)^{\alpha +s}}{2} \right ]} \, , \label {mu-sh}
\ee
and comparing the kernels (\ref {green-kernel}) $G^{(s,s')}_{\alpha ^{(s)}, \alpha ^{(s')}}$ with respect to expressions (\ref {tildeK}) of the tilded kernels $\tilde K$ and intertwining, keeping $\epsilon_{2,s}=\we^{(s)}_2$, the pseudorapidities

\ba\label{identificazione_eps}
&& \epsilon_{1,4k+1}=\we^{(4k+1)}_1 		\qquad\quad  \epsilon_{3,4k+1}=\we^{(4k+1)}_3 \nn\\
&& \epsilon_{1,4k+2}=\we^{(4k+2)}_1 		\qquad\quad  \epsilon_{3,4k+2}=\we^{(4k+2)}_3  \\
&& \epsilon_{1,4k+3}=\we^{(4k+3)}_3 		\qquad\quad  \epsilon_{3,4k+3}=\we^{(4k+3)}_1 \nn\\
&& \epsilon_{1,4k+4}=\we^{(4k+4)}_3 		\qquad\quad  \epsilon_{3,4k+4}=\we^{(4k+4)}_1 \nn
\ea

one shows that equations (\ref{epsbsv}) coincide with (\ref{tbaeps1},\ref{tbaeps2},\ref{tbaeps3}).

In a similar way, the extremal action (\ref {YY-n}) coincides with (\ref {YYc2}) times a factor $\frac{\sqrt{\lambda}}{2\pi}$, i.e. we have the equality
\be
S=\frac{\sqrt{\lambda}}{2\pi} YY_{c} \, .
\ee
We point out that the redefinitions of pseudorapidities were not needed for the hexagon and the heptagon and they become effective only for polygons with eight or more edges. Thus the OPE resummation, through the meson hypothesis, successfully reconstruct the TBA system for the Wls/amplitudes computed by the classical string, for any polygon.

\paragraph{Y-system}

In the previous paragraph we compared successfully the TBA equations for the amplitudes in appendix \ref{TBApp} to the relations (\ref{epsbsv}), resulting from the resummation of the OPE series. The interest now turns to the formulation, directly from the TBA equations (\ref{epsbsv}), of the corresponding set of functional equation or $Y$-system, in order to show the agreement with the scattering amplitude $Y$-system \cite{YSA}, which reads

\ba\label{YSA}
&& \frac{Y^-_{3,s}Y^+_{1,s}}{Y_{2,s}}=\frac{(1+Y_{3,s+1})(1+Y_{1,s-1})}{1+Y_{2,s}} \nn\\
&& \frac{Y^+_{2,s}Y^-_{2,s}}{Y_{1,s}Y_{3,s}}=\frac{(1+Y_{2,s-1})(1+Y_{2,s+1})}{(1+Y_{1,s})(1+Y_{3,s})}\\
&& \frac{Y^-_{1,s}Y^+_{3,s}}{Y_{2,s}}=\frac{(1+Y_{1,s+1})(1+Y_{3,s-1})}{1+Y_{2,s}}  \nn \ ,
\ea

We remark an unusual feature of the Y-system above, namely its crossed nature, which means the simultaneous presence in the LHS of two different functions (nodes) $Y_{a,s}$, in particular in the first and third equation. 


To begin with, we introduce the functions $\hat W_{a,s}$ in order to recast the equations of motion (\ref{epsbsv}) into a form resembling (\ref{hatTBA1})-(\ref{hatTBA2})
\ba
\we^{(s)}_{1}(\theta -i \hat \varphi _s)&=&-\ln \hat W_{1,s}\left (\theta - \frac{i\pi}{4}\frac{1+(-1)^{s+1}}{2} \right ) +i\phi_s \\
\we^{(s)}_{3}(\theta -i \hat \varphi _s)&=&-\ln \hat W_{3,s}\left (\theta - \frac{i\pi}{4}\frac{1+(-1)^{s+1}}{2} \right ) -i\phi_s \\
\we^{(s)}_{2}(\theta -i \hat \varphi _s)&=&-\ln \hat W_{2,s}\left (\theta - \frac{i\pi}{4}\frac{1+(-1)^{s}}{2} \right ) \ ,
\ea
and also the hatted version
\be
\hat W_{a,s}(\theta )= W_{a,s} \left ( \theta - \frac{i\pi}{4} \left [ \frac{1-(-1)^{a+s}}{2} \right ] \right ) \ ,
\ee
so that, thanks to the relations listed in Appendix \ref{ApSca} , the equations (\ref{epsbsv}) become

\footnotesize
\ba\label{TBA_hat1}
&& \ln \hat W_{2,s} (\theta )-{\cal E}_s(\theta)=-\int_{\textrm{Im} \theta ' =\varphi _{s}} d\theta' \biggl[\tilde K_2^{(s)}\left(\theta, \theta' + \frac{i\pi}{4}\frac{1-(-1)^{s+1}}{2}\right)\Lambda^+_s (\theta') + \nonumber \\
&+& 2\tilde K_1\left(\theta,\theta' + \frac{i\pi}{4}\frac{1-(-1)^s}{2}\right)\Lambda^0_s(\theta')\biggr] +  \int_{\textrm{Im} \theta ' =\varphi _{s-1}} d\theta ' \biggl[\tilde K_1\left(\theta, \theta '+ \frac{i\pi}{4}\frac{1-(-1)^s}{2}\right) \Lambda^+_{s-1}(\theta ') + \nonumber \\
&+& \tilde K_2^{(s)}\left(\theta, \theta' + \frac{i\pi}{4}\frac{1-(-1)^{s+1}}{2}\right) \Lambda^0_{s-1}(\theta')\biggr]  + \int_{\textrm{Im} \theta ' =\varphi _{s+1}} d\theta ' \biggl[\tilde K_1\left(\theta, \theta '+ \frac{i\pi}{4}\frac{1-(-1)^s}{2}\right) \Lambda^+_{s+1}(\theta ') + \nonumber \\
&+& \tilde K_2^{(s)}\left(\theta, \theta' + \frac{i\pi}{4}\frac{1-(-1)^{s+1}}{2}\right) \Lambda^0_{s+1}(\theta')\biggr] \, ,
\ea
\ba
&& \ln \hat W_{1,s} (\theta )+ \ln \hat W_{3,s} (\theta )-\sqrt{2}{\cal E}_s\left (\theta + \frac{i\pi}{4}(-1)^{s+1} \right ) = -\int_{\textrm{Im} \theta ' =\varphi _{s}} d\theta' \biggl[2\tilde K_2^{(s)}\left(\theta, \theta' + \frac{i\pi}{4}\frac{1-(-1)^{s}}{2}\right)\Lambda^0_s (\theta') + \nonumber \\
&+& 2\tilde K_1\left(\theta,\theta' + \frac{i\pi}{4}\frac{1-(-1)^{s+1}}{2}\right)\Lambda^+_s(\theta')\biggr] +  \int_{\textrm{Im} \theta ' =\varphi _{s-1}} d\theta ' \biggl[\tilde K^{(s)}_2\left(\theta, \theta '+ \frac{i\pi}{4}\frac{1-(-1)^s}{2}\right) \Lambda^+_{s-1}(\theta ') + \nonumber \\
&+& 2\tilde K_1\left(\theta, \theta' + \frac{i\pi}{4}\frac{1-(-1)^{s+1}}{2}\right) \Lambda^0_{s-1}(\theta')\biggr] + \int_{\textrm{Im} \theta ' =\varphi _{s+1}} d\theta ' \biggl[\tilde K^{(s)}_2\left(\theta, \theta '+ \frac{i\pi}{4}\frac{1-(-1)^s}{2}\right) \Lambda^+_{s+1}(\theta ') + \nonumber \\
&+& 2\tilde K_1\left(\theta, \theta' + \frac{i\pi}{4}\frac{1-(-1)^{s+1}}{2}\right) \Lambda^0_{s+1}(\theta')\biggr] \, , 
\ea

\ba\label{TBA_hat3}
\ln \hat W_{1,s} (\theta )- \ln \hat W_{3,s} (\theta ) -2i\phi_s
=(-1)^{s+1}\Biggl[\int_{\textrm{Im} \theta ' =\varphi _{s-1}} d\theta' \tilde K_3 \left(\theta , \theta'+ \frac{i\pi}{4}\frac{1-(-1)^s}{2}\right) \Lambda^-_{s-1}(\theta') + \nonumber \\
+ \int_{\textrm{Im} \theta ' =\varphi _{s+1}} d\theta' \tilde K_3 \left(\theta , \theta'+ \frac{i\pi}{4}\frac{1-(-1)^s}{2}\right) \Lambda^-_{s+1}(\theta')\Biggr] \, ,
\ea
\normalsize

where the functions $\Lambda_s^{\pm}(\q)$ and $\Lambda^0_s(\q)$ stand for

\small
\be
\Lambda^+_{s}(\theta)=\ln(1+W_{1,s}(\theta))(1+W_{3,s}(\theta)) \, , \quad \Lambda^0_{s}(\theta)=\ln(1+W_{2,s}(\theta)) \, , \quad \Lambda^-_{s}(\theta)=\ln\frac{(1+W_{1,s}(\theta))}{(1+W_{3,s}(\theta))} 
\ee
\normalsize



The hatted functions $\hat W_{a,s}$ computed in $\theta=0$ are linked  to the cross ratios $\tau_s$, $\sigma_s$ and $\phi_s$ through a set of equations analogous to (\ref{cross1}) and (\ref{cross2}), which are

\be
\ln \hat W_{2,s}(0)=-2\tau _s \, , \quad  \ln\frac{\hat W_{2,s}(0)}{\hat W_{1,s}(0)\hat W_{3,s}(0)}=2\sigma _s  \, , \quad
e^{i\phi _{s}}=\sqrt{\frac{\hat W_{1,s}(0)}{\hat W_{3,s}(0)}}  \, .
\ee

To ease  our task, we decide to rewrite equations (\ref{TBA_hat1})-(\ref{TBA_hat3}) in terms of the relativistic kernels $K_i$ and the parameters $m_s, C_s, \varphi _s$ appearing in (\ref{eq1})-(\ref{eq3}):
\small
\ba\label{TBA_rel1}
\ln W_{2,s}(\theta)&=& - |m_s| \sqrt{2} \cosh (\theta - i \varphi _s) -\int_{\textrm{Im} \theta ' =\varphi _{s}} d\theta'\biggl[K_2(\theta-\theta')
\Lambda^+_{s}(\theta) + \nonumber \\
&+& 2K_1(\theta-\theta')\Lambda^0_{s}(\theta')\biggr] + \int_{\textrm{Im} \theta ' =\varphi _{s-1}} d\theta'\biggl[K_2(\theta-\theta')
\Lambda^0_{s-1}(\theta') + \nonumber \\
&+& K_1(\theta-\theta')\Lambda^+_{s-1}(\theta')\biggr] + \int_{\textrm{Im} \theta ' =\varphi _{s+1}} d\theta'\biggl[K_2(\theta-\theta')
\Lambda^0_{s+1}(\theta') + \nonumber \\
&+& K_1(\theta-\theta')\Lambda^+_{s-1}(\theta')\biggr] \, , \\
\ln W_{1,s}(\theta)&=& - |m_s| \cosh (\theta - i \varphi _s) -C_s \left(\sin \frac{\pi s}{2} -\cos \frac{\pi s}{2}\right ) -\int_{\textrm{Im} \theta ' =\varphi _{s}} d\theta'\biggl[K_2(\theta-\theta')
\Lambda^0_{s}(\theta') + \nonumber \\
&+& K_1(\theta-\theta')\Lambda^+_{s}(\theta')\biggr] + \int_{\textrm{Im} \theta ' =\varphi _{s-1}}
d\theta'\biggl[K_1(\theta-\theta')\Lambda^0_{s-1}(\theta') + \nonumber \\
&+& \frac{1}{2}K_2(\theta-\theta')\Lambda^+_{s-1}(\theta') + (-1)^{s+1}\frac{1}{2}K_3(\theta -\theta')\Lambda^-_{s-1}(\theta')\biggr] + \nonumber \\
&+&  \int_{\textrm{Im} \theta ' =\varphi _{s+1}} d\theta'\biggl[K_1(\theta-\theta')\Lambda^0_{s+1}(\theta') + \frac{1}{2}K_2(\theta-\theta')\Lambda^+_{s+1}(\theta') + \nonumber \\
&+& (-1)^{s+1}\frac{1}{2}K_3(\theta -\theta')\Lambda^-_{s+1}(\theta')\biggr] \, ,
\ea
\ba\label{TBA_rel3}
\ln W_{3,s}(\theta)&=& - |m_s| \cosh (\theta - i \varphi _s) +C_s \left(\sin \frac{\pi s}{2} -\cos \frac{\pi s}{2}\right ) -\int_{\textrm{Im} \theta ' =\varphi _{s}} d\theta'\biggl[K_2(\theta-\theta')
\Lambda^0_{s}(\theta') + \nonumber \\
&+& K_1(\theta-\theta')\Lambda^+_{s}(\theta')\biggr] + \int_{\textrm{Im} \theta ' =\varphi _{s-1}}
d\theta'\biggl[K_1(\theta-\theta')\Lambda^0_{s-1}(\theta') + \nonumber \\
&+& \frac{1}{2}K_2(\theta-\theta')\Lambda^+_{s-1}(\theta') -(-1)^{s+1} \frac{1}{2}K_3(\theta -\theta')\Lambda^-_{s-1}(\theta')\biggr] + \nonumber \\
&+&  \int_{\textrm{Im} \theta ' =\varphi _{s+1}} d\theta'\biggl[K_1(\theta-\theta')\Lambda^0_{s+1}(\theta') + \frac{1}{2}K_2(\theta-\theta')\Lambda^+_{s+1}(\theta') - \nonumber \\
&-& (-1)^{s+1}\frac{1}{2}K_3(\theta -\theta')\Lambda^-_{s+1}(\theta')\biggr] \ .
\ea
\normalsize
We point out that all the differences between (\ref{TBA_rel1})-(\ref{TBA_rel3}) and (\ref{eq1})-(\ref{eq3})
are the signs multiplying the kernel $K_3$ and the constant $C_s$.

From equations (\ref{TBA_rel1})-(\ref{TBA_rel3}), the task of formulating the relative $Y$-system is easily achieved, by means of the bootstrap relations (\ref{boot_rel}). It turns out that the result explicitly depends on the parity of the label $s$ since: for odd values of $s$ we get
\ba\label{YsysOdd}
\mbox{[$s$ odd]:}&& \nn\\
&& \frac{W^-_{3,s}W^+_{1,s}}{W_{2,s}}=\frac{(1+W_{3,s+1})(1+W_{3,s-1})}{1+Y_{2,s}} \nn\\
&&\frac{W^+_{2,s}W^-_{2,s}}{W_{1,s}W_{3,s}}=\frac{(1+W_{2,s-1})(1+W_{2,s+1})}{(1+W_{1,s})(1+W_{3,s})} \\
&& \frac{W^-_{1,s}W^+_{3,s}}{W_{2,s}}=\frac{(1+W_{1,s+1})(1+W_{1,s-1})}{1+W_{2,s}} \nn
\ea
while the even case yields
\ba\label{YsysEven}
\mbox{[$s$ even]:}&& \nn\\
&& \frac{W^-_{3,s}W^+_{1,s}}{W_{2,s}}=\frac{(1+W_{1,s+1})(1+W_{1,s-1})}{1+W_{2,s}}  \nn\\
&&\frac{W^+_{2,s}W^-_{2,s}}{W_{1,s}Y_{3,s}}=\frac{(1+W_{2,s-1})(1+W_{2,s+1})}{(1+W_{1,s})(1+W_{3,s})} \\
&& \frac{W^-_{1,s}W^+_{3,s}}{W_{2,s}}=\frac{(1+W_{3,s+1})(1+W_{3,s-1})}{1+W_{2,s}} \nn \ .
\ea
The $Y$-system (\ref{YsysOdd}),(\ref{YsysEven}) is apparently different from that in \cite{YSA}, but we can recover the agreement by an identification analogous to (\ref{identificazione_eps})
\ba\label{ridefinizioneY}
&& Y_{2,s}=W_{2,s}  \\
&& Y_{1,4k+1}=W_{1,4k+1} 		\qquad\quad  Y_{3,4k+1}=W_{3,4k+1} \nn\\
&& Y_{1,4k+2}=W_{1,4k+2} 		\qquad\quad  Y_{3,4k+2}=W_{3,4k+2} \nn\\
&& Y_{1,4k+3}=W_{3,4k+3} 		\qquad\quad  Y_{3,4k+3}=W_{1,4k+3} \nn\\
&& Y_{1,4k+4}=W_{3,4k+4} 		\qquad\quad  Y_{3,4k+4}=W_{1,4k+4} \nn
\ea

so that (\ref{YsysOdd}),(\ref{YsysEven}) exactly matches with (\ref{YSA}).

An alternative way to get the $Y$-system (\ref{YsysOdd}),(\ref{YsysEven}) makes use of the pentagonal amplitudes and bootstrap relations among them, rather than the relativistic kernels. Indeed, this procedure moves the first step directly from the equations of motion (\ref{eq_moto}), which can be recast in the a more suitable shape 
\be\label{eq_moto2}
X^{(s)}_{\alpha}(\theta)+ \sum _{r=1}^{n-5} \sum _{\beta=1}^3(-1)^{r} \int _{\gamma_{r}}\frac{d\theta '}{2\pi} \mu _{\beta}(\theta ') G^{(s,r)}_{\alpha,\beta}(\theta , \theta ')
L_{\beta,r}\left(\q'-i\frac{\pi}{4}\,\frac{1-(-1)^r}{2}\right)=0 \nonumber
\ee
with $L_{\alpha,s}(\q)\equiv \ln[1+W_{\alpha,s}(\q)]$.
The $W$-functions are now defined as

\small
\be
(-1)^s\, X^{(s)}_{\alpha}(\q)=-\ln W_{\alpha,s}\left(\q-i\frac{\pi}{4}\,\frac{1-(-1)^s}{2}\right)
-\tau_s E_\alpha(\q) -i(-1)^s\sigma_s p_\alpha(\q)+(2-\alpha)\ln\sqrt{\frac{y_{1,s}}{y_{3,s}}}
\ee
\normalsize

in such a way the integral equations turn into

\footnotesize
\ba
&& \ln\left(\frac{W_{\alpha,s}^+(\q)W_{4-\alpha,s}^-(\q)}{W_{\alpha+1,s}(\q)W_{\alpha-1,s}(\q)}\right)=\sum_{r=1}^{n-5}\sum_{\beta=1}^3(-1)^{r+s}
\int_{\mbox{Im}\q'=\varphi_r} \frac{d\q'}{2\pi}\,\mu_\beta(\q'+i\frac{\pi}{4}\,\frac{1-(-1)^r}{2})L_{\beta,r}(\q')\times\\
&&\times\left[G^{(s,r)}_{\alpha,\beta}(\q+i\frac{\pi}{4}\,\frac{1-(-1)^s}{2}+\frac{i\pi}{4},\q'+i\frac{\pi}{4}\,\frac{1-(-1)^r}{2})+
G^{(s,r)}_{4-\alpha,\beta}(\q+i\frac{\pi}{4}\,\frac{1-(-1)^s}{2}-\frac{i\pi}{4},\q'+i\frac{\pi}{4}\,\frac{1-(-1)^r}{2})-\right. \nn\\
&& \left. -G^{(s,r)}_{\alpha+1,\beta}(\q+i\frac{\pi}{4}\,\frac{1-(-1)^s}{2},\q'+i\frac{\pi}{4}\,\frac{1-(-1)^r}{2})-G^{(s,r)}_{\alpha-1,\beta}(\q+i\frac{\pi}{4}\,\frac{1-(-1)^s}{2},\q'+i\frac{\pi}{4}\,\frac{1-(-1)^r}{2}) \right]  \nn
\ea
\normalsize

which becomes the set of functional equations (\ref{YsysOdd}),(\ref{YsysEven}) once we use the bootstrap relations (\ref{boot_pentagonal}).

\vspace{0.2cm}

\textbf{Uncrossing the $Y$-system:}\\

Finally, we can obtain an uncrossed version of the Y-system by means of the bootstrap formulae (\ref{boot_rel})(\ref{boot_rel2}). However, once again the result depends on the value of the $Y$-system column index $s$; for instance,  $s$ odd gives
\be\label{uncrossed_odd}
  \begin{split}
    W^{++}_{1,s}\,W^{--}_{1,s} &=\frac{(1+W^+_{3,s-1})(1+W^+_{3,s+1})(1+W^-_{1,s+1})(1+W^-_{1,s-1})}
      {(W_{3,s})^2\ (1+\frac{1}{W_{2,s}^+})(1+\frac{1}{W_{2,s}^-})}  \\
    W^{++}_{3,s}\,W^{--}_{3,s} &=\frac{(1+W^+_{1,s-1})(1+W^+_{1,s+1})(1+W^-_{3,s+1})(1+W^-_{3,s-1})}
      {(W_{1,s})^2\ (1+\frac{1}{W_{2,s}^+})(1+\frac{1}{W_{2,s}^-})} \\
    \frac{W^+_{2,s}W^-_{2,s}}{W_{1,s}W_{3,s}} &=\frac{(1+W_{2,s-1})(1+W_{2,s+1})}{(1+W_{1,s})(1+W_{3,s})}  \ ,
  \end{split}
\ee
while for an even value of $s$ reads
\be\label{uncrossed_even}
  \begin{split}
    W^{++}_{1,s}\,W^{--}_{1,s} &=\frac{(1+W^+_{1,s-1})(1+W^+_{1,s+1})(1+W^-_{3,s+1})(1+W^-_{3,s-1})}
      {(W_{3,s})^2\ (1+\frac{1}{W_{2,s}^+})(1+\frac{1}{W_{2,s}^-})}\\
    W^{++}_{3,s}\,W^{--}_{3,s} &=\frac{(1+W^+_{3,s-1})(1+W^+_{3,s+1})(1+W^-_{1,s+1})(1+W^-_{1,s-1})}
      {(W_{1,s})^2\ (1+\frac{1}{W_{2,s}^+})(1+\frac{1}{W_{2,s}^-})} \\
    \frac{W^+_{2,s}W^-_{2,s}}{W_{1,s}W_{3,s}} &=\frac{(1+W_{2,s-1})(1+W_{2,s+1})}{(1+W_{1,s})(1+W_{3,s})}  \ .
  \end{split}
\ee
To conclude, the same formulae (\ref{boot_rel})(\ref{boot_rel2}) may be used to recast the $Y$-system (\ref{YSA}) to the uncrossed form:
\be
  \begin{split}
    Y^{++}_{1,s}\,Y^{--}_{1,s} &=\frac{(1+Y^+_{1,s-1})(1+Y^-_{1,s+1})(1+Y^+_{3,s+1})(1+Y^-_{3,s-1})}
      {(Y_{3,s})^2\ (1+\frac{1}{Y_{2,s}^+})(1+\frac{1}{Y_{2,s}^-})}\\
    Y^{++}_{3,s}\,Y^{--}_{3,s} &=\frac{(1+Y^+_{3,s-1})(1+Y^-_{3,s+1})(1+Y^+_{1,s+1})(1+Y^-_{1,s-1})}
      {(Y_{1,s})^2\ (1+\frac{1}{Y_{2,s}^+})(1+\frac{1}{Y_{2,s}^-})} \\
    \frac{Y^+_{2,s}Y^-_{2,s}}{Y_{1,s}Y_{3,s}} &=\frac{(1+Y_{2,s-1})(1+Y_{2,s+1})}{(1+Y_{1,s})(1+Y_{3,s})}  \ .
  \end{split}
\ee

\chapter{The OPE scalars: semiclassical enhancement}
\label{ChSca}

In this chapter we deal with the scalars in the OPE series (\ref{OPEseries}), focusing on the strong coupling limit. In Section \ref{WLAmp} we anticipated a non-pertubative contributions, coming from the $O(6)$ dynamics on the string side, first proposed by \cite{BSV4}\footnote{For the NMHV amplitudes, this contribution has been analyzed by \cite{BEL}.}. Here we refine the method of form factors theory \cite{Smirnov}, depicted in the introductory chapter, to the case of the asymptotically free theories (here the $O(6)$ sigma-model). In this way, we easily show the expected exponential behaviour in $\sqrt\lambda$ for the null polygonal Wilson loop. We also find an expression for the coefficient in front of $\sqrt\lambda$, expandend in a series of multiparticle contributions. For the hexagon, some subleading corrections are discussed and numerically evaluated. A pivotal step is to show that the functions appearing in the integrals, product of dynamical and matrix part, enjoy the important property of the \emph{asymptotic factorisation}. We first deal with the simplest case, the hexagon, which is the perfect instance to show how the method works. The content of this part is largely based on the papers \cite{BFPR2,BFPR3}. Later, we give a sketch of how the argument can be extended to any polygon. In particular, an interesting recursive formula is obtained.

\section{Hexagonal Wilson loop} 
\label{ScalarHex}

Let us consider the hexagonal Wilson loop, whose OPE series is fully described in Section \ref{SecOPE}, and restrict the summation only over scalars. We remember the constraint of $SU(4)$ neutrality, which requires the number of particles to be even.

In the strong coupling regime the scalars decouple from the other particles and form a relativistic O(6) nlsm \cite{AM}: the OPE series reads
\be
\label{Wilson}
W=\sum_{n=0}^{\infty}W^{(2n)} \, , \quad W^{(2n)}=\frac{1}{(2n)!}\int\prod_{i=1}^{2n}\frac{d\theta_i}{2\pi}\,G^{(2n)}(\theta_1,\cdots,\theta_{2n})\, e^{- z\sum\limits_{i=1}^{2n}\cosh\theta_i} \, ,
\ee

where we indicated by $W$ the scalar contribution to the OPE series (\ref{OPEseries}).
The hyperbolic rapidities used to parametrize energy and momentum are related to the Bethe rapidities by $u_i=\frac{2}{\pi}\theta_i$. The dimensionless distance $z=m_{gap}\sqrt{\tau^2+\sigma^2}$ contains two conformal ratios $\sigma$, $\tau$ and is proportional to the mass gap, dynamically generated \cite {AM}, given in terms of the coupling as
\be\label{mgap}
m_{gap}(\lambda)=\frac{2^{1/4}}{\Gamma (5/4)} \lambda ^{1/8}e^{-\sqrt{\lambda}/4} (1+O(1/\sqrt{\lambda} ) ) \, .
\ee

We stress that formula (\ref{Wilson}) is nothing but the form factor expansion, compare to (\ref{CorrFF}), of the two-point function of the twist field operator

\be 
W(z)\equiv \langle\hat{P}(z)\hat{P}(0)\rangle
\ee

whose form factors are the pentagon transitions $P(\psi_i|\psi_j)$.
For this correlator we expect the short-distance behaviour \cite{BSV4}

\be\label{W(z)} 
W(z)\simeq c\frac{\log(1/z)^s}{z^J}
\ee

which, taking into account the definition of distance and the mass-coupling relation, gives the strong coupling limit anticipated in (\ref{strongW})

\be\label{W(lambda)} 
W(\lambda)=C(\tau,\sigma)\lambda^B e^{\sqrt{\lambda}A}\left[1+O\left(\frac{1}{\sqrt{\lambda}}\right)\right]
\ee

with the identifications

\be\label{lambda-z}
A=\frac{J}{4}, \quad  B=\frac{s}{2}-\frac{J}{8}, \quad C(\tau ,\sigma)=\frac{c}{4^s}\left[\frac{\Gamma(5/4)}{2^{1/4}\sqrt{\tau^2+\sigma^2}}\right]^{J} 
\ee

For generic coupling, the functions $G^{(2n)}$ factorise into a $\lambda$ dependent dynamical part, $\Pi_{dyn}^{(2n)}$, and the coupling-independent $\Pi_{mat}^{(2n)}$, encoding the internal $SO(6)$ structure of scalars \cite{BSV4}
\be\label{Gi2n}
G^{(2n)}(u_1,\cdots,u_{2n})=\Pi_{dyn}^{(2n)}(u_1,\cdots,u_{2n})\,\Pi_{mat}^{(2n)}(u_1,\cdots,u_{2n}) \, .
\ee

which, as mentioned previously, are the squared form factors summed over the internal indices, \emph{i.e.} compare with formula (\ref{FFsum}).

The dynamical part is a product of two-particle functions only, which becomes relativistic at strong coupling and reads

\small
\be\label{dynamical}
\Pi_{dyn}^{(2n)}(u_1,\cdots,u_{2n})=\mu^{2n}\prod_{i<j}^{2n}\Pi(u_i-u_j) \, , \quad
\Pi(u)=
\frac{8\theta\tanh \left(\frac{\theta}{2}\right)\Gamma \left (\frac{3}{4}+\frac{i\theta}{2\pi} \right)
\Gamma \left (\frac{3}{4}-\frac{i\theta}{2\pi} \right)}{\pi\Gamma \left (\frac{1}{4}+\frac{i\theta}{2\pi} \right) \Gamma \left (\frac{1}{4}-\frac{i\theta}{2\pi} \right)}\, ,
\quad  \mu=\frac{2\Gamma\left(\frac{3}{4}\right)}{\sqrt{\pi}\Gamma\left(\frac{1}{4}\right)} \, .
\ee
\normalsize

On the other hand, as discussed in Section \ref{SecMat}, the matrix part enjoys an integral representation \cite{BSV4, BFPR} given by the formula (\ref{ScaPiMat}). This matrix factor has been thoroughly analysed in the third chapter, where it has been given a Young tableaux representation. 

We are interested in the strong coupling regime, which, by means of (\ref{mgap}), corresponds to the short-distance regime of the Wl/$O(6)$ correlator. We thus employ the method outlined in subsection \ref{Int2QFT}: the main idea is to evaluate the logarithm of $W$, containing the connected counterparts $g^{(2n)}$ of the functions $G^{(2n)}$. For this purpose, we are going to prove another property of the matrix factor, the asymptotic factorisation. This is a crucial point for the method to work. There is a caveat, however: the asymptotic freedom of the $O(6)$ model gives a weaker (power-like) decay for the connected functions $g^{(2n)}$ than the more usual exponential \cite{Smirnov}, thus requiring a more careful analysis in the short-distance limit.

\subsection{Asymptotic factorisation}

With this program in mind, we are going to study the behaviour of $G^{(2n)}$ when $m$ rapidities are shifted by a large amount $\Lambda\rightarrow\infty$, while the remaining $2n-m$ ones are held fixed. We will obtain that for even $m$, $G^{(2n)}$ enjoys the {\it asymptotic factorisation} into two functions with fewer rapidities, schematically
\be\label{fact}
G^{(2n)}\rightarrow G^{(m)}\,G^{(2n-m)} \, , \quad 2\leq m\leq 2n-2 \, ;
\ee

On the other hand, for $m$ odd the function $G^{(2n)}$ goes to zero with the power-like behaviour $1/\Lambda ^2$. This typical power-like decay, ascribable to the asymptotic freedom of the $O(6)$, is the only significant difference with respect the case studied in \cite{Smirnov}. First, the dynamical part (\ref{dynamical}) (trivially extending the definition for odd $m$) enjoys the factorisation

\small
\be
\Pi_{dyn}^{(2n)}(u_1+\Lambda,\cdots,u_{m}+\Lambda,u_{m+1},\cdots,u_{2n})
\longrightarrow\Lambda^{2m(2n-m)}\Pi_{dyn}^{(m)}(u_1,\cdots,u_{m})\Pi_{dyn}^{(2n-m)}(u_{m+1},\cdots,u_{2n}) \ ,
\label {dynasy}
\ee
\normalsize
as a consequence of the asymptotic behaviour $\Pi(u)\simeq u^2$ for $u \rightarrow \infty$. The matrix part, given by the integrals in (\ref{ScaPiMat}), is more involved and we would rather tackle the simplest non trivial case first, \textit{i.e.} $\Pi_{mat}^{(4)}\rightarrow \Pi_{mat}^{(2)}\Pi_{mat}^{(2)}$ and then generalise. When we shift two rapidities, say $u_1\to u_1+\Lambda ,\,u_2\to u_2+\Lambda$, by a large amount $\Lambda$, the integrals in (\ref{ScaPiMat}) receive the main contribution from the region where one auxiliary variable $a$, two $b$'s, and one $c$ are large. Therefore, upon shifting, for instance, $a_1,b_1,b_2,c_1$ by $\Lambda$, we rewrite (\ref{ScaPiMat}) as
\ba
&&\Pi_{mat}^{(4)}(u_1+\Lambda,u_2+\Lambda,u_3,u_4)=\frac{1}{4!4}\int_{-\infty}^{+\infty}\frac{da_1db_1db_2dc_1}{(2\pi)^4}\frac{g(b_1-b_2)}{\displaystyle\prod_{i=1}^2f(a_1-b_i)f(c_1-b_i)\prod_{i,j=1}^{2}f(u_i-b_j)}\times \nn\\
&& \times \int_{-\infty}^{+\infty}\frac{da_2db_3db_4dc_2}{(2\pi)^4}\frac{g(b_3-b_4)}{\displaystyle\prod_{i=3}^4f(a_2-b_i)f(c_2-b_i)\prod_{i,j=3}^{4}f(u_i-b_j)}
\,\mathcal{R}^{(4,2)}(a_1,a_2,b_1,\dots ,b_4,c_1,c_2;\Lambda) \, ,\nn\\ \label {Pi_mat2}
\ea
where the function $\mathcal{R}^{(4,2)}$ stems for
\ba
&& \mathcal{R}^{(4,2)}(a_1,a_2,b_1,\dots ,b_4,c_1,c_2;\Lambda) =
\frac{\displaystyle\prod_{i=1}^2\prod_{j=3}^{4}g(b_i-b_j+\Lambda)}{\displaystyle\prod_{i=1}^{2}\prod_{j=3}^{4}f(u_i-b_j+\Lambda)f(u_j-b_i-\Lambda)} \times\nn\\
&& \times \frac{g(a_1-a_2+\Lambda)g(c_1-c_2+\Lambda)} {\displaystyle\prod_{i=3}^{4}f(a_1-b_i+\Lambda)f(c_1-b_i+\Lambda)\prod_{i=1}^{2}f(a_2-b_i-\Lambda)f(c_2-b_i-\Lambda)} \ .
\ea
and enjoys the large $\Lambda$ expansion
\be
\mathcal{R}^{(4,2)}(a_1,a_2,b_1,\dots ,b_4,c_1,c_2;\Lambda) =
\Lambda^{-8}\left[1+O\left(\frac{1}{\Lambda}\right)\right] \, , \quad \Lambda \rightarrow +\infty \, ,
\ee
Taking into account all the possible exchanges of isotopic rapidities (of the same type), we have an additional multiplicity factor of $24$, which yields
\ba\label{422Pimat}
\Pi_{mat}^{(4)}(u_1+\Lambda,u_2+\Lambda,u_3,u_4)\simeq 24\Lambda^{-8}\frac{1}{4!4}\int\frac{da_1db_1db_2dc_1}{(2\pi)^4}\frac{g(b_1-b_2)}{\displaystyle\prod_{i=1}^2f(a_1-b_i)f(c_1-b_i)\prod_{i,j=1}^{2}f(u_i-b_j)}\times \nn\\
\times \int\frac{da_2db_3db_4dc_2}{(2\pi)^4}\frac{g(b_3-b_4)}{\displaystyle\prod_{i=3}^4f(a_2-b_i)f(c_2-b_i)\prod_{i,j=3}^{4}f(u_i-b_j)}=\Lambda^{-8}\Pi_{mat}^{(2)}(u_1,u_2)\Pi_{mat}^{(2)}(u_3,u_4) \ . \nn\\
\ea
Assembling everything together and using (\ref{Gi2n}), the asymptotic factorisation is finally proven
\be\label{422fact}
G^{(4)}(u_1+\Lambda, u_2+\Lambda, u_3,u _4)\ \overset{\Lambda\rightarrow\infty}{\longrightarrow}\ G^{(2)}(u_1,u_2)G^{(2)}(u_3,u_4) + O(\Lambda^{-2})\, .
\ee
where the $O(1/\Lambda)$ term vanishes thanks to a refined cancellation coming from the matrix part and the dynamical one.

The most general case, $u_i\to u_i+\Lambda$ for $1\leq i\leq m$, goes along the same line. We need only to separate odd $m=2k-1$ from even $m=2k$. In a unified manner, we may describe the shifts $a_j\to a_j+\Lambda$ and $c_j\to c_j+\Lambda$ for $1\leq j\leq k$, along with $b_i\to b_i+\Lambda$ for $1\leq i\leq m$, namely

\small
\ba \label{Intpimat}
&&\Pi_{mat}^{(2n)}(u_1+\Lambda,\cdots,u_{m}+\Lambda,u_{m+1},\cdots,u_{2n})= \\
&& = \frac{1}{(2n)!(n!)^2}\int\prod_{i=1}^k\frac{da_i dc_i}{(2\pi)^2}\prod_{i=1}^m\frac{db_i}{2\pi}\,
\frac{\displaystyle\prod_{i<j,\,i=1}^{k}\left[g(a_i-a_j)g(c_i-c_j)\right]\displaystyle\prod_{i<j,\,i=1}^{m}g(b_i-b_j)}
{\displaystyle\prod_{j=1}^{m}\left[\prod_{i=1}^{k}f(a_i-b_j)f(c_i-b_j)\prod_{l=1}^{m}f(u_l-b_j)\right]}\times \nn\\
&& \times \int\prod_{i=k+1}^n\frac{da_i dc_i}{(2\pi)^2}\prod_{i=m+1}^{2n}\frac{db_i}{2\pi}\,
\frac{\displaystyle\prod_{i<j,\,i=k+1}^{n}\left[g(a_i-a_j)g(c_i-c_j)\right]\displaystyle\prod_{i<j,\,i=m+1}^{2n}g(b_i-b_j)}
{\displaystyle\prod_{j=m+1}^{2n}\left[\prod_{i=k+1}^{n}f(a_i-b_j)f(c_i-b_j)\prod_{l=m+1}^{2n}f(u_l-b_j)\right]}
\,\mathcal{R}^{(2n,m)}(a_1,\dots , c_{2n};\Lambda) \,\, , \nn
\ea
\normalsize

where 
\small
\ba
&& \mathcal{R}^{(2n,m)}(a_1,\dots , c_{2n};\Lambda)= \frac{\displaystyle\prod_{i=1}^{m}\prod_{j=m+1}^{2n}g(b_i-b_j+\Lambda)}{\displaystyle\prod_{i=1}^{m}\prod_{j=m+1}^{2n}f(u_j-b_i-\Lambda)f(u_i-b_j+\Lambda)} \times \nn\\
&& \times
\frac{\displaystyle\prod_{i=1}^{k}\prod_{j=k+1}^{n}g(a_i-a_j+\Lambda)g(c_i-c_j+\Lambda)}
{\displaystyle\prod_{j=1}^{m}\prod_{i=k+1}^{n}f(a_i-b_j-\Lambda)f(c_i-b_j-\Lambda)
\prod_{j=m+1}^{2n}\prod_{i=1}^{k}f(a_i-b_j+\Lambda)f(c_i-b_j+\Lambda)} \ .
\ea
\normalsize

The leading order is given by $\mathcal{R}^{(2n,m)} \simeq \Lambda^{4(n-k)(2k-m)}\Lambda^{-4k(2n-m)}$ and the factorisation of the matrix part (\ref {Intpimat}), with even $m=2k$, is achieved
\be
\Pi_{mat}^{(2n)}(u_1+\Lambda,\cdots,u_{2k}+\Lambda,u_{2k+1},\cdots,u_{2n})\longrightarrow  \Lambda^{-2m(2n-m)}\Pi_{mat}^{(2k)}(u_1,\cdots,u_{2k})\Pi_{mat}^{(2n-2k)}(u_{2k+1},\cdot\cdot,u_{2n}) \, 
\label{pimat-fact}
\ee
which, put together with (\ref{dynasy}), yields the result we aim for. This entails $G^{(2n)}(u_1+\Lambda,\cdots,u_{m}+\Lambda,u_{m+1},\cdots,u_{2n})$ weighted by a factor $\Lambda^{-2(m-2k)^2}$, which means, for odd $m=2k-1$, the expected suppression $G^{(2n)}\simeq \Lambda^{-2}$. On the other hand, for even $m=2k$, the task is accomplished and the asymptotic factorisation (\ref{fact}) follows
\be\label{factorization}
G^{(2n)}(u_1+\Lambda,\cdots,u_{2k}+\Lambda,u_{2k+1},\cdots,u_{2n})\ \overset{\Lambda\to\infty}{\longrightarrow}\ G^{(2k)}(u_1,\cdots,u_{2k})\,G^{(2n-2k)}(u_{2k+1},\cdots,u_{2n}) + O(\Lambda^{-2})\ .
\ee

We stress that, as in the case $4\to 2+2$, we considered all the possible shifts of the auxiliary variables within the integrand (\ref{Intpimat}), producing a multiplicity factor ${n \choose k}^2{2n \choose 2k}$, which, once combined with the present factorials $\frac{1}{(2n)!(n!)^2}{n \choose k}^2{2n \choose 2k}=\frac{1}{(2k)!(k!)^2}\frac{1}{(2n-2k)!((n-k)!)^2}$, reproduces the correct factorials of $G^{(2k)}$ and $G^{(2n-2k)}$.

We point out that (\ref{factorization}) is not a sufficient condition, although necessary, for our purpose: we need the connected functions to be integrable, see for instance formula (\ref{DeltaFF}), a fact which requires stronger conditions on the asymptotic behaviour. In fact, we need an extension of (\ref{factorization}) with different large shifts $\Lambda_i$: this is a consequence of the power like correction (due to asymptotic freedom) in place of the exponential one of \cite{Smirnov}. This issue, addressed in \cite{BFPR3}, is very technical and thus it is left to the Appendix \ref{FactConn}.

\subsection{Short-distance regime}

Now, we can profitably employ the technique depicted earlier in the text and switch to the connected functions $g^{(2n)}$, which characterize the series of the logarithm
\be\label{logW}
{\cal F}=\ln W = \sum_{n=1}^{\infty}{\cal F}^{(2n)}=\sum_{n=1}^{\infty}\frac{1}{(2n)!}\int\prod_{i=1}^{2n}\frac{d\theta_i}{2\pi}g^{(2n)}(\theta_1,\cdots,\theta_{2n})e^{-z\sum_{i=1}^{2n}\cosh\theta_i} \, .
\ee
A well-known fact tells us that the original functions $G^{(2n)}$ are expressed in terms of the connected $g^{(2k)}$: for instance, for few number of particles we have

\be 
G_{12}^{(2)}=g_{12}^{(2)}, \quad G_{1234}^{(4)}=g_{1234}^{(4)}+g_{12}^{(2)}g_{34}^{(2)}+g_{13}^{(2)}g_{24}^{(2)}+g_{14}^{(2)}g_{23}^{(2)}
\ee

where an obvious short-hand notation has been used. On the other hand, the reverse relations are:

\be 
g_{12}^{(2)}=G_{12}^{(2)}, \quad g_{1234}^{(4)}=G_{1234}^{(4)}-G_{12}^{(2)}G_{34}^{(2)}-G_{13}^{(2)}G_{24}^{(2)}-G_{14}^{(2)}G_{23}^{(2)}
\ee

The connected functions, included the general relation with the $G^{(2n)}$, are discussed in details in the Appendix \ref{FactConn}. 

The fundamental consequence of the factorisation (\ref{factorization}) is that the connected functions vanish whenever a subset of rapidities is sent far away from all the others by a large amount $\Lambda$, namely
\be
\lim_{\Lambda\to\infty}g^{(2n)}(\theta_1+\Lambda,\cdots,\theta_{m}+\Lambda,\theta_{m+1},\cdots,\theta_{2n})\simeq \frac{1}{\Lambda ^2} \rightarrow 0 \, , \qquad\mbox{for}\ \ m<2n \, . \label {g-asy}
\ee

This fact follows from the specific combinatorial relation between the $G$'s and the $g$'s, once manipulated by means of the factorisation (\ref{factorization}) to give rise to peculiar cancellations. Conversely, (\ref {g-asy}) entails the factorisation (\ref{factorization}), thus establishing the equivalence of the two properties. The connected functions $g^{(2n)}$ enjoy a plethora of computational advantages with respect to the $G^{(2m)}$ quite in general. In the present case, for instance, they make possible the large coupling expansion by allowing the limit $z\to 0$ inside ${\cal F}^{(2n)}$, which is not possible on the original terms $W^{(2n)}$ of the series (\ref{Wilson}). Therefore it is crucial that the functions $g^{(2n)}$ (differently from the $G^{(2m)}$) are integrable over the $2n-1$ variables they depend on.

Clearly, the limit (\ref {g-asy}) is crucial to decide the small $z$ behaviour of the logarithm of the Wilson loop\footnote{Do not forget, however, that a more stringent condition, necessarily involving different shifts, is required. This issue is discussed in the Appendix \ref{FactConn}.}. To derive the conformal/small $z$ limit, we repeat the procedure outlined in \ref{Int2QFT}: we shall consider the multi-integral $I^{(2n)}\equiv (2n)! (2\pi)^n{\cal F}^{(2n)}$ in the series (\ref {logW}), and, as the connected functions depend only on the differences $\theta_{ij}$, integrate over one rapidity $\theta _1$
\be
I^{(2n)}=\int d\theta _1 \prod _{i=1}^{2n-1}d \alpha _i \exp \Bigl [ -z \cosh \theta _1 -z \sum _{i=2}^{2n} \left ( \cosh \theta _1 \cosh \alpha _{i-1} + \sinh \theta _1 \sinh \alpha _{i-1} \right )  \Bigr ]  g^{(2n)}(\alpha _1, \ldots , \alpha _{2n-1}) \, ,
\ee
where se used the convenient definitions $a=1+\sum_{i=2}^{2n} \cosh  \alpha _{i-1}=\xi \cosh \eta$ and $b= \sum_{i=2}^{2n} \sinh  \alpha _{i-1}=\xi \sinh \eta$ for some real $\eta$ (depending on the $\alpha _i$), thanks of the identity $a^2-b^2=2n+2  \sum _{i=2}^{2n} \cosh  \alpha _{i-1} + 2 \sum _{i=2}^{2n} \sum _{j=i+1}^{2n} \cosh (\alpha _{i-1} - \alpha _{j-1})= \xi ^2 >0$. It follows that
\small
\be\label{I2n}
I^{(2n)}= \int \prod _{i=1}^{2n-1}d \alpha _i g^{(2n)}(\alpha _1, \ldots , \alpha _{2n-1})  \int d\theta _1 \exp \Bigl [ -z \xi \cosh (\theta _1 + \eta )  \Bigr ] = 2 \int \prod _{i=1}^{2n-1}d \alpha _i g^{(2n)}(\alpha _1, \ldots , \alpha _{2n-1})  K_0 (z \xi) \, .
\ee
\normalsize
This is the same formula we obtained earlier in the text, namely (\ref{IntK0}). Now the main difference, due to the asymptotic freedom, stands out. We are tempted to expand the integrand for small argument $K_0(z\xi)=-\ln z -\ln\xi + \ln 2-\gamma + O(z^2\ln z)$, with $\gamma=0.5772...$ the Euler-Mascheroni constant. However, because of the weak decay (\ref{g-asy}), we need to cut-off the integral to the region where the argument is small and expand later
\be\label{lead-subl}
I^{(2n)}= -2 \ln z  \int \prod _{i=1}^{2n-1}d \alpha _i g^{(2n)}(\alpha _1, \ldots , \alpha _{2n-1}) -2\int_{z\xi<1} \prod _{i=1}^{2n-1}d \alpha _i g^{(2n)}(\alpha _1, \ldots , \alpha _{2n-1})\ln\xi + O(1)\, ,
\ee

where we kept the cut-off in the second and removed it in the first one, as the function $g^{(2n)}$ is integrable, whilst the second one diverges (as $\ln\ln(1/z)$). This mechanism will be clearer when we analyse the two and four particles cases.

Eventually, the strong coupling expansion for $\mathcal{F}_n$  (\ref {logW}) can be systematically set down, whose first order is
\be\label{Wleading}
\ln W = \frac{\sqrt{\lambda}}{4\pi} \sum _{n=1}^{+\infty} \frac{1}{(2n)!}\int \prod _{i=1}^{2n-1} \frac{d\alpha _i}{2\pi}
g^{(2n)}(\alpha _1, \ldots , \alpha _{2n-1}) + O(\ln\sqrt{\lambda}) \, ,
\ee
where we used $\ln(1/z) \sim - \ln m_{gap} \sim \frac{\sqrt{\lambda}}{4}$. Thus the main claim is proven, as a factor $\sqrt{\lambda}$ is extracted in front of the series. Remarkably, the expression for the coefficient is a series of integrals which can be computed numerically with high precision. In the following we will discuss the first two contributions: the two and four particle cases.

In fact, also the (divergent) sub-leading term in (\ref{lead-subl}) can be obtained, even though without a closed formula. This gives rise to the unusual logarithmic $(\ln 1/z)^s$ factor in the two point 2D CFT correlation function, see formula (\ref{W(z)}). Therefore, we expect for the $2n$-particle contribution the expansion

\be\label{F2nExp}
\mathcal{F}^{(2n)}= J^{(2n)}\ln(1/z) + s^{(2n)}\ln\ln(1/z) + t^{(2n)} + O\left(\frac{1}{\ln z}\right)
\ee

\subsubsection{Two-particle case}

Considering (\ref{I2n}) for $n=1$, we have only one variable $\alpha_1\equiv \alpha$, with simply $\xi=2\cosh\frac{\alpha}{2}$
\be
\mathcal{F}^{(2)}=\frac{1}{(2\pi)^2}\int d\alpha g^{(2)}(\alpha)K_0\left (2z\cosh\frac{\alpha}{2}\right)=\frac{2}{(2\pi)^2}\int_0^{\infty} d\alpha g^{(2)}(\alpha)K_0\left (2z\cosh\frac{\alpha}{2}\right) \, ,
\ee
where the rescaled function $g^{(2)}(\alpha)=\frac{4}{\pi^2}g^{(2)}(u_1,u_2)$ is, explicitly
\be
g^{(2)}(\alpha)=\frac{\Gamma^2(3/4)}{\Gamma^2(1/4)}\frac{\alpha\tanh(\alpha/2)\Gamma\left (\frac{3}{4}-\frac{i\alpha }{2\pi} \right)\Gamma\left (\frac{3}{4}+\frac{i\alpha }{2\pi} \right)}{\Gamma\left (\frac{1}{4}-\frac{i\alpha }{2\pi} \right)\Gamma\left (\frac{1}{4}+\frac{i\alpha }{2\pi} \right)}\frac{12\pi^2}{\left(\alpha^2 + \frac{\pi^2}{4} \right)\left(\alpha^2 + \pi^2\right)}
\ee
and is endowed with the asymptotic behaviour $g(\alpha)=C\alpha^{-2} + O(\alpha^{-4})$ with $C=6\pi\frac{\Gamma^2(3/4)}{\Gamma^2(1/4)}$.
Now, we are willing to study the expansion (\ref{F2nExp}) for $n=1$
\be\label{F2}
\mathcal{F}^{(2)}= J^{(2)}\ln(1/z) + s^{(2)}\ln\ln(1/z) + t^{(2)} + O\left(\frac{1}{\ln z}\right) \, .
\ee
We split the integral in half
\be
\mathcal{F}^{(2)}=\int_0^{2\ln(1/z)} \frac{d\alpha}{2\pi^2} g^{(2)}(\alpha)K_0\left (2z\cosh\frac{\alpha}{2}\right)
+ \int_{2\ln(1/z)}^{\infty} \frac{d\alpha}{2\pi^2} g^{(2)}(\alpha)K_0\left (2z\cosh\frac{\alpha}{2}\right)=\mathcal{F}^{(2)}_1 + \mathcal{F}^{(2)}_2 \, .
\ee
In the limit $z\rightarrow 0$, $\mathcal{F}^{(2)}_2$ goes to zero as $K_0$ is bounded within the integration support and the function $g^{(2)}$ decays as $C\alpha^{-2}+O(\alpha ^{-4})$ for large rapidity, giving an $O(1/\ln z)$ contribution. For the first piece, in order to estimate the diverging and the finite contributions for $z\rightarrow 0$, we are allowed to expand $K_0(2z \cosh \frac{\alpha}{2})$ and, using $h(\alpha)\equiv \frac{1}{2\pi^2}g^{(2)}(\alpha)$, we get
\ba\label{leading}
\mathcal{F}^{(2)}&=&\ln\frac{1}{z} \int _{0}^{2\ln(1/z)} d\alpha h(\alpha)- \int _{0}^{2\ln(1/z)} d\alpha h(\alpha) \ln \left (\cosh \frac{\alpha}{2} \right ) - \gamma \int _{0}^{2\ln(1/z)} d\alpha h(\alpha) +  O\left(\frac{1}{\ln z}\right) \nonumber \\
&=& J^{(2)}\ln\frac{1}{z} - \int _{0}^{2\ln(1/z)} d\alpha h(\alpha) \ln \left (\cosh \frac{\alpha}{2} \right ) - J^{(2)}\gamma - \ln\frac{1}{z} \int _{2\ln(1/z)}^{\infty} d\alpha h(\alpha) + O\left(\frac{1}{\ln z}\right) \nn\\
\ea
with $\displaystyle J^{(2)}\equiv\int_{0}^{\infty} d\alpha h(\alpha)$ as the leading term of the series (\ref{F2}).
The second term in (\ref{leading}) is of order $\ln\ln(1/z)$ since the integrand is suppressed as $\sim\frac{1}{\alpha}$, while the remaining ones are finite, since
\be
-\ln(1/z) \int _{2\ln(1/z)}^{\infty} d\alpha h(\alpha) \simeq -\frac{C}{2\pi^2}\ln(1/z)\int _{2\ln(1/z)}^{\infty}\frac{d\alpha}{\alpha^2}=-\frac{C}{(2\pi)^2} \, .
\ee
In order to extract the $O(\ln\ln(1/z))$ contribution from the constant ones in
\be
- \int _{0}^{2\ln(1/z)} d\alpha h(\alpha) \ln \left (\cosh \frac{\alpha}{2} \right ) \, ,
\ee
we divide the integration domain into two pieces
\be
- \int _{0}^{1} d\alpha h(\alpha) \ln \left (\cosh \frac{\alpha}{2} \right )- \int _{1}^{2\ln(1/z)} d\alpha h(\alpha) \ln \left (\cosh \frac{\alpha}{2} \right ) \, ,
\ee
The divergence $\ln\ln(1/z)$ dwells in the first one: to extract it, we add and subtract the asymptotic behaviour
\be
- \int_{1}^{2\ln(1/z)} d\alpha\left[ h(\alpha) \ln \left (\cosh \frac{\alpha}{2} \right )-\frac{C}{(2\pi)^2\alpha}\right]-\frac{C}{(2\pi)^2}\int_{1}^{2\ln(1/z)}\frac{d\alpha}{\alpha} \, .
\ee
The first integral stays finite for $2\ln(1/z)\rightarrow\infty$ while the second gives the subleading $\ln\ln(1/z)$, up to an additional constant term
\be
-\frac{C}{(2\pi)^2}\int_{1}^{2\ln(1/z)}\frac{d\alpha}{\alpha}=-\frac{C}{(2\pi)^2}\ln \ln (1/z)-\frac{C}{(2\pi)^2}\ln 2 \, .
\ee
In the end, we obtain:
\be
J^{(2)}=\int  _{0}^{+\infty} d\alpha h(\alpha )= \frac{1}{2\pi^2}\int _{0}^{+\infty} d\alpha g^{(2)}(\alpha ) \simeq 0.03109
\ee
\be
s^{(2)}=-\frac{C}{(2\pi)^2}=- \frac{3}{2\pi } \frac{\Gamma ^2(3/4)}{\Gamma ^2(1/4)}\simeq -0.05454
\ee
\ba
t^{(2)} &=& - J^{(2)}\gamma - \frac{C}{(2\pi)^2}\left(1+\ln 2\right) - \frac{1}{2\pi^2}\int_{0}^{1}d\alpha g^{(2)}(\alpha) \ln \left (\cosh \frac{\alpha}{2} \right) + \nn\\
&+& \frac{1}{2\pi^2}\int_{1}^{\infty}d\alpha \left[\frac{C}{2\alpha}-g^{(2)}(\alpha) \ln \left ( \cosh \frac{\alpha}{2} \right ) \right ] \, .
\ea
Our numerical estimate for $t^{(2)}$ amounts to $t^{(2)}\simeq -0.00819$, agreeing with the Montecarlo evaluation by \cite{BSV4,BEL}.

\subsubsection{Four-particle case $n=2$}

For the leading order $J\ln(1/z)$, it is not difficult to evaluate the correction $\delta J^{(4)}$ from the explicit expression of the four scalar connected function $g^{(4)}$. Specializing (\ref{I2n}) for $n=2$, we get
\be
J^{(4)}=\frac{1}{12(2\pi)^4}\int d\alpha_1 d\alpha_2 d\alpha_3 g^{(4)}(\alpha_1,\alpha_2,\alpha_3) \, .
\ee
which we numerically integrate to obtain a correction to $J$ of $J^{(4)}= (-3.44\pm 0.01)\cdot 10^{-3}$, {\it i.e.} in total
\be
J^{(2)}+J^{(4)} \simeq 0.02765 \, .
\ee
This evaluation differs from the 2D-CFT prediction $J=\frac{1}{36}=0.02\bar{7}$ \cite{BSV4} by the small amount of $0.5\%$.

The correction $s^{(4)}$ is more complicated to evaluate, since it depends on the asymptotic behaviour of $g^{(4)}$ and there are many different regions to consider. We remind that the divergence $\ln\ln(1/z)$ is due to the combined action of the cutoff $z\xi<1$ and the term $g^{(4)}\ln\xi$ in the expansion of $K_0(z\xi)$. To be precise, it is inside the following integral
\be\label{deltas4int}
s^{(4)}\ln\ln(1/z)=-\frac{1}{12(2\pi)^4}\int_{z\xi<1} d\alpha_1 d\alpha_2 d\alpha_3 g^{(4)}(\alpha_1,\alpha_2,\alpha_3) \ln\xi + O(1) \, .
\ee
When one or more variables are large we have $\ln\xi\simeq\frac{|\alpha_i|}{2}$, with $\alpha_i$ the largest of them, and the cutoff condition becomes $|\alpha_i|<2\ln(1/z)$. Thanks to the linearity in $\alpha_i$, the only region where the integral becomes divergent corresponds to the split $4\to 3+1$, in which $g^{(4)}$ goes to zero with the minimum power required by convergence, see the Appendix \ref{FactConn}. The region has multiplicity four and, as they are equivalent, we choose to send $\alpha_1\to\infty$ and keep the other variables finite.
We parametrize the asymptotic behaviour as
\be
\lim_{\alpha_1\to\pm\infty}\alpha_1^2 g^{(4)}(\alpha_1,\alpha_2,\alpha_3)= g^{(4)}_{as}(\alpha_2,\alpha_3) \, .
\ee
from which the contribution of the regions $4\to 3+1$ is contained in
\be
-\frac{2}{3(2\pi)^4}\int^{2\ln(1/z)}d\alpha_1\frac{1}{\alpha_1^2}\frac{\alpha_1}{2}\int d\alpha_2 d\alpha_3 g^{(4)}_{as}(\alpha_2,\alpha_3) \, ,
\ee
where we considered the upper integration limit only, responsible for the divergent part. The additional factor $4\cdot 2$ is due to the number of regions (a particle can be sent either to $+\infty$ or $-\infty$).
Finally, we find the coefficient in (\ref{deltas4int}) expressed as
\be\label{deltas4}
s^{(4)}=-\frac{1}{3(2\pi)^4}\int d\alpha_2 d\alpha_3 g^{(4)}_{as}(\alpha_2,\alpha_3) \, ,
\ee
where the explicit expression of the integrand is
\ba
 &&g^{(4)}_{as}(\alpha_2,\alpha_3)=-6\mu^2\left[g^{(2)}(\alpha_2-\alpha_3) + g^{(2)}(\alpha_2)+g^{(2)}(\alpha_3)\right] +\mu^4 \left(\frac{2}{\pi}\right)^2 36\Pi(u_3)\Pi(u_4)\Pi(u_{34}) \cdot \nn\\
&\cdot & \frac{(u_3^2+4)(u_4^2+4) + (u_3^2+4)(u_{34}^2+4)+(u_4^2+4)(u_{34}^2+4) + \frac{3}{2}(u_3^2+u_4^2+u_{34}^2 +24)}{(u_3^2+1)(u_3^2+4)(u_4^2+1)(u_4^2+4)(u_{34}^2+1)(u_{34}^2+4)} \, ,
\ea
where we used the variables $u_{3,4}=\frac{2}{\pi}\alpha_{2,3}$ for brevity. The numerical integration yields $s^{(4)}\simeq 0.017650$, thus the four particles prediction sums up to
\be
s^{(2)} + s^{(4)}\simeq -0.036894 \, .
\ee
The discrepancy with respect to the expected value  $s=-1/24=-0.041\bar{6}$ \cite{BSV4} is about $11\%$ ({\it cf.} also \cite{BEL}), not as good as that of $J^{(4)}$ but still valuable.

The finite contribution $t^{(4)}$ comes from three different terms $t^{(4)}=t^{(4)}_1+t^{(4)}_2+ t^{(4)}_3$, contained in
\be
\mathcal{F}^{(4)}=\frac{1}{12(2\pi)^4}\int_{z\xi<1} d\alpha_1 d\alpha_2 d\alpha_3 g^{(4)}(\alpha_1,\alpha_2,\alpha_3)K_0(z\xi)
+ O\left(\frac{1}{\ln z}\right) \, ,
\ee
once we subtract both the divergent terms, $J^{(4)}\ln(1/z)$ and $s^{(4)}\ln\ln(1/z)$, previously obtained.
We immediately see that a finite contribution comes from the constant term in the expansion of the Bessel function $\ln 2-\gamma$, which is
\be
t^{(4)}_1=\frac{(\ln 2-\gamma)}{12(2\pi)^4}\int d\alpha_1 d\alpha_2 d\alpha_3 g^{(4)}(\alpha_1,\alpha_2,\alpha_3)=(\ln 2-\gamma)J^{(4)} \, .
\ee
Another one is due to the removal of the cutoff $z\xi < 1$, in the computation of $J^{(4)}$
\be
t^{(4)}_2=\lim_{z\to 0}\left[-\frac{\ln(1/z)}{12(2\pi)^4}\int_{z\xi>1}d\alpha_1 d\alpha_2 d\alpha_3 g^{(4)}(\alpha_1,\alpha_2,\alpha_3)\right] \, .
\ee
Using the same argument as for the subleading $s^{(4)}$, only the region $4\to 3+1$ matters and we get
\be
t^{(4)}_2=-\frac{2\ln(1/z)}{3(2\pi)^4}\int_{2\ln(1/z)}^{\infty}\frac{d\alpha_1}{\alpha_1^2}\int d\alpha_2 d\alpha_3 g_{as}^{(4)}(\alpha_2,\alpha_3) =s^{(4)} \, .
\ee
The last piece, $t^{(4)}_3$, is more involved and comes from the integral (\ref{deltas4int})
\be
-\frac{1}{12(2\pi)^4}\int_{z\xi<1} d\alpha_1 d\alpha_2 d\alpha_3 g^{(4)}(\alpha_1,\alpha_2,\alpha_3) \ln\xi \simeq s^{(4)}\ln\ln(1/z) + t^{(4)}_3 ,
\ee
which yielded the $\ln\ln(1/z)$ contribution in (\ref{deltas4}).

As for the $n=1$ case, we subtract the asymptotic behaviours and get the finite integral (we are now allowed to remove the cutoff $z\xi<1$)
\ba
&& -\frac{1}{12(2\pi)^4}\int d\alpha_1d\alpha_2d\alpha_3 \Bigl[g^{(4)}(\alpha_1,\alpha_2,\alpha_3)\ln\xi-\frac{g^{(4)}_{as}(\alpha_2,\alpha_3)}{2(|\alpha_1|+ a)}-\frac{g^{(4)}_{as}(\alpha_1,\alpha_3)}{2(|\alpha_2|+a)}-\nn\\
&&-\frac{g^{(4)}_{as}(\alpha_1,\alpha_2)}{2(|\alpha_3|+a)} -
 \frac{g^{(4)}_{as}(\alpha_2 -\alpha_1,\alpha_3 -\alpha_1)}{2(|\alpha_1| + a)}\Bigr] \, ,
\ea
where the parameter $a>0$, which does not spoil the large $\alpha_i$ limit, is introduced to prevent the singularities for $\alpha_i=0$. Here, on the contrary of the two particle case, we do not split the integration in parts, as it would be a rather cumbersome procedure. The insertion of a parameter $a$ is much more easily employable.
The divergence $s^{(4)}\ln\ln(1/z)$ is confined to the simple integral
\ba
&& -\frac{1}{12(2\pi)^4}\int_{z\xi<1}d\alpha_1d\alpha_2d\alpha_3 \Bigl[\frac{1}{2(|\alpha_1|+ a)}g^{(4)}_{as}(\alpha_2,\alpha_3)+\frac{1}{2(|\alpha_2|+a)}g^{(4)}_{as}(\alpha_1,\alpha_3)+\nn\\
&&+ \frac{1}{2(|\alpha_3|+a)}g^{(4)}_{as}(\alpha_1,\alpha_2) +
 \frac{1}{2(|\alpha_1| + a)}g^{(4)}_{as}(\alpha_2 -\alpha_1,\alpha_3 -\alpha_1)\Bigr] \, ,
\ea
which also contains the correction $t_3^{(4)}$ we are aiming for.
The four terms contribute the same, therefore we are left with
\be
-\frac{1}{6(2\pi)^4}\int_{z\xi <1}d\alpha_1d\alpha_2d\alpha_3 \frac{1}{|\alpha_1|+ a}\,g^{(4)}_{as}(\alpha_2,\alpha_3) \, .
\ee
Neglecting the vanishing terms, the integral becomes
\be
-\frac{1}{3(2\pi)^4}\int_{0}^{2\ln(1/z)}d\alpha_1 \frac{1}{\alpha_1+ a} \int_{\mathbb{R}^2}d\alpha_2d\alpha_3\, g^{(4)}_{as}(\alpha_2,\alpha_3) \, ,
\ee
since the divergence appears only where $|\alpha_1|$ is large and we can remove the cutoff in the other directions. Integrating over $\alpha_1$ gives
\be
-\frac{1}{3(2\pi)^4}\left[\ln\ln(1/z) + \ln\frac{2}{a}\right]\int_{\mathbb{R}^2}d\alpha_2d\alpha_3 \,g^{(4)}_{as}(\alpha_2,\alpha_3) 
\, ,
\ee
which reproduces $s^{(4)}$ plus a finite correction proportional to it. Everything sums up to
\ba
&& t^{(4)}_3 = -\frac{1}{12(2\pi)^4}\int d\alpha_1d\alpha_2d\alpha_3 \Bigl[g^{(4)}(\alpha_1,\alpha_2,\alpha_3)\ln\xi-\frac{1}{2(|\alpha_1|+ a)}g^{(4)}_{as}(\alpha_2,\alpha_3)-\nn\\
&&-\frac{1}{2(|\alpha_2|+a)}g^{(4)}_{as}(\alpha_1,\alpha_3)-
 \frac{1}{2(|\alpha_3|+a)}g^{(4)}_{as}(\alpha_1,\alpha_2) -
 \frac{1}{2(|\alpha_1| + a)}g^{(4)}_{as}(\alpha_2 -\alpha_1,\alpha_3 -\alpha_1)\Bigr] + \nn\\
 && + s^{(4)}\ln\frac{2}{a} ,
\ea
where the dependence on $a$ disappears, as $\int_{0}^{\infty}d\alpha\left(\frac{1}{\alpha + a}-\frac{1}{\alpha + a'}\right)=\ln\frac{a'}{a}$. For simplicity, we take $a=2$ and get the final answer
\ba
&& t^{(4)} = -\frac{1}{12(2\pi)^4}\int d\alpha_1d\alpha_2d\alpha_3 \Bigl[g^{(4)}(\alpha_1,\alpha_2,\alpha_3)\ln\xi-\frac{1}{2(|\alpha_1|+ 2)}g^{(4)}_{as}(\alpha_2,\alpha_3)-\nn\\
&&-\frac{1}{2(|\alpha_2|+2)}g^{(4)}_{as}(\alpha_1,\alpha_3)-
 \frac{1}{2(|\alpha_3|+2)}g^{(4)}_{as}(\alpha_1,\alpha_2) -
 \frac{1}{2(|\alpha_1| + 2)}g^{(4)}_{as}(\alpha_2 -\alpha_1,\alpha_3 -\alpha_1)\Bigr] + \nn\\
 && + (\ln 2 -\gamma) J^{(4)} +  s^{(4)} \, .
\ea
A rough numerical estimate returns $t^{(4)}\simeq -0.006133$,
which added to the two-particle contribution yields
\be
t^{(2)} +  t^{(4)}\simeq -0.01432 \, .
\ee

We notice that, on the contrary of $J^{(4)}$ and $s^{(4)}$, $t^{(4)}$ is of the same order as the two-particle term $t^{(2)}\simeq -0.00819$. This suggests that we might need a larger $n$ to evaluate this coefficient with a good accuracy. An optimal estimate of $t$ is still missing, as the Montecarlo evaluations by \cite{BSV4,BEL} furnish $t\simeq -0.01$ with only one significant digit.

\subsubsection{$2n$ scalars}

Referring to the notation
\be
\ln W = \mathcal{F}\simeq J \ln (1/z) + s \ln \ln (1/z) + t
\ee
we get the following expressions of the $2n$ particle contributions to $J$, $s$ and $t$: the leading divergence $J$ is just simply given by
\be
J^{(2n)}=-\frac{2}{(2n)! (2\pi)^{2n}} \int \prod _{i=1}^{2n-1} d\alpha _i g^{(2n)}(\alpha _1,\ldots , \alpha _{2n-1}) \, ,
\ee
while $s^{(2n)}$ comes from the integral
\be\label{deltas2n}
-\frac{2}{(2n)! (2\pi)^{2n}} \int _{z\xi <1}\prod _{i=1}^{2n-1} d\alpha _i g^{(2n)}(\alpha _1,\ldots , \alpha _{2n-1})
\ln \xi \simeq s^{(2n)} \ln \ln (1/z) + t^{(2n)}_3 \, ,
\ee
which also contains the finite piece $t^{(2n)}_3$. As concerns $t$, we have three contributions $t^{(2n)}=t^{(2n)}_1+ t^{(2n)}_2+ t^{(2n)}_3$. The first is the simplest and is given by the constant term in the expansion of $K_0$
\be
 t^{(2n)}_1=(\ln 2 - \gamma)J^{(2n)} \, ,
\ee
while the second comes from the removal of the cutoff in the computation of $J^{(2n)}$ and reads
\be
t^{(2n)}_2=\lim_{z\to 0}\left[-\frac{2\ln(1/z)}{(2n)!(2\pi)^{2n}}\int_{z\xi >1}\prod _{i=1}^{2n-1} d\alpha _i g^{(2n)}(\alpha _1,\ldots , \alpha _{2n-1})\right] \, .
\ee
Following the $n=1,2$ cases, it is shown to be equal\footnote{The regions are the same, where the decay is just enough for the function $g^{(2n)}$ to be $L^{1}(\mathbb{R}^{2n-1})$.} to $s^{(2n)}$. Collecting everything, we obtain
\be
t^{(2n)}=(\ln 2 - \gamma)J^{(2n)} + s^{(2n)} + t^{(2n)}_3 \, .
\ee

To summarize, we provided many explicit formul{\ae} for the coefficients $J$, $s$ and $t$ parametrising the small $z$ limit of $W$, see the expansion of the logarithm (\ref{F2nExp}). The series representation of the coefficients is a very effective procedure, as their contributions $J^{(2n)}$, $s^{(2n)}$ and $t^{(2n)}$ can be extracted, in most cases analytically, from the integral in (\ref{I2n}).  For $n=2$, the expected values for $J$ and $s$ are already reproduced with a good accuracy. On the other hand, the difficult constant $t$ is still calling for a better evaluation. A better numerical/analytical analysis of (\ref{I2n}) would yield their values with more precision, without the numerical subtleties of the direct evaluation of $W$, as in \cite{BSV4,BEL}. Eventually, by means of (\ref{lambda-z}), the coefficients $J$, $s$ and $t$ can be used to parametrise the scalar contribution $W$ in terms of the coupling constant $\lambda$.

\section{Polygonal Wilson loop $N>6$}
\label{ScalarPol}

This section deals with the more general case, a null polygonal Wilson loop composed by $N>6$ edges. We stick to the scalar contribution in the strong coupling limit, \emph{i.e.} we have $N-4$-point function in the $O(6)$. The author \cite{BSV4} pushed forward, as far the leading order is concerned, a proposal for the strong coupling limit for the general polygon, which reads

\be\label{BSVJProp} 
W_N\simeq e^{A_N\sqrt{\lambda}}, \quad A_N=\frac{(N-4)(N-5)}{48N}\equiv \frac{J_N}{4}
\ee

This follows from the application of the standard Operator Product Expansion to the twist field $\hat{P}$, which suggests the scaling above.

The method exposed and applied in the previous section still works, with minor modifications \cite{BFPR5}.

Thinking in terms of the $O(6)$ correlation functions, the strong coupling limit of the loop $W_N$ corresponds to the $(N-4)$-point function 

\be 
W_N(\tau_1,\sigma_1;\cdots;\tau_{N-5},\sigma_{N-5};m)=\langle 0|\mathcal{P}(z_1)\cdots\mathcal{P}(z_{N-4})|0\rangle
\ee

The cross ratios\footnote{Along with $\phi_i$ which do not appear in the formula.} $\tau_i$, $\sigma_i$, where $i=1,..,N-5$, fix the geometry of the polygon. The $O(6)$ dimensionless coordinates are defined by extending the definition to $N>6$, according to

\be\label{zcross}
z_i-z_{i+1}=(m\tau_i,m\sigma_i)
\ee

where $m$ is the mass (\ref{mgap}) of the scalars. Again, the strong coupling limit corresponds to the short-distance regime $z_i-z_{i+1}\to 0$ in the correlator.

Instead of just one as in the hexagon case, now we insert $N-5$ identities inside the correlator, getting the form factor series for the multi-point function/Wilson loop $W_N$, which reads
\small
\ba\label{WN}
W_N&=&\sum_{n_1,\cdots,n_{N-5}=0}^{\infty}\prod_{l=1}^{N-5}\frac{1}{(2n_l)!}\int\prod_{l=1}^{N-5}\prod_{i_l=1}^{2n_l}\left(\frac{d\theta_{i_l}^{(l)}}{2\pi}e^{-m\tau_l\sum_{i_l=1}^{2n_l}\cosh\theta_{i_l}^{(l)}}e^{+im\sigma_l\sum_{i_l=1}^{2n_l}\sinh\theta_{i_l}^{(l)}}\right)\cdot \\
&&\cdot G^{(2n_1,\cdots ,2n_{N-5})}(\vec{\theta}^{(1)};\vec{\theta}^{(2)};\cdots ;\vec{\theta}^{(N-6)};\vec{\theta}^{(N-5)})  \nn
\ea
\normalsize
where the short-hand vector notation has been introduced $\vec{\theta}^{(l)}=(\theta^{(l)}_1,\cdots,\theta^{(l)}_{2n_l})$. The functions $G^{(2n_1,\cdots ,2n_{N-5})}$ are the generalization of the $G^{(2n)}$ introduced for $N=6$ and they depend on the form factors of the twist operator according to

\be\label{multif}
G^{(2n_1,\cdots ,2n_l)}=\sum_{j^{(1)}_1,\cdots ,j^{(1)}_{2n_1}}\cdots\sum_{j^{(l)}_1,\cdots ,j^{(l)}_{2n_l}}\langle 0|\mathcal{P}|\phi_{j^{(1)}_1}(\theta^{(1)}_1)\cdots \phi_{j^{(1)}_{2n_1}}(\theta^{(1)}_{2n_1})\rangle\cdots \langle \phi_{j^{(l)}_1}(\theta^{(l)}_1)\cdots \phi_{j^{(l)}_{2n_l}}(\theta^{(l)}_{2n_l})|\mathcal{P}|0\rangle
\ee

This already leads us to an important property: thanks to $\langle 0|\mathcal{P}|0\rangle=1$, whenever the vacuum appears in one or more intermediate states the function $G$ decouples and can be expressed in terms of those of smaller polygons. For instance, for the heptagon $N=7$ 

\be\label{dec7}
G^{(2n,0)}(\theta_1,\cdots,\theta_{2n};\emptyset)=G^{(0,2n)}(\emptyset;\theta_1,\cdots,\theta_{2n})=G^{(2n)}(\theta_1,\cdots,\theta_{2n})
\ee

while for $N=8$ we have

\small
\be\label{dec8}
G^{(2n,0,0)}=G^{(0,2n,0)}=G^{(0,0,2n)}=G^{(2n)} , \quad G^{(2n,2m,0)}=G^{(0,2n,2m)}=G^{(2n,2m)}, \quad G^{(2n,0,2m)}=G^{(2n)}G^{(2m)}
\ee
\normalsize

where we omitted the rapidities for brevity. This decoupling holds in more generality and it is described in details in the Appendix \ref{FactConn}.

The leivmotiv of the method is the same, \emph{i.e.} we switch to the logarithm $\mathcal{F}_N\equiv\log W_N$ which enjoys the sum

\ba\label{FN}
\mathcal{F}_N&=&\sum^{\infty}_{\stackrel{(n_1,\cdots,n_{N-5})\neq (0,\cdots,0)}{n_1,\cdots,n_{N-5}=0}}\prod_{l=1}^{N-5}\frac{1}{(2n_l)!}\int\prod_{l=1}^{N-5}\prod_{i_l=1}^{2n_l}\left(\frac{d\theta_{i_l}^{(l)}}{2\pi}e^{-m\tau_l\sum_{i_l=1}^{2n_l}\cosh\theta_{i_l}^{(l)}}e^{+im\sigma_l\sum_{i_l=1}^{2n_l}\sinh\theta_{i_l}^{(l)}}\right)\cdot \nn \\
&&\cdot g^{(2n_1,\cdots ,2n_{N-5})}(\vec{\theta}^{(1)};\vec{\theta}^{(2)};\cdots ;\vec{\theta}^{(N-6)};\vec{\theta}^{(N-5)}) \equiv \sum^{\infty}_{\stackrel{(n_1,\cdots,n_{N-5})\neq (0,\cdots,0)}{n_1,\cdots,n_{N-5}=0}} \mathcal{F}_N^{(2n_1,....,2n_{N-5})}  
\ea

where the connected counterparts $g^{(2n_1,\cdots ,2n_{N-5})}$ appear. Along the same line of the hexagon, they are related to the $G$'s in a combinatorial fashion. The exact relation between them is displayed in the Appendix \ref{FactConn}.

From the decoupling (\ref{dec7},\ref{dec8}), a crucial property of the connected functions follows. When the vacuum is external, \emph{i.e.} in the first or the last state, the formula is the same as for the $G$'s, for instance

\be\label{decg7}
g^{(2n,0)}=g^{(0,2n)}=g^{(2n)} , \quad g^{(2n,0,0)}=g^{(2n)} , \quad g^{(2n,2m,0)}=g^{(2n,2m)}
\ee

whereas, on the other hand, the properties of $g^{(2n,0,2m)}$ are more interesting: it is possible to prove that it is vanishing for $m,n \neq 0$. 

This identitiy follow from (\ref{dec1}), (\ref{dec2}) and can be further generalised to the important
\be\label{zero_int}
g^{(\cdots , 2n, 0,0,\cdots ,0,0,2m,\cdots)}=0 ,  \quad m,n\neq 0
\ee
that is, the connected function vanishes whenever we have a string of zeroes in the internal intermediate states, bounded by two non-zero $2m,2n$. 
The reason to this feature is pretty much the same as that only connected graphs contribute to the logarithm of the partition function. As we are going to see in the next paragraph, the property (\ref{zero_int}) allows us to find an interesting recursion formula, partially reconstructing the $N$-gon from the smaller polygons.

\subsection{The recursion formula}

Here we show that, thanks to the special decoupling properties of the $G$'s and, consequently, formulae (\ref{decg7},\ref{zero_int}) for the $g$'s, a recursion formula among polygons exists. This allows us to describe the polygon $W_N$ in terms of smaller ones up to corrections, which become subleading as $N$ grows. 

To see how it works, we start with the easiest case, the heptagon, whose series contains two sums over the intermediate states

\be
\mathcal{F}_7(\tau_1,\sigma_1;\tau_2,\sigma_2)=\sum_{(n,m)\neq (0,0)}^{\infty}\mathcal{F}_7^{(2n,2m)}(\tau_1,\sigma_1;\tau_2,\sigma_2)
\ee

The key point is to notice that the decoupling property (\ref{decg7}) entails

\be
\mathcal{F}_7^{(2n,0)}(\tau_1,\sigma_1;\tau_2,\sigma_2)=\mathcal{F}_6^{(2n)}(\tau_1,\sigma_1), \quad \mathcal{F}_7^{(0,2n)}(\tau_1,\sigma_1;\tau_2,\sigma_2)=\mathcal{F}_6^{(2n)}(\tau_2,\sigma_2)
\ee

so that, as the sum can be split into three contributions

\be
\mathcal{F}_7=\sum_{(n,m)\neq (0,0)}^{\infty}\mathcal{F}_7^{(2n,2m)}=\sum_{n=1}^{\infty}\mathcal{F}_7^{(2n,0)}(\tau_1,\sigma_1) + \sum_{n=1}^{\infty}\mathcal{F}_7^{(0,2n)}(\tau_2,\sigma_2) + \sum_{n,m=1}^{\infty}\mathcal{F}_7^{(2n,2m)}(\tau_1,\sigma_1;\tau_2,\sigma_2)
\ee

implies the relation 

\be
\mathcal{F}_7(\tau_1,\sigma_1;\tau_2,\sigma_2)=\mathcal{F}_6(\tau_1,\sigma_1) + \mathcal{F}_6(\tau_2,\sigma_2) + \sum_{n,m=1}^{\infty}\mathcal{F}_7^{(2n,2m)}(\tau_1,\sigma_1;\tau_2,\sigma_2)
\ee

which describes the heptagon as composed by the two inner hexagons plus corrections.

To see the special property (\ref{zero_int}) at work, we move to the octagon which reads (omitting the obvious cross ratios dependence)

\be\label{oct}
\mathcal{F}_8 = \sum_{n=1}^{\infty}\left(\mathcal{F}_8^{(2n,0,0)} + \mathcal{F}_8^{(0,2n,0)} + \mathcal{F}_8^{(0,0,2n)}\right) + \sum_{n,m=1}^{\infty}\left(\mathcal{F}_8^{(2n,2m,0)}+\mathcal{F}_8^{(0,2n,2m)}\right)  +\sum_{n,m,l=1}^{\infty}\mathcal{F}_8^{(2n,2m,2l)}
\ee

where it is important the absence of $\mathcal{F}_8^{(2n,0,2m)}$, thanks to (\ref{zero_int}).

With a simple rearrangement of the various pieces, we get

\ba 
\mathcal{F}_8(\tau_1,\sigma_1;\tau_2,\sigma_2;\tau_3,\sigma_3) &=&\mathcal{F}_7(\tau_1,\sigma_1;\tau_2,\sigma_2) + \mathcal{F}_7(\tau_2,\sigma_2;\tau_3,\sigma_3) - \mathcal{F}_6(\tau_2,\sigma_2) + \nn \\
&+& \sum_{n,m,l=1}^{\infty}\mathcal{F}_8^{(2n,2m,2l)} (\tau_1,\sigma_1;\tau_2,\sigma_2;\tau_3,\sigma_3)
\ea 

\emph{i.e.} the octagon is mainly composed by the two inner heptagons, from which we substract their overlap, namely the middle hexagon. This formula is straightforwardly extended for any $N$, to the beautiful recursion relation

\ba\label{RecF-ex}
&&\mathcal{F}_{N}(\tau_1,\sigma_1;\dots;\tau_{N-5},\sigma_{N-5}) = \mathcal{F}_{N-1}(\tau_1,\sigma_1;\dots;\tau_{N-6},\sigma_{N-6})+\mathcal{F}_{N-1}(\tau_2,\sigma_2;\dots;\tau_{N-5},\sigma_{N-5})- \nn\\
&& -\mathcal{F}_{N-2}(\tau_2,\sigma_2;\dots;\tau_{N-6},\sigma_{N-6}) + \sum_{n_1,\cdots,n_{N-5}=1}^{\infty}\mathcal{F}_{N}^{(2n_1,\cdots,2n_{N-5})}(\tau_1,\sigma_1;\dots;\tau_{N-5},\sigma_{N-5}) 
\ea

where, again, the interpretation is clear: up to corrections ($2(N-5)$ particles or more), a $N$-gon is composed by the two inner $(N-1)$-gons, to which we subtract the middle $(N-2)$-gon.

\subsection{The strong coupling limit}

As we are interested in the strong coupling limit, we need to address the short-distance regime. We expect, taking the hint from the hexagon, that the purely $N$-gonal terms $\mathcal{F}_N^{(2n_1,...,2n_{N-5})}$ behave as

\be\label{F_Nexp}
\mathcal{F}_N^{(2n_1,...,2n_{N-5})}=J_N^{(2n_1,...,2n_{N-5})}\log (1/m) + s_N^{(2n_1,...,2n_{N-5})}\log\log (1/m) + O(1)
\ee

which would allow us to apply the recursion formula to the coefficients as well. This is actually what we are going to show in the following.

\subsubsection{N-gonal corrections}

In this paragraph we analyse the contribution $\mathcal{F}_N^{(2n_1,....,2n_{N-5})}$, showing that it enjoys the expansion (\ref{F_Nexp}). We also give a formula for the leading coefficient $J_N^{(2n_1,....,2n_{N-5})}$. 

\paragraph{Heptagon}

We first deal with the heptagon, extending the argument in \cite{BFPR2} and depicted in the Section \ref{ScalarHex} for $N=6$. The starting expression is
\be
\mathcal{F}_7^{(2n,2m)}=\frac{1}{(2n)!(2m)!}\int\prod_{i=1}^{2n}\frac{d\theta_i}{2\pi}\prod_{j=1}^{2m}\frac{d\theta'_j}{2\pi}e^{-m\tau_1\sum_i\cosh\theta_i}e^{-m\tau_2\sum_j\cosh\theta'_j+im\sigma_2\sum_j\sinh\theta'_j}g^{(2n,2m)}
\ee

where, for simplicity, we got rid of the cross ratio $\sigma_1$ by a rotation.\footnote{Similarly to the hexagon, where the cross ratio $\sigma$ disappears.}

The procedure follows the same pattern of the previous section, where we dealt with the hexagon. The connected function $g^{(2n,2m)}$ depends on the differences $\theta_{ij}$, $\theta'_{ij}$ and $\theta_i-\theta'_j$, thus we define $\alpha_i\equiv \theta_i-\theta_1$, $\alpha'_j\equiv\theta'_j-\theta_1$, with $i=2,\cdots,2n$ and $j=1,\cdots,2m$ so that our variables are now $\theta_1$, $\alpha_i$ and $\alpha'_j$.
As $g^{(2n,2m)}$ does not depend on $\theta_1\equiv\theta$, we can integrate over it to get
\ba\label{F7theta}
\mathcal{F}_7^{(2n,2m)}&=&\frac{1}{(2n)!(2m)!}\int\prod_{i=2}^{2n}\frac{d\alpha_i}{2\pi}\prod_{j=1}^{2m}\frac{d\alpha'_j}{2\pi}g^{(2n,2m)}(\alpha_2,\cdots,\alpha_{2n};\alpha'_1,\cdots,\alpha'_{2m})\cdot \\
&\cdot & \int d\theta \exp\left[-m\tau_1\xi\cosh\left(\theta+\eta \right) - m\tau_2\xi'\cosh\left(\theta+\eta' \right) + im\sigma_2\xi'\sinh\left(\theta+\eta'\right)\right]   \nn
\ea
where $\xi$, $\xi'$, $\eta$ and $\eta'$ are functions of $\alpha_i$, $\alpha'_j$ through
\ba
 1 &+&\sum_{i=2}^{2n}\cosh\alpha_i=\xi\cosh\eta , \quad \sum_{i=2}^{2n}\sinh\alpha_i=\xi\sinh\eta \\
&& \sum_{j=1}^{2m}\cosh\alpha'_j=\xi'\cosh\eta' , \quad \sum_{j=1}^{2m}\sinh\alpha'_j=\xi'\sinh\eta' \nn
\ea

The integral over $\theta$ in (\ref{F7theta}) is the modified version of what we had for the hexagon, $2 K_0(z\xi)\equiv\int d\theta e^{-z\xi\cosh\theta}$. It depends on four variables (we can shift by $\eta$ and the integral depends only on $\eta'-\eta$), but in the limit $m\to 0$ the leading (divergent) term can be extracted by trading the exponentials for a finite integration volume $-\log(1/m)<\theta<\log (1/m)$

\ba\label{mto0}
&&\int d\theta \exp\left[-m\tau_1\xi\cosh\theta - m\tau_2\xi'\cosh\left(\theta+\eta' -\eta \right) + im\sigma_2\xi'\sinh\left(\theta+\eta' -\eta\right)\right]\simeq \nn \\
&& \simeq \int_{-\log(1/m)}^{\log(1/m)}d\theta=2\log(1/m)
\ea

which gives, as coefficient for the leading term $\log(1/m)$, the integral

\be
J_7^{(2n,2m)}=\frac{2}{(2n)!(2m)!}\int\prod_{i=2}^{2n}\frac{d\alpha_i}{2\pi}\prod_{j=1}^{2m}\frac{d\alpha'_j}{2\pi}g^{(2n,2m)}(\alpha_2,\cdots,\alpha_{2n};\alpha'_1,\cdots,\alpha'_{2m})
\ee

The subleading divergence, parametrised here by $s_N^{(2n,2m)}$, appears when considering the subleading contribution in the integral (\ref{mto0}). It requires, as in the hexagon case, the introduction of a cutoff, whose removal should yield a term proportional to $\log\log(1/m)$. However, this procedure is more involved for $N>6$ and we will stick to the leading order only.

\paragraph{General case $N>7$}

For the most general polygon, we need to address
\ba\label{FNnnn}
\mathcal{F}_N^{(2n_1,\dots,2n_{N-5})} &=& \prod_{l=1}^{N-5}\frac{1}{(2n_l)!}\int\prod_{l=1}^{N-5}\prod_{i_l=1}^{2n_l}\left(\frac{d\theta_{i_l}^{(l)}}{2\pi}e^{\displaystyle-m\tau_l\sum_{i_l=1}^{2n_l}\cosh\theta_{i_l}^{(l)}+im\sigma_l\sum_{i_l=1}^{2n_l}\sinh\theta_{i_l}^{(l)}}\right)\cdot \nn\\
&&\cdot g^{(2n_1,\cdots ,2n_{N-5})}(\vec{\theta}^{(1)};\cdots;\vec{\theta}^{(N-5)})  \nn
\ea
We proceed in the same way as for the heptagon case, in the first place by suppressing one cross ratio
\be
\mathcal{F}_N^{(2n_1,\dots,2n_{N-5})}(\tau_1,\sigma_1;\dots;\tau_{N-5},\sigma_{N-5})=
\mathcal{F}_N^{(2n_1,\dots,2n_{N-5})}(\tau_1',0;\tau_{2}',\sigma_{2}';\dots;\tau_{N-5}',\sigma_{N-5}')
\ee
where $\tau_1'=\sqrt{\tau_1^2+\sigma_1^2}$, while $\tau_k'=\frac{\tau_1\tau_k+\sigma_1\sigma_k}{\sqrt{\tau_1^2+\sigma_1^2}}$ and
$\sigma_k'=\frac{-\sigma_1\tau_k+\tau_1\sigma_k}{\sqrt{\tau_1^2+\sigma_1^2}}$ for $k\neq 1$\,, then omitting the prime for simplicity.
We make use of the variables $\alpha_i^{(l)}\equiv \theta_i^{(l)}-\theta_1^{(1)}$ and $\theta_1^{(1)}\equiv\theta$ and introduce
the quantities $\xi,\,\eta,\,\xi^{(l)},\,\eta^{(l)}$, depending on the $\alpha_j^{(l)}$'s through to the relations
\ba
 1 &+&\sum_{i=2}^{2n_1}\cosh\alpha_i^{(1)}=\xi\cosh\eta \ , \qquad \sum_{i=2}^{2n_1}\sinh\alpha_i^{(1)}=\xi\sinh\eta \\
&& \sum_{j=1}^{2n_l}\cosh\alpha_j^{(l)}=\xi^{(l)}\cosh\eta^{(l)} \ ,\qquad \sum_{j=1}^{2n_l}\sinh\alpha^{(l)}_j=\xi^{(l)}\sinh\eta^{(l)}\ . \nn
\ea
The quantity (\ref{FNnnn}) can be thus recast into
\small
\ba
&&\mathcal{F}_N^{(2n_1,\dots,2n_{N-5})} = \prod_{l=1}^{N-5}\frac{1}{(2n_l)!}\int\frac{d\q}{2\pi}\int\prod_{l=1}^{N-5}\prod_{i_l=1}^{2n_l}
\left[\frac{d\alpha_{i_l}^{(l)}}{2\pi}\,e^{\displaystyle-m\tau_l\xi^{(l)}\cosh(\q+\eta^{(l)})+im\sigma_l\sinh(\q+\eta^{(l)})}\right]\cdot \nn\\
&&\cdot\ e^{-m\tau_1\xi\cosh\left(\theta+\eta \right)}\,
g^{(2n_1,\cdots ,2n_{N-5})}(\alpha_2^{(1)},\cdots,\alpha_{2n_1}^{(1)};\alpha_1^{(2)}\cdots\alpha_{2n_{N-5}}^{(N-5)}) \ .
\ea
\normalsize

which eventually leads, making use of the generalized version of (\ref{mto0}), to the leading correction $J_N^{(2n_1,\cdots ,2n_{N-5})}$ as an integral over all the variables $\alpha_i^{(j)}$ of the connected function $g^{(2n_1,\cdots ,2n_{N-5})}$

\be 
J_N^{(2n_1,\cdots ,2n_{N-5})}=2\prod_{l=1}^{N-5}\frac{1}{(2n_l)!}\int\prod_{l=1}^{N-5}\prod_{i_l=1}^{2n_l}\frac{d\alpha_{i_l}^{(l)}}{2\pi}g^{(2n_1,\cdots ,2n_{N-5})}(\alpha_2^{(1)},\cdots,\alpha_{2n_1}^{(1)};\alpha_1^{(2)}\cdots\alpha_{2n_{N-5}}^{(N-5)})
\ee

In the expansion (\ref{F_Nexp}), the cross ratios contribute only to the finite term $O(1)$ and not to the divergence contributions: thus $J$ and $s$ are independent on the geometry of the loop. This fact is very important, as it entails a simpler recursion formula for the coefficients $J_N$, $s_N$

\ba\label{RecJs}
J_{N}&=& 2J_{N-1}-J_{N-2} + \sum_{n_1,\cdots,n_{N-5}=1}^{\infty}J_{N}^{(2n_1,\cdots,2n_{N-5})} \nn \\
s_{N}&=& 2s_{N-1}-s_{N-2} + \sum_{n_1,\cdots,n_{N-5}=1}^{\infty}s_{N}^{(2n_1,\cdots,2n_{N-5})}
\ea

where, of course, they enjoy the expansion

\be
J_N=\sum^{\infty}_{\stackrel{(n_1,\cdots,n_{N-5})\neq (0,\cdots,0)}{n_1,\cdots,n_{N-5}=0}}J_N^{(2n_1,...,2n_{N-5})}, \quad s_N=\sum^{\infty}_{\stackrel{(n_1,\cdots,n_{N-5})\neq (0,\cdots,0)}{n_1,\cdots,n_{N-5}=0}}s_N^{(2n_1,...,2n_{N-5})}
\ee

From the parametrization of the $N$-gon in terms of the mass

\be\label{FNpar}
\mathcal{F}_N=J_N\log(1/m) + s_N\log\log(1/m) + O(1)
\ee

we can find, using formula (\ref{mgap}), its strong coupling expansion

\be
\mathcal{F}_N = J_N\log(1/m) + s_N\log\log(1/m) + O(1) = \frac{J_N}{4}\sqrt{\lambda} + \left(s_N-\frac{J_N}{4}\right)\log\sqrt{\lambda} + O(1)
\ee

\subsubsection{Solution of the recursion formula}

Here we analyze in details the recursion formula for the coefficients $J,s$, finding the general solution and comparing with the prediction of \cite{BSV4}. Everything we say is valid for $s_N$ as well. We write it as

\be\label{Recdelta}
J_{N} = 2J_{N-1}-J_{N-2} + \delta_N 
\ee

where the purely $N$-gonal contributions are 

\be\label{deltaNJ}
\delta_N\equiv \sum_{n_1,\cdots,n_{N-5}=1}^{\infty}J_{N}^{(2n_1,\cdots,2n_{N-5})} 
\ee

We can solve it iteratively with the initial conditions\footnote{The square and the pentagon are trivial, $W_4=W_5=1$.} $J_4=J_5=0$ and express the solution as

\be\label{Fdelta}
J_N = \sum_{n=6}^{N}(N+1-n)\delta_n
\ee

We suppose $\delta_N\to 0$ for large $N$ and study the homogeneous recursion formula: we guess a linear large $N$ behaviour

\be\label{largeN} 
J_N = aN+b +o(1)
\ee

which is the general solution to $J_N = 2J_{N-1}-J_{N-2}$, corresponding to the discrete version of $\partial^2_NJ_N=0$. 

From (\ref{largeN}) and with the expansion (\ref{Fdelta}) we get an expression for our coefficients $a,b$

\be\label{ab}
a=\sum_{n=6}^{\infty} \delta_n, \quad b=\sum_{n=6}^{\infty} (1-n)\delta_n
\ee

We point out that, a posteriori, (\ref{largeN}) and (\ref{ab}) make sense only if $\delta_n$ goes to zero fast enough for the series of $a,b$ to be convergent. The solution (\ref{Fdelta}) is, on the contrary, more general and does not rely on any particular feature of $\delta_n$. 
If we require a minimal corrections to (\ref{largeN}) of the type $N^{-1}$

\be 
J_N = aN+b + \frac{c}{N}
\ee
we fix, from the recursion formula (\ref{Recdelta}), $\delta_N$ up to a prefactor

\be\label{deltaN}
\delta_N=\frac{2c}{N(N-1)(N-2)}
\ee
which shows a cubic decay, that is the minimum required for the series of $b$ to be convergent.

From the $\delta_n$ found in (\ref{deltaN}) we can evaluate the series (\ref{ab}) to obtain

\be 
b=-\frac{9}{20}c, \quad a=\frac{1}{20}c
\ee

while the summation of (\ref{Fdelta}) gives

\be\label{BSVJ} 
J_N=\frac{c}{20}\frac{(N-4)(N-5)}{N}
\ee

reproducing, up to a prefactor, the expected result (\ref{BSVJProp}) \cite{BSV4} for the leading order. In addition, it allows us to push forward the following proposal: the behaviour (\ref{BSVJ}) may hold for the subleading coefficient $s_N$ as well, with of course a different prefactor so that $s_6=-1/24$.

A simpler way to derive the same result, which does not involve any series, is the following: we require the linear large $N$ limit (\ref{largeN}) and ask that the only zeros of $J_N$ are $J_4=J_5=0$; the simplest rational solution is then

\be
J_N=a\frac{(N-4)(N-5)}{N}
\ee
where $a$ is determined by the hexagon as $a=3J_6$, thus we write, extending also the argument to $s_N$

\be 
J_N=3\frac{(N-4)(N-5)}{N}J_6, \quad s_N=3\frac{(N-4)(N-5)}{N}s_6
\ee

\paragraph{Numerics} We conclude with some comments on the numerical computations. The non-trivial corrections to the leading coeffient $J_N$ are the purely $N$-gonal ones $\delta_N$, formula (\ref{deltaNJ}). Using formula (\ref{deltaN}) and fixing the prefactor $c=5/3$ (by means of $J_6=1/36$), gives the expected value for the heptagon $\delta_7=1/63$. We can evaluate the first term $J^{(2,2)}_7$, which should be the main contribution to $\delta_7$. The numerical evaluation is roughly in agreement with the expected result. For the computation of the heptagonal function $g^{(2,2)}$, following from $G^{(2,2)}$ defined through (\ref{multif}), we made use of the general pentagon transitions found in \cite{Bel1607}.

\chapter*{Conclusions}
\addcontentsline{toc}{chapter}{Conclusions}

In this thesis we computed the null polygonal Wilson loops in planar $\mathcal{N}=4$ Super Yang-Mills at strong 't Hooft coupling $\lambda\to\infty$, taking advantage of the pentagon approach based on the Operator Product Expansion. This method employs the integrability underlying the gauge theory under examination.  These Wilson loops are dual to the $4d$ gluons scattering amplitudes, therefore the OPE series provides tools for a non-perturbative evaluation of them. This is a unique case in the realm of interacting four dimensional gauge theories.  

Two different contributions stand out in the strong coupling limit. The first corresponds, in the language of the $AdS/CFT$ duality, to the classical string living in $AdS_5$. Its leading order is described in terms of a string whose worldsheet is attached to the polygonal contour on the boundary, the $4d$ Minkowski spacetime. The minimal area problem assumes the form of a set of non-linear integral equations and intriguingly recalls in form that of the Thermodynamic Bethe Ansatz, typical of integrable models. On the gauge side, we reproduced this result by a resummation of the fermions and gluons contributions to the OPE series. Interestingly, the former bind up to generate effective excitations in the strong coupling limit. From the point of view of the Bethe equations they are not physical particles, however they turn out to be useful to describe the strong coupling behaviour of the OPE series. This physical picture is confirmed by comparing with the string side, where these excitations are naturally present in the spectrum. This result constitutes an important check on the validity of the OPE series for the Wilson loop. 

The other regime corresponds to the non-perturbative string dynamics on the sphere $S^5$ and it has not been computed yet in that framework. However, the OPE series provides an easy evaluation of this effect, as the Wilson loop becomes a correlation function in the $O(6)$ non-linear $\sigma$-model. The strong coupling limit corresponds to the short-distance regime for the correlator, reducing the problem to find the scaling dimension of the twist operator appearing in the OPE series. Using a standard technique in the form factor theory, namely passing to the series of the logarithm and studying the connected integrals, we have been able to confirm this proposed correction, which remarkably turns out to be of the same order as the classical one, \emph{i.e.} exponential in $\sqrt{\lambda}$.

At any coupling, we studied the $SU(4)$ structure of the form factors of the operator $\hat{P}$. Its square, once summed over the internal indices, is described by a multiple integral over the auxiliary rapidities of the underlying $SU(4)$ chain. We focused on two cases of particular interest: when we have $n$ couples fermion-antifermion and the part due to an even number of scalars.  We managed to solve sistematically these integrals by residues, giving them an interesting representation in terms of Young diagrams. An expression as a sum over rational functions is thus obtained. In addition, the polar structure in the complex plane has been unveiled, recasting the remaining informations in certain polynomials. We must stress that the simplified form of the squared form factor, \emph{i.e.} the split in a dynamical contribution and a matrix part, is rather peculiar in the form factor theory and it should be related to the special features of the twist field under consideration. Finding a detailed explanation of this effect is one of the main investigation to do in the near future.

Intriguingly, along our path we encountered some analogies with the Nekrasov function for $\mathcal{N}=2$ theories. The role of the coupling $g$ is played by the spacetime deformation parameter $\epsilon$, which goes to zero as the coupling increases towards infinity as $\epsilon\sim i/g$. The formation of bound states between mesons in the $\mathcal{N}=4$ flux-tube follows the same pattern as for the instantons, leading to the emergence of the typical dilogarithm function and the associated TBA-like equation. This surely deserves more attention, as integrability also pops out, although differently, in $\mathcal{N}=2$ and there could be additional hidden connections between the two theories, yet to be unravelled.

\section*{Aknowledgements}
\addcontentsline{toc}{chapter}{Aknowledgements}

My first thanks goes to my scientific advisor Davide Fioravanti, who has been my mentor for the PhD and guided me throughout these years of research. For the various results achieved during these years, I am very grateful to Marco Rossi and Simone Piscaglia for their fundamental contributions. I must thank my PhD supervisor Francesco Ravanini for the many advices during these years. I cannot forget the warm hospitality that Didina Serban and Ivan Kostov gave me during my stay at the IpHT in Saclay. I also thank the University of Bologna and INFN for financial support, along with the CEA which eased my accomodation during my stay in Saclay. The biggest thanks goes to my parents, for their constant support and encouragement during my studies.


\begin{appendices}
\noappendicestocpagenum

\chapter{The $\mathcal{N}=2$ Nekrasov function}
\label{NekApp}

In this appendix we discuss in details some features of the Nekrasov instanton partition function $\mathcal{Z}$ for $\mathcal{N}=2$ $SU(N)$ gauge theories, previously introduced in Section \ref{IntSUSY}. It has been proposed in \cite{Nekrasov} to encode the non-perturbative effects to the partition function of the theory. In order to perform the sum over instantons, it needs to be defined in a regularized spacetime, deformed by two parameters $\epsilon_1,\epsilon_2$, called $\Omega$-background. 

The function $\mathcal{Z}$, besides the aforementioned deformations, depends also on the instanton parameter $q$, on the scalar fields VEVs $a_i$ and possibly on the masses of the matter fields. However, for our purpose, we explicit only the dependence on one deformation, say $\epsilon_2\equiv\epsilon$, so that the partition function is a multiple integral over the instanton coordinates $u_i$

\be\label{Nekrasov} 
\mathcal{Z}=\sum_{n=0}^{\infty}\frac{q^n}{n!\epsilon ^n}\oint_{\Gamma}\displaystyle\prod_{i=1}^n\frac{du_i}{2\pi i}Q(u_i)\displaystyle\prod_{i<j}^n e^{\epsilon G(u_{ij})}\displaystyle\prod_{i<j}^n\frac{u_{ij}^2}{u_{ij}^2-\epsilon^2}
\ee

which resembles very much the grand canonical partition function of a one dimensional two-body interacting gas. 

The dependence on the many parameters is encoded in the functions $Q(x),G(x)$. The kernel $G(x)$ is universal, whereas the polynomial $Q(x)$ contains the informations on the gauge group rank and the matter content of the theory. The VEVs $a_i$ of the scalar fields  appear inside this polynomial. The instanton parameter $q$ contains che coupling constant of the theory. We highlighted the dependence on $\epsilon$ as we are expecially interested in the correspondence with the meson series $W^{(M)}$: we will see that $\epsilon$ roughly plays the role of the inverse coupling constant $g^{-1}$. Therefore, the strong coupling limit corresponds to $\epsilon\to 0$, which is called the Nekrasov-Shatashvili limit.

The interacting part has been split in two different contributions. The first, $e^{\epsilon G(u_{ij})}$, is regular\footnote{Its particular form is not important for our analysis, as long as it remains regular in the limit $\epsilon\to 0$.} in the limit $\epsilon\to 0$ and it is named long-range potential as it assumes the maximum value for a finite distance. The short-range potential

\be\label{shortrange} 
\displaystyle\prod_{i<j}^n\frac{u_{ij}^2}{u_{ij}^2-\epsilon^2}
\ee

will play a major role in forming the bound states between instantons.

\paragraph{Relation with $W^{(M)}$}

The similarities between the two series $\mathcal{Z}$, $W^{(M)}$ have been first noticed in \cite{BFPR}, although only for the two particle contribution. The series (\ref{SingMes}), obtained after integrating over the antifermionic variables and employing the strong coupling approximation, has the same form of the Nekrasov function (\ref{Nekrasov}). To make this statement more precise, the function $Q$ mimics the measure and the propagation phase, while the role of the regular part $P^{MM}_{reg}$ is played by the long-range potential. The polar parts of the two series, which are responsible for the bound states and the emergence of the dilogarithm, coincide once we identify $\epsilon\sim i/g$. 

In spite of the many similarities, there are two significant mathematical differences between $W^{(M)}$ and $\mathcal{Z}$, which however do not spoil the duality in the limits $\lambda\to\infty/\epsilon\to 0$. The contour $\Gamma$ in (\ref{Nekrasov}) is closed\footnote{We choose the upper half plane.}, while the path in the small fermion sheet $C_S$ in (\ref{SingMes}) is open. On the other hand, the functions $Q(x),G(x)$ are endowed with poles inside the contour, whereas the only poles in the small fermion sheet in (\ref{SingMes}) come from the polar part. These differences imply that the sources of the subleading corrections are different for the two cases. In addition, we must remember that $W^{(M)}$ is \emph{already} a strong coupling approximation of the OPE series (\ref{OPEFer}), as mesons are composite objects and free fermions should start to contribute for generic $\lambda$. On the other hand, instantons are fundamentals and the series $\mathcal{Z}$ is valid for any $\epsilon$. In fact, the subleading corrections to the NS limit have been computed in \cite{BoF}: those for the Wilson loop are more involved but a part of them should follow the same pattern. The content of this appendix is largely devoted to the study of the NS limit and the exposition, in more details, of the methods applied in Section (\ref{FerHex}) to the meson series.

\section{Path integral, Fredholm determinant and the NS limit}

In this part we study the partition function $\mathcal{Z}$ in details, eventually reproducing the TBA-like equation in the NS limit $\epsilon\to 0$. Even though already obtained in \cite{MenYang}, it is useful to get the result in a different and more straightforward manner. We employ the techniques intorduced the main text and applied to the meson series $W^{(M)}$, giving some more details on the methods. We study separately the two kinds of interactions, long and short range, and then merge the results. 

\subsection{Short-range interaction}

The short-range partition function reads

\be\label{Zshort}
\mathcal{Z}_s=\sum_{n=0}^{\infty}\frac{q^n}{n!\epsilon ^n}\oint_{\Gamma}\displaystyle\prod_{i=1}^n\frac{du_i}{2\pi i}Q(u_i)\displaystyle\prod_{i<j}^n\frac{u_{ij}^2}{u_{ij}^2-\epsilon^2}
\ee
where the integrals are closed in the upper half plane so that we can evaluate them by residues. The parameter $\epsilon$ has a positive imaginary part, therefore there are no poles along the contour and the function is well-defined. We now employ two different techniques to get the leading order in the $\epsilon\to 0$ limit.

\subsubsection{Mayer expansion}

The Mayer expansion was originally introduced to study a classical gas interacting through a two-body potential. The Boltzmann factor is split according to $e^{-V(x)}=1+f(x)$, where $f(x)$ can be pictorially represented as a link between the two particles, representing the nodes of the cluster. It was applied to the Nekrasov function by \cite{MenYang} to obtain the leading oder in the NS limit. In our case, we write the short range as
\be\label{short-range}
\frac{u_{ij}^2}{u_{ij}^2-\epsilon^2} = 1+ \frac{\epsilon^2}{u_{ij}^2-\epsilon^2}
\ee
and expand the product in a sum over all the different $n$-clusters $C_n$
\be
\displaystyle\prod_{i<j}^n\frac{u_{ij}^2}{u_{ij}^2-\epsilon^2} = \sum_{C_n}\displaystyle\prod_{(i,j)\in C_n}\frac{\epsilon^2}{u_{ij}^2-\epsilon^2}
\ee
where the couple $(i,j)$ represents the link between the nodes $i$, $j$ of the cluster. We have then the product over all the links belonging to the specific cluster $C_n$.
The statement of the Mayer expansion is that the logarithm of the grand canonical partition function enjoys the same form with the sum restricted over the connected clusters $C^c_n$. This expansion reads
\be
F_s\equiv \ln \mathcal{Z}_s=\sum_{n=1}^{\infty}\frac{q^n}{n!\epsilon ^n}\oint_{\Gamma}\displaystyle\prod_{i=1}^n\frac{du_i}{2\pi i}Q(u_i)\sum_{C^c_n}\displaystyle\prod_{(i,j)\in C^c_n}\frac{\epsilon^2}{u_{ij}^2-\epsilon^2}
\ee
A connected cluster $C_n^c$ is one in which every node is connected to any other through, at least, one path of links. The connected cluster $C_n^c$ contains at least $n-1$ links, and those are called tree clusters $T_n$. In the Mayer expansion for the weakly interacting gas, the leading order is given by the tree contributions, as the addition of one link increases the order by $\epsilon$. This is actually what happens for the long-range contributions, studied in the next subsection. The issue is more subtle for the short-range, as it is naively of order $\epsilon^2$ from (\ref{short-range}) but it becomes of order $O(1)$ for small distances. What happens is that all the connected clusters contribute to the same order, see the discussion in \cite{MenYang}.
The main point to notice is that the poles of $Q(u)$ give a subleading contribution\footnote{Except for the last integral on $u_n$, for which they are necessary otherwise the whole thing vanishes.}, thus we can extract $Q^n(u_n)$ and write

\be
F_s = \sum_{n=1}^{\infty}\frac{q^n}{n!\epsilon ^n}\oint_{\Gamma} \frac{du_n}{2\pi i}Q^n(u_n)\oint_{\Gamma} \displaystyle\prod_{i=1}^{n-1}\frac{du_i}{2\pi i}\sum_{C^c_n}\displaystyle\prod_{(i,j)\in C^c_n}\frac{\epsilon^2}{u_{ij}^2-\epsilon^2} + O(1)
\ee

The effect of the short-range is recast in the $n-1$ closed integrals, to which we can add all the disconnected clusters, for their contribution is vanishing. We eventually get

\be
F_s = \frac{1}{\epsilon}\sum_{n=1}^{\infty}q^n\oint_{\Gamma} \frac{du}{2\pi i}Q^n(u)J_n(u) + O(1)
\ee

where we defined the multiple integral

\be\label{In}
J_n(u_n) \equiv \frac{1}{n!\epsilon^{n-1}}\oint_{\Gamma} \displaystyle\prod_{i=1}^{n-1}\frac{du_i}{2\pi i}\displaystyle\prod_{i<j}^n\frac{u_{ij}^2}{u_{ij}^2-\epsilon^2} = \frac{1}{n^2}
\ee

As expected, (\ref{In}) does not depend on $u_n$ (the center of the cluster) and it represents a measure factor, due to the short-range potential among the constituents, of the bound state of $n$ instantons. The bound state interpretation comes from the fact that, as concerns the external potential $Q(u_i)$ acting on the instantons, all the constituents enjoy the same coordinate since $Q(u_i+n\epsilon)\simeq Q(u_i)$ at leading order. 


We can sum over $n$ to see the emergence of the dilogarithm at the leading order

\be\label{Fshort}
F_s = \frac{1}{\epsilon}\sum_{n=1}^{\infty}\frac{q^n}{n^2}\int \frac{du}{2\pi i}Q^n(u)+ O(1) = \frac{1}{\epsilon}\int\frac{du}{2\pi i}Li_2\left[qQ(u)\right] + O(1)
\ee

\subsubsection{Fredholm determinant}

The short-range partition function $\mathcal{Z}_s$ enjoys a very nice alternative representation, which allows us to find the leading order (\ref{Fshort}) without any cluster expansion. This representation, valid for any $\epsilon$, is interesting by itself and could also be used to go beyond the leading order or even analyse $\mathcal{Z}_s$ for finite $\epsilon$. This technique has been applied to the meson series in the maix text, here we give some more details.

The key property comes from the Cauchy formula specialized to the short-range interaction

\be
\frac{1}{\epsilon^n}\displaystyle\prod_{i<j}^n\frac{u_{ij}^2}{u_{ij}^2-\epsilon^2} = (-1)^n \det\left(\frac{1}{u_i-u_j-\epsilon}\right)
\ee

from which we can write the whole integrand as a determinant

\be\label{Fredholm}
\mathcal{Z}_s=\sum_{n=0}^{\infty}\frac{(-q)^n}{n!}\int\displaystyle\prod_{i=1}^n\frac{du_i}{2\pi i}\det_{ij} M(u_i,u_j)
\ee

where the kernel $M(u_i,u_j)$ includes the potential $Q$ and reads

\be\label{M}
M(u_i,u_j)=\frac{Q^{1/2}(u_i)Q^{1/2}(u_j)}{u_i-u_j-\epsilon}
\ee

The expression (\ref{Fredholm}) is the definition of the Fredholm determinant

\be\label{Fred}
\mathcal{Z}_s= \det(1 - qM)
\ee

This representation holds for any $\epsilon$, regardless of the functional form of $Q(u)$, as the only property we employed is the Cauchy identity for the short-range interaction. Formula (\ref{Fred}) comes in handy when we consider the logarithm $F_s$ which, by means of the identity $\log\det=\textit{Tr}\log$, becomes

\be\label{SerTrace}
F_s=\log \mathcal{Z}_s =\log\det (1-qM)=\textit{Tr}\log(1-qM)=-\sum_{n=1}^{\infty}\frac{q^n}{n}\textit{Tr}M^n
\ee

The trace of an integral operator is defined as

\be
\textit{Tr}M^n \equiv \int\displaystyle\prod_{i=1}^n\frac{du_i}{2\pi i} \displaystyle\prod_{i=1}^n M(u_i,u_{i+1}) = \int\displaystyle\prod_{i=1}^n\frac{du_i}{2\pi i}Q(u_i) \displaystyle\prod_{i=1}^n \frac{1}{u_i-u_{i+1}-\epsilon}, \quad u_{n+1}\equiv u_1
\ee

Now we employ the small $\epsilon$ limit: the main contribution to the trace is given by the residues of the polar part $\frac{1}{u_i-u_{i+1}-\epsilon}$. Performing the $n-1$ integrations, we obtain

\be\label{TrLead}
\textit{Tr}M^n = -\frac{1}{n\epsilon}\int\frac{du}{2\pi i}Q^n(u) + O(1)
\ee

where the shifts inside the functions $Q(u+k\epsilon)$ have been neglected, as usual. The summation of (\ref{SerTrace}) with (\ref{TrLead}) yields the result (\ref{Fshort}) previously obtained from the cluster expansion, \emph{i.e.} the dilogarithm appears straightforwardly. 

\paragraph{Bound states}

In this paragraph we see how the leading order (\ref{Fshort}), which contains the dilogarithm function, is equivalent to a sum over bound states of instantons. 

Let us recall the partition function $\mathcal{Z}_s$ in the small $\epsilon$ limit

\be
\mathcal{Z}_s = \exp\left[\frac{1}{\epsilon}\int\frac{du}{2\pi i}Li_2\left[qQ(u)\right] + O(1)\right]
\ee

and, expanding both the dilogarithm and the exponential 

\be
\mathcal{Z}_s\simeq \sum_{N=0}^{\infty}\frac{1}{N!\epsilon^N}\left[\int\frac{du}{2\pi i}Li_2[qQ(u)]\right]^N=\sum_{N=0}^{\infty}\frac{1}{N!\epsilon^N}\left[\sum_{a=1}^{\infty}\int\frac{du}{2\pi i}\frac{q^a Q^a(u)}{a^2}\right]^N
\ee

we can write a multiple sum over $a_i$

\be
\mathcal{Z}_s\simeq \sum_{N=0}^{\infty}\frac{1}{N!\epsilon^N}\sum_{a_1=1}^{\infty}\cdots\sum_{a_N=1}^{\infty}\int\displaystyle\prod
_{i=1}^N\frac{du_i}{2\pi i}\frac{q^{a_i}Q^{a_i}(u_i)}{a_i^2}
\ee

Here, $N$ represents the number of composite particles, while the indices $a_i$ tell us how many instantons are bound inside the $i$-th particle.

It is worth to point out that the typical dilogarithm function appears thanks to the particular measure $1/a_i^2$ of the bound states, see the integral (\ref{In}). 

\subsection{Long-range interaction}

In this subsection we deal with the other simplified case, where only the long-range interaction is present. This time, the partition function reads

\be\label{Zlong}
\mathcal{Z}_L=\sum_{n=0}^{\infty}\frac{q^n}{n!\epsilon ^n}\int\displaystyle\prod_{i=1}^n\frac{du_i}{2\pi i}Q(u_i)\displaystyle\prod_{i<j}^n e^{\epsilon G(u_{ij})}
\ee

The difference with respect to $\mathcal{Z}_S$ is that the two-body potential is smooth in the limit $\epsilon\to 0$ and we can push the Mayer expansion all the way through. As before, we define $e^{\epsilon G(u)}\equiv 1 + \epsilon f(u)$, from which the free energy $F_L$ as a sum over all the connected clusters follows

\be
F_L\equiv \ln \mathcal{Z}_L=\sum_{n=1}^{\infty}\frac{q^n}{n!\epsilon ^n}\int\displaystyle\prod_{i=1}^n\frac{du_i}{2\pi i}Q(u_i)\sum_{C^c_n}\displaystyle\prod_{(i,j)\in C^c_n}\epsilon f(u_{ij})
\ee

This time, since we do not have a singular behaviour for $\epsilon\to 0$, there are no subtleties and the leading order is simply given by the tree clusters, which contain $n-1$ links

\be\label{tree}
F_L = \sum_{n=1}^{\infty}\frac{q^n}{n!\epsilon }\int\displaystyle\prod_{i=1}^n\frac{du_i}{2\pi i}Q(u_i)\sum_{T_n}\displaystyle\prod_{(i,j)\in T_n} f(u_{ij}) + O(1)
\ee

We remind that this statement was not true for the short-range potential, where all the connected clusters contribute at the leading order and their combined effect is summarized in the integral (\ref{In}). In the following we apply a more immediate method, which makes use of the Hubbard-Stratonovich transformation as depicted in the maix text for the meson series (\ref{SingMes}).

\subsubsection{Path integral representation}

An efficient method to study $\mathcal{Z}_L$ makes use of a path integral representation \cite{FPR2,BoF}. The long-range potential admits the natural interpretation of the propagator of a quantum field $X(u)$, according to

\be
\left\langle X(u)X(v)\right\rangle \equiv \epsilon G(u-v)
\ee

The Gaussian identity, extended to the functional integration, leads to the important equivalence

\be\label{Hubb}
\displaystyle\prod_{i<j}^n e^{\epsilon G(u_{ij})}=\displaystyle\prod_{i<j}^n e^{\left\langle X(u_i)X(u_j)\right\rangle} = e^{-\frac{1}{2}n\epsilon G(0)} \left\langle\displaystyle\prod_{i=1}^n e^{X(u_i)}\right\rangle
\ee

which enables us to represent the two-body interaction through an average of a product of single particle terms. This procedure is known in literature as the Hubbard-Stratonovich transformation. We define the renormalized instanton parameter as $q'=q e^{-\frac{\epsilon}{2}G(0)}$, to account for the diagonal term in (\ref{Hubb}). The partition function is thus written as the expectation value

\be\label{ZLaverage}
\mathcal{Z}_L=\left\langle \exp\left[\frac{q'}{\epsilon}\int\frac{du}{2\pi i} Q(u)e^{X(u)}\right]\right\rangle
\ee




where the average of a generic functional $\mathcal{F}[X]$ is defined by the path integral

\be
\left\langle \mathcal{F}[X] \right\rangle \equiv \int DX \mathcal{F}[X] \exp\left[\frac{1}{\epsilon} S_0[X]\right] , \quad S_0[X]=- \frac{1}{2}\int \frac{dudv}{(2\pi i)^2}X(u)G^{-1}(u-v)X(v)
\ee

The inverse propagator is defined through

\be
\int \frac{dv}{2\pi i} G^{-1}(u-v)G(v-w)=\delta(u-w)
\ee

with the $\delta(x)$ function is normalized as

\be
\int \frac{du}{2\pi i} f(u)\delta(u-v)=f(v)
\ee

The partition function (\ref{Zlong}) is thus recast as a path integral

\be
\mathcal{Z}_L=\int DX \exp\left[\frac{1}{\epsilon}S[X]\right]
\ee

where the action $S[X]$ is given by

\be
S[X]=S_0[X] + q'\int\frac{du}{2\pi i} Q(u)e^{X(u)}
\ee

We remark that the path integral representation (\ref{ZLaverage}) for (\ref{Zlong}) is valid for any $\epsilon$. However, having extracted a factor $\epsilon^{-1}$ in the action, the limit $\epsilon\to 0$ follows immediately by the saddle point approximation

\be
F_L \simeq \frac{1}{\epsilon}S[X_c]\equiv \frac{1}{\epsilon}S_c
\ee

The saddle point equation $\frac{\delta S[X]}{\delta X(u)}=0$ reads\footnote{At the leading order, the instanton parameter is not corrected.}

\be
qQ(u)e^{X(u)}=\int\frac{dv}{2\pi i}G^{-1}(u-v)X(v)
\ee

which can be expressed in term of the direct kernel $G(x)$ as

\be\label{eom}
X(u)=q\int\frac{dv}{2\pi i}G(u-v)Q(v)e^{X(v)}
\ee

The critical action is then

\be\label{Scr}
S_c = q\int\frac{du}{2\pi i}Q(u)\left[1-\frac{1}{2}X(u)\right]e^{X(u)}
\ee

where $X(u)$ satisfies the classical equation of motion (\ref{eom}).

To check the validity of the method, we can expand the solution (\ref{eom}) in powers of $q$ and then substitute in (\ref{Scr}): considering that $G \simeq f$ at the leading order, what we obtain is the standard expansion over the connected tree clusters (\ref{tree}) for $F_L$. 

\subsection{The full partition function}

Equipped with two efficient techniques to deal with long and short range parts respectively, we are now ready to tackle the whole partition function $\mathcal{Z}$ (\ref{Nekrasov}). We remind that the leading order in the NS limit has already been unravelled in \cite{MenYang}, mainly by means of the Mayer expansion. The approach here is different and much faster, which combines both the techniques discussed above.

To address the problem we apply in sequence the Hubbard-Stratonovich transformation and the Fredholm formula. First, we use the fluctuating field $X(u)$ to obtain a path integral representation

\be\label{Zaverage}
\mathcal{Z}=\left\langle \mathcal{Z}_s[q\to q', Q\to Qe^X]\right\rangle
\ee

which differs from (\ref{ZLaverage}), since we still have the short-range interaction to deal with. As a matter of fact, (\ref{Zaverage}) is the expectation value of a short-range partition function $\mathcal{Z}_s$, where the potential is modified by the fluctuating field through $Q(u) \to Q(u)e^{X(u)}$. We can work out the short-range part with the Fredholm technique, so that we have a fluctuating matrix $M'[X]$, related to $M$ in (\ref{M}) through

\be
M'_{ij}[X]=M_{ij}\exp\left[\frac{1}{2}X(u_i) + \frac{1}{2}X(u_j)\right]
\ee

The full partition function, for any $\epsilon$, is then the expectation value of a Fredholm determinant

\be\label{ZFre}
\mathcal{Z}=\left\langle\det(1-q'M'[X])\right\rangle
\ee

In the NS limit we obtain the same result as for $\mathcal{Z}_s$ and the dilogarithm appears inside the average

\be\label{Zpath}
\mathcal{Z}\simeq \left\langle\exp\left[\frac{1}{\epsilon}\int\frac{du}{2\pi i}Li_2[qQ(u)e^{X(u)}]\right]\right\rangle = \int DX \exp\left[\frac{1}{\epsilon}S[X]\right]
\ee

The total action is

\be
S[X]=-\frac{1}{2}\int\frac{dudv}{(2\pi i)^2}X(u)G^{-1}(u-v)X(v) + \int\frac{du}{2\pi i}Li_2[qQ(u)e^{X(u)}]
\ee

and for small $\epsilon$ the path integral is dominated by its critical value

\be
F=\ln \mathcal{Z} \simeq \frac{1}{\epsilon}S[X_c]=\frac{1}{\epsilon}S_c
\ee

coming from the saddle point, which is the TBA-like equation

\be\label{TBAlikeZ}
X(u) + \int\frac{dv}{2\pi i}G(u-v)\ln\left[1-qQ(v)e^{X(v)}\right]=0
\ee

The critical action $S_c$ is thus given by

\be
S_c=\frac{1}{2}\int\frac{du}{2\pi i}X(u)\ln\left[1-qQ(u)e^{X(u)}\right] + \int\frac{du}{2\pi i}Li_2\left[qQ(u)e^{X(u)}\right]
\ee

that matches the critical Yang-Yang functional obtained in \cite{Bou,MenYang}.

\paragraph{Bound states}

As discussed for the short-range function $\mathcal{Z}_s$, we can get the sum over bound states by expanding the dilogarithm and the exponential inside the average in (\ref{Zpath}). The generalized Gaussian identity (\ref{Hubb}) can be extended and, neglecting the diagonal term, reads

\be
\displaystyle\prod_{i<j}^N e^{a_i a_j \left\langle X(u_i)X(u_j)\right\rangle} \simeq \left\langle\displaystyle\prod_{i=1}^N e^{a_i X(u_i)}\right\rangle
\ee

allowing us to find the alternative expression of the partition function in the $\epsilon\to 0$ limit, as a sum over bound states

\be
\mathcal{Z}\simeq \sum_{N=0}^{\infty}\frac{1}{N!\epsilon^N}\sum_{a_1=1}^{\infty}\cdots\sum_{a_N=1}^{\infty}\int
\displaystyle\prod_{i=1}^N\frac{du_i}{2\pi i}\frac{q^{a_i}Q^{a_i}(u_i)}{a_i^2}\displaystyle\prod_{i<j}^N e^{\epsilon a_i a_j G(u_{ij})}
\ee

This sum runs over the instantons and bound states thereof, emerging from the short-range interaction in the same way as the mesons in (\ref{SingMes}). The interaction between these composite particles is given by the long range part $\epsilon a_i a_jG(u_{ij})$. The numbers $a_i$ represent the number of elementary constituents of the bound state, whose measure is proportional to $a_i^{-2}$: the long range interaction acts between any couple of elementary constituents, so the total effect contains the multiplicity factor $a_i a_j$. 

The representation (\ref{ZFre}) is interesting for several reasons: as we have seen, it yields the leading order in the NS limit without any cluster approach. Furthermore, it could be used to analyse the partition function beyond the leading order and even for any value of the parameter $\epsilon$. As a conclusive analysis, the analogies with the series of mesons (\ref{SingMes}) are even clearer now. The short-range interaction is responsible, in both cases, for the emergence of bound states among them with the associated dilogarithm function.

\chapter{TBA-like equations for $\mathcal{N}=4$ amplitudes}
\label{TBApp}

Here we will review the TBA-like equations for the scattering amplitudes, following the discussion in the main text of subsection \ref{StCou}. They read, for the polygon with $n$ sides \cite{YSA,Anope}

\ba
\ln Y_{2,s}(\theta)&=& - |m_s| \sqrt{2} \cosh (\theta - i \varphi _s) -\int_{\textrm{Im} \theta ' =\varphi _{s}} d\theta'\biggl[K_2(\theta-\theta')
{\cal L}_{s}(\theta ') + \nonumber \\
&+& 2K_1(\theta-\theta')\tilde {\cal L}_{s}(\theta')\biggr] + \int_{\textrm{Im} \theta ' =\varphi _{s-1}} d\theta'\biggl[K_2(\theta-\theta')
\tilde {\cal L}_{s-1}(\theta') + \nonumber \\
&+& K_1(\theta-\theta'){\cal L}_{s-1}(\theta')\biggr] + \int_{\textrm{Im} \theta ' =\varphi _{s+1}} d\theta'\biggl[K_2(\theta-\theta')
\tilde {\cal L}_{s+1}(\theta') + \nonumber \\
&+& K_1(\theta-\theta'){\cal L}_{s-1}(\theta')\biggr] \, , \label {eq1} \\
\ln Y_{1,s}(\theta)&=& - |m_s| \cosh (\theta - i \varphi _s) -C_s -\int_{\textrm{Im} \theta ' =\varphi _{s}} d\theta'\biggl[K_2(\theta-\theta')
\tilde {\cal L}_{s}(\theta') + \nonumber \\
&+& K_1(\theta-\theta'){\cal L}_{s}(\theta')\biggr] + \int_{\textrm{Im} \theta ' =\varphi _{s-1}}
d\theta'\biggl[K_1(\theta-\theta')\tilde {\cal L}_{s-1}(\theta') + \nonumber \\
&+& \frac{1}{2}K_2(\theta-\theta'){\cal L}_{s-1}(\theta') - \frac{1}{2}K_3(\theta -\theta'){\cal M}_{s-1}(\theta')\biggr] + \nonumber \\
&+&  \int_{\textrm{Im} \theta ' =\varphi _{s+1}} d\theta'\biggl[K_1(\theta-\theta')\tilde {\cal L}_{s+1}(\theta') + \frac{1}{2}K_2(\theta-\theta'){\cal L}_{s+1}(\theta') + \nonumber \\
&+& \frac{1}{2}K_3(\theta -\theta'){\cal M}_{s+1}(\theta')\biggr] \, , \label {eq2} \\
\ln Y_{3,s}(\theta)&=& - |m_s| \cosh (\theta - i \varphi _s) +C_s -\int_{\textrm{Im} \theta ' =\varphi _{s}} d\theta'\biggl[K_2(\theta-\theta')
\tilde {\cal L}_{s}(\theta') + \nonumber \\
&+& K_1(\theta-\theta'){\cal L}_{s}(\theta')\biggr] + \int_{\textrm{Im} \theta ' =\varphi _{s-1}}
d\theta'\biggl[K_1(\theta-\theta')\tilde {\cal L}_{s-1}(\theta') + \nonumber \\
&+& \frac{1}{2}K_2(\theta-\theta'){\cal L}_{s-1}(\theta') + \frac{1}{2}K_3(\theta -\theta'){\cal M}_{s-1}(\theta')\biggr] + \nonumber \\
&+&  \int_{\textrm{Im} \theta ' =\varphi _{s+1}} d\theta'\biggl[K_1(\theta-\theta')\tilde {\cal L}_{s+1}(\theta') + \frac{1}{2}K_2(\theta-\theta'){\cal L}_{s+1}(\theta') - \nonumber \\
&-& \frac{1}{2}K_3(\theta -\theta'){\cal M}_{s+1}(\theta')\biggr] \label {eq3} \, .
\ea

where $s=1,...,n-5$ and the geometric informations of the loop are encoded in the parameters $m_s,C_2,\varphi_s$ which are, in a complicated way\footnote{For instance, see (\ref{cross1}) and the various definitions in this appendix.}, related to the cross ratios $\tau_s,\sigma_s,\phi_s$. 

In the following we write them in a form suitable for comparisons with the strong coupling resummation of the OPE series in Chapter \ref{ChCla}. To begin with, we define the kernels \cite{YSA,Anope}
\be
K_1(\theta )= \frac{1}{2\pi \cosh \theta} \, , \quad K_2(\theta )= \frac{\sqrt{2}\cosh \theta}{\pi \cosh 2\theta}
\, , \quad K_3(\theta )= \frac{i}{\pi} \tanh 2\theta  \label {ker} \, ,
\ee
and introduce the short-hand for the nonlinear functions of $\ln Y$'s:
\begin{equation}
 {\cal L}_{s}(\theta)=\ln(1+Y_{1,s}(\theta))(1+Y_{3,s}(\theta)) \, , \quad \tilde {\cal L}_{s}(\theta)=\ln(1+Y_{2,s}(\theta)) \, , \quad {\cal M}_{s}(\theta)=\ln\frac{(1+Y_{1,s}(\theta))}{(1+Y_{3,s}(\theta))} \label {nlinY} \, .
\end{equation}

The hexagon is obtained when $s$ is fixed to $1$. In that case equations (\ref {eq1}, \ref {eq2}, \ref {eq3}) become those of \cite{TBuA} after the identifications $2Z=|m|$, $\mu =e^{-C}$ and
\be
\epsilon (\theta - i\varphi)=-\ln Y_{1,1}(\theta ) - C \, , \quad \tilde \epsilon (\theta - i\varphi)=-\ln Y_{2,1}(\theta )
\ee
with $\ln Y_{1,1}(\theta )=\ln Y_{3,1}(\theta ) -2C$.

In the more general case $s$ goes from $1$ to $n-5$, with $n$ the number of edges of our polygon. Let us introduce the hatted Y-functions $\hat Y_{\alpha,s} (\theta )$:
\be
\hat Y_{\alpha,s}(\theta )= Y_{\alpha,s} \left ( \theta - \frac{i\pi}{4}b_{\alpha + s + 1} \right )
\label {hatY}
\ee

and the physical cross-ratios $y_{\alpha,s}= \hat Y_{\alpha,s}(0) \label {cr-rat}$ where the symbol $b_s$ equals 1 or 0 for respectively even and odd values of $s$.

Following \cite{Anope}, we introduce the tilded kernels as
\ba
&& \tilde K_1(\theta , \theta ')= - \frac{1}{2\pi} \frac{\sinh 2\theta}{\sinh 2\theta ' \cosh (\theta -\theta ')}
\, , \quad \tilde K_3(\theta , \theta ')=\frac{i}{\pi} \frac{\sinh 2\theta}{\sinh 2\theta ' \sinh (2\theta -2\theta ')} \nonumber \\
&& \tilde K_2^{(s)} (\theta , \theta ')=- \frac{\sqrt{2}}{\pi}\sinh \left (\theta -\theta ' + \frac{i\pi}{4}(-1)^s \right )
\frac{\sinh 2\theta}{\sinh 2\theta ' \sinh (2\theta -2\theta ')}  \, . \label {tildeK}
\ea
With these definitions, the hatted $Y$ functions satisfy the TBA-like equations

\footnotesize
\ba\label{hatTBA1}
&& \ln \hat Y_{2,s} (\theta )-{\cal E}_s(\theta)=-\int_{\textrm{Im} \theta ' =\varphi _{s}} d\theta' \biggl[\tilde K_2^{(s)}\left(\theta, \theta' + \frac{i\pi}{4}b_s\right){\cal L}_s (\theta') + \nonumber \\
&+& 2\tilde K_1\left(\theta,\theta' + \frac{i\pi}{4}b_{s+1}\right)\tilde {\cal L}_s(\theta')\biggr] +  \int_{\textrm{Im} \theta ' =\varphi _{s-1}} d\theta ' \biggl[\tilde K_1\left(\theta, \theta '+ \frac{i\pi}{4}b_{s+1}\right) {\cal L}_{s-1}(\theta ') + \nonumber \\
&+& \tilde K_2^{(s)}\left(\theta, \theta' + \frac{i\pi}{4}b_s\right) \tilde {\cal L}_{s-1}(\theta')\biggr]  + \int_{\textrm{Im} \theta ' =\varphi _{s+1}} d\theta ' \biggl[\tilde K_1\left(\theta, \theta '+ \frac{i\pi}{4}b_{s+1}\right) {\cal L}_{s+1}(\theta ') + \nonumber \\
&+& \tilde K_2^{(s)}\left(\theta, \theta' + \frac{i\pi}{4}b_s\right) \tilde {\cal L}_{s+1}(\theta')\biggr] \, , 
\ea
\ba
&& \ln \hat Y_{1,s} (\theta )+ \ln \hat Y_{3,s} (\theta )-\sqrt{2}{\cal E}_s\left (\theta + \frac{i\pi}{4}(-1)^{s+1} \right ) = -\int_{\textrm{Im} \theta ' =\varphi _{s}} d\theta' \biggl[2\tilde K_2^{(s)}\left(\theta, \theta' + \frac{i\pi}{4}b_{s+1}\right)\tilde {\cal L}_s (\theta') + \nonumber \\
&+& 2\tilde K_1\left(\theta,\theta' + \frac{i\pi}{4}b_s\right){\cal L}_s(\theta')\biggr] +  \int_{\textrm{Im} \theta ' =\varphi _{s-1}} d\theta ' \biggl[\tilde K^{(s)}_2\left(\theta, \theta '+ \frac{i\pi}{4}b_{s+1}\right) {\cal L}_{s-1}(\theta ') + \nonumber \\
&+& 2\tilde K_1\left(\theta, \theta' + \frac{i\pi}{4}b_s\right) \tilde {\cal L}_{s-1}(\theta')\biggr] + \int_{\textrm{Im} \theta ' =\varphi _{s+1}} d\theta ' \biggl[\tilde K^{(s)}_2\left(\theta, \theta '+ \frac{i\pi}{4}b_{s+1}\right) {\cal L}_{s+1}(\theta ') + \nonumber \\
&+& 2\tilde K_1\left(\theta, \theta' + \frac{i\pi}{4}b_s\right) \tilde {\cal L}_{s+1}(\theta')\biggr] \, ,  \\
&&  \ln \hat Y_{1,s} (\theta )- \ln \hat Y_{3,s} (\theta ) -\ln y_{1,s} + \ln y_{3,s} =-\int_{\textrm{Im} \theta ' =\varphi _{s-1}} d\theta' \tilde K_3 \left(\theta , \theta'+ \frac{i\pi}{4}b_{s+1}\right) {\cal M}_{s-1}(\theta') + \nonumber \\
&+& \int_{\textrm{Im} \theta ' =\varphi _{s+1}} d\theta' \tilde K_3 \left(\theta , \theta'+ \frac{i\pi}{4}b_{s+1}\right) {\cal M}_{s+1}(\theta')
\label{hatTBA2} \, , 
\ea
\normalsize

where the driving term ${\cal E}_s(\theta )$ is given by
\ba
{\cal E}_s(\theta )&=& -i (-1)^s \left [\sqrt{2}\sinh \left ( \theta + \frac{i\pi}{4}(-1)^s \right ) \ln y_{2,s}-\sinh \theta \ln y_{1,s}y_{3,s} \right ]  = \label  {Estheta} \, \\
&=& \cosh \theta \ln y_{2,s} + i (-1)^{s+1}\sinh \theta \ln \frac{y_{2,s}}{y_{1,s}y_{3,s}} \, . \nonumber
\ea
For future purpose, we now propose the following definitions, generalising the hexagon case:
\ba
\epsilon _{1,s}(\theta -i \varphi _s)&=&-\ln Y_{1,s}(\theta) - \frac{1}{2}\ln \frac{y_{3,s}}{y_{1,s}} \, ,  \\
\epsilon _{3,s}(\theta -i \varphi _s)&=&-\ln Y_{3,s}(\theta) + \frac{1}{2}\ln \frac{y_{3,s}}{y_{1,s}} \, , \\
\epsilon _{2,s}(\theta -i \varphi _s)&=&-\ln Y_{2,s}(\theta) \, .
\ea

The ''pseudoenergies'' $\epsilon $ are related to the hatted-Y through
\ba
\epsilon _{1,s}(\theta -i \hat \varphi _s)&=&-\ln \hat Y_{1,s}\left (\theta - \frac{i\pi}{4}b_{s+1} \right ) - \frac{1}{2}\ln \frac{y_{3,s}}{y_{1,s}} \, , \\
\epsilon _{3,s}(\theta -i \hat \varphi _s)&=&-\ln \hat Y_{3,s}\left (\theta - \frac{i\pi}{4}b_{s+1} \right ) + \frac{1}{2}\ln \frac{y_{3,s}}{y_{1,s}} \, , \\
\epsilon _{2,s}(\theta -i \hat \varphi _s)&=&-\ln \hat Y_{2,s}\left (\theta - \frac{i\pi}{4}b_s \right ) \, , 
\ea
where 
\be
\hat \varphi _s= \varphi _s + \frac{\pi}{4} \, . 
\ee
To write the new TBA equations in a more compact way, we define
\be
L_s(\theta )=\ln \left [\left ( 1+\sqrt{\frac{y_{1,s}}{y_{3,s}}}e^{-\epsilon _{1,s}(\theta -i\hat \varphi _s)}\right )\left (1+\sqrt{\frac{y_{3,s}}{y_{1,s}}}e^{-\epsilon _{3,s}(\theta -i\hat \varphi _s)}\right ) \right ]={\cal L}_s\left(\theta -\frac{i\pi}{4}\right) \, , 
\ee
\be
\tilde L_s(\theta )= \ln \left (1+e^{-\epsilon _{2,s}(\theta -i\hat \varphi _{s})}\right )=\tilde {\cal L}_s\left(\theta -\frac{i\pi}{4}\right) \, , 
\ee
\be
M_s(\theta )=\ln \left ( \frac{1+\sqrt{\frac{y_{1,s}}{y_{3,s}}}e^{-\epsilon _{1,s}(\theta -i\hat \varphi _s)}}{ 1+\sqrt{\frac{y_{3,s}}{y_{1,s}}}e^{-\epsilon _{3,s}(\theta -i\hat \varphi _s)}} \right )={\cal M}_s\left(\theta -\frac{i\pi}{4}\right) \, . 
\ee

which lead, for $\epsilon_{\alpha,s}$, to the TBA-like

\footnotesize
\ba
&& \epsilon _{2,s}(\theta -i \hat \varphi _s)=-{\cal E}_s \left ( \theta - \frac{i\pi}{4}b_s \right ) + \label{tbaeps1}  \\
&+& \int_{\textrm{Im} \theta ' =\hat\varphi _{s}}d\theta'\biggl[\tilde K_2^{(s)} \left ( \theta - \frac{i\pi}{4}b_s, \theta '- \frac{i\pi}{4}b_{s+1} \right ) L_s(\theta') + 2\tilde K_1 \left ( \theta - \frac{i\pi}{4}b_s, \theta ' - \frac{i\pi}{4}b_s\right )\tilde L_s(\theta')\biggr] - \nonumber \\
&-& \int_{\textrm{Im} \theta ' =\hat\varphi _{s-1}}d\theta'\biggl[\tilde K_2^{(s)} \left ( \theta - \frac{i\pi}{4}b_s, \theta '- \frac{i\pi}{4}b_{s+1} \right )\tilde L_{s-1}(\theta') + \tilde K_1 \left ( \theta - \frac{i\pi}{4}b_s, \theta ' - \frac{i\pi}{4}b_s\right)L_{s-1}(\theta')\biggr] - \nonumber \\
&-& \int_{\textrm{Im} \theta ' =\hat\varphi _{s+1}}d\theta'\biggl[\tilde K_2^{(s)} \left ( \theta - \frac{i\pi}{4}b_s, \theta '- \frac{i\pi}{4}b_{s+1} \right )\tilde L_{s+1}(\theta') + \tilde K_1 \left ( \theta - \frac{i\pi}{4}b_s, \theta ' - \frac{i\pi}{4}b_s\right)L_{s+1}(\theta')\biggr] \nonumber \\
&& \epsilon _{3,s}(\theta -i \hat \varphi _s)-\epsilon _{1,s}(\theta -i \hat \varphi _s)= -\int_{\textrm{Im} \theta ' =\hat\varphi _{s-1}}d\theta'\biggl[ \tilde K_3 \left ( \theta - \frac{i\pi}{4}b_{s+1}, \theta ' - \frac{i\pi}{4}b_s\right ) M_{s-1}(\theta')\biggr] + \label{tbaeps2}  \\
&+& \int_{\textrm{Im} \theta ' =\hat\varphi _{s+1}}d\theta'\biggl[ \tilde K_3 \left ( \theta - \frac{i\pi}{4}b_{s+1}, \theta ' - \frac{i\pi}{4}b_s\right ) M_{s+1}(\theta')\biggr] \nonumber \\
&& \epsilon _{3,s}(\theta -i \hat \varphi _s)+\epsilon _{1,s}(\theta -i \hat \varphi _s)= -\sqrt{2} {\cal E}_s \left ( \theta - \frac{i\pi}{4}b_s \right ) + \label{tbaeps3} \\
&+& 2\int_{\textrm{Im} \theta ' =\hat\varphi _{s}}d\theta'\biggl[ \tilde K_1 \left ( \theta - \frac{i\pi}{4}b_{s+1}, \theta ' - \frac{i\pi}{4}b_{s+1}\right )L_s(\theta') +
 \tilde K_2^{(s)} \left ( \theta - \frac{i\pi}{4}(-1)^s-\frac{i\pi}{4}b_s, \theta ' - \frac{i\pi}{4}b_s \right )\tilde L_s(\theta')\biggr] - \nonumber \\
&-&  \int_{\textrm{Im} \theta ' =\hat\varphi _{s-1}}d\theta'\biggl[2 \tilde K_1 \left ( \theta - \frac{i\pi}{4}b_{s+1}, \theta ' - \frac{i\pi}{4}b_{s+1}\right )\tilde L_{s-1}(\theta') + \tilde K_2^{(s)} \left ( \theta - \frac{i\pi}{4}(-1)^s -\frac{i\pi}{4}b_s, \theta ' - \frac{i\pi}{4}b_s \right ) L_{s-1}(\theta')\biggr] - \nonumber \\
&-&  \int_{\textrm{Im} \theta ' =\hat\varphi _{s+1}}d\theta'\biggl[2 \tilde K_1 \left ( \theta - \frac{i\pi}{4}b_{s+1}, \theta ' - \frac{i\pi}{4}b_{s+1}\right )\tilde L_{s+1}(\theta') + \tilde K_2^{(s)} \left ( \theta - \frac{i\pi}{4}(-1)^s -\frac{i\pi}{4}b_s, \theta ' - \frac{i\pi}{4}b_s \right ) L_{s+1}(\theta')\biggr] \nonumber \, .
\ea
\normalsize

Following \cite {Anope}, we write equations (\ref {tbaeps1}, \ref {tbaeps2}, \ref {tbaeps3}) as
\ba
&& \epsilon _{2,s}(\theta -i \hat \varphi _s) + {\cal E}_s \left ( \theta - \frac{i\pi}{4}b_s \right )
= -\hat A_{2,s}(\theta) \, , 
\ea
\ba
&& \epsilon _{3,s}(\theta -i \hat \varphi _s)-\epsilon _{1,s}(\theta -i \hat \varphi _s)=
\hat A_{1,s}(\theta) - \hat A_{3,s}(\theta) \, , 
\ea
\ba
&& \epsilon _{3,s}(\theta -i \hat \varphi _s)+\epsilon _{1,s}(\theta -i \hat \varphi _s)
 + \sqrt{2} {\cal E}_s \left ( \theta - \frac{i\pi}{4}b_s \right )
 =  -\hat A_{3,s}(\theta) - \hat A_{1,s}(\theta) \, ,
\ea
where the quantities $\hat A_{\alpha,s}(\theta)$ includes the terms in the RHS. Then, the critical Yang-Yang functional, which describes the strong coupling limit (\ref{SYYc}), is given by \cite{Anope}
\ba
YY_c&=&\sum _{\alpha=1}^3 \sum _{s=1}^{n-5} \int _{\textrm{Im} \theta =\hat \varphi _s}\frac{d\theta}{\pi \sinh ^2 \left [2\theta-
\frac{i\pi}{2}b_{\alpha + s}\right ] }  \Bigl [ \textrm{Li}_2 \left (-e^{-\epsilon _{\alpha,s}(\theta -i\hat \varphi _s)} \left ( \frac{y_{1,s}}{y_{3,s}} \right )^{1-\frac{\alpha}{2}} \right )+ \nonumber \\
&+&  \frac{1}{2}\ln \left ( 1+e^{-\epsilon _{\alpha,s}(\theta -i\hat \varphi _s)} \left ( \frac{y_{1,s}}{y_{3,s}}\right )^{1-\frac{\alpha}{2}}  \right ) \hat A_{\alpha,s} (\theta ) \Bigr ] \label {YYc2} \, .
\ea

\chapter{Scattering matrices and Pentagon transitions}
\label{ApSca}

In this appendix we list several formulae for the S matrices and the pentagon transitions concerning fermions and their effective bound states, the mesons. 

\section{All couplings}

The exact relation, \textit{i.e.} valid at any coupling, between the S matrix of two mesons and the S matrices of their constituents reads
\be
S^{(MM)}(u,v)=\frac{u-v+i}{u-v-i} S^{(ff)}(u+i,v+i)S^{(ff)}(u-i,v+i)S^{(ff)}(u+i,v-i)S^{(ff)}(u-i,v-i) \, . 
\ee
Inspired by this formula, the pentagon transition between mesons is proposed as follows \cite{BFPR}
\be\label{PMM}
P^{(MM)}(u,v)=(u-v)(u-v+i) P^{(ff)}(u+i,v+i) P^{(ff)}(u-i,v-i) P^{(f\bar f)}(u-i,v+i)
P^{(f\bar f)}(u+i,v-i) \, , 
\ee
which reads, in terms of fermionic S-matrices,
\ba\label{PiMM}
P^{(MM)}(u,v)&=&\sqrt{\frac{f_{\psi \psi}(u+i,v+i) f_{\psi \psi}(u-i,v-i)}{f_{\psi \psi}(u-i,v+i) f_{\psi \psi}(u+i,v-i)}} \times \label {P(MM)}\\
&\times & \sqrt{\frac{S^{(ff)}(u+i,v+i) S^{(ff)}(u-i,v-i)S^{(ff)}(u+i,v-i)S^{(ff)}(u-i,v+i)}{S^{(\ast ff)}(u+i,v+i) S^{(\ast ff)}(u-i,v-i)S^{(\ast f\bar f)}(u+i,v-i)S^{(\ast f\bar f)}(u-i,v+i)}} \nonumber \, ,
\ea
with
\be
f_{\psi \psi}(u,v)=\frac{x(u)x(v)}{g^{2}}\left (1-\frac{g^2}{x(u)x(v)} \right) \, , \quad x(u)=\frac{u}{2}\left (1+\sqrt{1-\frac{4g^2}{u^2}} \right ) \, ,
 \quad x^{\pm}(u)=x(u\pm\frac{i}{2})
\ee
and $S^{(\ast ff)}$,  $S^{(\ast f\bar f)}$ are mirror S-matrices. The latter follow from formul{\ae}\ \cite{BSV3}
\ba
\ln S^{(\ast f\bar f)}(u,v) &=& \ln S^{(\ast \bar f f)}(u,v)=
\ln S^{(Ff)}(u,v)-\ln S^{(s f)}(u-\frac{i}{2},v) \\
\ln S^{(\ast f f)}(u,v) &=& \ln\left(\frac{u-v}{u-v+i}\right)\,+\ln S^{(\ast \bar f f)}(u,v) \label{psipsi} \, .
\ea
Importantly, $S^{(\ast f\bar f)}$  is free from zeroes and poles. Using relation (\ref{psipsi}) 
, we can express the meson-meson pentagon amplitude as:
\ba
P^{(MM)}(u,v)&=&\sqrt{\frac{f_{\psi \psi}(u+i,v+i) f_{\psi \psi}(u-i,v-i)}{f_{\psi \psi}(u-i,v+i) f_{\psi \psi}(u+i,v-i)}} \frac{u-v+i}{u-v}\times \label {P(MM)2}\\
&\times & \sqrt{\frac{S^{(ff)}(u+i,v+i) S^{(ff)}(u-i,v-i)S^{(ff)}(u+i,v-i)S^{(ff)}(u-i,v+i)}{S^{(\ast f\bar f)}(u+i,v+i) S^{(\ast f\bar f)}(u-i,v-i)S^{(\ast f\bar f)}(u+i,v-i)S^{(\ast f\bar f)}(u-i,v+i)}} \nonumber \, .
\ea
Since $f_{\psi \psi}$ and $S^{(ff)}$ have no zeros nor poles in the small fermion sheet, this expression shows that $P^{(MM)}(u,v)$ has a pole for $u=v$ and a zero for $u=v-i$. No other poles or zeroes are present. From this property, the important split (\ref{PMMreg}) has been proposed in the main text, which is of fundamental importance for the formation of bound states between mesons.

 The proposal (\ref {P(MM)2}) is confirmed by a strong coupling limit analysis. In fact, when $\lambda \rightarrow \infty$ it reproduces formula (10.15) of \cite {FPR2}, which has been obtained by solving the axioms for the meson pentagonal amplitudes directly at strong coupling.

\section{Pentagonal amplitudes at strong coupling}\label{appA2}

Here is a collection of the functions $P_{\alpha, \beta}(\theta, \theta ')$, appearing in the resummation of the OPE series for the polygonal Wilson loop in Section \ref{ResumPol}:
\ba
P_{11}(\theta | \theta ') &=& P_{33}(\theta | \theta ')=1+\frac{i\pi}{\sqrt{\lambda}} \frac{\cosh 2\theta \cosh 2\theta '}{\sinh (2\theta -2\theta ')}\left [1+\cosh (\theta -\theta ')-i\sinh (\theta -\theta ') \right ]=  \label {Pgg} \nn \\
&=& 1+ \frac{2\pi}{\sqrt{\lambda}}\,K^{(gg)}(\q,\q')=1+ \frac{2\pi}{\sqrt{\lambda}}\,K^{(\bar g\bar g)}(\q,\q')  \, , \ea
\ba
P_{13}(\theta | \theta ') &=& P_{31}(\theta | \theta ')=1+\frac{i\pi}{\sqrt{\lambda}} \frac{\cosh 2\theta \cosh 2\theta '}{\sinh (2\theta -2\theta ')}\left [-1+\cosh (\theta -\theta ')-i\sinh (\theta -\theta ') \right ] =  \label {Pgbarg} \nn \\
&=& 1+ \frac{2\pi}{\sqrt{\lambda}}\,K^{(g\bar g)}(\q,\q')=1+ \frac{2\pi}{\sqrt{\lambda}}\,K^{(\bar g g)}(\q,\q') 
\, ,
\ea
\ba
P_{22}(\theta | \theta ') &=& 1-\frac{2\pi}{\sqrt{\lambda}}\frac{i \sinh 2\theta \sinh 2\theta '}{ \sinh (2\theta -2\theta ')} \sqrt{2}\cosh \left (\theta -\theta ' -i\frac{\pi}{4} \right ) =  \label {PmMM} \nn \\
&=& 1+ \frac{2\pi}{\sqrt{\lambda}}\,K^{(MM)}(\q,\q')  \, ,
\ea
\ba
P_{21}(\theta | \theta ') &=& P_{23}(\theta | \theta ')=1+\frac{2\pi}{\sqrt{\lambda}}\frac{\sinh 2\theta \cosh 2\theta '}{\sqrt{2} \cosh (2\theta -2\theta ')}[ \sinh (\theta -\theta ')+i \cosh (\theta -\theta ') ] = \nn \\
&=& 1+ \frac{2\pi}{\sqrt{\lambda}}\,K^{(Mg)}(\q,\q')=1+ \frac{2\pi}{\sqrt{\lambda}}\,K^{(M\bar g)}(\q,\q')  \, ,
\ea
\ba
P_{12}(\theta | \theta ') &=& P_{32}(\theta | \theta ')=1+\frac{2\pi}{\sqrt{\lambda}} \frac{\sinh 2\theta ' \cosh 2\theta }{\sqrt{2} \cosh (2\theta ' -2\theta )}[ \sinh (\theta ' -\theta )-i \cosh (\theta '-\theta ) ] = \nn \\
&=& 1+ \frac{2\pi}{\sqrt{\lambda}}\,K^{(gM)}(\q,\q')=1+ \frac{2\pi}{\sqrt{\lambda}}\,K^{(\bar gM)}(\q,\q')  \, .
\ea

where we remind that the indices $1,3$ stand for the gluons, respectively with positive and negative helicity, whereas $2$ indicates the meson. 

\subsection{Relations between kernels}\label{appA3}

From the definition (\ref{green-kernel}) and the formul{\ae} above, several relations follow 

\be
G^{(s,s)}_{1,1}(\theta,\theta')=G^{(s,s)}_{1,3}(\theta,\theta')=G^{(s,s)}_{3,1}(\theta,\theta')=
G^{(s,s)}_{3,3}(\theta,\theta') \, ,
\ee
\be
G^{(s,s)}_{1,2}(\theta,\theta')=G^{(s,s)}_{3,2}(\theta,\theta') \, , \quad G^{(s,s)}_{2,1}(\theta,\theta')=G^{(s,s)}_{2,3}(\theta,\theta') \, ,
\ee
\be
G^{(s,s+1)}_{1,2}(\theta,\theta')=G^{(s,s+1)}_{3,2}(\theta,\theta') \, , \quad G^{(s,s+1)}_{2,1}(\theta,\theta')=G^{(s,s+1)}_{2,3}(\theta,\theta') \, ,
\ee
\be
G^{(s,s+1)}_{1,1}(\theta,\theta')=G^{(s,s+1)}_{3,3}(\theta,\theta') \, , \quad G^{(s,s+1)}_{1,3}(\theta,\theta')=G^{(s,s+1)}_{3,1}(\theta,\theta') \, ,
\ee

in addition to the obvious $G^{(s,s+1)}_{\alpha,\beta}(\theta,\theta')=G^{(s,s-1)}_{\alpha,\beta}(\theta,\theta')$.

We display also some of the relations between the kernels $G$ and the tilded ones $\tilde K$:
\be
\frac{\mu _2(\theta ')}{2\pi} G^{(s,s)}_{2,2}(\theta , \theta ') =-2 \tilde K_1 (\theta , \theta ') \, ,
\ee
\be
\frac{\mu _2(\theta ')}{2\pi} G^{(s,s+1)}_{2,2}(\theta , \theta ') =
-\tilde K_2^{(s)}(\theta , \theta ') \, ,
\ee
\begin{equation}
\frac{\mu_1(\theta')}{2\pi}G^{(s,s)}_{2,1}(\theta,\theta')=-\tilde K^{(s)}_2\left(\theta,\theta'+i\frac{\pi}{4}(-1)^s\right) \, ,
\end{equation}

\begin{equation}
\frac{\mu_1(\theta')}{2\pi}G^{(s,s+1)}_{2,1}(\theta,\theta')=-\tilde K_1\left(\theta,\theta'-i\frac{\pi}{4}(-1)^s\right) \, ,
\end{equation}


\begin{equation}
\frac{\mu_2(\theta')}{2\pi}G^{(s,s)}_{1,2}(\theta,\theta')=-\tilde K^{(s)}_2\left(\theta-i\frac{\pi}{4}(-1)^s,\theta'\right) \, ,
\end{equation}

\begin{equation}
\frac{\mu_1(\theta')}{2\pi}G^{(s,s)}_{1,1}(\theta,\theta')=-\tilde K_1\left(\theta+i\frac{\pi}{4}(-1)^s,\theta'+i\frac{\pi}{4}(-1)^s\right) \, ,
\end{equation}


\begin{equation}
\frac{\mu_2(\theta')}{2\pi}G^{(s,s+1)}_{1,2}(\theta,\theta')=-\tilde K_1\left(\theta+i\frac{\pi}{4}(-1)^s,\theta'\right) \, ,
\end{equation}


\begin{equation}
\frac{\mu_1(\theta')}{2\pi}\left[G^{(s,s+1)}_{1,1}(\theta,\theta')+G^{(s,s+1)}_{3,1}(\theta,\theta')\right]=-\tilde K^{(s)}_2\left(\theta-i\frac{\pi}{4}(-1)^s,\theta'-i\frac{\pi}{4}(-1)^s\right) \, ,
\end{equation}



\begin{equation}
\frac{\mu_1(\theta')}{2\pi}\left[G^{(s,s+1)}_{1,1}(\theta,\theta')-G^{(s,s+1)}_{3,1}(\theta,\theta')\right]=(-1)^s \tilde K_3\left(\theta+i\frac{\pi}{4}(-1)^s,\theta'-i\frac{\pi}{4}(-1)^s\right) \, .
\end{equation}



\subsection{Bootstrap relations} \label{bootstrap}

In this part, some bootstrap relations involving the relativistic kernels $K_i$ are displayed. We use the shorthand notation for the shifts $K_a(\q^\pm)=K^\pm_a(\q)=K_a(\q \pm\frac{i\pi}{4})$ and also $K_a(\q^{\pm\pm})=K^{\pm\pm}_a(\q)=K_a(\q \pm\frac{i\pi}{2})$.
\ba\label{boot_rel}
&& K_1^{+}+K_1^{-}=K_2 \, , \nn\\
&& K_2^{+}+K_2^{-}=2K_1+\delta(\q) \, ,\\
&& K_3^{-}-K_3^{+}=\delta(\q) \nn \, ,
\ea
\ba\label{boot_rel2}
&& K_1^{++}+K_1^{--}=\delta(\q) \nn \, , \\
&& K_2^{++}+K_2^{--}=\delta(\q^+)+\delta(\q^-) \, , \\
&& K_3^{++}+K_3^{--}=2K_3-\delta(\q^+)+\delta(\q^-) \ . \nn
\ea
The bootstrap relations concerning the pentagonal amplitudes are:
\ba\label{boot_pentagonal}
&& K^{(MM)}_{sym}\left(\q^{++},\q'^+\right)+ K^{(MM)}_{sym}\left(\q,\q'^+\right)
   -2K^{(gM)}_{sym}\left(\q^+,\q'^+\right) =0 \, , \\
&& K^{(Mg)}_{sym}\left(\q^{++},\q'^+\right)+ K^{(Mg)}_{sym}\left(\q,\q'^+\right)
   -2K^{(gg)}_{sym}\left(\q^+,\q'^+\right) =\pi \sinh^2(2\q)\,\delta(\q-\q') \, , \nn\\
&& K^{(gM)}\left(\q',\q^{++}\right)+ K^{(gM)}\left(\q',\q\right)
   -K^{(gg)}\left(\q',\q^+\right)-K^{(g\bar g)}\left(\q',\q^+\right) =0 \nn \, , \\
&& K^{(MM)}\left(\q',\q^{++}\right)+ K^{(MM)}\left(\q',\q\right)
   -2K^{(Mg)}\left(\q',\q^+\right) = \pi \sinh^2(2\q)\,\delta(\q-\q') \nn \, , \\
&& K^{(gM)}_{sym}\left(\q^{++},\q'^+\right)+ K^{(gM)}_{sym}\left(\q,\q'^+\right)
   -K^{(MM)}_{sym}\left(\q^+,\q'^+\right) = -\pi \cosh^2(2\q)\,\delta(\q-\q') \nn \, , \\
&& K^{(gg)}_{sym}\left(\q^{++},\q'^+\right)+ K^{(gg)}_{sym}\left(\q,\q'^+\right)
   -K^{(Mg)}_{sym}\left(\q^+,\q'^+\right) = 0  \nn \, , \\
&& K^{(Mg)}\left(\q',\q^{++}\right)+ K^{(Mg)}\left(\q',\q\right)
   -K^{(MM)}\left(\q',\q^+\right) = 0  \nn \, , \\
&& K^{(gg)}\left(\q',\q^{++}\right)+ K^{(g\bar g)}\left(\q',\q\right)
   -K^{(gM)}\left(\q',\q^+\right) = 0  \nn \, , \\
&& K^{(g\bar g)}\left(\q',\q^{++}\right)+ K^{(gg)}\left(\q',\q\right)
   -K^{(gM)}\left(\q',\q^+\right) = -\pi \cosh^2(2\q)\,\delta(\q-\q')  \nn \, , \\
&& K^{(Mg)}\left(\q^+,\q'^+\right)+ K^{(Mg)}\left(\q^-,\q'^+\right)
   -K^{(g\bar g)}\left(\q,\q'^+\right)- K^{(gg)}\left(\q,\q'^+\right) = 0 \nn \, , \\
&& K^{(MM)}\left(\q^+,\q'^+\right)+ K^{(MM)}\left(\q^-,\q'^+\right)
   -2K^{(gM)}\left(\q,\q'^+\right) = -\pi \cosh^2(2\q)\,\delta(\q-\q')  \nn \, , \\
&& K^{(Mg)}_{sym}\left(\q^+,\q'\right)+ K^{(Mg)}_{sym}\left(\q^-,\q'\right)
   -2K^{(gg)}_{sym}\left(\q,\q'\right) = -\pi \cosh^2(2\q)\,\delta(\q-\q') \nn \, , \\
&& K^{(MM)}_{sym}\left(\q^+,\q'\right)+ K^{(MM)}_{sym}\left(\q^-,\q'\right)
   -2K^{(gM)}_{sym}\left(\q,\q'\right) = 0 \nn \, , \\
&& K^{(g\bar g)}\left(\q^-,\q'^+\right)+ K^{(gg)}\left(\q^+,\q'^+\right)
   -K^{(Mg)}\left(\q,\q'^+\right) = \pi \sinh^2(2\q)\,\delta(\q-\q') \nn \, , \\
&& K^{(gg)}\left(\q^-,\q'^+\right)+ K^{(g\bar g)}\left(\q^+,\q'^+\right)
   -K^{(Mg)}\left(\q,\q'^+\right) = 0 \nn \, , \\
&& K^{(gM)}\left(\q^-,\q'^+\right)+ K^{(gM)}\left(\q^+,\q'^+\right)
   -K^{(MM)}\left(\q,\q'^+\right) = 0 \nn \, , \\
&& K^{(gg)}_{sym}\left(\q^-,\q'\right)+ K^{(gg)}_{sym}\left(\q^+,\q'\right)
   -K^{(Mg)}_{sym}\left(\q,\q'\right) = 0 \nn \, , \\
&& K^{(gM)}_{sym}\left(\q^-,\q'\right)+ K^{(gM)}_{sym}\left(\q^+,\q'\right)
   -K^{(MM)}_{sym}\left(\q,\q'\right) = \pi\sinh^2(2\q)\, \delta(\q-\q') \nn \, ,
\ea
where the shifts have to be read as $\q^\pm=\q\pm\frac{i\pi}{4}$ and $\q^{\pm\pm}=\q\pm\frac{i\pi}{2}$. In terms of the tensor (\ref{green-kernel}), the relations above can be summarised as:
\small
\ba
 G^{(s,s)}_{\alpha,\beta}(\q^{+},\q')+G^{(s,s)}_{4-\alpha,\beta}(\q^{-},\q')
  -G^{(s,s)}_{\alpha+1,\beta}(\q,\q')-G^{(s,s)}_{\alpha-1,\beta}(\q,\q') =
  -2\pi \delta_{\alpha+\beta,odd}\,\frac{\delta(\q-\q')}{\mu_\beta(\q')} \nn \, , \\
 G^{(2k\pm 1,2k)}_{\alpha,\beta}(\q^{++},\q')+G^{(2k\pm 1,2k)}_{4-\alpha,\beta}(\q,\q')
  -G^{(2k\pm 1,2k)}_{\alpha+1,\beta}(\q^+,\q')-G^{(2k\pm 1,2k)}_{\alpha-1,\beta}(\q^+,\q') =
  -2\pi \delta_{4-\alpha,\beta}  \,\frac{\delta(\q-\q')}{\mu_\beta(\q')}  \nn \, , \\
 G^{(2k,2k\pm 1)}_{\alpha,\beta}(\q^{+},\q'^+)+G^{(2k,2k\pm 1)}_{4-\alpha,\beta}(\q^-,\q'^+)
  -G^{(2k,2k\pm 1)}_{\alpha+1,\beta}(\q,\q'^+)-G^{(2k,2k\pm 1)}_{\alpha-1,\beta}(\q,\q'^+) =
  -2\pi \delta_{\alpha,\beta}  \,\frac{\delta(\q-\q')}{\mu_\beta(\q'^+)} \nn \, ,\\
\ea
where $k=1,2,3,\dots$ \,.
\normalsize

\chapter{Polynomials}
\label{Pol}

In this appendix we display several properties of the many polynomials appearing throghout the text, expecially in Chapter \ref{ChMat}.

\section{$\delta_{2n}$ functions}

This part is devoted to the polynomials $\delta_{2n}$, appearing in the multiple integrals of the matrix part after integrating over the auxiliary variables $a_i,c_i$.

We recall the defining expression as a sum over partitions 
\ba\label{deltapart}
\delta _{2n}(b_1,\ldots,b_{2n})&\equiv &\frac{n!}{2n} \sum _{\alpha _1<\alpha _2< \ldots  < \alpha _{n}=1}^{2n} \left (  \prod _{\stackrel {i\in S_{\vec{\alpha}},
j\in S_{\vec{\alpha}}, i<j} {i\in \bar S_{\vec{\alpha}},
j\in \bar S_{\vec{\alpha}}, i<j}} [ (b_i-b_j)^2+1] \right ) \cdot \nonumber \\
&\cdot & \prod _{k=1}^n \prod _{\beta \in  \bar S_{\vec{\alpha}}} \frac{b_{\alpha _k}-b_{\beta}-i}{b_{\alpha _k}-b_{\beta}} \, .
\label{def-deltaA}
\ea
which implies the symmetry in all the arguments, {\it i.e.} 
\be
\delta_{2n}(b_1,\ldots,b_i,b_{i+1},\ldots,b_{2n})=\delta_{2n}(b_1,\ldots,b_{i+1},b_i,\ldots,b_{2n})
\ee

From the representation (\ref{deltapart}), it is possible to show that
$\delta_{2n}$ is vanishing whenever three or more variables are aligned in the complex plane with a displacement of $i$
\be
\delta_{2n}(b_1,b_1+i,b_1+2i,b_4,\ldots,b_{2n})=0 \ .
\ee
The sum over partition (\ref {def-deltaA}) can be written in a more compact form: we recognise, in its highest degree $\delta^{(0)} _{2n}$, the Moore-Read wave function\footnote{I am indebted to Ivan Kostov and Didina Serban for pointing out this interesting representation.}
\be
\delta^{(0)} _{2n}(b_1,\ldots,b_{2n})\equiv\frac{n!}{2n} \sum _{\alpha _1<\alpha _2< \ldots < \alpha _{n}=1}^{2n}   \prod _{\stackrel {i\in S_{\vec{\alpha}},
j\in S_{\vec{\alpha}}, i<j} {i\in \bar S_{\vec{\alpha}},
j\in \bar S_{\vec{\alpha}}, i<j}}  (b_i-b_j)^2 \ ,
\ee
which is equivalent to the Pfaffian 
\be\label{delta0det}
\delta^{(0)} _{2n}(b_1,\ldots,b_{2n})=\frac{n!}{2n}2^n\displaystyle\prod_{i<j}b_{ij}\textrm{Pf}\left(\frac{1}{b_{ij}}\right) \, .
\ee
Formula (\ref{delta0det}) can be extended to the full $\delta_{2n}$ by means of the substitution $b_{ij}\to \frac{b_{ij}^2+1}{b_{ij}}$, finding the appealing
\be\label{deltadet}
\delta_{2n}(b_1,\ldots ,b_{2n})=
\frac{n!}{2n}2^n\displaystyle\prod_{i<j}\frac{b_{ij}^2+1}{b_{ij}}\textrm{Pf} \, D \, , \quad D_{ij}=\left(\frac{b_{ij}}{b_{ij}^2+1}\right) \, ,
\ee
This Pfaffian representation also gives a recursion relation for the $\delta$-polynomials:
\ba
\delta_2(b_1,b_2) &=& 1 \nn\\
\delta_{2n}(b_1,\ldots,b_{2n}) &=& 2(n-1)\sum_{\stackrel{l=1}{l\neq k}}^{2n}\prod_{\stackrel{i=1}{i\neq k,l}}^{2n}
\frac{b_{ik}^2+1}{b_{ik}}\frac{b_{il}^2+1}{b_{il}}\,\delta_{2n-2}(b_1,\ldots,\underline{b_k},\ldots,\underline{b_l}\ldots,b_{2n})
\label {delta-rec} \, ,
\ea

where the notation $\underline{b_k}$ means that $b_k$ is removed as variable of the function $\delta_{2n-2}$\,.\\
Referring to a specific configuration as in (\ref{diagr}), the functions $\delta_{2n}(b_1,\ldots,b_{2n})$ take simple forms. We use of shorthand notation of Section \ref{ScaMat}, as for instance $\delta_{2n}(Y)$ to indicate that the variables of $\delta_{2n}$ are computed on the residue configuration $Y=(l_1,\ldots,l_{2n})$\,.\\
The first formula is 
\ba
\delta_{2n}(2,0,1,\ldots, 1)&\equiv &\delta_{2n}(u_1, u_1+i, u_3,\ldots , u_{2n})= \nn\\
&=& 2(n-1)\displaystyle\prod_{j=3}^{2n}(u_1-u_j-i)(u_1-u_j+2i)\delta_{2n-2}(u_3,\ldots , u_{2n}) \, ,
\ea
which, upon iteration, yields the most general one
\ba
&&\delta_{2n}(2,0,..,2,0_{2k+2},1,\ldots ,1)\equiv\delta_{2n}(u_1, u_1+i,u_3, u_3+i,.., u_{2k+1}, u_{2k+1}+i, u_{2k+3},\ldots , u_{2n})= \nn\\
&& =2^{k+1}\frac{(n-1)!}{(n-2-k)!} \displaystyle\prod_{i<j=0}^{k}[(u_{2i+1}-u_{2j+1})^2+1][(u_{2i+1}-u_{2j+1})^2+4]\cdot \nn\\ &&\cdot\displaystyle\prod_{j=2k+3}^{2n}\displaystyle\prod_{l=0}^{k}(u_{1+2l}-u_j-i)(u_{1+2l}-u_j+2i)\delta_{2n-2-2k}(u_{2k+3}, \ldots , u_{2n}) \, .
\ea

where $0\leq k \leq n-2$.

It allows us to express any $\delta_{2n}(Y)$ in terms of $\delta_{2k}(1,\ldots ,1)$, with less particles: for instance, we consider the particular case $k=n-2$ and choose $u_{2n}=u_{2n-1}+i$ to find
\small
\ba\label{delta20}
\delta_{2n}(2,0,\ldots ,2,0)&\equiv &\delta_{2n}(u_1, u_1+i, \ldots , u_{2n-1}, u_{2n-1}+i)= \nn\\
&=& 2^{n-1}(n-1)!\displaystyle\prod_{i<j=0}^{n-1}[(u_{2i+1}-u_{2j+1})^2+1][(u_{2i+1}-u_{2j+1})^2+4] \, .
\ea
\normalsize
Merging the last two equations allows us to express $\delta_{2n} (Y=(Y_1,Y_2))$, where $Y_1=(2,0,\cdots ,2,0)$ and $Y_2=(1,\cdots ,1)$, in terms of the product $\delta_{2k+2}(Y_1)\delta_{2n-2k-2}(Y_2)$ 
\ba\label{delta-fact}
&&\delta_{2n}(2,0,\ldots ,2,0_{2k+2},1,\ldots ,1)=2\displaystyle\prod_{j=2k+3}^{2n}\displaystyle\prod_{l=0}^{k}(u_{1+2l}-u_j-i)(u_{1+2l}-u_j+2i)\cdot \nn\\
&& \cdot \frac{(n-1)!}{(n-2-k)!k!}\delta_{2k+2}(2,0,\ldots ,2,0)\delta_{2n-2-2k}(1,1,\ldots ,1,1) \, ,
\ea

where a mixing part is present. Formula (\ref{delta-fact}) holds for $0\leq k \leq n-2$. In (\ref{delta20}), we can move all the columns to the left to obtain
\ba
\delta_{2n}(2,2,\ldots ,0,0)&\equiv &\delta_{2n}(u_1, u_1+i, u_2, u_2 +i, \ldots , u_{n}, u_{n}+i)= \nn\\
&=& 2^{n-1}(n-1)!\displaystyle\prod_{i<j}^{n}[(u_{i}-u_{j})^2+1][(u_{i}-u_{j})^2+4] \, ,
\ea
\emph{i.e.} the configuration considered in the main text.

A factorisation similar to that of the functions $G^{(2n)}$ holds for the $\delta_{2n}$ functions as well. When a set of $2k$ particle rapidities is sent to infinity, $\delta_{2n}$ splits into  the product
\be
\delta_{2n}(u_1+\Lambda,\cdots ,u_{2k}+\Lambda, u_{2k+1},\cdots ,u_{2n})= \Lambda^{4k(n-k)}\frac{(n-1)!}{(k-1)!(n-k-1)!}\delta_{2k}\delta_{2n-2k}\left[1+O(\Lambda^{-1})\right] \, .
\ee

where we omitted the obvious arguments of the polynomials with less particles.

We end the appendix by giving a compact formula for the case $n=2$

\ba
\delta_4(b_1,\ldots,b_4) &=&
14 + \frac{1}{2}(b_1 - b_2)^2 [(b_4 - b_3)^2+4] +
 \frac{1}{2}(b_1 - b_3)^2 [(b_2 - b_4)^2+4] + \nn\\
& +& \frac{1}{2}(b_1 - b_4)^2 [(b_2 - b_3)^2+4] +
 \frac{1}{2}(b_2 - b_3)^2 [(b_1 - b_4)^2+4] + \nn\\
&+& \frac{1}{2}(b_2 - b_4)^2 [(b_1 - b_3)^2+4] + \frac{1}{2} (b_3 - b_4)^2 [(b_1 - b_2)^2+4] = \\
&=& 2+[(b_1-b_2)^2+2][(b_3-b_4)^2+2] +[(b_1-b_3)^2+2][(b_2-b_4)^2+2]+ \nn \\
&+& [(b_1-b_4)^2+2][(b_2-b_3)^2+2] \, . \nn
\ea

\section{Scalars}

In Section \ref{ScaMat} we proved the polar structure (\ref {P2n}), recalled here by convenience
\be
\label{P2nbis}
\Pi_{mat}^{(2n)}(u_1,\ldots ,u_{2n})=\frac{P_{2n}(u_1,\ldots,u_{2n})}{\displaystyle\prod_{i<j}^{2n}(u_{ij}^2+1)(u_{ij}^2+4)} \, ,
\ee
where $P_{2n}$ is a polynomial of degree $4 n(n-1)$. Now we list some useful properties of these functions and give explicit expressions for the first cases $P_2$ and $P_4$. In addition, by means of the factorisation, we derive a simple formula for the highest degree $P_{2n}^{(0)}$ for any $n$.

To begin with, making use of the residue formula (\ref{ResPimatSca}), we express the full polynomial $P_{2n}$ computed in specific values in terms of smaller polynomials $P_{2k}$, with $k<n$
\be\label{RecSca}
P_{2n}(u_1-i,u_1+i,u_3,\ldots ,u_{2n})=6P_{2n-2}(u_3,\ldots ,u_{2n})\displaystyle\prod_{j=3}^{2n}(u_{1j}^2+4)(u_{1j}^2+9)\, ,
\ee
that gives, once iterated, the most general one valid for $k=0,\ldots, n$
\ba
&&P_{2n}(u_1-i,u_1+i,\ldots ,u_k-i,u_k+i,u_{2k+1},\ldots, u_{2n})=6^k P_{2n-2k}(u_{2k+1},\ldots ,u_{2n})\cdot \nn\\
&& \cdot\displaystyle\prod_{i<j}^k (u_{ij}^2+1)(u_{ij}^2+4)(u_{ij}^2+9)(u_{ij}^2+16)\displaystyle\prod_{i=1}^k\displaystyle\prod_{j=2k+1}^{2n}(u_{ij}^2+4)(u_{ij}^2+9) \, .
\ea
The full iteration $k=n$ gives the expression of the polynomials $P_{2n}$ in the specific configuration:
\small
\be
P_{2n}(u_1-i,u_1 +i,u_2-i,u_2 + i,\ldots ,u_{n}-i,u_{n}+i)= 6^n \displaystyle\prod_{i<j}^n(u_{ij}^2+1)(u_{ij}^2+4)(u_{ij}^2+9)(u_{ij}^2+16) \, .
\ee
\normalsize
Furthermore, the recursion formula (\ref{RecSca}) entails that the polynomial vanishes in some particular configurations, such as
\ba
&& P_{2n}(u_1,u_1+i,u_1+3i,u_4,\ldots ,u_{2n})=0 \, , \nn\\
&& P_{2n}(u_1,u_1+2i,u_1+4i,u_4,\ldots ,u_{2n})=0 \, . \label {van-conf}
\ea

\paragraph{Factorisation}

The factorisability of the functions $G^{(2n)}$ also affects the behaviour of the polynomials $P_{2n}$. 
Indeed, they satisfy a factorisation property as well, which reads 
\ba\label{factP}
&& P_{2n} (u_1 + \Lambda,\ldots,u_{2k}+\Lambda,u_{2k+1},\ldots,u_{2n})= \Lambda ^{8(n-k)k}
P_{2k}(u_1,\ldots\,u_{2k})P_{2n-2k}(u_{2k+1},\ldots\,u_{2n})\cdot \nonumber \\
&\cdot & \left [ 1+2\Lambda^{-1} \sum _{i=1}^{2k} \sum _{j=2k+1}^{2n} (u_i-u_j)  + \Delta_{2n,2k}^{(2)}(u_1,\ldots ,u_{2n})\Lambda^{-2} + O(\Lambda^{-3})\right] \, ,
\ea
where function $\Delta_{2n,2k}^{(2)}$ parametrizes the quadratic subleading. 
On the other hand, by shifting an odd number of particles, we get instead the power law
\be
P_{2n} (u_1 + \Lambda,\ldots,u_{2k+1}+\Lambda,u_{2k+2},\ldots,u_{2n})= O(\Lambda ^{2(2k+1)(2n-2k-1)-2}) \, ,
\ee

\paragraph{Explicit expressions}

We now provide the polynomials appearing in (\ref {P2n}) in the cases $n=1, n=2$: beside to the simple $P_2(u_1,u_2) =6$, we have
\footnotesize
\begin{align}\label{P4}
P_4(u_1,u_2,u_3,u_4) &= 36 \left[9((u_1-u_2)^2+4)((u_3-u_4)^2+4)+9((u_1-u_3)^2+4)((u_2-u_4)^2+4)+\right. \nn\\
&+ 9((u_1-u_4)^2+4)((u_2-u_3)^2+4) +\nn\\
&+ ((u_1-u_3)^2+4)((u_1-u_4)^2+4)((u_2-u_3)^2+4)((u_2-u_4)^2+4)+\nn\\
&+ ((u_1-u_2)^2+4)((u_1-u_4)^2+4)((u_3-u_2)^2+4)((u_3-u_4)^2+4)+\nn\\
&+ ((u_1-u_2)^2+4)((u_1-u_3)^2+4)((u_4-u_2)^2+4)((u_4-u_3)^2+4)+\nn\\
&+ \frac{3}{2}((u_1 - u_2)^2+4)((u_2 - u_3)^2+4)((u_1- u_4)^2+4) +\nn\\
&+ \frac{3}{2}((u_1 - u_3)^2+4)((u_2-u_3)^2+4)((u_1 - u_4)^2+4) + \nn\\
&+ \frac{3}{2}((u_1 - u_2)^2+4)((u_1-u_3)^2+4)((u_2 - u_4)^2+4) + \nn\\
&+ \frac{3}{2}((u_1 - u_3)^2+4) ((u_2 - u_3)^2+4) ((u_2 - u_4)^2+4) + \nn\\
&+ \frac{3}{2}((u_1 - u_3)^2+4) ((u_1 - u_4)^2+4) ((u_2 - u_4)^2+4) + \nn\\
&+ \frac{3}{2}((u_2 - u_3)^2+4) ((u_1 - u_4)^2+4) ((u_2 - u_4)^2+4) + \nn\\
&+ \frac{3}{2}((u_1 - u_2)^2+4) ((u_1 - u_3)^2+4) ((u_3 - u_4)^2+4) + \nn\\
&+ \frac{3}{2}((u_1 - u_2)^2+4) ((u_2 - u_3)^2+4) ((u_3 - u_4)^2+4) + \nn\\
&+ \frac{3}{2}((u_1 - u_2)^2+4) ((u_1 - u_4)^2+4) ((u_3 - u_4)^2+4) + \nn\\
&+ \frac{3}{2}((u_2 - u_3)^2+4) ((u_1 - u_4)^2+4) ((u_3 - u_4)^2+4) + \nn\\
&+ \frac{3}{2}((u_1 - u_2)^2+4) ((u_2 - u_4)^2+4) ((u_3 - u_4)^2+4) + \nn\\
&+ \frac{3}{2}((u_1 - u_3)^2+4) ((u_2 - u_4)^2+4) ((u_3 - u_4)^2+4)-\nn\\
&- \frac{3}{2}((u_1 - u_2)^2(u_2 - u_3)^2(u_1 - u_3)^2 + (u_1 - u_2)^2(u_2 - u_4)^2(u_1-u_4)^2 +\nn\\
&+ (u_1 - u_4)^2(u_4 - u_3)^2(u_1 - u_3)^2 +(u_4 - u_2)^2(u_2 - u_3)^2(u_4 -u_3)^2)+ \nn\\
&+ 48(u_1 - u_2)^2 + 48(u_1 - u_3)^2 + 48(u_1 - u_4)^2 + 48(u_3 - u_2)^2 + 48(u_4 - u_2)^2 +\nn\\
&+ 48(u_3 -u_4)^2 +\frac{3}{2}((u_1 - u_2)^2(u_3 - u_4)^2 +(u_1 - u_3)^2(u_2 - u_4)^2 +\nn\\
&+ (u_1 - u_4)^2(u_3 - u_2)^2) + \left. 1152 \right]
\end{align}
\normalsize
From the expression of $P_4$ and the factorisation property (\ref{factP}), we can make a guess
for the highest degree term $P^{(0)}_{2n}$. For $n=2$ the exact formula (\ref{P4}) yields
\be
P_{4}^{(0)}(u_1,u_2,u_3,u_4)=6^2\displaystyle\prod_{i<j}^4u_{ij}^2\left(\frac{1}{u_{12}^2 u_{34}^2} + \frac{1}{u_{13}^2 u_{24}^2} + \frac{1}{u_{14}^2 u_{23}^2}\right) \, .
\ee
This formula has a nice interpretation as a sum over the pairings, recalling the Wick theorem for bosonic particles: prefactor aside, we can think of $P^{(0)}_4$ as the four point function of a free boson with propagator $u^{-2}_{ij}$. The generalization of this formula to the $2n$ goes through an expression that, in the factorisation limit, reproduces exactly the property (\ref{factP}) for $P_{2n}^{(0)}$, with any $n,k$.
We thus conjecture
\be\label{P0}
P_{2n}^{(0)}(u_1,\ldots ,u_{2n})=6^n\displaystyle\prod_{i<j}^{2n} u_{ij}^2 \sum'_p\displaystyle\prod_{i=1}^{n}\frac{1}{(u_{p(2i-1)} - u_{p(2i)})^2}  \, ,
\ee
where the sum is restricted over the pairings, such that the total number of terms is $(2n-1)!!$.
A careful analysis shows that (\ref{P0}) is the only (symmetric) polynomial solution satisfying the factorisation (\ref{factP}). Formula (\ref{P0}) is confirmed for $n=3$, directly from the sum over Young tableaux (\ref{MatYoung}) depicted in Chapter \ref{ChMat}.

There is an interesting link with the polynomials $\delta_{2n}$: we use the identity\footnote{From the physical point of view, this identity is a sort of bosonisation, as the LHS can be thought of as a correlator of a free fermion with propagator $u^{-1}_{ij}$.} for the special $2n\times 2n$ antisymmetric matrix
\be\label{Pf-Wick}
\textrm{Det} \left(\frac{1}{u_{ij}}\right)=\left[\textit{Pf}\left(\frac{1}{u_{ij}}\right)\right]^2 = \sum'_p\displaystyle\prod_{i=1}^{n}\frac{1}{(u_{p(2i-1)} - u_{p(2i)})^2} \, ,
\ee
to relate the highest degrees of the polynomials $P_{2n}$ and $\delta_{2n}$. Merging formulae (\ref{delta0det}),(\ref{P0}) and (\ref{Pf-Wick}) together, we can express the highest degree as a determinant
\be
P^{(0)}_{2n}(u_1,\ldots ,u_{2n})=6^n\displaystyle\prod_{i<j}^{2n} u_{ij}^2\textrm{Det}\left(\frac{1}{u_{ij}}\right)=\frac{6^n4n^2}{4^n(n!)^2}\left[\delta^{(0)}_{2n}(u_1,\ldots ,u_{2n})\right]^2 \, .
\ee
Unfortunately, this remarkable equality does not survive when we consider the full polynomial $P_{2n}$, as we can check for $n=2$ with the explicit formula (\ref{P4}).  We do not know if a determinant representation of the full $P_{2n}$ exists: however, it is an interesting idea to pursue since it would allow to find a nice representation of $W$.

\section{Fermions}

Here we provide a list of properties for the polynomials $P^{(n)}$ which helped us to prove some of the assertions contained in the main text.\\
The formula (\ref{ResPimat}) for the residue of the matrix factor entails a sort of recursion relation among the polynomials, when computed in the specific configuration
\ba\label{RecConj}
&& P^{(n)}(u_1,\cdots , u_n,u_1-2i,v_2, \cdots , v_n)=4P^{(n-1)}(u_2,\cdots , u_n,v_2, \cdots , v_n)\cdot \\
&& \cdot \displaystyle\prod_{j=2}^n (u_{1j}+i)(u_{1j}-4i)(u_1-v_j+2i)(u_1-v_j-3i) \nn
\ea
which can be iterated $k$ times ($k\leq n$), to  get
\ba\label{Reck}
&& P^{(n)}(u_1,\cdots , u_n,u_1-2i,,\cdots ,u_k -2i, v_{k+1}, \cdots , v_n)=4^k P^{(n-k)}(u_{k+1},\cdots , u_n,v_{k+1}, \cdots , v_n)\cdot \nn \\
&& \cdot \displaystyle\prod_{i=1}^k\displaystyle\prod_{j=k+1}^n (u_{ij}+i)(u_{ij}-4i)(u_i-v_j+2i)(u_i-v_j-3i)\displaystyle\prod_{i<j}^k(u_{ij}^2+1)(u_{ij}^2+16) \ ;
\ea
The complete iteration $k=n$ gives the property
\be\label{fundP}
P^{(n)}(u_1,\cdots , u_n,u_1-2i, \cdots , u_n-2i)=4^n\displaystyle\prod_{i<j}^n(u^2_{ij}+1)(u^2_{ij}+16) ,
\ee
pivotal for the achievements in the main text.
As an application of the recursion relation (\ref{RecConj}), one finds that the polynomials vanish under some special configurations:
\ba\label{Pfer=0}
&& P^{(n)}(u_1,\cdots , u_n,u_1-2i,u_1-3i,v_3,\cdots ,v_n)=0 \nn\\
&& P^{(n)}(u_1,\cdots , u_n,u_1-2i,u_1 +2i,v_3,\cdots ,v_n)=0 \nn\\
&& P^{(n)}(u_1,u_1+i,u_3,\cdots , u_n,u_1-2i,v_2,\cdots ,v_n)=0 \nn\\
&& P^{(n)}(u_1,u_1-4i,u_3,\cdots , u_n,u_1-2i,v_2,\cdots ,v_n)=0 \ .
\ea

Furthermore, as in the scalar case, we observe the factorisation of the polynomials $P^{(n)}$ as a straightforward consequence of (\ref{Mat-fer}) and the factorisation of the matrix part\footnote{It works in the same way as the scalar matrix part.}
\be\label{factPfer}
P^{(n)}(\{u_{i=1}^{k} +\Lambda , u_{i=k+1}^n \},\{v_{i=1}^{k} +\Lambda , v_{i=k+1}^n  \}) \simeq \Lambda^{4k(n-k)}P^{(k)}(\{u_{i=1}^k\},\{v_{i=1}^k\})P^{(n-k)}(\{u_{i=k+1}^n\},\{v_{i=k+1}^n\}) \ .
\ee

\paragraph{Explicit expressions}

The polynomials get involved very quickly, so that we have explicit compact expression only up to $n=2$, which are:

\ba\label{P1P2}
P^{(1)}(u_1,v_1) &=& 4 \\
P^{(2)}(u_1,u_2,v_1,v_2) &=& 4[24 + 3((u_2-v_1)^2+6)((u_1-v_2)^2+6) + \nn\\
&+&   3((u_1-v_1)^2+6)((u_2-v_2)^2+6) +((u_1-u_2)^2+4)((v_1-v_2)^2+4)] \ . \nn
\ea

\chapter{Factorisation and Connected functions}
\label{FactConn}

In this appendix we discuss in details the connected functions introduced in Chapter \ref{ChSca}, both for the hexagon and for $N>6$. In the former case, we also discuss their asymptotic properties by extending the argument of the main text, with different large shifts $\Lambda_i$. They are crucial to prove the $\sim\sqrt{\lambda}$ behaviour of the logarithm of the Wilson loop at strong coupling.

\section{Asymptotic factorisation}

In this part we are going to extend the argument of the maix text, where we proved the factorisation $G^{(2n)}\to G^{(2k)}G^{(2n-2k)}$ when the $2n$ particles are split in two groups composed respectively by $2k$ and $2n-2k$ particles, by considering the case with different large shifts. To be specific, for even $m$, we prove the factorisation arises when considering
\be
G^{(2n)}(u_1+\Lambda _1, u_2 +\Lambda _2,\ldots , u_m+\Lambda _m, u_{m+1},\ldots ,u_{2n}) \, ,
\ee
where the shifts $\Lambda _i$ are parametrised as $\Lambda _i=c_i R +O(R^0)$, with $c_i$ constants and $R$ a large quantity. We need to unravel the structure of $G^{(2n)}$ in this particular limit.

\medskip

To begin with, the dynamical factor (\ref{dynamical}) enjoys the behaviour (valid also for $m$ odd)
\ba
&& \Pi_{dyn}^{(2n)}(u_1+\Lambda _1,\ldots,u_{m}+\Lambda_m,u_{m+1},\ldots,u_{2n})\longrightarrow  \nonumber \\
&& \Pi_{dyn}^{(m)}(u_{1}+\Lambda _1,\ldots,u_{m}+\Lambda _m) \left ( \prod _{i=1}^{m} \Lambda _i ^2 \right ) ^{2n-m} \Pi_{dyn}^{(2n-m)}(u_{m+1},\ldots,u_{2n}) \label {as-dyn} \cdot \nonumber \\
&& \cdot \left [ 1+2 \sum _{i=1}^m\sum _{j=m+1}^{2n}\frac{u_i-u_j}{\Lambda _i} +O\left (R^{-2} \right ) \right ] \, ,
\ea
as a consequence of the asymptotic behaviour of the two-particle part
\be\label{asymptF}
u \rightarrow \infty \quad \Rightarrow \quad \Pi(u)= u^2-\frac{1}{2}-\frac{9}{8u^2}+O\left (u^{-4}\right ) \ .
\ee
We point out that $\Pi_{dyn}^{(m)}(u_{1}+\Lambda _1,\ldots,u_{m}+\Lambda _m)$ is actually divergent, if at least one of the $\Lambda _i$ is different from the others
\be
\Pi_{dyn}^{(m)}(u_{1}+\Lambda _1,\ldots,u_{m}+\Lambda _m) =
\mu ^m \prod _{\underset{i<j}{i,j=1}}^{m} (\Lambda _i-\Lambda _j)^2 + \ldots  \, ,
\ee

\medskip

For the matrix part (\ref{ScaPiMat}) we follow the same procedure of the main text: we tackle the simplest case first, {\it i.e.} $\Pi_{mat}^{(4)}\rightarrow \Pi_{mat}^{(2)}\Pi_{mat}^{(2)}$: eventually, the procedure can be extended to the general case $\Pi_{mat}^{(2n)}\rightarrow \Pi_{mat}^{(2k)}\Pi_{mat}^{(2n-2k)}$\,.
To start with, we perform the shifts $u_1\to u_1+\Lambda _1 ,\,u_2\to u_2+\Lambda _2$; for large $\Lambda_ 1$, $\Lambda _2$ the integrals (\ref{ScaPiMat}) receive the main contribution from the region where two roots $b$, one $a$ and one $c$ are shifted by $\Lambda _i$ as well. We decide to shift, for instance, $a_1$ by $\Lambda _1^{a}$, $c_1$ by $\Lambda _1^{c}$ and $b_1,b_2$ by $\Lambda _1^{b}$, $\Lambda _2^{b}$, respectively, where the large shifts of isotopic variables can be equal to $\Lambda _1$ or $\Lambda _2$.
In addition, we have to sum over all the possible choices for the shifts $\Lambda _i^{\alpha}$, $\alpha=a,b,c,$: for this sum we use the short-hand notation $\sum \limits_{\textrm{shifts}}$. Moreover, the usual combinatorial factor appears
\be
24= \binom{4}{2} \cdot 2 \cdot 2
\ee
which takes into account the $\binom{4}{2}$ independent choices of the auxiliary rapidities, all giving the same result. More details of the calculations are reported in \cite{BFPR3}.
In the end, for the matrix part we obtain the factorisation
\ba\label{4-fact-PiMat}
&&\Pi_{mat}^{(4)}(u_1+\Lambda_1,u_2+\Lambda_2,u_3,u_4)=
\Lambda_1^{-4}\Lambda _2^{-4}\Bigl [1+2(u_3+u_4)\left (\frac{1}{\Lambda _1}+\frac{1}{\Lambda _2} \right )- \nn \\
&&- 4\left (\frac{u_1}{\Lambda _1}+\frac{u_2}{\Lambda _2} \right ) +O(R^{-2}) \Bigr ]  \Pi_{mat}^{(2)}(u_1+\Lambda _1,u_2+\Lambda _2)\Pi_{mat}^{(2)}(u_3,u_4) \ , \label {Pi4-fact}
\ea

Since the dynamical factor behaves as
\ba
&& \Pi_{dyn}^{(4)}(u_1+\Lambda_1,u_2+\Lambda_2,u_3,u_4)\rightarrow \Pi_{dyn}^{(2)}(u_1+\Lambda_1,u_2+\Lambda_2)\Pi_{dyn}^{(2)}(u_3,u_4)
\Lambda _1^4 \Lambda _2^4 \cdot \nn \\
&& \cdot \left [1-2(u_3+u_4)\left (\frac{1}{\Lambda _1}+\frac{1}{\Lambda _2} \right )+4\left (\frac{u_1}{\Lambda _1}+\frac{u_2}{\Lambda _2} \right ) +O(R^{-2}) \right ]  \, ,
\ea

for the complete function (\ref{Gi2n}) we obtain the factorisation we were aiming for:
\be
G^{(4)}(u_1+\Lambda_1,u_2+\Lambda_2,u_3,u_4)\ \overset{\Lambda_i\rightarrow\infty}{\longrightarrow}\ G^{(2)}(u_1+\Lambda _1,u_2+\Lambda _2)G^{(2)}(u_3,u_4) [1+O(R^{-2}) ]\, , \label{4-fact}
\ee
where, if $c_1 \not= c_2$, $G^{(2)}(u_1+\Lambda _1,u_2+\Lambda _2)=6\mu ^2 /\Lambda _{12}^2+\ldots $, if $c_1=c_2$, $G^{(2)}$ is finite.
Of course, as $G^{(4)}(u_1,u_2,u_3,u_4)$ is symmetric under exchange of rapidities, the property (\ref {4-fact}) is indeed valid for any couple of rapidities.

\medskip

We are now ready to address the most general case: we shift $m=2k$ rapidities by amounts $\Lambda _i$. Thanks to the symmetry of the function $G$, we can shift the first $m$ rapidities without losing generality: $u_i\to u_i+\Lambda _i$ for $1\leq i\leq m$. We also shift $a_i\to a_i+\Lambda _i^a$ and $c_i\to c_i+\Lambda_i^c$ for $1\leq i\leq k$, along with $b_i\to b_i+\Lambda _i^b$ for $1\leq i\leq m$. 

The final result is:
\small
\ba
&& \Pi_{mat}^{(2n)}(u_1+\Lambda _1,\ldots,u_{2k}+\Lambda _{2k},u_{2k+1},\ldots,u_{2n}) \rightarrow
\frac{1}{\left ( \prod \limits _{i=1}^m \Lambda _i \right )^{4n-4k}}\Bigl [1+2 \sum _{i=1}^{2k}\frac{1}{\Lambda _i}\sum _{j=2k+1}^{2n}u_j- \nonumber \\
&-& 2(2n-2k)\sum _{i=1}^{2k}\frac{u_i}{\Lambda _i} + O\left (R^{-2}\right ) \Bigr ] \Pi_{mat}^{(2k)}(u_1+\Lambda _1,\ldots,u_{2k}+\Lambda_{2k})\Pi_{mat}^{(2n-2k)}(u_{2k+1},\ldots,u_{2n}) \, ,
\ea
\normalsize
which extends that of the main text to the case with different $\Lambda_i$. We remark that $\Pi_{mat}^{(2k)}(u_1+\Lambda _1,\ldots,u_{2k}+\Lambda_{2k})$, in the general case with different $\Lambda _i$, goes to zero as
\be
\Pi_{mat}^{(2k)}(u_1+\Lambda _1,\ldots,u_{2k}+\Lambda_{2k})\sim
\prod _{i<j=1}^{2k} (\Lambda _i-\Lambda _j)^{-2} \left [ \Lambda _{12}^{-2}\Lambda _{34}^{-2}\ldots \Lambda _{2k-1,2k}^{-2}+pairings \right ]
\, . \label{pimatzero}
\ee

We already know the dynamical part behaviour, which added to our result yields the sought factorisation property:
\ba\label{2n-fact}
&& G^{(2n)}(u_1+\Lambda _1,\ldots,u_{2k}+\Lambda _{2k},u_{2k+1},\ldots,u_{2n}) \rightarrow \nonumber \\
&& \rightarrow G^{(2k)}(u_1+\Lambda _1,\ldots,u_{2k}+\Lambda_{2k})G^{(2n-2k)}(u_{2k+1},\ldots,u_{2n})\Bigl [1+ O\left (R^{-2}\right ) \Bigr ] \, ,
\ea

We remark that the function $G^{(2k)}(u_1+\Lambda _1,\ldots,u_{2k}+\Lambda_{2k})$, when $\Lambda _i$ are all different, behaves like
\be
G^{(2k)}(u_1+\Lambda _1,\ldots,u_{2k}+\Lambda_{2k})
\sim \left [ \Lambda _{12}^{-2}\Lambda _{34}^{-2}\ldots \Lambda _{2k-1,2k}^{-2}+pairings \right ] \, .
\ee
Therefore, putting $\Lambda _i =c_i R +O(R^0)$, with different $c_i$ and $R\rightarrow +\infty$, the behaviour of $G^{(2k)}(u_1+\Lambda _1,\ldots,u_{2k}+\Lambda_{2k})$ and of $G^{(2n)}(u_1+\Lambda _1,\ldots,u_{2k}+\Lambda _{2k},u_{2k+1},\ldots,u_{2n})$ is respectively given by
\be\label{G-even}
G^{(2k)}(u_1+\Lambda _1,\ldots,u_{2k}+\Lambda_{2k}) \sim R^{-2k} \, , \quad G^{(2n)}(u_1+\Lambda _1,\ldots,u_{2k}+\Lambda _{2k},u_{2k+1},\ldots,u_{2n})
\sim R^{-2k} \, .
\ee
If $p$ of the $c_i$ coincide, {\it i.e.} $\Lambda _1=\Lambda _2=\ldots =\Lambda _p \not = \Lambda _{p+1} \ldots \not= \Lambda _{2k}$, $G^{(2k)}$ and consequently $G^{(2n)}$ go to zero as $R^{-2k+2\left [\frac{p}{2}\right ]}$: some of these configurations will be examined in the next part, when we discuss the connected functions. 

Now, we spend a few words to analyse the asymptotic of $G^{(2n)}$ when we shift an odd number $m$ of rapidities.
In this case there is no factorisation, since there are no functions $G$ with an odd number of arguments.

When $m$ is odd, we find convenient to define $m=2k-1$. Then, we shift $a_i\to a_i+\Lambda _i^a$ and $c_i\to c_i+\Lambda_i^c$ for $1\leq i\leq k$, along with $b_i\to b_i+\Lambda _i^b$ for $1\leq i\leq m$. With these choices, formula (\ref {Intpimat}) still holds. Sending $\Lambda _i \rightarrow +\infty$ inside the integrals, we get
\ba
&& \m{R}^{(2n,m)} \rightarrow  \frac{1}{\left ( \prod \limits _{i=1}^m \Lambda _i \right )^{4n-2m}}
\prod _{i=1}^m (\Lambda _i^b)^2 \prod _{i=1}^k (\Lambda _i^a \Lambda _i ^c )^{-2}
\Bigl \{ 1+2 \sum _{i=1}^{m}\frac{1}{\Lambda _i^b}\sum _{j=m+1}^{2n}u_j-2(2n-m)\sum _{i=1}^{m}\frac{u_i}{\Lambda _i} + \nonumber \\
&& + \sum _{j=m+1}^{2n} 2b_j \left [ \sum _{i=1}^{m}\frac{1}{\Lambda _i}-2\sum _{i=1}^{m} \frac{1}{\Lambda _i^b} + \sum _{i=1}^k\left (\frac{1}{\Lambda _i^a}+\frac{1}{\Lambda _i^c} \right ) \right ] +  \sum _{j=k+1}^{n} 2a_j \left ( \sum _{i=1}^{m}\frac{1}{\Lambda _i^b}-2 \sum _{i=1}^{k}\frac{1}{\Lambda _i^a} \right ) + \nn \\
&& +  \sum _{j=k+1}^{n} 2c_j \left ( \sum _{i=1}^{m}\frac{1}{\Lambda _i^b}-2 \sum _{i=1}^{k}\frac{1}{\Lambda _i^c} \right ) + 2 \sum _{i=1}^m \frac{b_i}{\Lambda _i^b}-2\sum _{i=1}^k \frac{a_i}{\Lambda _i^a}-2\sum _{i=1}^k \frac{c_i}{\Lambda _i^c}+ O\left (R^{-2} \right ) \Bigr \} \, . \label {R-int2}
\ea
Therefore, keeping only the leading term, when $m=2k-1$ the matrix part behaves as
\ba
&& \Pi_{mat}^{(2n)}(u_1+\Lambda _1,\ldots,u_{2k-1}+\Lambda _{2k-1},u_{2k},\ldots,u_{2n}) \sim
\frac{(\prod \limits _{i=1}^k \Lambda _i ^{-4}+perm.)}{\left ( \prod \limits _{i=1}^{2k-1} \Lambda _i \right )^{4n-4k}} \cdot\nn\\
&& \cdot\prod \limits _{\underset{i<j}{i,j=1}}^{2k-1} (\Lambda _i-\Lambda _j)^{-2}
\left [ \Lambda _{12}^{-2}\Lambda _{34}^{-2}\ldots \Lambda _{2k-3,2k-2}^{-2}+pairings \right ]
\cdot \textrm{finite function} (u_{2k},\ldots,u_{2n}) \, , \nonumber
\ea
Considering tha dynamical factor, we get
\ba
&& G^{(2n)}(u_1+\Lambda _1,\ldots,u_{2k-1}+\Lambda _{2k-1},u_{2k},\ldots,u_{2n}) \sim \nonumber \\
&&
\left [ \Lambda _{12}^{-2}\Lambda _{34}^{-2}\ldots \Lambda _{2k-3,2k-2}^{-2}+pairings \right ]
\prod \limits _{i=1}^{2k-1}\Lambda _i ^2 \left ( \prod \limits _{i=1}^k \Lambda _i ^{-4} +perm. \right ) \cdot  \textrm{finite function} (u_{2k},\ldots,u_{2n}) \, . \nonumber
\ea
In the end, we conclude that, if all the $\Lambda_i$ are different and scale with $R$, the behaviour of $G^{(2n)}(u_1+\Lambda _1,\ldots,u_{2k-1}+\Lambda _{2k-1},u_{2k},\ldots,u_{2n}) $  is
\be
G^{(2n)}(u_1+\Lambda _1,\ldots,u_{2k-1}+\Lambda _{2k-1},u_{2k},\ldots,u_{2n})  \sim R^{-2k}=R^{-m-1} \, . \label{G-odd}
\ee
On the other hand, if $p$ of the shifts $\Lambda _i$ coincide, the function $G^{(2n)}$ vanishes as $R^{-2k+2\left [ \frac{p}{2}\right ]}$.

To summarize, we addressed some different splits of the rapidities and obtained the corresponding asymptotic behaviours of the $G^{(2n)}$. The main formulae, (\ref{2n-fact}), (\ref{G-even}) and (\ref{G-odd}), will be of  fundamental importance to study the asymptotic properties of the connected functions $g^{(2n)}$. We have to say, however, that our analysis is not complete, as there are other asymptotic regions to be considered.

\paragraph{Young tableaux}

It is useful to mention an alternative method to prove the factorisation properties obtained so far, which makes use of the Young tableaux expansion (\ref{MatYoung}). For simplicity here we stick to the simplest case, the split $4\to 2+2$ which contains only one shift $\Lambda$ and has been already analysed in the main text, resulting in (\ref{422Pimat}). 
We observe that many diagrams of $\Pi_{mat}^{(4)}$ split into a product of two diagrams belonging to the expansion of $\Pi_{mat}^{(2)}$, weighted by the prefactor $\Lambda^{-8}$, as follows

\footnotesize
\ba\label{422Young}
(1,1,1,1)&&\to \Lambda^{-8}(1,1)_{12}\times(1,1)_{34} \\
(2,0,2,0)+(2,0,0,2)+(0,2,2,0)+(0,2,0,2)&&\to \Lambda^{-8}[(2,0)_{12}+(0,2)_{12}]\times[(2,0)_{34}+(0,2)_{34}]\nn\\
(1,1,2,0)+(1,1,0,2)+(2,0,1,1)+(0,2,1,1)&&\to \Lambda^{-8}(1,1)_{12}\times[(2,0)_{34}+(0,2)_{34}] + 12 \leftrightarrow 34 \nn
\ea
\normalsize

The RHS members sum up to $\Lambda^{-8}\Pi_{mat}^{(2)}(u_1,u_2)\Pi_{mat}^{(2)}(u_3,u_4)$, in agreement with (\ref{422Pimat}). It is interesting to note that the other diagrams of $\Pi_{mat}^{(4)}$, not present in (\ref{422Young}), decay faster than $O(\Lambda^{-8})$ and they do not contribute to the factorisation. This procedure can be extended to the general split $2n\to 2(n-k) + 2k$ with also the possibility to have different shifts $\Lambda_i$.

\section{Connected functions}

In this appendix we elaborate on the connected functions, both for the hexagonal Wilson loop and the bigger polygons $N>6$. In the former case we also analyse in details their asymptotic properties, which are of paramount importance for the method in Section \ref{ScalarHex} to work.

\subsection{Hexagon}

Here we discuss the connected functions $g^{(2n)}$ for the hexagon. We first analyse the relation with the  functions $G^{(2n)}$, sketched for the cases $n=2,3$ in the main text. We also give evidence that $g^{(2n)}\in L^1(\mathbb{R}^{2n-1})$, \emph{i.e.} that the integral over the $2n-1$ variables of $g^{(2n)}$ is well-defined.

The expansion of $G^{(2n)}$
in terms of the connected parts is well-known and involves a sum over all the possible arrangements of $2n$ particles in subgroups of even particles
\be\label{direct}
G^{(2n)}=\sum_{\left\{m\right\}}\sum_{\textit{pair.}}\displaystyle\prod_{k=1}^{n}\underbrace{g^{(2k)}\ldots g^{(2k)}\,}_\text{$m_k$ terms} \, ,
\ee

where, for the sake of compactness, we omitted the dependence on the rapidities.

The set $\left\{m\right\}$ represents the integers $m_k$ with $k=1,\ldots,n$, identifying a specific cluster configuration for the $2n$ particles and fulfilling the constraint $\displaystyle\sum_{k=1}^{n}2k \, m_k=2n$\footnote{A similar formula holds in general, when also odd numbers of particles are allowed.}. For any definite set $\{m\}$, the number of inequivalent ways of clustering is given by

\be 
\left(\prod_{k=1}^{n}\frac{1}{m_k!}\right)\frac{(2n)!}{\prod_{k=1}^{n}((2k)!)^{m_k}}
\ee

The inverse relation can be obtained, resulting in the general expansion
\be\label{inverse}
g^{(2n)}=\sum_{\left\{m\right\}}f(\left\{m\right\})\sum_{\textit{pair.}}\displaystyle\prod_{k=1}^{n}\underbrace{G^{(2k)}\ldots G^{(2k)}\,}_\text{$m_k$ terms} \, ,
\ee
where, in contrast to (\ref {direct}), the products of functions are weighted by the prefactor
\be
f(\left\{m\right\})=(-1)^{p}p! \, , \quad p\equiv\sum_{k=1}^{n}m_k - 1 \, ,
\ee
containing also an oscillating sign. In an alternative manner, it is possible to sum over all the permutations and account for the overcounting with the specific prefactor, rewriting (\ref{direct}) and (\ref{inverse}) as \cite{Smirnov}
\ba
&&G^{(n)}(u_1,\ldots,u_n)=\sum_{q=1}^{n}\frac{1}{q!}\sum_{k_1+\ldots+k_q=n}\frac{1}{k_1!\ldots k_q!} \cdot \nn\\
&&\cdot\sum_P g^{(k_1)}(u_{P_1},\ldots,u_{P_{k_1}})\ldots g^{(k_q)}(u_{P_{n-k_q+1}}\ldots,u_{P_{n}}) \, ,
\ea
\ba
&&g^{(n)}(u_1,\ldots,u_n)=\sum_{q=1}^{n}\frac{(-1)^{q-1}}{q}\sum_{k_1+\ldots+k_q=n}\frac{1}{k_1!\ldots k_q!}\cdot \nn\\
&&\cdot\sum_P G^{(k_1)}(u_{P_1},\ldots,u_{P_{k_1}})\ldots G^{(k_q)}(u_{P_{n-k_q+1}}\ldots,u_{P_{n}}) \, ,
\ea

which actually holds for $n$ odd as well.

In the following we discuss the asymptotic properties of $g^{(2n)}$, giving evidence of their integrability $g^{(2n)}\in L^1(\mathbb{R}^{2n-1})$. This is a fundamental condition for the method in the main text to work, see formula (\ref{Wleading}).

\subsubsection{Asymptotic properties}

To this aim, it is sufficient to prove that $g^{(2n)}$ belongs to the class $L^1(\mathbb{R}^{2n-1})$, a stronger condition as it involves the integral of $|g^{(2n)}|$. To show $g^{(2n)}\in L^1(\mathbb{R}^{2n-1})$, we need to address all the possible asymptotic behaviours in the integration space. The most general situation concerns $l$ subsets composed by $k_i \ (i=1,\ldots,l)$ variables going to infinity with the shifts $\Lambda_i=c_i R$, $i=1,\ldots ,l$. A sufficient condition is 
\small
\be\label{convg2n}
g^{(2n)}(u_1 + \Lambda_1, \ldots ,u_{k_1}+\Lambda_1, u_{k_1 +1} + \Lambda_2, \ldots ,u_{k_1 +k_2}+\Lambda_2 ,\ldots , u_{\sum_i k_i} +\Lambda_l, \ldots ,u_{2n} ) \simeq O(R^{a\leq -l-1}) \,.
\ee
\normalsize
This is the generalisation of the one dimensional case $R^{a\leq -2}$, once we take into account the  growth of the integration volume $R^{l-1}$.
A rigorous proof of (\ref{convg2n}) is not easy, as the number of regions is very large as $n$ increases. However, there are many indications that all the functions belong to $L^1(\mathbb{R}^{2n-1})$. A thorough analysis of the condition (\ref{convg2n}) for the first cases $g^{(4)}$ and $g^{(6)}$ will be more convincing and it is done later. Here we will also address some of them for any $n$, giving more hints for $g^{(2n)}\in L^1(\mathbb{R}^{2n-1})$. The analysis here is expoited in details in the appendix of \cite{BFPR3}.

Let us start with the analysis of $g^{(2n)}$: we observe that the function $G^{(2n)}$ behaves as
\be
G^{(2n)}(u_1+\Lambda _1,u_2+\Lambda_2, \ldots,u_{m}+\Lambda_{m},u_{m+1},\ldots,u_{2n}) \sim R^{-2k+2\left [ \frac{p}{2}\right ]} \label{Goddeven} \, ,
\ee
where $2k=m$ for even $m$, $2k=m+1$ for $m$ odd and $p\leq m$ is the number of $\Lambda _i$ which mutually coincide. As a direct consequence of (\ref{inverse}), the connected functions exhibit instead a different asymptotic behaviour: in fact, when $m$ is odd, {\it i.e.} $m=2k-1$, we find
\be\label{gRodd}
g^{(2n)}(u_1+\Lambda _1,u_2+\Lambda_2,\ldots,u_{2k-1}+\Lambda_{2k-1},u_{2k},\ldots,u_{2n}) \sim R^{-2k+2\left [ \frac{p-1}{2}\right ]} \,;
\ee
otherwise for even $m=2k$, with $1<m\leq 2n-2$, a faster decay shows up
\be\label{gReven}
g^{(2n)}(u_1+\Lambda _1,u_2+\Lambda_2,\ldots,u_{2k}+\Lambda_{2k},u_{2k+1},\ldots,u_{2n}) \sim R^{-2k-2+2\left [ \frac{p}{2}\right ]} \,.
\ee

So far, we considered the case with $m$ particles to infinity, shifted by $m$ (or by $m-p+1$) different quantities $\Lambda_i$, $i=1,\ldots ,m$: the function $g^{(2n)}$ decays fast enough to be integrable. This asymptotic region corresponds to the split $2n\to p + m' +1+ \ldots + 1$, where $m'\equiv 2n-m$. However, these cases are just a small subset of all the different ways of grouping particles. The most general case, represented by the formula (\ref{convg2n}), describe the split $2n\to k_1 + \cdots + k_{l+1}$. Even though the general proof is very complicated, being the number of asymptotic regions rapidly growing with $n$, a thorough analysis of the simplest cases $g^{(4)}$, $g^{(6)}$ strongly hints that $g^{(2n)}$ belongs to $L^1(\mathbb{R}^{2n-1})$ for any $n$. 

\medskip

\noindent\textbf{$\bullet$ Four particles case $g^{(4)}$}

\medskip

For the four point function $g^{(4)}$, the conditions above are sufficient to guarantee $g^{(4)}\in L^1(\mathbb{R}^3)$, as all the asymptotic regions belong to the type $2n\to p+m'+1+\ldots + 1$.
However, as the polynomials (\ref{P4}) provides a compact formula for $g^{(4)}$, we can push further the analysis and find the actual decay in the different regimes.
We use the variables $\alpha_i\equiv \theta_{i+1}-\theta_1$, for which the invariance under exchange of rapidities implies
\be
g^{(4)}(\alpha_1,\alpha_2,\alpha_3)=g^{(4)}(-\alpha_1,\alpha_2-\alpha_1,\alpha_3-\alpha_1)=g^{(4)}(\alpha_1-\alpha_2,-\alpha_2,\alpha_3-\alpha_2)=g^{(4)}(\alpha_1-\alpha_3,\alpha_2-\alpha_3,-\alpha_3) \,.
\ee

When one of the $\alpha_i$ grows to infinity, which corresponds to the split $4\to 3+1$,
we obtain
\be\label{1p-3p}
g^{(4)}(\alpha_1 +\Lambda,\alpha_2,\alpha_3) = \frac{g^{(4)}_{as}(\alpha_2,\alpha_3) }{\Lambda^2} + O(\Lambda^{-3})\, ,
\ee
which is the minimum decay assuring integrability at infinity. A physically different limit occurs when we split into $4\to 2+2$, realized by sending two $\alpha_i$ to infinity together. It results in a faster decay\footnote {The result above means that the correction to the factorisation (\ref{422fact}) are in fact of order $O(\Lambda^{-4})$.} than the expected $O(\Lambda^{-2})$, namely
\be\label{g4faster}
g^{(4)}(\alpha_1 +\Lambda,\alpha_2 +\Lambda,\alpha_3) = O(\Lambda^{-4}) \,.
\ee
Now we introduce different shifts $\Lambda_i$, all of order $R$: this is the split $4\to 2+1+1$, where the function behaves as
\be
g^{(4)}(\alpha_1 +\Lambda_1,\alpha_2 +\Lambda_2,\alpha_3) = O(R^{-4}) \, ,
\ee
that, again, turns out to be faster than the required $O(R^{-3})$.
The last region to analyse is $4\to 1+1+1+1$, where our function decays as
\be
g^{(4)}(\alpha_1 +\Lambda_1,\alpha_2 +\Lambda_2,\alpha_3+\Lambda_3) =
O(R^{-6}) \, ,
\ee
faster than the minimum $O(R^{-4})$. 

Summarising, in all regions except $4\to 3+1$, the connected function $g^{(4)}$ goes to zero more rapidly than the minimum assuring integrability at infinity. This fact has practical effects on the computation of the subleading term $s^{(4)}\ln\ln(1/z)$. In particular, the function $g^{(4)}_{as}(\alpha_2,\alpha_3)$ belongs to $L^1(\mathbb{R}^2)$, thanks to
\be
g^{(4)}_{as}(\alpha_2+\Lambda,\alpha_3) = O(\Lambda^{-2}), \quad g^{(4)}_{as}(\alpha_2+\Lambda_1,\alpha_3+\Lambda_2) = O(R^{-4}) \, .
\ee

assuring the validity of formula (\ref{deltas4}) for the subleading correction $s^{(4)}$. 

\medskip

\noindent\textbf{$\bullet$ Six particles case $g^{(6)}$}

\medskip

With six scalars ($n=3$) the analysis is more involved due to the several different asymptotic regions. Fortunately, all of them except one are already included in the subset $2n\to m + p + 1 + \cdots + 1$ analysed before.
Let us recall the connected function $g^{(6)}$
\be\label{inverse-bis}
g_{123456}=G_{123456}-(G_{12}G_{3456} + 14 \textit{ d.e.}) + 2(G_{12}G_{34}G_{56} + 14 \textit{ d.e.}) \, .
\ee

where we made use of an obvious short-hand notation.
The only new process we need to address is $6\to 2+2+2$, which is not trivial as it involves only groups composed by an even number of particles, making the RHS of (\ref{inverse-bis}) of order $O(1)$. As a consequence, in addition to the finite part $O(1)$, a refined cancellation of the subleading terms $O(R^{-2})$ needs to occur, a fact not a priori guaranteed by (\ref{2n-fact}).
We choose to group the particles according to $(12 \quad 34 \quad 56)$, leaving us with
\be\label{g6left}
g_{123456} = G_{123456} - G_{12}G_{3456} - G_{34}G_{1256} - G_{56}G_{1234} + 2G_{12}G_{34}G_{56} + O(R^{-4}) \, ,
\ee
By means of (\ref{4-fact}) and (\ref{g4faster}), the general condition (\ref{convg2n}) becomes
\be\label{cond6222}
G^{(6)}(u_1+\Lambda_1, u_2+\Lambda_1, u_3+\Lambda_2, u_4+\Lambda_2, u_5,u_6)= G^{(2)}(u_1,u_2)G^{(2)}(u_3,u_4)G^{(2)}(u_5,u_6) + O(R^{-3}) \, ,
\ee
which is not a straightforward extension of the previous results and represents a sort of multiple factorisation.\\
To clarify the condition (\ref{cond6222}), we introduce the corrections to the single factorisation process $2n\to 2k + 2(n-k)$ as
\be
G^{(2n)}\to G^{(2k)}G^{(2n-2k)} + \sum_l \Lambda^{-l}S_{2n,2k}^{(l)}(u_1, \ldots , u_{2n}) \, .
\ee
For the sake of the physical picture, we temporarily ignore that the quadratic subleading $S^{(2)}_{(4,2)}$ is actually vanishing. The cancellation of the terms $O(R^{-2})$ in (\ref{g6left}) occurs if
\ba\label{mult6222}
&& G^{(6)}(u_1 + \Lambda_1 , u_2 + \Lambda_1 , u_3 + \Lambda_2, u_4 + \Lambda_2 , u_5, u_6)= G^{(2)}(u_1, u_2)G^{(2)}(u_3, u_4)G^{(2)}(u_5, u_6) + \nn\\
&& + \Lambda_1^{-2}G^{(2)}(u_3, u_4)S^{(2)}_{4,2}(u_1, u_2, u_5, u_6) + \Lambda_2^{-2}G^{(2)}(u_1, u_2)S^{(2)}_{4,2}(u_3, u_4, u_5, u_6) + \nn\\
&& + \Lambda_{12}^{-2}G^{(2)}(u_5, u_6)S^{(2)}_{4,2}(u_1, u_2, u_3, u_4) + O(R^{-3})
\ea
If we use $S^{(2)}_{(4,2)}=0$, we recover the previous formula (\ref{cond6222}): nevertheless, (\ref{mult6222}) is interesting on its own for its clear physical meaning and it is easily extendible to any process of the type $2n\to 2k_1 + \cdots + 2k_l$.
Formula (\ref{mult6222}) is a relation among the subleading corrections for different factorisation processes: in simple words, the multi-factorisation process gets corrected by all the subleading terms associated to the various sub-factorisations involved, which are three in the case $6\to 2+2+2$.
The constraint (\ref{mult6222}) translates into, for the polynomial $P_6$
\ba\label{sublP6}
&& P_6(12_{\Lambda_1} \quad 34_{\Lambda_2} \quad 56)=\Lambda_1^{8}\Lambda_2^{8}\Lambda_{12}^{8}P_2P_2P_2 \biggl[ 1 + 2(u_{13}+u_{14}+u_{23}+u_{24})\Lambda_{12}^{-1} + \nn\\
&+& 2(u_{15}+u_{16}+u_{25}+u_{26})\Lambda_{1}^{-1} + 2(u_{35}+u_{36}+u_{45}+u_{46})\Lambda_{2}^{-1}  + \nn\\
&+& 4(u_{13} + u_{14} +u_{23} +u_{24} )(u_{15} + u_{16} +u_{25} +u_{26} )\Lambda_{12}^{-1}\Lambda_1^{-1}  + \nn\\
&+& 4(u_{13} + u_{14} +u_{23} +u_{24} )(u_{35} + u_{36} +u_{45} +u_{46} )\Lambda_{12}^{-1}\Lambda_2^{-1}  + \nn\\
&+& 4(u_{15} + u_{16} +u_{25} +u_{26} )(u_{35} + u_{36} +u_{45} +u_{46} )\Lambda_{2}^{-1}\Lambda_1^{-1}  + \nn\\
&+& \Delta_{4,2}^{(2)}(u_1,u_2,u_3,u_4)\Lambda_{12}^{-2} + \Delta_{4,2}^{(2)}(u_1,u_2,u_5,u_6)\Lambda_{1}^{-2} + \Delta_{4,2}^{(2)}(u_3,u_4,u_5,u_6)\Lambda_{2}^{-2} +  O(R^{-3}) \biggr] \nn \, , \\
\ea

In plain words, the quadratic corrections to the process $6 \to 2+2+2$ of $P_6$ shall be fixed by $\Delta_{4,2}^{(2)}$, which parametrises the correction to the factorization $4 \to 2+2$, see (\ref{factP}). It is easy to extend (\ref{sublP6}) to the general process $2n\to 2k_1 + \cdots 2k_l$ and check that the highest degree $P_{2n}^{(0)}$ satisfies these constraints. To ensure the integrability condition (\ref{convg2n}) for any $n$ and any split, the generalizations of (\ref{cond6222}), (\ref{mult6222}) and (\ref{sublP6}) must hold.
A deeper analysis of $g^{(6)}$, employing the sum over Young tableaux (\ref{PiMat-Young}) with Mathematica, confirms that in the limit $6 \to 2+2+2$ the function $g^{(6)}$ decays as $O(R^{-4})$ and thus belongs to the class $L^1(\mathbb{R}^5)$.

In conclusion, in spite of the lack of a general argument, we collected additional evidence for $g^{(2n)}\in L^1(\mathbb{R}^{2n-1})$. We stress that the convergence of the integral in (\ref{DeltaFF}) is obvious for theories without asymptotic freedom, as the functions $g^{(n)}$ decay exponentially. In our case, the power-like decay not only gives the logarithmic correction $\log(1/z)^s$ to the correlator but also makes the convergence of the integrals more subtle. 
 
\subsection{Polygons $N>6$}

This part is dedicated to the connected functions appearing for the bigger polygons, \emph{i.e.} $N>6$.

We begin by writing the general formula 
\be\label{MultiConn}
g^{(2n_1,\cdots,2n_k)}=\sum_{l=1}^{n_1+ \cdots + n_k}(-1)^{l-1}(l-1)!\sum_{\left \{ n^{(j)}_m \right \}}\sum_{d.e.}\prod_{j=1}^{l}G^{(2n^{(j)}_1,\cdots,2n^{(j)}_{k})}
\ee

which relates them to the original $G^{(2n_1,\cdots,2n_{k})}$ characterising the expansion of the Wilson loop.

Formula (\ref{MultiConn}) is an extension of (\ref{inverse}), valid for the hexagon. The number $l$ counts the functions $G$ appearing in the product, $\left \{ n^{(j)}_m \right \}$ is the set of different products of $l$ functions and the last sum contains all the permutations (different exchanges) among the equivalent rapidities. Of course, we have the constraints $\sum_{j=1}^{l}n^{(j)}_m=n_m$. The total number of terms in the sum  $\sum_{d.e.}$ is

\be
\frac{\displaystyle\prod_{m=1}^{k}(2n_m)!}{\displaystyle\prod_{j=1}^{l}\displaystyle\prod_{m=1}^{k}(2n^{(j)}_m)!}\frac{1}{\displaystyle\prod_{j'=1}^{l'}(m_{j'})!}
\ee

where $l'$ is the number of different $G$ in the product and $m_{j'}$ is the multiplicity of the $j'$-th function. For instance, in $G^{(2,0)}G^{(2,0)}G^{(0,4)}$ we have $l=3$ but $l'=2$ and $\displaystyle\prod_{j'=1}^{l'}(m_{j'})!=2!1!=2$.

From the decoupling of the $G$'s introduced in the main text
\be\label{dec1}
G^{(2n_1,\cdots,2n_k,0,0,\cdots,0,0,2m_1,\cdots,2m_l)}=G^{(2n_1,\cdots,2n_k)}G^{(2m_1,\cdots,2m_l)}
\ee
\be\label{dec2}
G^{(2n_1,\cdots,2n_k,0,0,\cdots,0,0)}=G^{(2n_1,\cdots,2n_k)}
\ee
the same follows for the connected part
\be\label{propg1}
g^{(2n_1,\cdots,2n_k,0,0,\cdots,0,0)}=g^{(2n_1,\cdots,2n_k)}
\ee
while, for intermediate vacuum states, they are vanishing
\be\label{propg2}
g^{(.....,2n,0,0,.....,0,0,2m,.....)}=0, \quad m,n\neq 0
\ee

The last two formulae (\ref{propg1},\ref{propg2}) are crucial to give rise to the recursion formula for the polygons. In the following we put formula (\ref{MultiConn}) to work for the simplest cases $N=7,8$ and for a small number of particles $n_i$.

\paragraph{Heptagon}

Up to eight particles, the non trivial heptagonal functions are $g^{(2,2)}$, $g^{(4,2)}$, $g^{(2,4)}$, $g^{(4,4)}$ and $g^{(6,2)}$, all the others are hexagonal thanks to (\ref{propg1}).

The simplest is

\ba
g^{(2,2)}(\theta_1,\theta_2;\theta'_1,\theta'_2)&=&G^{(2,2)}(\theta_1,\theta_2;\theta'_1,\theta'_2) -G^{(2,0)}(\theta_1,\theta_2;\emptyset)G^{(0,2)}(\emptyset;\theta'_1,\theta'_2) = \nn \\
&=& G^{(2,2)}(\theta_1,\theta_2;\theta'_1,\theta'_2) -G^{(2)}(\theta_1,\theta_2)G^{(2)}(\theta'_1,\theta'_2) 
\ea

The six particles function $g^{(4,2)}$, $g^{(2,4)}$ are related by symmetry and are given by

\ba
&& g^{(4,2)}(\theta_1,\theta_2,\theta_3,\theta_4;\theta'_1,\theta'_2)=G^{(4,2)}(\theta_1,\theta_2,\theta_3,\theta_4;\theta'_1,\theta'_2)- G^{(2)}(\theta'_1,\theta'_2) G^{(4)}(\theta_1,\theta_2,\theta_3,\theta_4) - \\
&-& \left(G^{(2)}(\theta_1,\theta_2)G^{(2,2)}(\theta_3,\theta_4;\theta'_1,\theta'_2) + \textit{5 terms}\right) + 2G^{(2)}(\theta'_1,\theta'_2)\left(G^{(2)}(\theta_1,\theta_2)G^{(2)}(\theta_3,\theta_4) + \textit{2 terms}\right) \nn
\ea

where, inside the brackets, we have all the different permutations among the equivalent rapidities $\theta_i$ in order to obtain a $g^{(4,2)}$ which is symmetric under exchange of any of them.

We write down $g^{(4,4)}$ omitting the dependence on the rapidities

\ba
g^{(4,4)} &=& G^{(4,4)}-G^{(4,0)}G^{(0,4)} -\left(G^{(2,2)}G^{(2,2)} + \textit{17 terms}\right) - \left(G^{(4,2)}G^{(0,2)} + \textit{5 terms}\right) - \nn \\ 
&-& \left(G^{(2,0)}G^{(2,4)} + \textit{5 terms}\right) + 2\left(G^{(2,0)}G^{(2,0)}G^{(0,4)} + \textit{2 terms}\right) + \nn \\ 
&+& 2\left(G^{(0,2)}G^{(0,2)}G^{(4,0)} + \textit{2 terms}\right)- 6\left(G^{(2,0)}G^{(2,0)}G^{(0,2)}G^{(0,2)}+\textit{8 terms}\right) + \nn \\ 
&+& 2\left(G^{(2,0)}G^{(0,2)}G^{(2,2)} + \textit{35 terms}\right) 
\ea

\paragraph{Octagon}

From $N=8$ the property (\ref{propg2}) starts to play an important role, allowing us to cancel many contributions.
Since $g^{(2,0,2)}=g^{(4,0,2)}=g^{(2,0,4)}=0$, the first non-zero octagonal function is then
\be
g^{(2,2,2)}=G^{(2,2,2)}-G^{(2,2,0)}G^{(0,0,2)}-G^{(2,0,2)}G^{(0,2,0)}-G^{(0,2,2)}G^{(2,0,0)} + 2G^{(2,0,0)}G^{(0,2,0)}G^{(0,0,2)}
\ee
in which we do not have any permutation as there are at most only two rapidities in each set.
More explicitly, making use of (\ref{dec1}) and (\ref{dec2}), we have

\ba
&& g^{(2,2,2)}(\theta_1,\theta_2;\theta'_1,\theta'_2;\theta''_1,\theta''_2) = G^{(2,2,2)}(\theta_1,\theta_2;\theta'_1,\theta'_2;\theta''_1,\theta''_2) - G^{(2,2)}(\theta_1,\theta_2;\theta'_1,\theta'_2)G^{(2)}(\theta''_1,\theta''_2) - \nn \\ 
&& - G^{(2)}(\theta_1,\theta_2)G^{(2,2)}(\theta'_1,\theta'_2;\theta''_1,\theta''_2) + G^{(2)}(\theta_1,\theta_2)G^{(2)}(\theta'_1,\theta'_2)G^{(2)}(\theta''_1,\theta''_2)
\ea

\end{appendices}

\end{document}